\documentclass{article}

\pdfoutput=1

\usepackage{arxiv}

\usepackage[utf8]{inputenc} 
\usepackage[T1]{fontenc}    
\usepackage{hyperref}       
\usepackage{url}            
\usepackage{booktabs}       
\usepackage{amsfonts}       
\usepackage{nicefrac}       
\usepackage{microtype}      
\usepackage{lipsum}		
\usepackage{graphicx}
\usepackage{natbib}
\usepackage{doi}

\usepackage{appendix}

\usepackage{amsmath}
\usepackage{algorithm}
\usepackage{textcomp}
\usepackage{xcolor}
\usepackage{multirow}
\usepackage{algpseudocode}
\usepackage{subcaption}
\usepackage[para,online,flushleft]{threeparttable} 
\usepackage{tikz}
\usetikzlibrary{shapes, arrows.meta, positioning}
\usepackage{lscape}
\usepackage{amsthm}

\newcommand{\HT}[1]{\textcolor{black}{#1}}  

\title{Benchmarking Quantum(-inspired) Annealing Hardware on Practical Use Cases}

\author{ 	\href{https://orcid.org/0000-0002-8765-1783}{\includegraphics[scale=0.06]{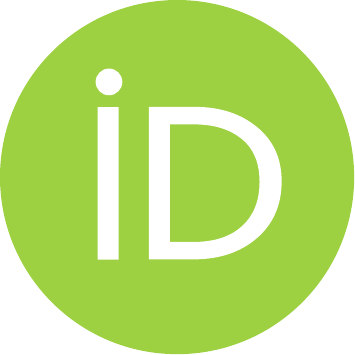}\hspace{1mm}Tian Huang} \\
	Institute of High Performance Computing \\
	Agency for Science Technology and Research \\
	Singapore, 138632 \\
	\texttt{huangtian44@hotmail.com} \\
	\And
	Jun Xu \\
    Department of Computer Science \\
    National University of Singapore \\
    13 Computing Drive, Singapore, 117417 \\
	\texttt{xu.jun@u.nus.edu} \\
	\And
	\href{https://orcid.org/0000-0002-3415-3676}{\includegraphics[scale=0.06]{figures_tc/orcid.pdf}\hspace{1mm}Tao Luo} \thanks{Corresponding author.} \\
	Institute of High Performance Computing \\
	Agency for Science Technology and Research \\
	Singapore, 138632 \\
	\texttt{leto.luo@gmail.com} \\
	\And
	\href{https://orcid.org/0000-0001-7948-7672}{\includegraphics[scale=0.06]{figures_tc/orcid.pdf}\hspace{1mm}Xiaozhe Gu} \\
	The Chinese University of Hong Kong \\
	Shenzhen, China \\
	\texttt{guxi0002@e.ntu.edu.sg} \\
	\And
	Rick Goh \\
	Institute of High Performance Computing \\
	Agency for Science Technology and Research \\
	Singapore, 138632 \\
	\texttt{gohsm@ihpc.a-star.edu.sg} \\
	\And
	\href{https://orcid.org/0000-0002-4281-2053}{\includegraphics[scale=0.06]{figures_tc/orcid.pdf}\hspace{1mm}Weng-Fai Wong} \\
    Department of Computer Science \\
    National University of Singapore \\
    13 Computing Drive, Singapore, 117417 \\
	\texttt{wongwf@nus.edu.sg} \\
}

\renewcommand{\shorttitle}{Benchmarking Quantum(-inspired) Annealers}

\hypersetup{
pdftitle={Benchmarking Quantum(-inspired) Annealing Hardware on Practical Use Cases},
pdfsubject={quant-ph},
pdfauthor={Tian Huang, Jun Xu, Tao Luo, Xiaozhe Gu, Rick Goh, and Weng-Fai Wong},
pdfkeywords={Quantum Annealer, Digital Annealer, Combinatorial optimisation, Benchmark},
}

\begin{document}
\maketitle

\begin{abstract}
Quantum(-inspired) annealers show promise in solving combinatorial optimisation problems in practice. There has been extensive researches demonstrating the utility of D-Wave quantum annealer and quantum-inspired annealer, i.e., Fujitsu Digital Annealer on various applications, but few works are comparing these platforms. In this paper, we benchmark quantum(-inspired) annealers with three combinatorial optimisation problems ranging from generic scientific problems to complex problems in practical use. In the case where the problem size goes beyond the capacity of a quantum(-inspired) computer, we evaluate them in the context of decomposition. Experiments suggest that both annealers are effective on problems with small size and simple settings, but lose their utility when facing problems in practical size and settings. Decomposition methods extend the scalability of annealers, but they are still far away from practical use. Based on the experiments and comparison, we discuss the advantages and limitations of quantum(-inspired) annealers, as well as the research directions that may improve the utility and scalability of the these emerging computing technologies. 
\end{abstract}


\section{Introduction}\label{sec:intro}

Qannealing is a generic method for solving combinatorial optimisation problems by exploiting the \HT{quantum effects}. The motivation for solving the problem with quantum annealer is speed. Many combinatorial optimisation problems, despite their simple problem settings, are computationally difficult. The problem-solving process utilises quantum fluctuation-based computation instead of classical computation. A combinatorial optimisation problem can be modelled as Quadratic Unconstrained Binary Optimisation (QUBO) form \cite{glover2019tutorial}, which corresponds naturally to the transverse Ising model and benefits from the speed up by quantum annealing \cite{kadowaki1998quantum}. Some problems, e.g. max-cut problem \cite{poljak1995maximum} and max-sat problem \cite{bian2017solving}, are especially popular among the quantum computing research community because of the simplicity of the problem settings. Applications with more advanced problem settings have attracted more attention recently. Examples are portfolio optimisation in finance \cite{grant2021benchmarking}, traffic flow management \cite{inoue2020traffic}, warehouse management problem \cite{sao2019application}.

Although there are plenty of works demonstrating the capability of quantum annealing, the utility of quantum annealers is restricted by today's manufacturing technology. For example, D-Wave Quantum Annealer (QA) has a limited number of qubits. Furthermore, the architecture of a Quantum Processing Unit (QPU) puts restrictions on the connectivity of the graph that represents the problem. On the other hand, Fujitsu Digital Annealer (DA) is a quantum-inspired CMOS-based ASIC, which simulates the annealing process efficiently. It enjoys speedup over general-purpose classical computers and has fewer restrictions than QA does.

One may ask which one, QA or DA, is better for combinatorial optimisation problems. As far as we know there is no comprehensive benchmark between them. The answer to the question depends on the problem settings. In this paper, we benchmark QA and DA with different combinatorial optimisation problems, ranging from simple to complex problem settings in practical use. Through analysis on experiments, we discuss the advantages and disadvantages of quantum(-inspired) annealers under different problems settings.

Due to the limitations of today's manufacturing techniques, the capacity of a quantum(-inspired) annealer is often far away from the problem scale in practical use. Decomposition techniques are essential when the size of a problem goes beyond the capacity of a computer. This paper also benchmarks the quantum(-inspired) annealers in the context of decomposition. We use warehouse assignment problem as a case study and propose a new decomposition heuristic and compare it with various other methods. In this paper, We have the following contributions:

\begin{itemize}
    \item To our best knowledge, this is the first work to systematically compare the performance between quantum annealer and digital annealer. We include combinatorial optimisation problems in different levels of complexity and quantitatively evaluate how these annealers respond to different problem settings.
    \item We use the warehouse assignment problem as a case study and propose a new heuristic that enables quantum-inspired annealers to solve the warehouse management optimisation of size linear to the number of qubits, provided a certain block-structural property is satisfied.
    \item We identify the advantages and missing pieces for QA and DA through a comprehensive analysis of the experiments.
\end{itemize}

We have the following observations from experiments: First, both QA and DA are very fast and efficient if the problem has a small scale and sparse connectivity between decision variables. Although QA and DA have different working principles, both of the performance, in terms of quality of solutions, decrease when the size of the problem scales up. Introducing constraints into the problem will add difficulty for QA in finding optimal or even feasible solutions. Dense connectivity reduces the utility of QA. DA is more robust to these challenges. DA has a better scaling factor compared with CPU-based simulated annealing. It is effective on constrained and densely connected problems, which are more common in real-world applications.

Second, our new decomposition heuristic for warehouse management is based on the assumption of a block-structural property. Such property is frequently observed in many practical settings including warehouse allocation. Theoretical proof provides a boundary for the performance of the problem-solving process. Experiments suggest that our new decomposition heuristic produces more promising solutions with less memory consumption than QBSolv does and enjoys speedup over simulated annealing on generic purpose classical CPUs. However, experiments show that both hybrid methods lose their utility as the problem size continues to scale up.

Through the experiments, we conclude the following directions to improving the performance of quantum(-inspired) annealers: 1. Error and noise mitigation; 2. Constraints simplification and post-readout processing; 3. Smart encoding; 4. Smart decomposition.

\section{Related Work}
\label{sec:related}

\subsection{Annealing-based Computers}

\textbf{Quantum Annealing} (QA) is a method that minimises the energy of an objective function. It takes advantage of quantum mechanics. Quantum bit or qubit is the basic unit in a quantum computing system, which resembles the binary bit in classical digital circuits. A coupler in quantum annealer bonds two qubits together and allows quantum mechanics to work its magic. Qubits and couplers in a quantum annealer can be seen as the nodes and edges in a graph. To use a quantum annealer, One has to convert a problem into a Quadratic Unconstrained Binary Optimisation (QUBO) \cite{glover2019tutorial} form and map the QUBO problem onto the quantum annealer.

D-Wave is a world-leading company that designs and builds quantum annealers. The Quantum Processing Unit (QPU) in the D-Wave's quantum annealing system is a lattice of interconnected qubits. Each qubit is made of a superconducting loop. D-Wave released the D-Wave 2000Q system in 2017, the QPU in which employs a Chimera graph architecture \cite{vert2019limitations}, equipped with 2048 qubits and 6016 couplers. In 2019, D-Wave released the Advantage system, the QPU of which employs a \emph{Pegasus} \cite{dwave_arch} graph architecture, which has 5640 qubits and 40,484 couplers. 

The Fujitsu \textbf{Digital Annealer} (DA) \cite{aramon2019physics} is a hardware implementation of an enhanced variant of the Simulated Annealing (SA) algorithm. DA employs Metropolis-Hastings update with Parallel Tempering. In each cycle, it adjusts the temperature, proposes bit flips, and accepts/rejects the proposals according to the temperature and periodically swaps configurations between systems. Apart from the sophisticated techniques in proposing and selecting updates, the competence of DA also comes from the efficient hardware implementation that exploits the parallelism in the algorithm.

\subsection{Benchmark for Quantum(-inspired) Annealing}

There are many works demonstrating the utility of quantum and digital annealers in various applications\cite{poljak1995maximum,bian2017solving,inoue2020traffic,grant2021benchmarking,naghsh2019digitally,miasnikof2020graph,sao2019application,maruo2020optimization}. Bass et al. \cite{bass2021optimizing} did a performance benchmark for four hybrid methods, in which the largest problem requires 10,000 binary decision variables. Vert et. al. \cite{vert2019limitations} investigated the relationship between the quality of solutions and the connectivity of decision variables for hybrid methods. Ohzeki et.al. \cite{ohzeki2019control} compares the performance of quantum annealer and digital annealer on automated guided vehicles (AGV) problems. Albash et. al. \cite{albash2018demonstration} compares the scalability of quantum and classical annealing. Seker et. al. \cite{cseker2020digital} systematically investigate the performance of DA on various problems. Kowalsky et.al. \cite{kowalsky20223} perform a scaling and prefactor analysis of the performance of four quantum(-inspired) optimizers on 3-regular 3-XORSAT problems.

None of these works provides a comprehensive comparison between a quantum annealer and a digital annealer. In this paper, we evaluate the two annealers on three representative combinatorial optimisation problems and compare their performance qualitatively and quantitatively. In the context of decomposition, the largest problem instance in this paper involves $8100^2=65.61\times10^6$ binary decision variables, which is close to the size of the application in practical use.

\subsection{QUBO decomposition}
\label{sec:QUBO_decomp}



In a large neighbourhood local search, a random feasible solution is usually taken as the starting point and a small subset of variables solvable on hardware are chosen in each iteration. The subproblem generated by this subset is solved while variables outside of the subset are conditioned. This yields an exploration of a very large neighbourhood, whose result can be optionally plugged into a complementary software local search that gets to a local minimum. Overall, this approach has two complementary steps: the perturbation/jump step that find a point with a good neighbourhood to work with, and the local search step for finding the local optimum within such a neighbourhood. \emph{QBSolv} is the canonical D-Wave implementation \cite{booth2017partitioning} that uses subsets of variables for perturbation. A comprehensive evaluation \cite{gideon2020optimizing} shows \emph{QBSolv} outperforms many other decomposition methods. There is another implementation by Oracle \cite{mihic2018randomized} which inverts the way by solving random subsets of variables for local search and updating randomly chosen variables for perturbation. The approach described in this paper falls into this category but is a special case without perturbation. A neighbourhood is a-priori determined and subsets of variables are chosen heuristically on a higher abstraction level instead of at random.


Much of the quantum annealer application work cited in Section \ref{sec:intro} is experimental in scale and does not deal directly with Quadratic Assignment Problem (QAP). QAP asks for optimal assignment of $n$ facilities to $n$ locations that minimise the logistic efforts between them. There is only one instance of QAP being solved on a quantum device with application in flight gate assignment \cite{stollenwerk2019flight}. Just like what we mentioned in Section \ref{sec:intro}, the QAP instance in \cite{stollenwerk2019flight} is too big for DA. The scaling method they used is to randomly divide a graph into disconnected components if it is too large, and solve each component separately. This approach, while scalable, is a brute-force one based on randomness, and does not give significant insight as to how a QAP can be meaningfully decomposed.

\section{On Direct Problems}
\label{sec:simple}

In this section, we demonstrate the advantages and limitations of quantum(-inspired) computers by applying them to three representative combinatorial optimisation problems. The demonstration of problems is arranged progressively, such that a latter problem setting poses more challenges, which further weaken the utility of a quantum(-inspired) annealer. We try to answer the following three questions empirically:

\begin{itemize}
    \item What problem size can a quantum(-inspired) annealer handle?
    \item How much performance improvement can a quantum annealer achieve compared to a classical one?
    \item How is the quality/optimality of the solutions found by a quantum annealer compared to those of a classical one?
\end{itemize}

We include the D-Wave QPU of \textbf{Chimera} and \textbf{Pegasus} architecture, and Fujitsu \textbf{Digital Annealer} (DA) in the evaluation. We also include \textbf{Simulated Annealing} (SA) by D-Wave, which is an open-source CPU-based implementation of the Metropolis-Hastings algorithm. We include \textbf{QBSolv} from D-Wave, which is a heuristic hybrid solver that incorporates classical computer and quantum annealer. We also include \textbf{Gurobi}, which represents the SOTA commercial optimisation solver. We use Gurobi to find the global optima of the synthetic datasets in this paper.

The annealing process is crucial to the performance of annealing-based solvers. We try out different hyper-parameters for the annealing process. More specifically, we are varying \emph{annealing time}, or $T_{neal}$, in D-Wave QA, \emph{number of iterations}, or $\#iter$, in Fujitsu DA and \emph{number of sweeps}, or $\#sweep$, in SA. We only include the results with the best hyper-parameters in the comparison in the main text. Our definition of ``best'' prioritises energy functions and take timing\footnote{\HT{In optimisation field, the convention is to increase the run-time (through hyper-parameters, e.g., number of iterations) of a solver along with the complexity/scale of a problem. Given that classical annealers (CA) are generally a few orders of magnitude slower than a quantum annealer (QA), the comparability between CA and QA would be impaired if we further increasing run-time of CA. To conduct a more reasonable comparison, we evaluate a solver with each individual hyper-parameter on a range of problem complexity/scale.}} into consideration as well. Readers can find the details of the experimental settings in \HT{Appendix B} and complete results in \HT{Appendix C, D and E}.

\subsection{max-cut}
\label{sec:maxcut}

The max-cut problem is a well-known combinatorial optimisation problem that aims to find a partition of a node set into two parts, maximising the sum of weights over all edges across the two node subsets of a graph. Such a partition is called a maximum cut. Consider an undirected graph $G = (V, E)$, where $|V| = n$ with edge weights $w_{i,j}>0$, $w_{i,j}=w_{j,i}$, for $(i,j) \in E$. We partition $V$ into two subsets. The cost function to be optimized is the sum of the weights of edges between two subsets of $V$. The definition of the cost function is given by:

\begin{equation}
    \sum_{(i,j)\in E}w_{i,j}x_i(1-x_j)
    \label{equ:maxcut_obj}
\end{equation}

$x_i$ and $x_j$ are binary decision variables. $x_i$ is set to 0 or 1 depending on whether the node $i$ is in the first or second subset, respectively. The goal is to find a partition that maximises the cost. One solution would be in the form of $x_0, \ldots, x_{n-1}$. An annealer must have at least $n$ qubits to handle a max-cut problem with $n$ nodes.

\subsubsection{Pegasus-like max-cut problems}
\label{sec:maxcut_pegasus}

We formulate ten max-cut problems of different sizes, each of which is a sub-graph of the Pegasus graph architecture. The number of nodes of a graph $\left | V \right |$ ranges from 543 to 5430.

The reason for generating Pegasus-like graphs is two folds. First, we want to understand the performance of a quantum annealer when we make full use of the resources on the QPU. The largest problem that can be handled by a QPU is related to the match between the problem and QPU's architecture. If the graph of a problem is the same as or is a subset of the architecture, then the problem can be easily mapped onto the QPU. Otherwise, extra qubits are needed to map the problem. Using Pegasus-like graphs minimises the extra cost and maximises the utilisation of resources on the QPU.

Second, we want to minimise the uncertainties in measuring the performance of the QPU. To handle a problem that does not match with the architecture very well, an extra step called \emph{minor embedding} \cite{yang2016graph} is needed to map the problem onto the QPU. Minor embedding itself is an NP-hard problem and usually introduces extra qubit cost and extra noises to the system, which weaken the performance of the QPU. Pegasus-like problems can be directly mapped onto a Pegasus architecture without a resort to minor embedding, and therefore minimises the uncertainties in the measurement.

\begin{figure}[htb]
     \centering
     \begin{subfigure}[b]{0.45\linewidth}
        \centering
    	\includegraphics[width=\textwidth]{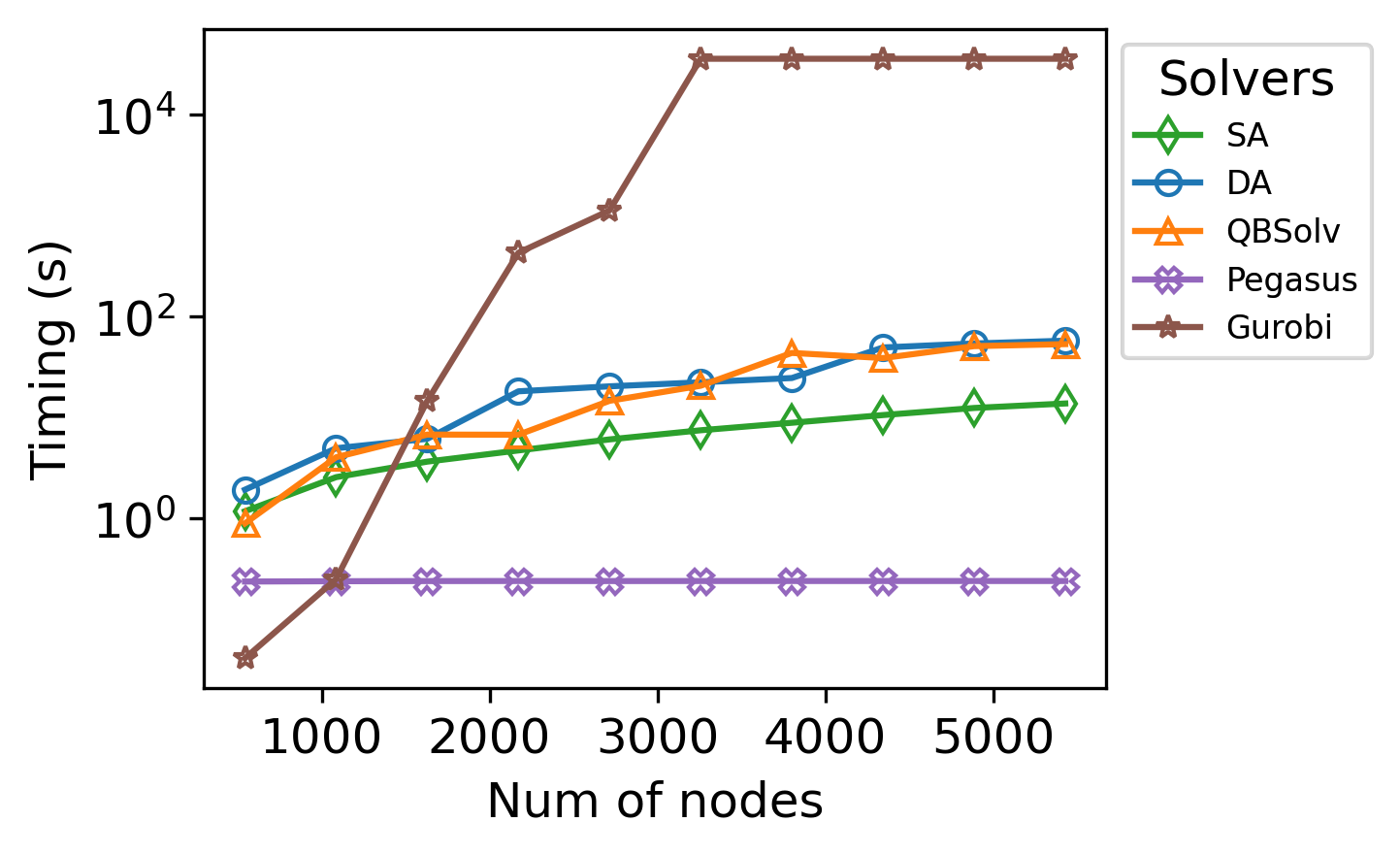}
    	\caption{Timing}
    	\label{fig:maxcut_pegasus_timing}
     \end{subfigure}
     \begin{subfigure}[b]{0.45\linewidth}
        \centering
    	\includegraphics[width=\textwidth]{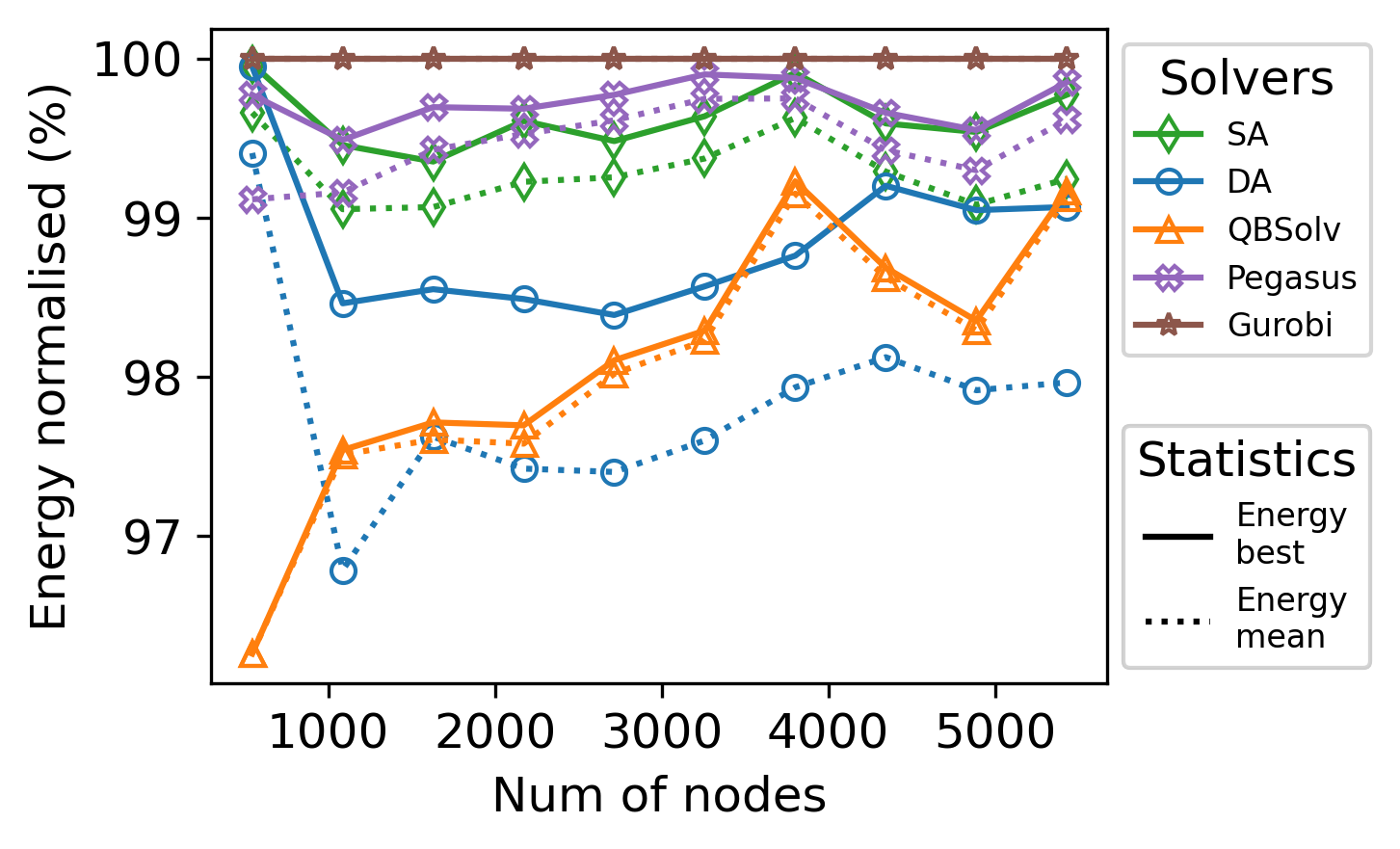}
    	\caption{Energy}
    	\label{fig:maxcut_pegasus_energy}
     \end{subfigure}
    \caption{Performance on Pegasus-like max-cut problems. X axis represents the number of nodes. (a) Timing. Y axis represents seconds in log scale. (b) Objective energy, normalised to that of Gurobi. Two statistics, mean energy (dotted lines) and best energy (solid lines), are included. For hyper-parameters, we use \HT{$T_{neal}=2000\mu s$} for D-Wave annealer,  $\#iter=10^6$ for DA, and $\#sweep=10^4$ for SA}
    \label{fig:maxcut_pegasus}
\end{figure}

Fig. \ref{fig:maxcut_pegasus_timing} shows the timing of the solvers on the Pegasus-like max-cut problems. Pegasus QPU spends about $200 ms$ on all these problems. The total execution time of the QPU can be decomposed into programming time and sampling time. The sampling time is fixed because the QPU is based on the same annealing process. The programming time increases along with the problem size. Pegasus is generally 2-3 orders of magnitude faster then the classical annealing-based solvers. The difference in timing mostly comes from the fundamental difference in the working principles. The solving time of classical solvers depend on the size of a problem. Larger problem usually requires longer running time. The solving time of Gurobi increases sub-exponentially along with the size of the problem. Because we set the Gurobi program to terminate when its running exceeds ten hours, the second half of the Gurobi curve clips. 

Fig.\ref{fig:maxcut_pegasus_energy} shows the objective energy, normalised to that of Gurobi. To describe the statistics of the results, we use solid curves to represent the energy of the best solution, and use the dotted curves to represent the mean energy of the solutions.\footnote{The mean-best plot setting is a replacement of an error-bar setting. The latter one introduces overlapping issues and is hard to read. Readers can find per-solver error-bar plot setting for a range of hyper-parameters in \HT{Appendix C, D, and E}.} Since max-cut is a maximisation problem, higher energy is better. The energy achieved by classical annealers and quantum annealer is not closely related with problem size. This experiment demonstrates the scalability of classical and quantum annealers. Pegasus outperforms all classical annealers on most of the problem instances by a margin of $1\%-3\%$ energy. DA performs better on the problem instances of size over 4000, which have higher connectivity than what smaller instances have. 


We also have experiments on Chimera-like max-cut problems. Please find the details in the \HT{Appendix C.2}.

\subsubsection{Connectivity-varied max-cut problems}
\label{sec:maxcut_connect}

The Pegasus-like graph in the previous section is sparsely connected graphs and are not in favour of DA. According to Fujitsu's official documentation \cite{aramon2019physics}, DA is designed for solving optimisation problems with densely connected graphs. In this section, we examine how connectivity affects the performance of an annealer. We make 32 randomly generated max-cut graphs, each of which has 145 nodes\footnote{This is the largest complete graph Pegasus accepts. For DA, the up-limit is 8192. To compare Pegasus and DA, we have to use 145.}. The average degree of these graphs ranges from 1 to 140. For the Pegasus QPU, there will be auxiliary qubits and connections involved in the final embedding, since the connectivity-varying problems are mostly not a sub-graph of the Pegasus graph architecture.

\begin{figure}[htb]
     \centering
     \begin{subfigure}[b]{0.45\linewidth}
        \centering
    	\includegraphics[width=\textwidth]{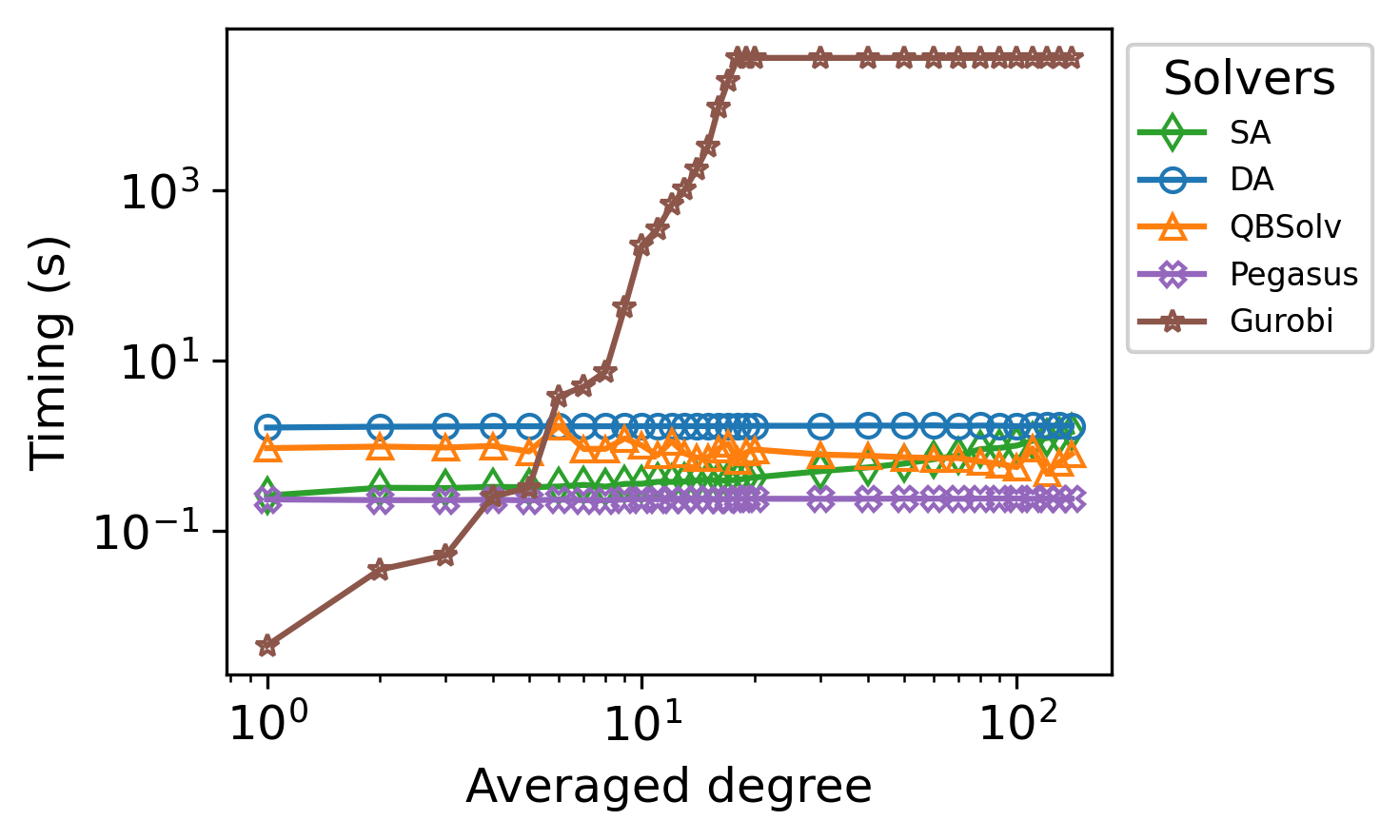}
    	\caption{Timing}
    	\label{fig:maxcut_connectivity_timing}
     \end{subfigure}
     \begin{subfigure}[b]{0.45\linewidth}
        \centering
    	\includegraphics[width=\textwidth]{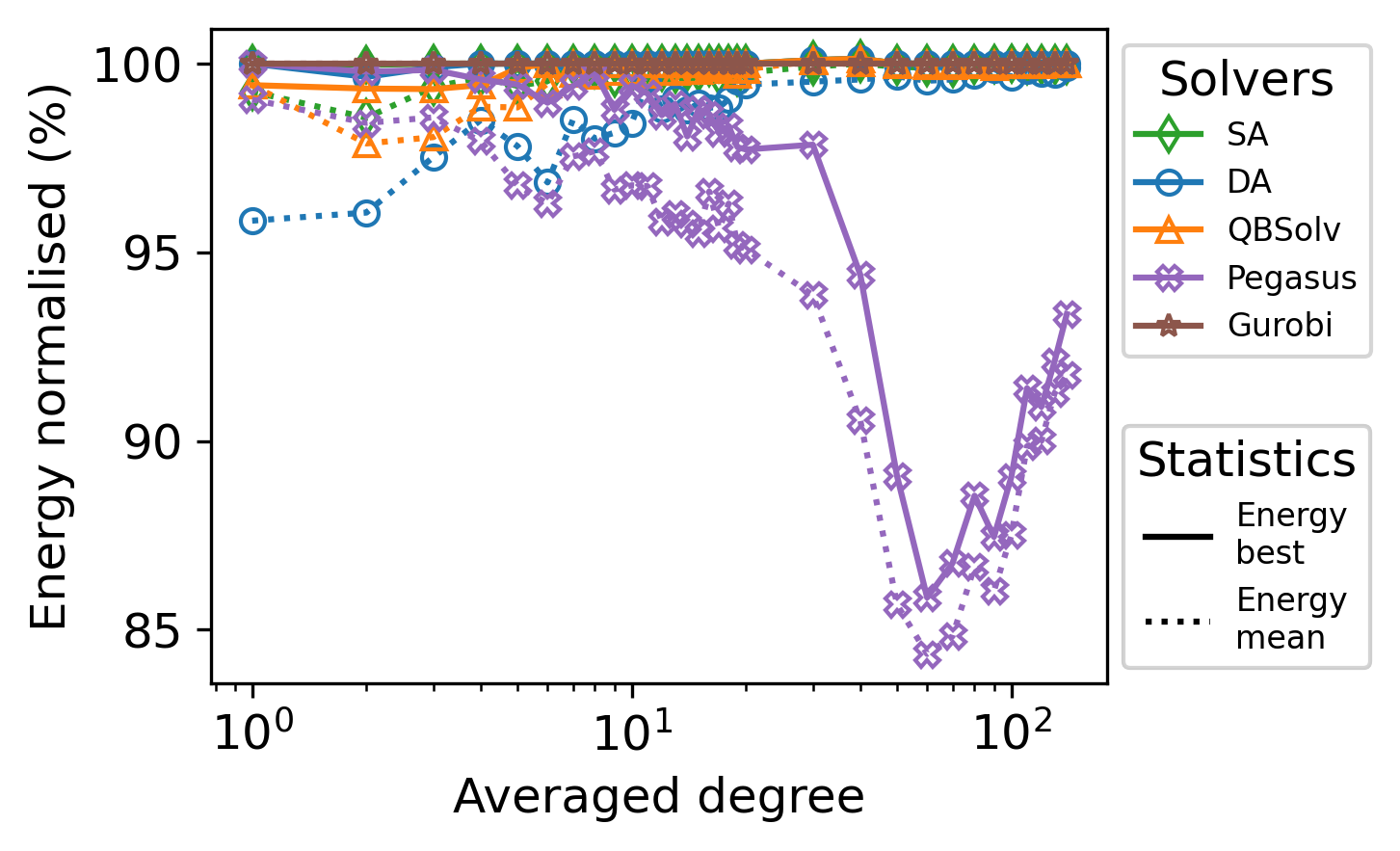}
    	\caption{Normalised energy}
    	\label{fig:maxcut_connectivity_energy}
     \end{subfigure}
    \caption{Performance on connectivity-varied max-cut problems. The plot setting is similar to that in Fig.\ref{fig:maxcut_pegasus}. For hyper-parameters, we use \HT{$T_{neal}=2000\mu s$} for D-Wave annealer, $\#iter=10^6$ for DA, and $\#sweep=10^4$ for SA}
    \label{fig:maxcut_connectivity}
\end{figure}

Fig.\ref{fig:maxcut_connectivity_timing} shows the time consumption of all solvers. Problems with degree over ten are challenging such that Gurobi cannot quickly traverse the solution spaces, so we set a timeout limit of ten hours. On the other hand, there are only 145 nodes in these graphs, which is ``trivial'' for classical annealers. DA, SA and QBSolv finish searching within one second, which is very close to the timing of Pegasus.

Fig.\ref{fig:maxcut_connectivity_energy} shows the objective energy of the solutions. All solvers, except Pegasus, can achieve 100\% energy on almost all problem instances. The mean energy of DA improves as the connectivity increases. We have similar observations in Section \ref{sec:maxcut_pegasus}, where the energy of DA improves as problem size increases. This suggest that DA performs better on densely connected problems. In comparison, the energy of Pegasus decreases when averaged degree is over 12. The mean energy could drop down to as low as 85\%, which corresponds to averaged degree of 16. This suggests the Pegasus performs better on sparsely connected problems.

The problem setting of max-cut is straightforward. Many other problems have constraints that cannot be directly represented in existing annealers. An annealer may find a solution that violate the constraints. We shall see how the situation changes for such problem settings.

\subsection{Minimum Vertex Cover}
\label{sec:mvc}

Given an undirected graph with a set of nodes $V$ and edges $E$, a vertex cover of a graph is a set of nodes that includes at least one endpoint of every edge of the graph. The \emph{Minimum Vertex Cover} (MVC) problem is an optimisation problem that finds the smallest vertex cover of a given graph. MVC problem can be formulated as eq.\ref{equ:mvc_obj}. 

\begin{equation}
\mathrm{Minimize} \sum_{i\in V} x_i \quad \text{subject to} \quad x_i+x_j\geq 1, \; \forall (i,j)\in E
\label{equ:mvc_obj}
\end{equation}

$x_i=1$ indicates node $i$ is in the cover and $x_i=0$ otherwise. The objective function minimises the number of nodes in the vertex cover, whereas the constraints ensure that at least one of the endpoints of each edge $(i,j)$ will be in the cover. In a quantum(-inspired) annealer, there is no corresponding mechanism to ensure the satisfaction of the constraints. A workaround is to lift the energy of configurations that violates the constraints and such that it becomes a less favourable solutions. This method is widely known as ``penalty''. The penalty of MVC can be formulated in the form of $\alpha \cdot (1-x_i-x_j+x_ix_j)$, where $\alpha$ represents a positive scalar penalty. A QUBO form for MVC is given by:

\begin{equation}
    \mathrm{Minimize} \; \sum_{i\in V}x_i+\alpha\sum_{(i, j)\in E}(1-x_i-x_j+x_ix_j)
    \label{equ:mvc_QUBO}
\end{equation}

The penalty term does not introduce new connections to the graph. The QUBO form has the same topology as that of the original problem. The choice of penalty weight $\alpha$ is application-specific. For eq.\ref{equ:mvc_QUBO}, adding a node to a minimum vertex cover will increase the objective energy by one. Removing any node from a minimum vertex cover will increase the objective energy by $\alpha-1$. \cite{glover2019tutorial} suggest any $\alpha>1$, for example, $\alpha=2$ would ensure that a solver can find solutions that satisfy the constraints of a MVC problem. 

\subsubsection{Connectivity-varied MVC problems}
\label{sec:mvc_connect}

We reuse the connectivity-varied graphs in Section \ref{sec:maxcut_connect} for constructing MVC problems. We discover that some solvers have difficulty in finding feasible solutions with penalty weight $\alpha=2$. We set $\alpha = \max_{i \in V}\deg(i)$, which represents the highest degree in a graph. With this setting, we can understand the comparison between the annealers more clearly.

\begin{figure}[htb]
     \centering
     \begin{subfigure}[b]{0.45\linewidth}
        \centering
    	\includegraphics[width=\textwidth]{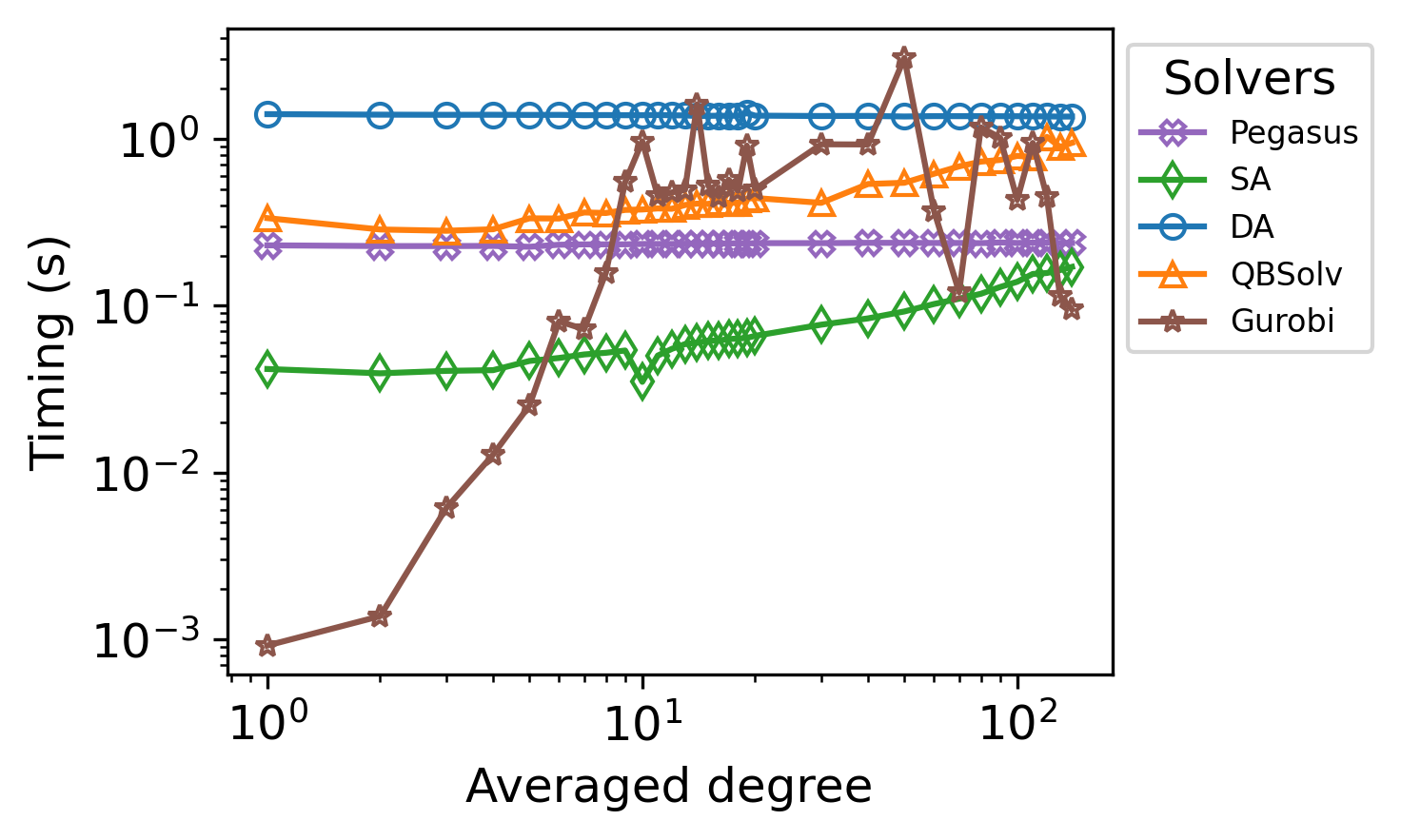}
    	\caption{Timing}
    	\label{fig:mvc_connectivity_timing}
     \end{subfigure}
     \begin{subfigure}[b]{0.45\linewidth}
        \centering
    	\includegraphics[width=\textwidth]{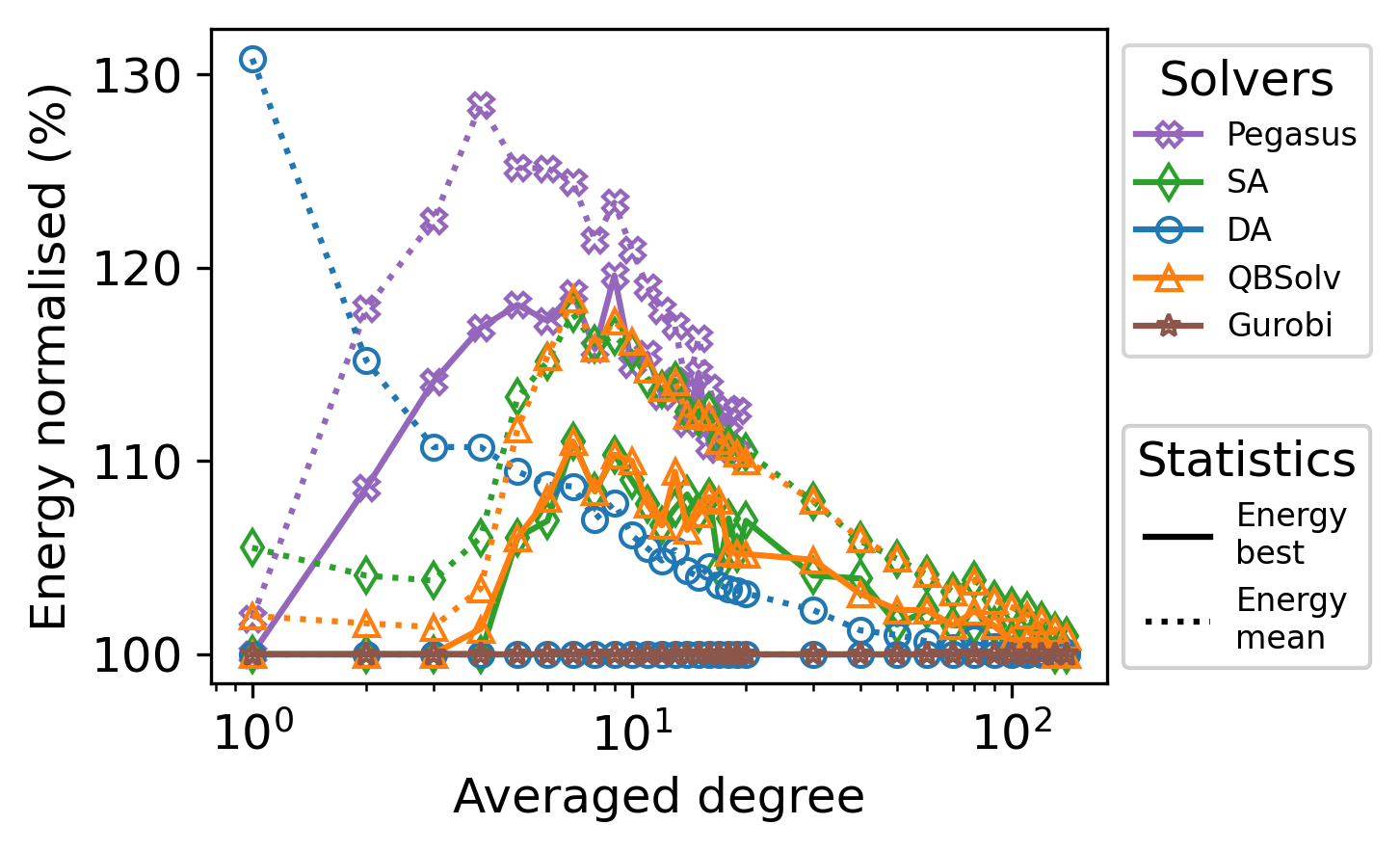}
    	\caption{Normalised energy}
    	\label{fig:mvc_connectivity_energy}
     \end{subfigure}
    \caption{Performance on connectivity-varied MVC problems. The plot setting is similar to that in Fig.\ref{fig:maxcut_pegasus}, except that infeasible solutions are not counted in Fig.\ref{fig:mvc_connectivity_energy}. For hyper-parameters, we use $T_{neal}=2000\mu s$ for D-Wave annealer,  $\#iter=10^5$ for DA, and $\#sweep=10^2$ for SA}
    \label{fig:mvc_connectivity}
\end{figure}

Fig.\ref{fig:mvc_connectivity_timing} shows the timing of the solver on the connectivity-varied MVC problems. Since we are using $T_{neal}=2000\mu s$ for Pegasus, it loses its leading position in terms of timing. This figure is similar to Fig.\ref{fig:maxcut_connectivity_timing}, except the timing of Gurobi stops increasing when the averaged degree is above 10. All solutions obtained by Gurobi are global optimal solutions.

Fig.\ref{fig:mvc_connectivity_energy} shows the objective energy of solutions, normalised to that of Gurobi. Since MVC is a minimisation problem, lower energy is better. The best energy curve of DA is always at 100\%, which means it finds optimal solutions for all problems. It is overlapped with Gurobi. One can spot the markers of DA and Gurobi by zooming in. Pegasus find the optimal solution only when the average degree is equal to one. Then there is more than 20\% performance drop as the degree increases. The Pegasus curve does not continue beyond a degree of 11 as it fails to find any feasible solution. 

In annealing-based methods, we sample multiple times for a problem instance. If a problem has constraints, some samples of the QUBO form may not be feasible, i.e., does not satisfy with constraints. We introduce a metric called \emph{probability of feasibility}, or $P_f$:

\begin{equation}
    P_f = \frac{\text{Feasible \#samples}}{\text{Total \#samples}} \times 100\%
    \label{equ:pf}
\end{equation}

\begin{figure}[htb]
    \centering
    \begin{subfigure}[b]{0.45\linewidth}
        \centering
    	\includegraphics[width=\textwidth]{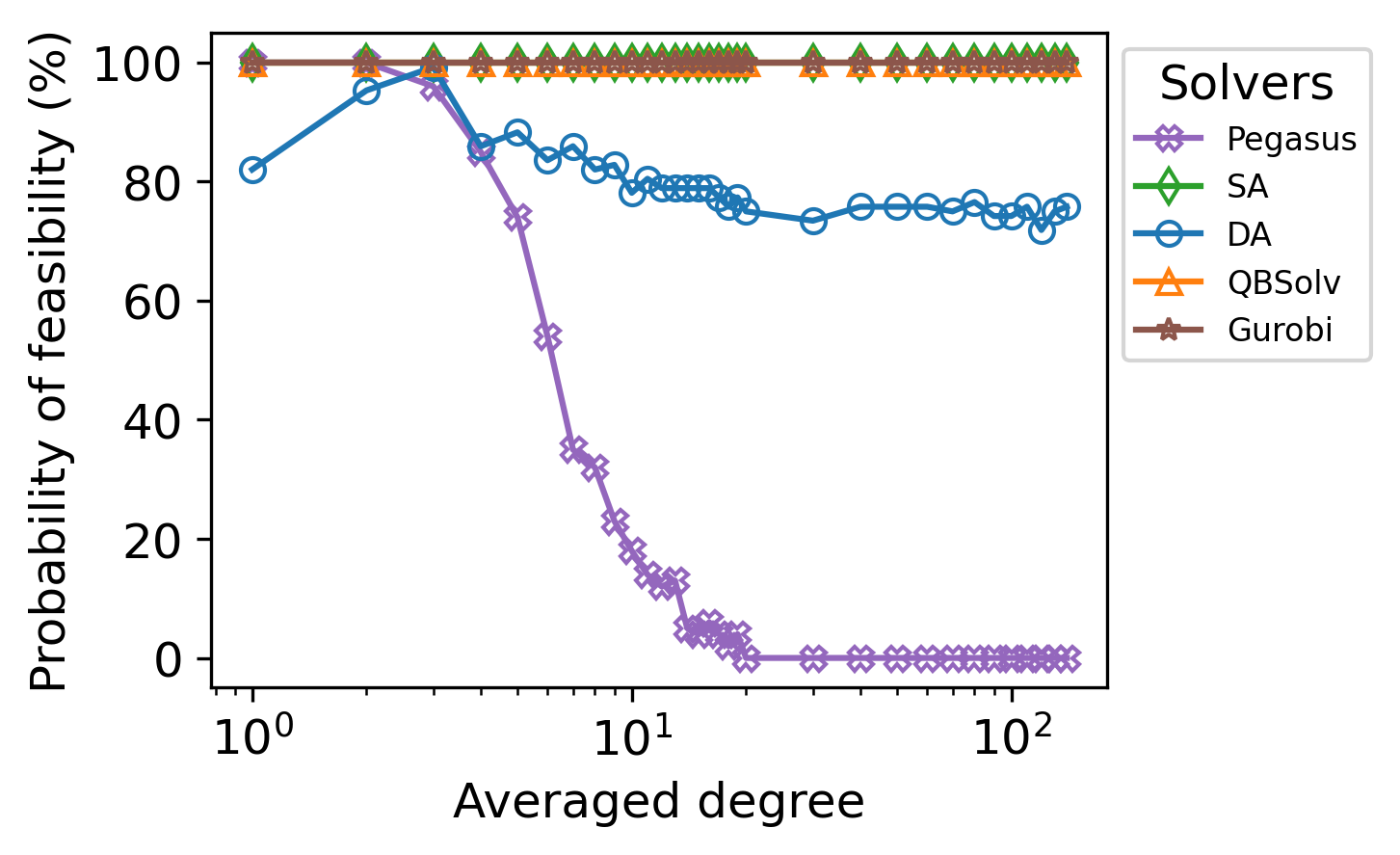}
    	\caption{Probability of feasibility}
    	\label{fig:mvc_connectivity_pf}
    \end{subfigure}
    \begin{subfigure}[b]{0.45\linewidth}
        \centering
        \includegraphics[width=\textwidth]{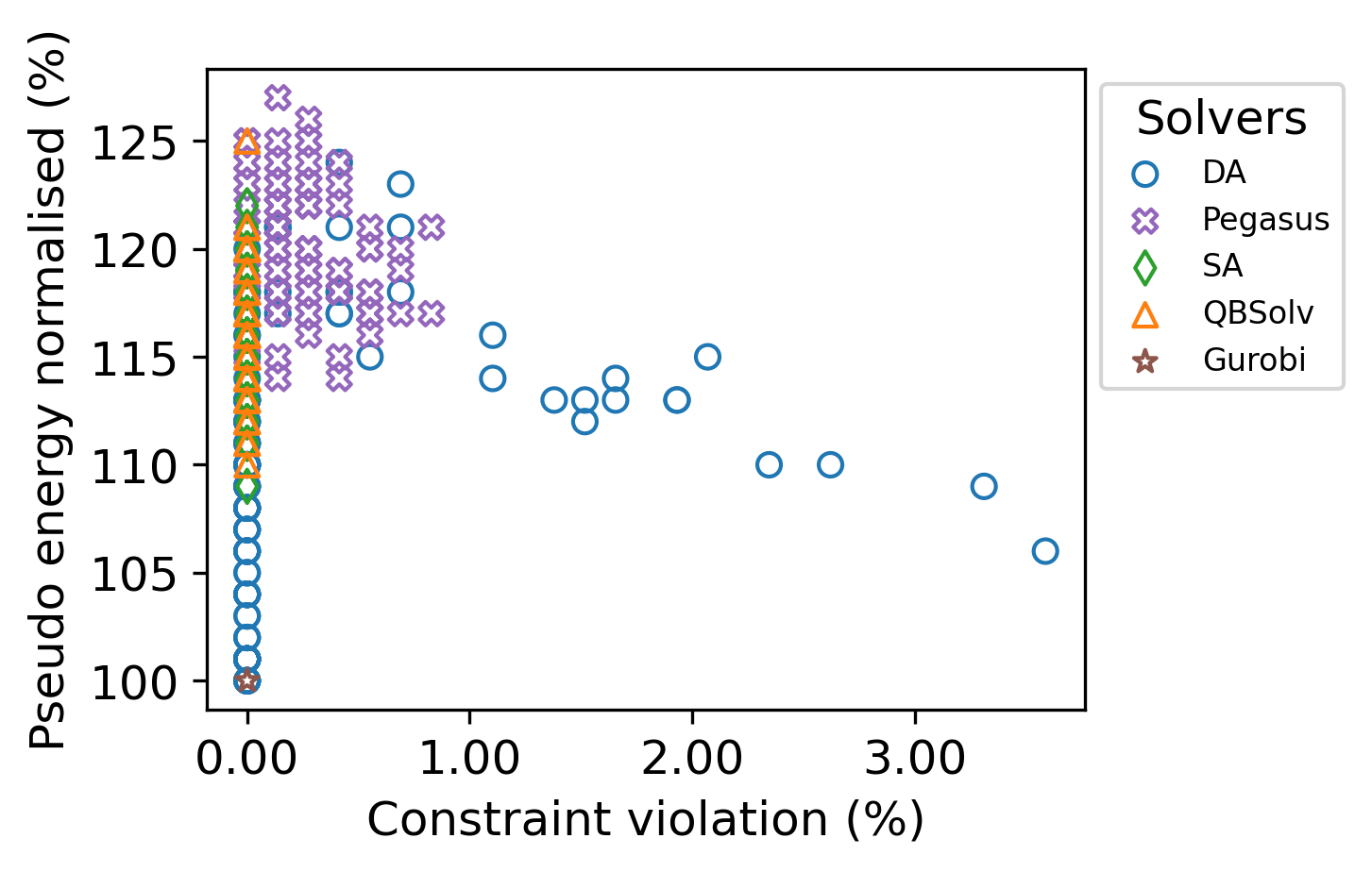}
        \caption{Constraint violation}
        \label{fig:mvc_connectivity_violation}
    \end{subfigure}
    \caption{Feasibility in connectivity-varied MVC problems. a) X axis is averaged degree in log scale. Y axis is $P_f$ in percentage. Gurobi, SA and QBSolv always achieve 100\% $P_f$, the curves of which are overlapped with each other and cannot be seen, but one can spot their markers by zooming in. b) The distribution of constraint violation on the problem instance with an averaged degree of 10. X axis is the constraint violation in percentage. Y axis is pseudo objective energy, normalised to that of Gurobi.}
    \label{fig:mvc_connectivity_indepth}
\end{figure}

Fig.\ref{fig:mvc_connectivity_pf} shows the relation between averaged degree and $P_f$. The SA curve is always 100\%, which provides a sanity check of the choice of the penalty coefficient $\alpha$. The Pegasus curve drops to zero when the degree goes beyond 11. This is where the Pegasus curve stops in Fig.\ref{fig:mvc_connectivity_energy}. The Pegasus curve suggests it is struggling to find feasible solutions as the average degree increases. In fact, for the problem with 140 degrees, we have fine-tuned $\alpha$, or even remove the objective and solve the constraints solely, and still cannot find any feasible solution. DA in Fig.\ref{fig:mvc_connectivity_pf} achieves about 80\% $P_f$. With higher $\#iter$, e.g. $10^6$, it achieves 100\% all the time. Gurobi always find feasible solutions because the constraints can be explicitly implemented in Gurobi and directly regulate the searching process. 

We further investigate the constraint violation in a solution. This is for us to understand how easy it is to fix the violation and salvage promising solutions from broken ones. For MVC, the number of constraints is equal to the number of edges in a graph, according to eq.\ref{equ:mvc_obj}. If two decision variables on two sides of an edge are all zeros, we count it a violation. We divide the number of violations in a solution by the number of constraints, and name the result as the percentage of constraint violation. We use pseudo energy to measure the quality of infeasible solutions, which is the original objective energy plus the non-zero penalty terms.

We examine the problem of degree equals to 10, because this is one of the most challenging instances, according to the time consumption of Gurobi. The relation between energy and constraint violation is shown in Fig.\ref{fig:mvc_connectivity_violation}. Each marker represents a solution. The left bottom corner, i.e., the coordinate (0.0, 1.0), is the global optima. The solutions of QBSolv and SA reside on the line of violation=0.0\%, which means these solutions are feasible. On the other hand, many solutions from Pegasus distribute across 0\%-1\%, which are infeasible solutions. The distribution of the infeasible solutions of DA is similar to a Pareto frontier, in which if you emphasize more on fewer violations, you get higher energy. In some cases, one can see a similar frontier in quantum annealer's constraint violations, which is available in \HT{Appendix D and E}. Fixing broken constraints is another broad topic, which falls out side the scope of this paper. If fixation is not an option, one can also use these infeasible solutions as initialisation of another search.


\subsubsection{DIMACS 10th Challenge}
\label{sec:mvc_dimacs10}

We further evaluate the annealers on real and random datasets extracted from the 10th DIMACS Challenge \cite{bader2013graph}. Eleven graphs are included in this benchmark. The number of nodes ranges from 34 to 22963. Please find the detailed of dataset in \HT{Appendix D.3}.

\begin{figure}[htb]
    \centering
    \begin{subfigure}[b]{0.45\linewidth}
        \centering
    	\includegraphics[width=\textwidth]{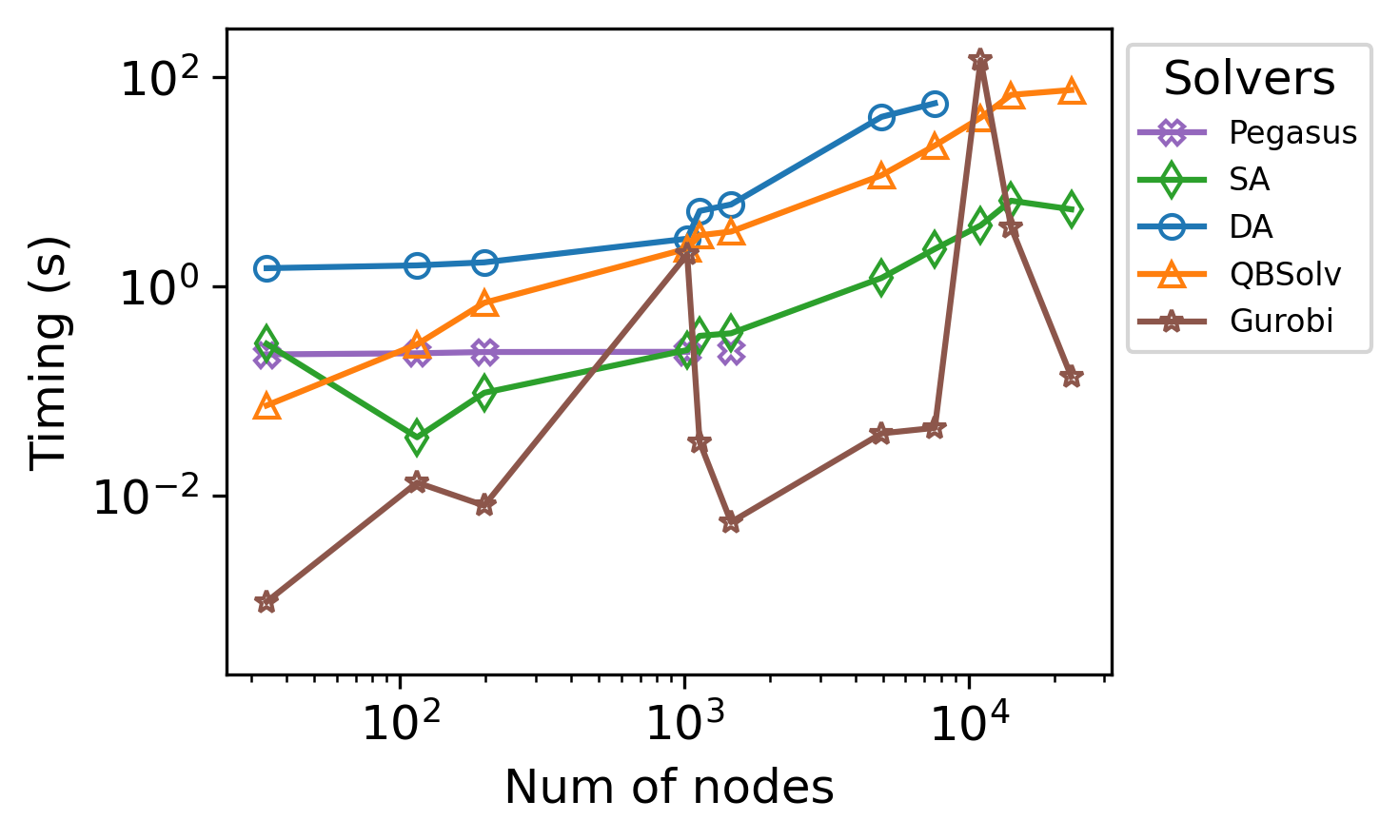}
    	\caption{Timing}
    	\label{fig:mvc_dimacs10_timing}
    \end{subfigure}
    \begin{subfigure}[b]{0.45\linewidth}
        \centering
    	\includegraphics[width=\textwidth]{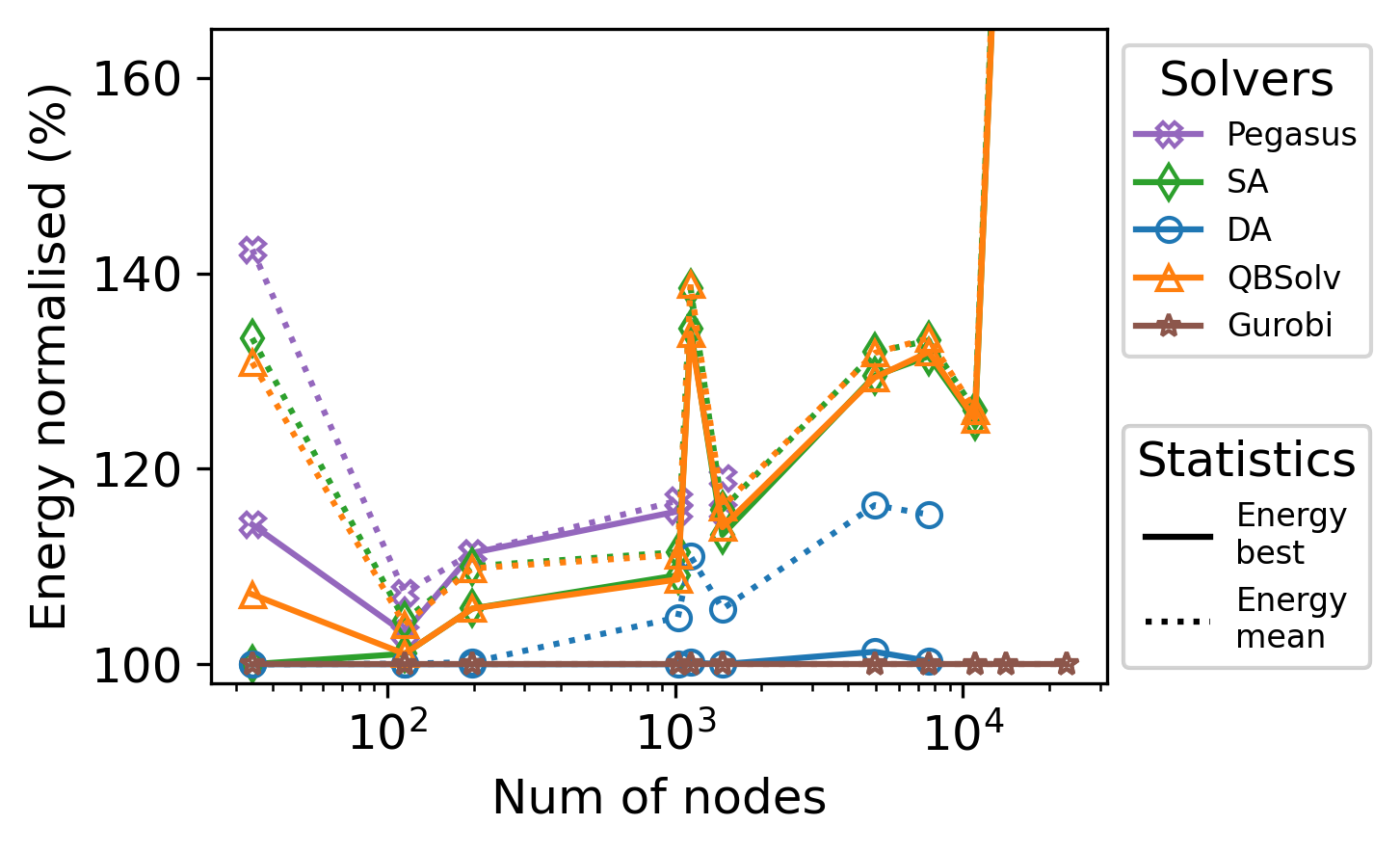}
    	\caption{Normalised energy}
    	\label{fig:mvc_dimacs10_energy}
    \end{subfigure}
    \caption{Performance on MVC problems from DIMACS 10th Challenge. The plot setting is similar to that in Fig.\ref{fig:mvc_connectivity}. For hyper-parameters, we use $T_{neal}=2000\mu s$ for D-Wave annealer,  $\#iter=10^6$ for DA, and $\#sweep=10^2$ for SA}
    \label{fig:mvc_dimacs10}
\end{figure}

In Fig.\ref{fig:mvc_dimacs10_timing}, the timing of the solvers remains a similar trend compared to previous experiments. The Pegasus curve stops when the size of the problem is over 1024 because D-Wave's minor embedding heuristic failed to map larger problems onto Pegasus. DA cannot handle the problem size over 8192.

\begin{figure}[htb]
    \centering
    \begin{subfigure}[b]{0.45\linewidth}
        \centering
    	\includegraphics[width=\textwidth]{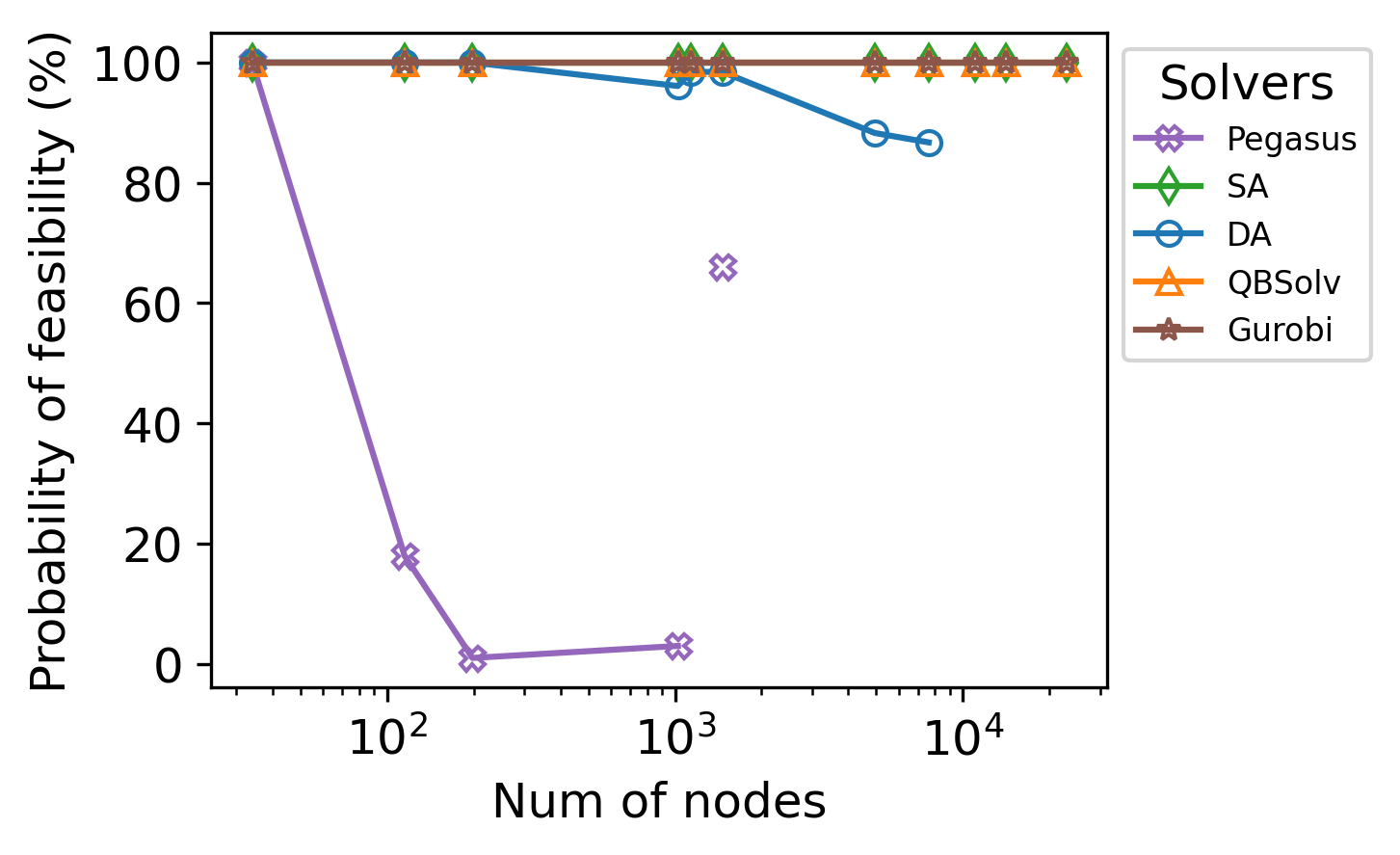}
    	\caption{Probability of feasibility}
    	\label{fig:mvc_dimacs10_pf}
    \end{subfigure}
    \begin{subfigure}[b]{0.45\linewidth}
        \centering
        \includegraphics[width=\textwidth]{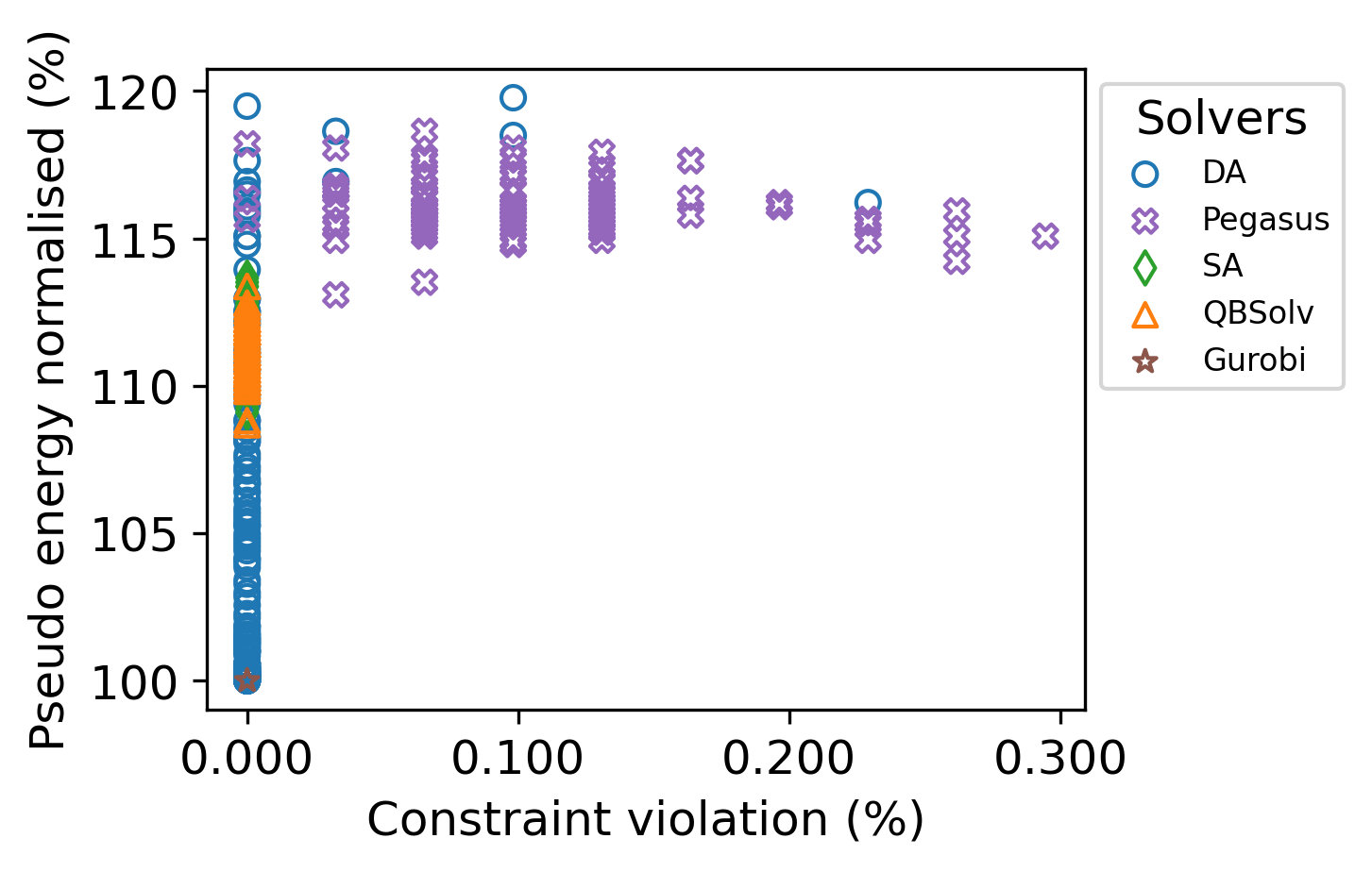}
        \caption{Constraint violation}
        \label{fig:mvc_dimacs10_violation}
    \end{subfigure}
    \caption{Feasibility on MVC problems from DIMACS 10th Challenge. The plot setting is similar to that in Fig.\ref{fig:mvc_connectivity_indepth}. b) The constraint violation on the ``delaunay\_n10'' problem}
    \label{fig:mvc_dimacs10_indepth}
\end{figure}

According to Fig.\ref{fig:mvc_dimacs10_energy}, DA outperforms Pegasus in terms of the quality of solutions, as well as problem size. If we check with Fig.\ref{fig:mvc_dimacs10_pf}, we know that Pegasus is struggling to find feasible solutions, when the problem size is over 100. This matches with the results in previous section. In comparison, the solutions found by DA are mostly feasible. 

Fig.\ref{fig:mvc_dimacs10_violation} shows the relation between normalised energy and constraint violation on the famous ``delaunay\_n10'' problem. We choose it because this is the largest that Pegasus can handle in this dataset. Most of the solvers have similar distributions of solutions, compared with Fig.\ref{fig:mvc_connectivity_violation}. For Pegasus, the distribution is not similar to a Pareto frontier. The normalised energy for Pegasus is around 1.18 and is independent of the constraint violation. We see this as a sign that the noises and errors overwhelm in the quantum annealer, such that the solutions are no longer related to the original problem.

We use the benchmark from DIMACS 10th Challenge in 2012, because we believe it closely follows the trend of requirements nowadays. However, most of the previous works on heuristic MVC methods are based on the benchmark from DIMACS 2nd Challenge in 1992. We evaluate benchmark from DIMACS 2nd as well and include the results in \HT{Appendix D.4}. We also have experiments on a set of Pegasus-like MVC problems in \HT{Appendix D.1}. 

\subsection{Quadratic Assignment Problem (QAP)}
\label{sec:qap}

The \emph{Quadratic Assignment Problem} (QAP) is a well-known combinatorial optimisation problem with a wide range of applications, such as backboard wiring, statistical analysis, placement of electronic components, etc.

QAP can be visualised as the problem of assigning $n$ facilities to $n$ locations. Between any two facilities, there are products transported back and forth, and the amount transported is called the \emph{flow}. Between two locations there is some distance. \emph{Traffic} is defined as the product of flow and distance. The problem is to assign facilities to locations such that the sum of traffic is minimised. The aim is to find an assignment of the facilities to the locations such that the sum of all products of flows and their corresponding distances is minimised, subject to the constraints that each facility is assigned to exactly one location and each location contains exactly one facility. 

Formally, we assume $n$ is the number of facilities which is also the number of locations. $F = (f_{ij})$ is the flow matrix for facilities. Each entry represents the flow between facilities $i$ and $j$. $D = (d_{kl})$ is the distance matrix. Each entry represents the distance between location $i$ and $j$. $X = (x_{ij})$ is the binary decision matrix. Each entry represents whether facility $i$ is assigned to location $j$. Note that the dimension of $X$ is $n \times n$, which implies that there are $O(n^2)$ binary decision variables. If the flow is given as the matrix $(f_{ij})$ and distance is given as the matrix $(d_{kl})$, then the objective function to minimise is:

\begin{align}
\text{Minimise \;} & \sum_{i=1}^{n}\sum_{j=1}^{n}\sum_{k=1}^{n}\sum_{l=1}^{n}f_{ij}d_{kl}x_{ik}x_{jl}
\label{equ:qap_obj}\\
\text{subject to \;} & \sum_{k=1}^{n}x_{ik}=1 \qquad \forall 1\leq i \leq n 
\label{equ:qap_ct1}\\
\text{and \;} & \sum_{i=1}^{n}x_{ik}=1 \qquad \forall 1\leq k \leq n
\label{equ:qap_ct2}
\end{align} 

$x_{ik}$ is the binary decision variable representing whether facility $i$ goes to location $k$. The constraints ensure that one facility is assigned to a location exactly once.

To enable the annealing of a QAP, we have to subsume constraints eq.\ref{equ:qap_ct1} and \ref{equ:qap_ct2} as a part of the objective function eq.\ref{equ:qap_obj}. Conventionally, we can encode the constraints as penalty terms which augment the objective function.

\begin{equation} \label{equ:qap_QUBO}
\begin{split}
\text{Minimise} \quad & \sum_{i=1}^{n}\sum_{j=1}^{n}\sum_{k=1}^{n}\sum_{l=1}^{n}f_{ij}d_{kl}x_{ik}x_{jl} \\
+ & \alpha(\sum_{k=1}^{n}x_{ik}-1)^2 + \alpha(\sum_{i=1}^{n}x_{ik}-1)^2
\end{split}
\end{equation}

Eq.\ref{equ:qap_QUBO} is a standard QUBO form of QAP. $\alpha$ is a penalty weight, which connects the original objective function eq.\ref{equ:qap_obj} and the constraints eq.\ref{equ:qap_ct1} and \ref{equ:qap_ct2}. It can be shown that for QUBOs, the optimal solution to the augmented objective function also minimises the original objective function.

\subsubsection{TinyQAP}
\label{sec:tinyqap}

A QAP with $n$ facilities and $n$ locations have $n^2$ variables, which often yields large models in practical settings. Compared to the max-cut and MVC problems, QAP is more demanding on the number of qubits. All problems in QAPLIB \cite{burkard1997qaplib} are beyond the capability of a quantum annealer.

We generate a set of tiny QAP problems to facilitate the evaluation of quantum annealers. The problem size ranges from 3 to 12. We set penalty weight $\alpha = n\times \mathrm{max}\left ( \left | C \right | \right )$, where $C$ represents a list of the coefficients of all linear and quadratic terms in eq.\ref{equ:qap_obj}. This setting follows the suggestion in \cite{glover2019tutorial} about the ``Scalar Penalty $P$'', with a little more emphasis on the feasibility. The settings are applied to all experiments in Section \ref{sec:qap}.

\begin{figure}[htb]
    \centering
    \begin{subfigure}[b]{0.45\linewidth}
        \centering
    	\includegraphics[width=\textwidth]{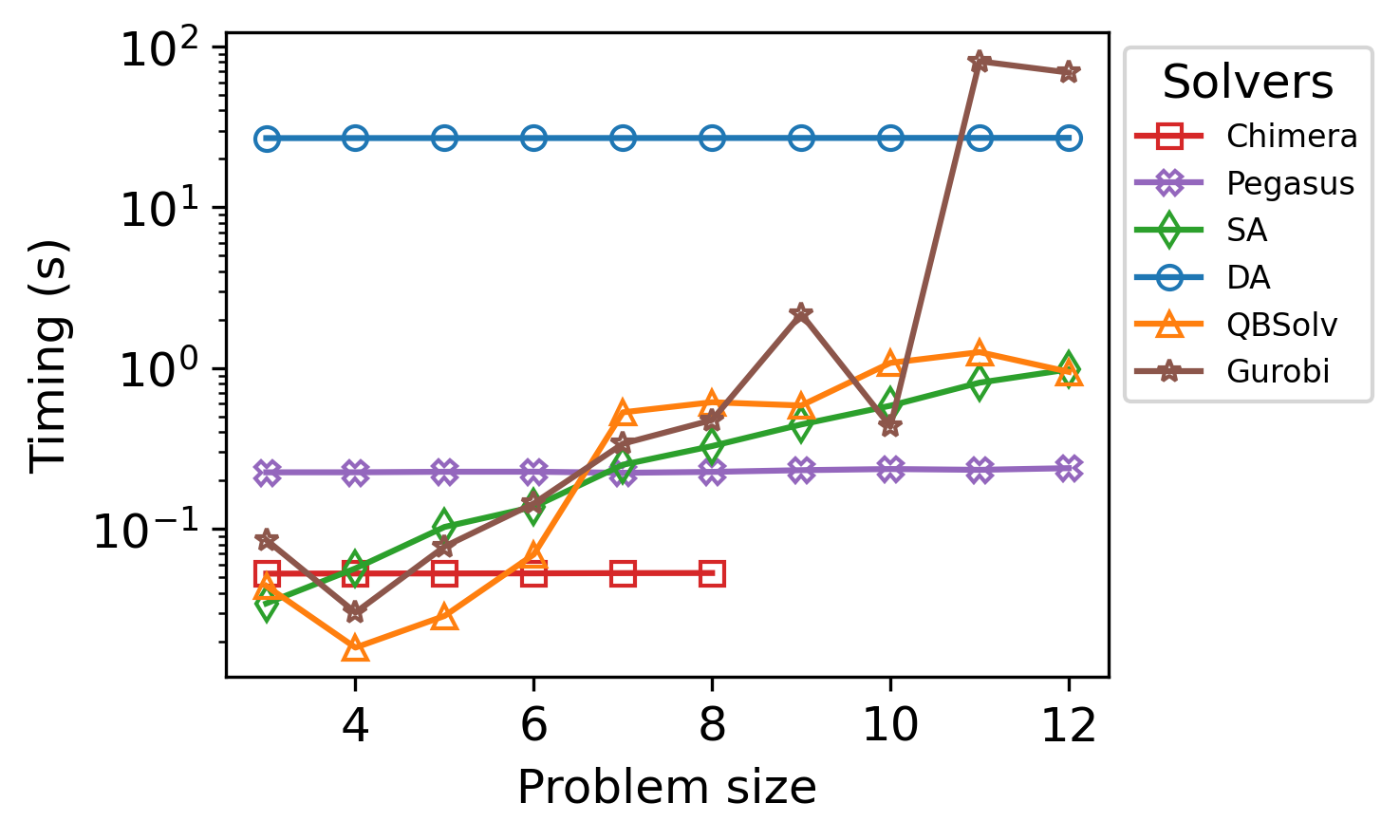}
    	\caption{Timing}
    	\label{fig:tinyqap_timing}
    \end{subfigure}
    \begin{subfigure}[b]{0.45\linewidth}
        \centering
    	\includegraphics[width=\textwidth]{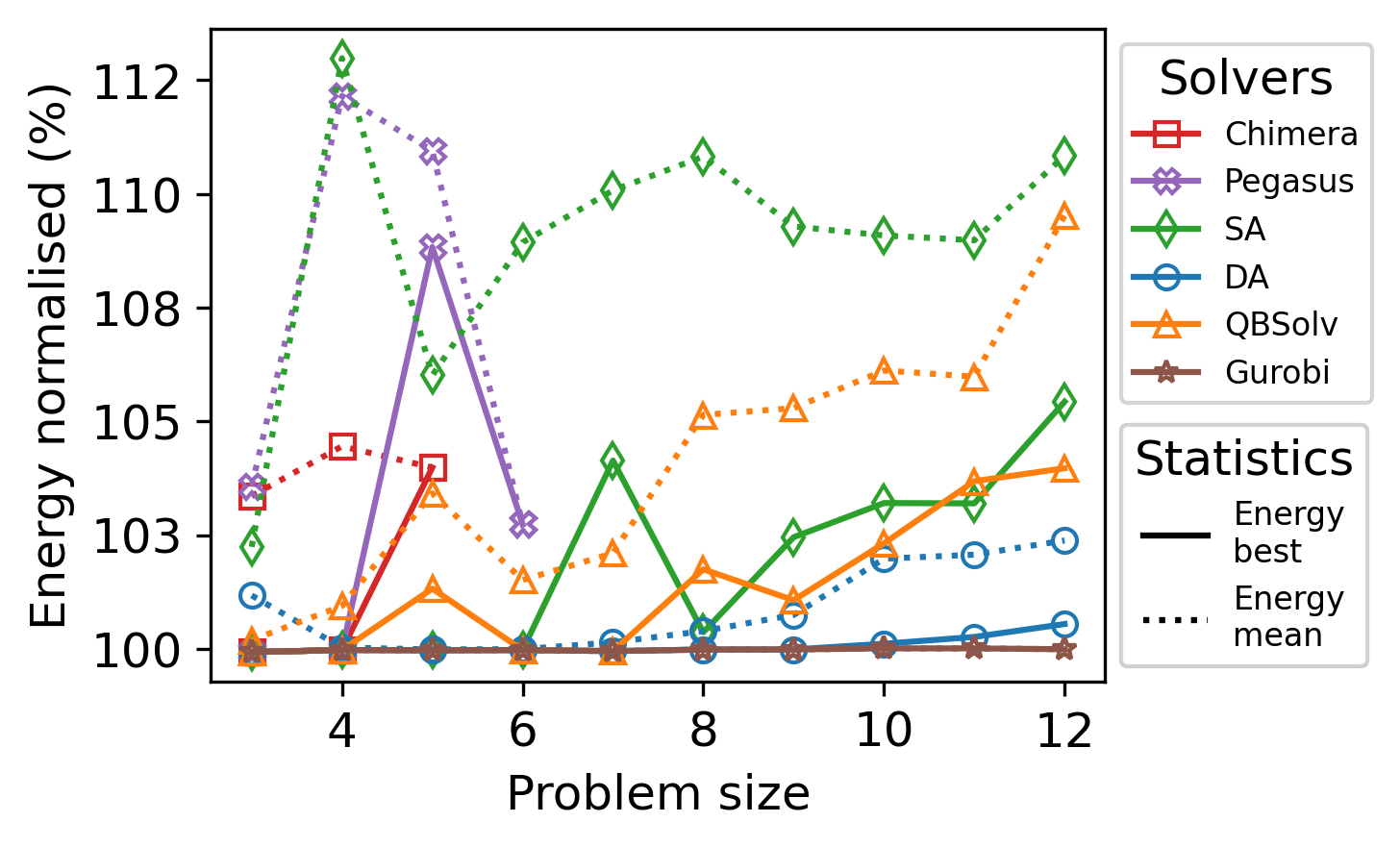}
    	\caption{Normalised energy}
    	\label{fig:tinyqap_energy}
    \end{subfigure}
    \caption{Performance on TinyQAP. The plot setting is similar to that in Fig.\ref{fig:mvc_connectivity}. For hyper-parameters, we use $T_{neal}=200\mu s$ for D-Wave Chimera, $T_{neal}=2000\mu s$ for D-Wave Pegasus, $\#iter=10^8$ for DA, and $\#sweep=10^3$ for SA}
    \label{fig:tinyqap}
\end{figure}

In Fig.\ref{fig:tinyqap_timing}, Chimera and Pegasus are not the fastest. The Chimera curve stops at 8, which corresponds to 64 decision variables. The largest complete graph that Chimera accepts is 65. QAP of $n=12$ is the up-limit of a Pegasus QPU can handle. DA is not designed for such trivial optimisation problems. We can see that the timing of DA does not scales up as the problem size increases.

\begin{figure}[htb]
    \centering
    \begin{subfigure}[b]{0.45\linewidth}
        \centering
    	\includegraphics[width=\textwidth]{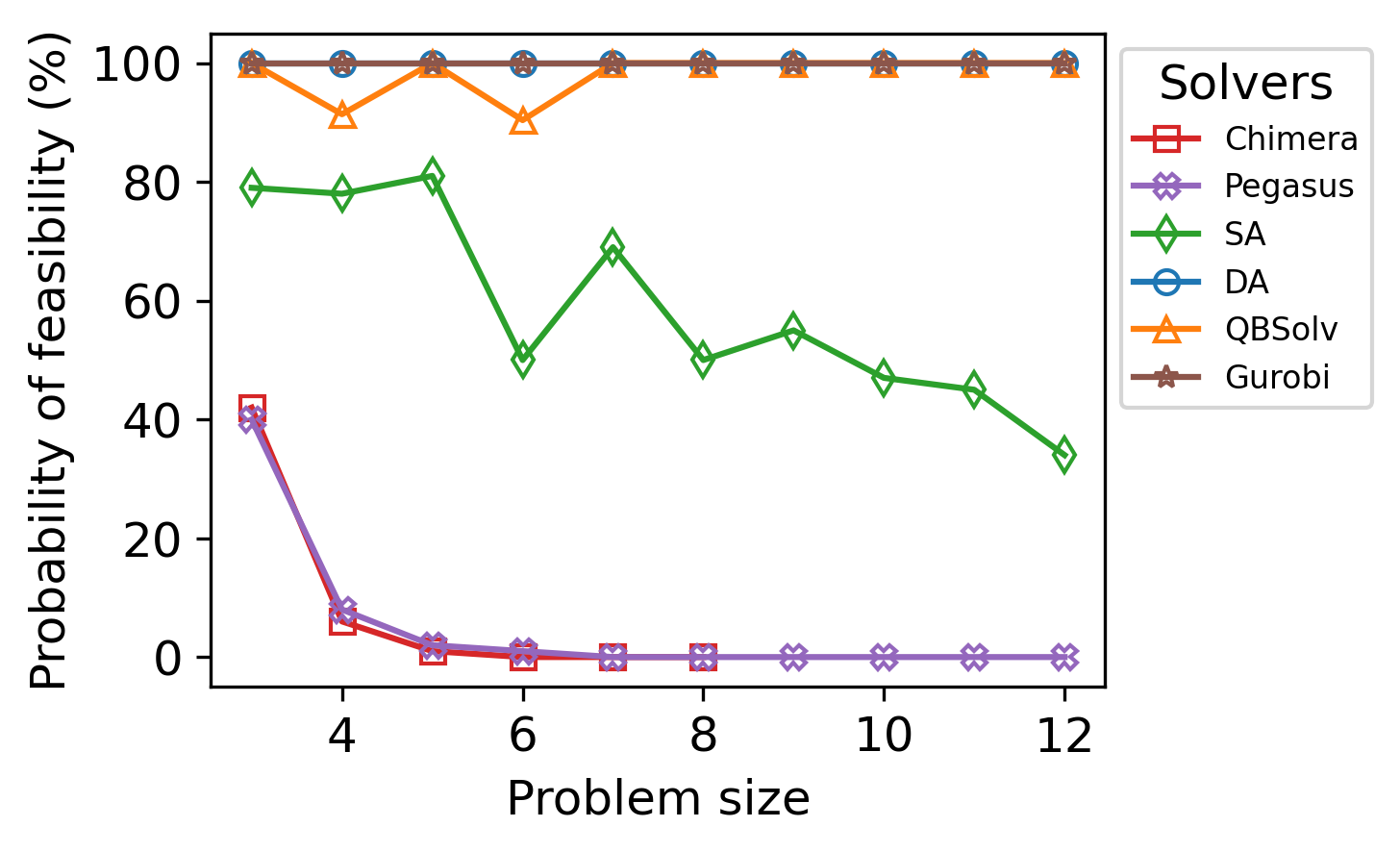}
    	\caption{Probability of feasibility}
    	\label{fig:tinyqap_pf}
    \end{subfigure}
    \begin{subfigure}[b]{0.45\linewidth}
        \centering
        \includegraphics[width=\textwidth]{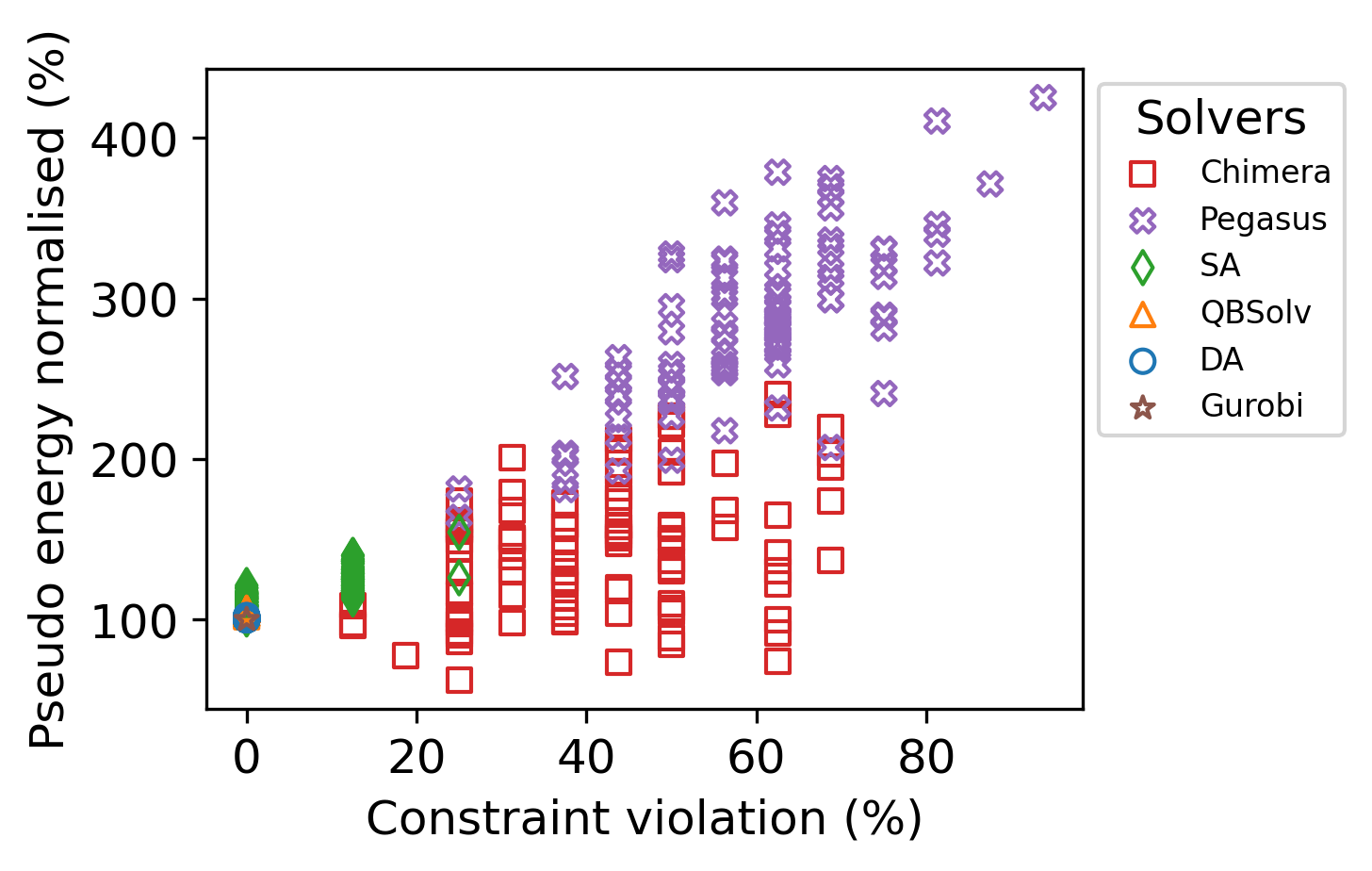}
        \caption{Constraint violation}
        \label{fig:tinyqap_violation}
    \end{subfigure}
    \caption{Feasibility on TinyQAP. The plot setting is similar to that in Fig.\ref{fig:mvc_connectivity_indepth} b) The constraint violation distribution on ``tiny-08''.}
    \label{fig:tinyqap_indepth}
\end{figure}

As shown in Fig.\ref{fig:tinyqap_energy}, Chimera only finds feasible solutions when $n \le 4$. From Fig.\ref{fig:tinyqap_pf} We know that the $P_f$ of Chimera quickly drops to 0\% when $n>4$. The Pegasus QPU managed to find feasible solutions for $n=5$, but its $P_f$ is very close to that of Chimera. This suggests Pegasus architecture does not have a clear advantage over the Chimera architecture. Please note that we are using quite large $T_{anneal}$ for Chimera and Pegasus, which generally leads to better feasibility. For Chimera, increasing $T_{anneal}$ by a factor of 10 does not give any improvement in terms of $P_f$, so we just demonstrate the results of $T_{neal}=200\mu s$. DA can find very promising solutions, which are all feasible and very close to the global optima.

We also investigate the constraint violation of the solutions to ``tiny-08'', which allows us to compare Chimera and Pegasus. The constraints in QAP are one-to-one mapping constraints. For example, in the tiny-03 problem, we have 6 constraints, i.e., 3 constraints to ensure each facility is only assigned to one location, and another 3 constraints to ensure each location only accommodates one facility. From Fig.\ref{fig:tinyqap_violation} we see the solutions of QBSolv and DA reside at coordinate (0.0, 1.0), which means these solutions are all feasible and optimal. Some of the solutions of SA are infeasible, but with only one or two violations.

In comparison, the two quantum annealers are suffering from many constraint violations. From Fig.\ref{fig:tinyqap_violation} we see the energy distribution of Chimera and Pegasus is roughly proportional to the number of violations. This is partly because we put more emphasis on the feasibility of a solution and put high penalty in the objective energy. With the same penalty settings, the annealing-based classical solvers find feasible and competitive solutions most times, but neither of the two quantum annealers can find any feasible solutions. In fact, we have fine-tuned $\alpha$. We even try to solve the penalty term solely. Neither of the quantum annealers find any feasible solutions. We believe this is partly due to the mis-implementation of the problems on the quantum annealers because of the analog control errors in QPUs. \cite{pearson2019analog}. In this experiment the solutions of Chimera is closer to the optimal solution compared with Pegasus. However we cannot assert Chimera architecture is better than Pegasus, because we only have access to one QPU instance of each architecture and cannot rule out the impact of individual differences in the performance evaluation.

TinyQAP is too small for DA. We are going to push towards the limit of DA in the next experiment.

\subsubsection{QAPLIB}
\label{sec:tai}

We choose the Tai dataset from QAPLIB \cite{burkard1997qaplib} because the size fits well with the purpose of our evaluation. We only use the instances Taixxa, which are uniformly generated. We do not include quantum annealers in this experiment because most of the problems are too large for them.

\begin{figure}[htb]
    \centering
    \begin{subfigure}[b]{0.45\linewidth}
        \centering
    	\includegraphics[width=\textwidth]{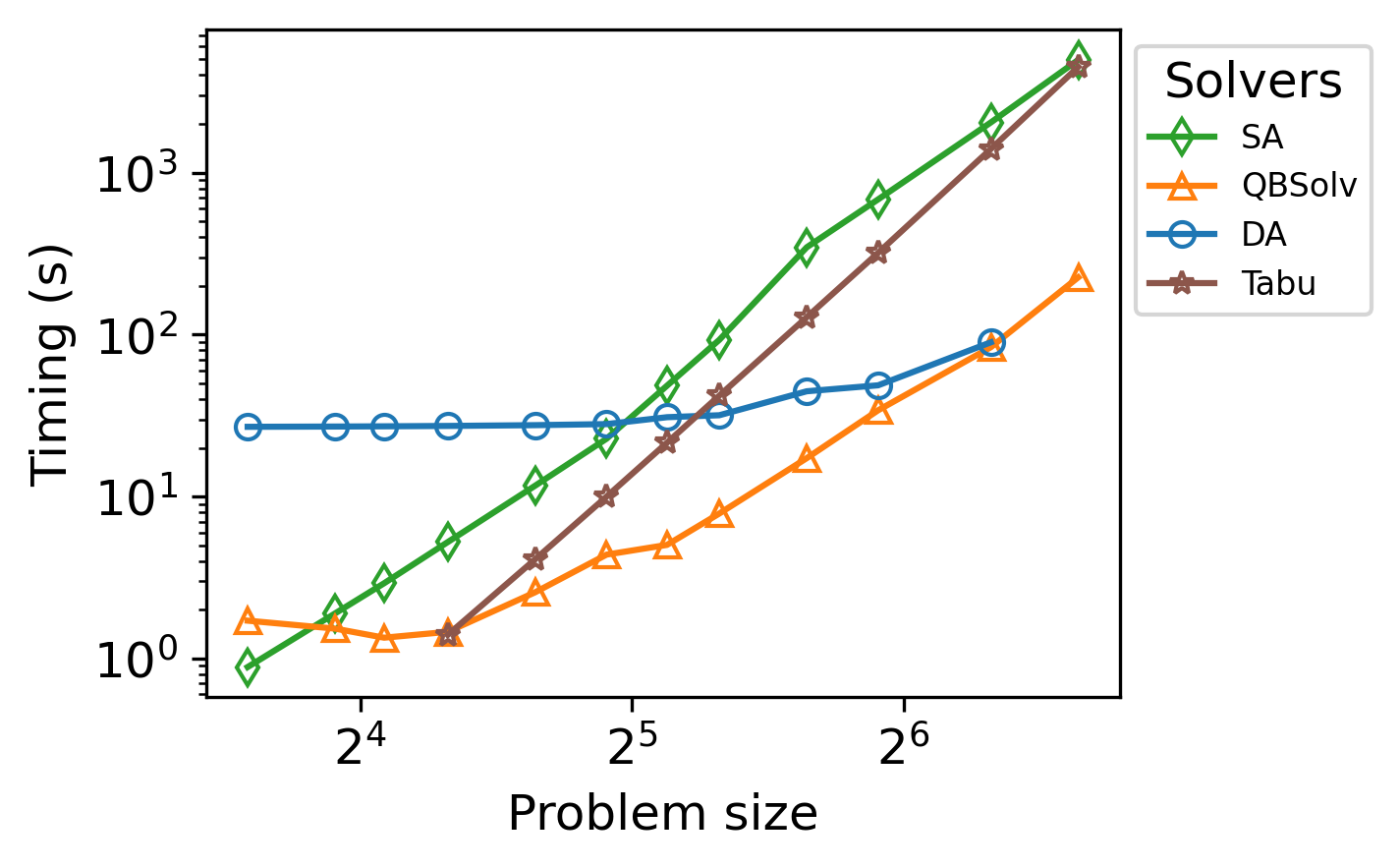}
    	\caption{Timing}
    	\label{fig:qap_tai_timing}
    \end{subfigure}
    \begin{subfigure}[b]{0.45\linewidth}
        \centering
    	\includegraphics[width=\textwidth]{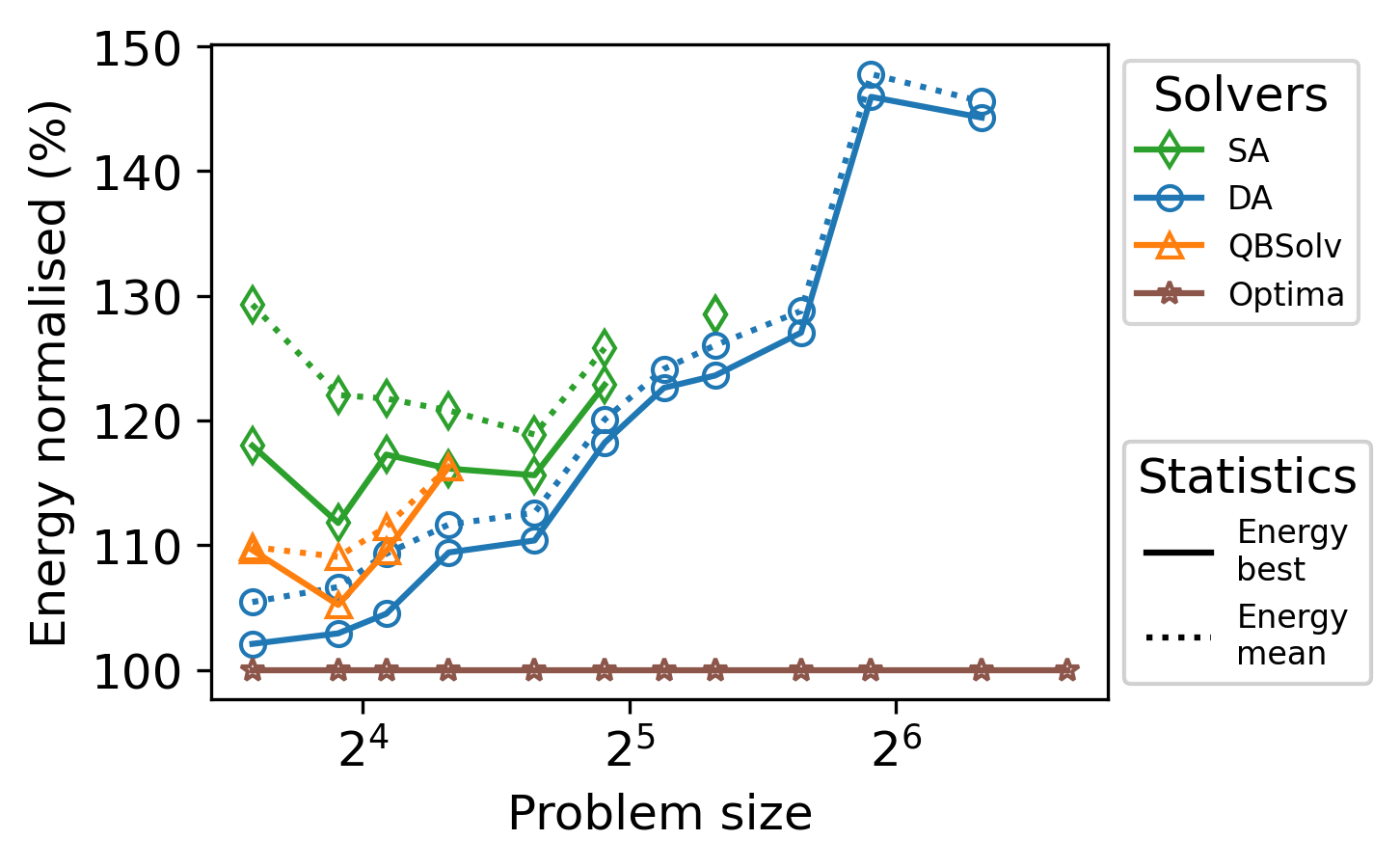}
    	\caption{Normalised energy}
    	\label{fig:qap_tai_energy}
    \end{subfigure}
    \caption{Performance on Tai benchmark. The plot setting is similar to that in Fig.\ref{fig:mvc_connectivity}. For hyper-parameters, we use $T_{neal}=2000\mu s$ for D-Wave Pegasus, $\#iter=10^8$ for DA, and $\#sweep=3\times 10^3$ for SA. We obtain global optima from \cite{burkard1997qaplib}. The timing of Tabu algorithm is from \cite{misevicius2005tabu}}
    \label{fig:qap_tai}
\end{figure}

\begin{figure}[htb]
    \centering
    \begin{subfigure}[b]{0.45\linewidth}
        \centering
    	\includegraphics[width=\textwidth]{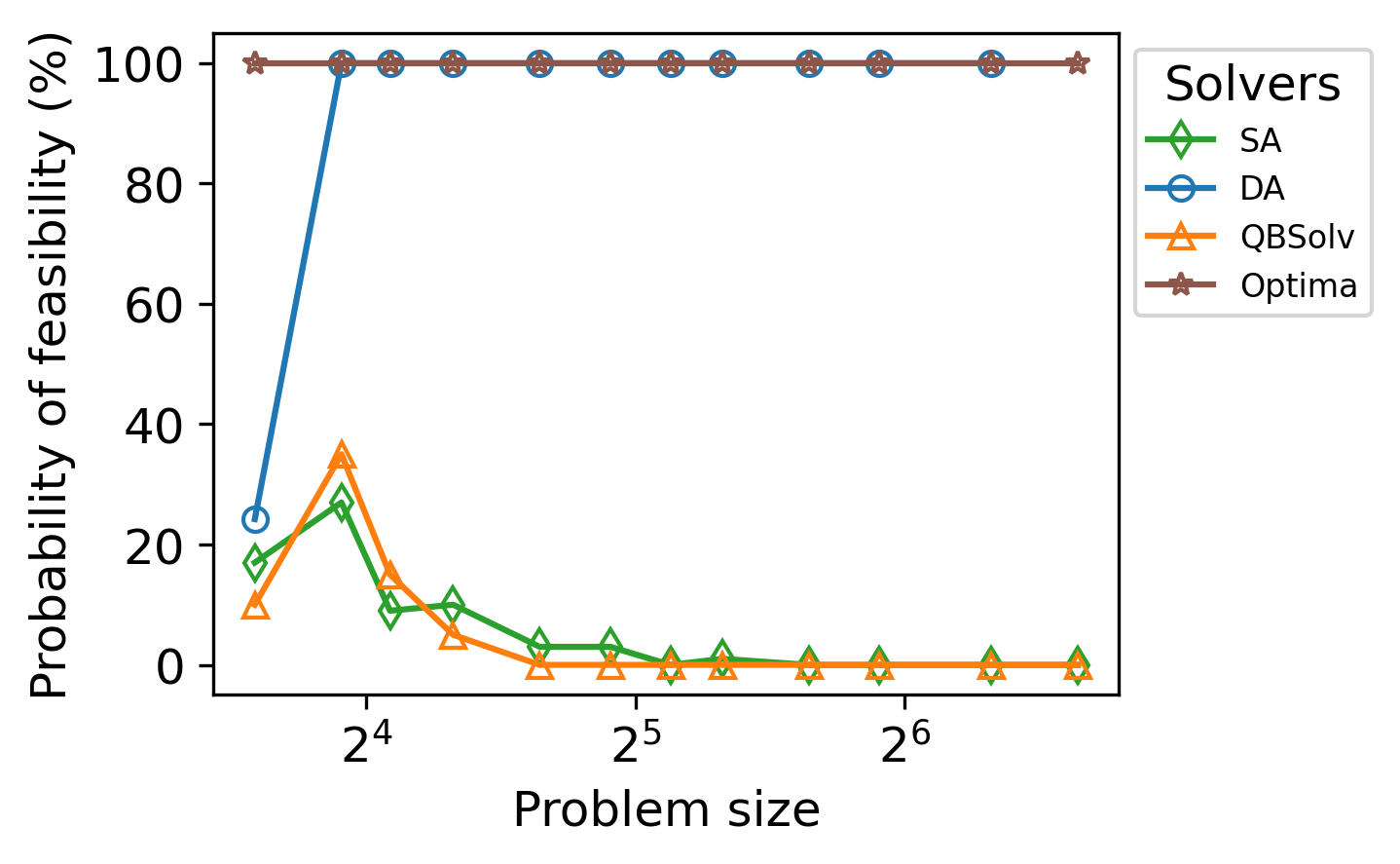}
    	\caption{Probability of feasibility}
    	\label{fig:qap_tai_pf}
    \end{subfigure}
    \begin{subfigure}[b]{0.45\linewidth}
        \centering
    	\includegraphics[width=\textwidth]{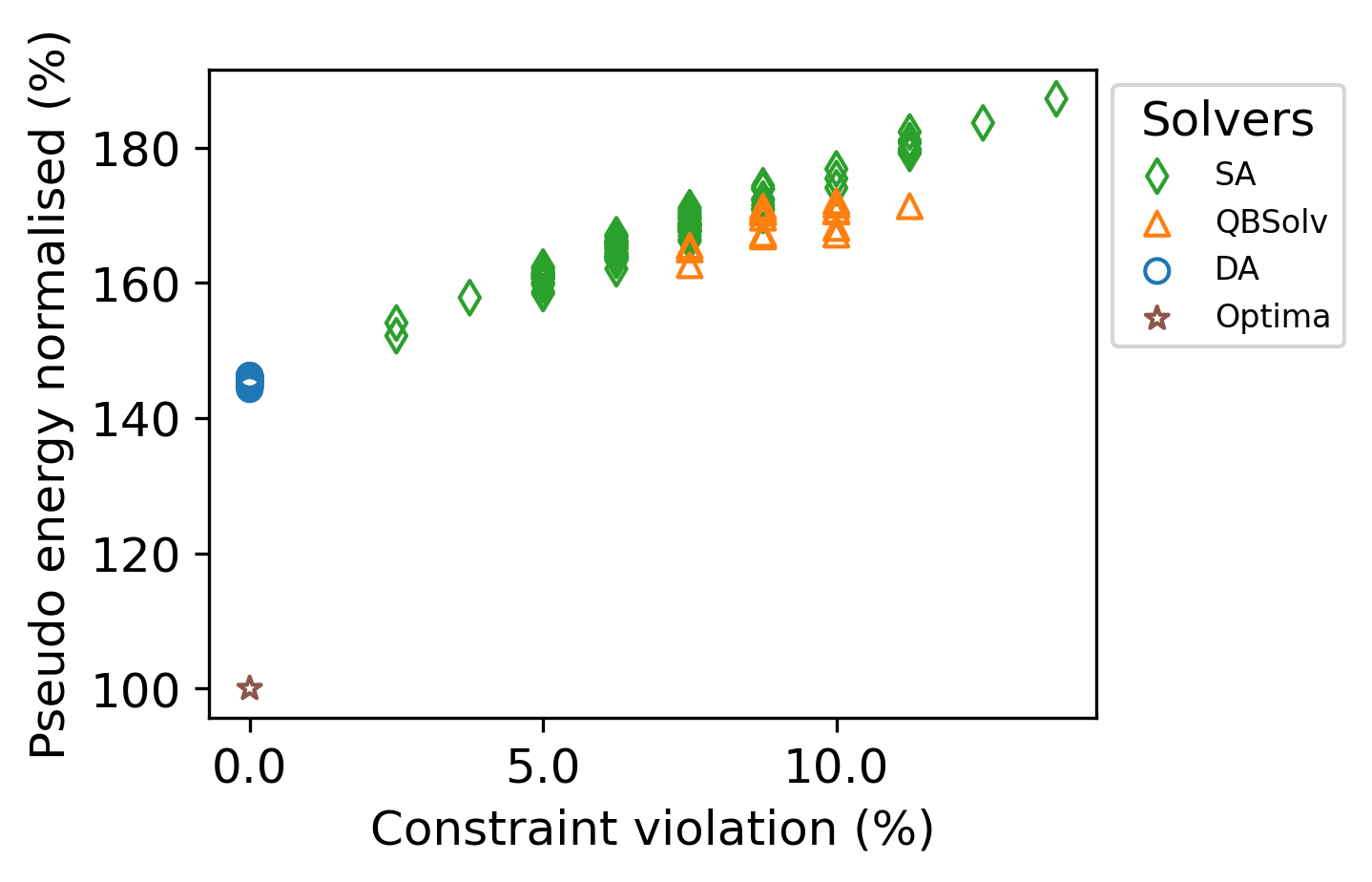}
    	\caption{Constraint Violation}
    	\label{fig:qap_tai_violation}
    \end{subfigure}
    \caption{Feasibility on Tai benchmark from QAPLIB. The plot setting is similar to that in Fig.\ref{fig:mvc_connectivity_indepth} b) The constraint violation distribution on ``tai80a''.}
    \label{fig:qap_tai_indepth}
\end{figure}

According to Fig.\ref{fig:qap_tai_timing} DA has its advantage over SA when the problem size is over 30. It also outperforms SA and QBSolv in terms of objective energy in Fig.\ref{fig:qap_tai_energy}. The hybrid method QBSolv find better solutions than SA, but not as good as DA. We also observe that both SA and QBSolv have difficulty in finding feasible solutions, according to Fig.\ref{fig:qap_tai_pf}

Fig.\ref{fig:qap_tai_violation} is the constraint violation of solutions to ``tai80a''. We choose this problem because it is the largest one in the dataset that can be handled by DA. All solutions of DA feasible. For SA and QBSolv, although none of their solutions is feasible, there is only about 3-15\% of constraint violation, which means the violation fixation task is not very challenging.

Warehouse management is a representative application of QAP. A warehouse usually has items and locations that are much larger than the problem size in previous QAP datasets. We aim to develop a scalable method and facilitate Warehouse management problems with the power of quantum(-inspired) computers.

\section{The Warehouse Assignment Problem}
\label{sec:warehouse}

In warehouse management, order picking is a labour-intensive and costly activity. \cite{tsige2013improving} Consider the problem of assigning items to locations in a warehouse. How should the storage assignment be done, so that when orders come in and items are picked, the travel distance of the picker is at a minimum? For a company, the reduction of such distance translates to the reduction of labour costs. The objective is to find a decent assignment that minimises the travel distance for a given set of orders. 

\begin{figure*}[!t]
     \centering
     \begin{subfigure}[b]{0.30\textwidth}
         \centering
         \includegraphics[width=\textwidth]{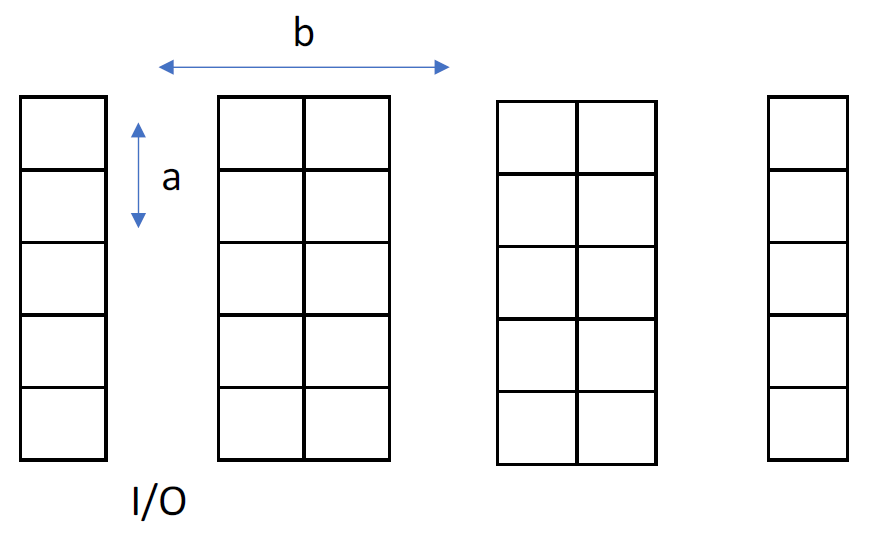}
         \caption{Warehouse layout example}
         \label{fig:layout_example}
     \end{subfigure}
     \hfill
     \begin{subfigure}[b]{0.30\textwidth}
         \centering
         \includegraphics[width=\textwidth]{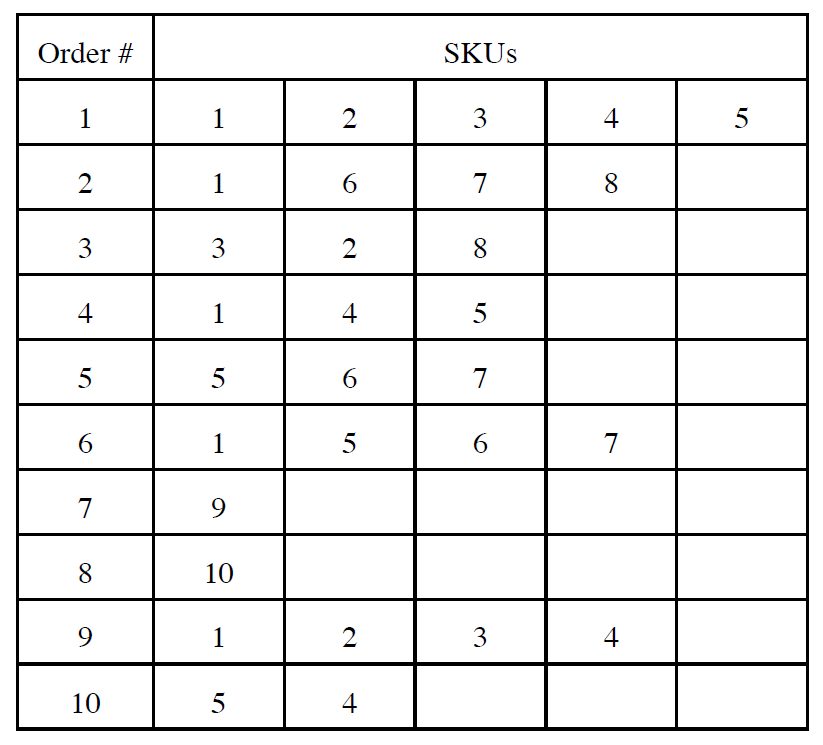}
         \caption{Example of a set of orders}
         \label{fig:order_example}
     \end{subfigure}
     \hfill
     \begin{subfigure}[b]{0.30\textwidth}
         \centering
         \includegraphics[width=\textwidth]{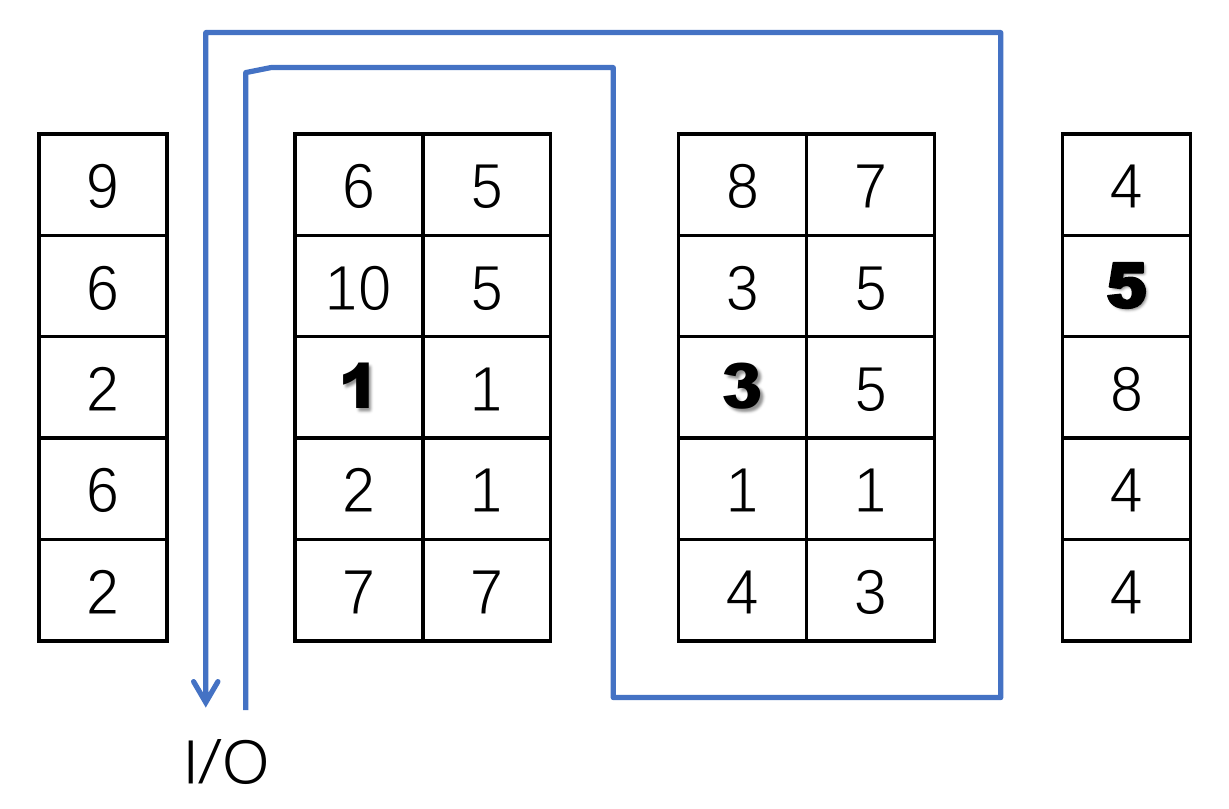}
         \caption{S-shaped routing policy}
         \label{fig:sshape}
     \end{subfigure}
        \caption{Warehouse Assignment Problem}
        \label{fig:warehouse}
\end{figure*}

Consider a warehouse with a layout of Fig.\ref{fig:layout_example}. It has three aisles and 30 locations in total. Each aisle can reach the two columns on its left and right. There is an Input/Output (I/O) point at the bottom left corner through which items are transported in and out of the warehouse.

A set of orders is shown in Fig.\ref{fig:order_example}. There are 10 orders and 10 {\em storage keep units} (SKUs). An SKU represents a unique type of item. The 10 SKUs are labelled 1 to 10. There are a total of 30 items, each of which belongs to a particular SKU.

To pick an order, the picker always starts from the I/O point. It may follow a different route to collect all items. Among various routing strategies, S-shaped (traversal) is the most common and simplest routing policy used in practice. \cite{tsige2013improving} With this policy, the order picker begins by entering the aisle closest to the I/O point. If an aisle contains at least one item, the picker traverses the aisle over the entire length, otherwise, the picker skips that aisle. After picking the last item, the order picker takes the shortest route back to the I/O point. In this paper, we applied a modified S-shaped routing policy, in which a picker will not skip an aisle if there is no item in it. The modification provides a better match between the objective function of the warehouse assignment problem and QAP. An example of the S-shaped routing policy is illustrated in Fig.\ref{fig:sshape}. The picker collects the highlighted item in the first and last aisle by following an S-shaped route even if there is no item to pick in the second aisle.

It is easy to apply QAP to the setting of warehouse assignments. \cite{ruijter2007improved} To do so, simply replace ``facilities'' with ``items''. The flow between items is defined to be the frequency at which their respective SKUs appear in the same order. This can be computed based on the incoming order set. The distance between locations is routing-specific. With our revised S-shaped routing policy, the distance between two items is roughly proportional to the number of columns between them, since the picker would in principle traverse those columns in a zig-zag manner.

\subsection{Heuristics: Block Structure}
\label{sec:warehouse_heuristics}

What inspires the application of QAP in warehouse assignment is the idea to assign items with high flow closer to each other. Intuitively, this would reduce the amount of traffic in picking orders because the picker is more likely to travel a shorter distance in between items appearing in the order.

The block structure of QAP occurs in the distance matrix $(d_{ij})$. Suppose $V(H)$ is the nodes in a graph of the locations. There is a partition of $V(H)$ into $K$ subsets and a function that maps each location to its respective subset: $c: V(H) \rightarrow K$, where $K={1,2,\ldots, k}$, such that

\begin{align}
d(x,y)=
\begin{cases}
\delta(x,y) & \text{ if } c(x)=c(y)      \\ 
M(x,y)      & \text{ if } c(x) \neq c(y)
\end{cases}
\label{equ:cond_dist}
\end{align}

where $\delta$ and $M$ are both functions. The concrete functions would depend on $d$, but with block structure $\delta$ would ``usually'' return smaller values than $M$. A special case is when $\delta$ and $M$ are constant with $\delta<M$, and $(d_{ij})$ will be a ``block matrix''.

\begin{figure}[!t]
     \centering
     \begin{subfigure}[b]{0.33\columnwidth}
         \centering
         \includegraphics[width=\linewidth]{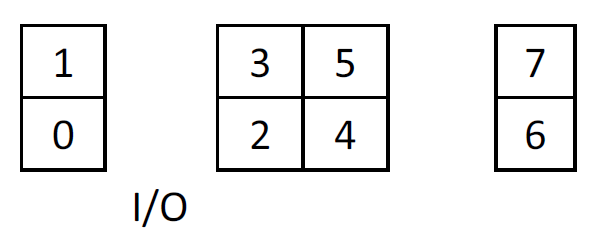}
         \caption{Toy layout}
         \label{fig:toy_layout}
     \end{subfigure}
     \begin{subfigure}[b]{0.33\columnwidth}
         \centering
         \includegraphics[width=\linewidth]{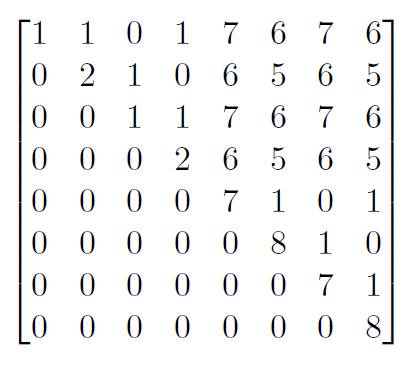}
         \caption{Toy layout distance matrix}
         \label{fig:toy_layout_adj}
     \end{subfigure}
     \hfill
     \begin{subfigure}[b]{0.33\columnwidth}
         \centering
         \includegraphics[width=\linewidth]{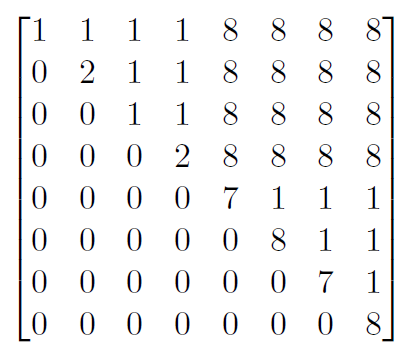}
         \caption{Simplified distance matrix}
         \label{fig:toy_layout_adj_simplified}
     \end{subfigure}
        \caption{Demonstration of block structure}
        \label{fig:block_structure}
\end{figure}

Consider an example with the following warehouse of 8 locations, with each location labelled as in Fig.\ref{fig:toy_layout}. Assume that the vertical distance is 1 and the horizontal distance is 3. The distance matrix looks like the one in Fig.\ref{fig:toy_layout_adj} (only the upper triangular part is shown).

Noticeably, the $4\times 4$ sub-matrix in the upper right corner, corresponding to the distance between the 1st and the 2nd aisles, is uniformly larger in number than the sub-matrices near the diagonal. In general, we could relabel the indices of locations such that the columns of a warehouse form intervals, and that when we move away from the diagonal, the matrix entries become larger because the columns are further away.

Since the above block-structure is observed, one simplification is that all locations within the same column have distance 1 (i.e. $\delta \equiv 1$) in between, and all locations in different columns have a larger distance, say 8 (i.e. $M \equiv 8$). Then the block structure becomes more pronounced as shown in Fig.\ref{fig:toy_layout_adj_simplified}. Please note that  the diagonal of the distance matrix represents the distance between the entry and a certain location. The calculation of the diagonal does not follow the block structure. That's why the diagonal is the same in Fig.\ref{fig:toy_layout_adj} and Fig.\ref{fig:toy_layout_adj_simplified}.

The simplification would normally be expected to flatten the energy landscape and remove tall and thick energy barriers. This benefits annealing-based computers, since energy barriers are notoriously challenging to both quantum and classical annealers. Quantum annealers use quantum mechanics to overcome energy barriers. For classical annealing, the parallel tempering method in DA and many other methods are proposed to overcome energy barriers. A flattened energy landscape is generally easier for annealing-based solvers to work with.

Similar to the way we can perform partitioning on locations, the item set can also be partitioned into $K$ of equal size, where $K$ is fixed to be $|K|$. Intuitively, we want items to appear together in high frequency to be assigned closer to each other. Therefore, we maximise the sum of interaction frequencies within the subsets and then assign each subset to a subset in the partition of $V(H)$, the set of locations. Note that $K$ divides $n$ is assumed, so that locations and items can be divided evenly. 

\subsection{Decomposition}
\label{sec:decomp}

Next, we describe the decomposition formally and provide proof of the theoretical boundary. To minimise the travelling distance, intuitively, we want to maximise the interaction frequencies of items within a block. This objective can be formally described as follows:

\begin{equation}
\text{Maximise} \quad \sum_{i,j=1}^{n}\sum_{l=1}^{k}f_{ij}x_{il}x_{jl}
\label{equ:graph_part}
\end{equation}

Eq.\ref{equ:graph_part} denotes the sum of interaction frequencies among items within their respective subsets. $(x_{ij})$ is an $n \times k$ decision matrix. $x_{ij}$ denotes whether item $i$ goes to subset $j$. The objective comes with the following constraints:

\begin{align}
\sum_{l=1}^{k}x_{il} = 1 \quad \forall i, 1\leq i\leq n
\label{equ:graph_part_ct1}\\
\sum_{i=1}^{n}x_{il} \leq s \quad \forall l, 1\leq l \leq k
\label{equ:graph_part_ct2}
\end{align}

Eq.\ref{equ:graph_part_ct1} means an item belongs to exactly one subset. Note that each subset can have at most $s=\frac{n}{k}$ items. Therefore, eq.\ref{equ:graph_part_ct2} means each subset must not exceed its capacity $s$.

Note that eq.\ref{equ:graph_part} is actually another QAP with a particular decision matrix (but with different constraints). Thus, it can be translated into a graph formulation. Suppose $V(G)$ and $E(G)$ are the nodes and edges in a graph of items. In the graph formulation, $(x_{ij})$ is equivalent to a function $g: V(G)\to K$, which maps an element of $V(G)$ into its subset, such that $g(i) = a \iff x_{ia} = 1$. This function is well-defined due to constraint eq.\ref{equ:graph_part_ct1}. Eq.\ref{equ:graph_part} can be converted to the equivalent graph formulation as follows:

\begin{equation}
\sum_{i,j=1}^{n}\sum_{l=1}^{k}f_{ij}x_{il}x_{jl} =  \sum_{\substack{(i,j)\in E(G)\\g(i) = g(j)}}f(i,j) \label{equ:graph_part_g}
\end{equation}

which, intuitively, is the sum of flows within subsets. This is the setup for the theorem below, which states that with a solution to eq.\ref{equ:graph_part}, an optimal solution to the overall QAP eq.\ref{equ:qap_obj} can be constructed. The proof is available from \HT{Appendix F.5}.

\newtheorem{theorem1}{Theorem}
\begin{theorem1}
Let $d(x,y) = \begin{cases}
		\delta & c(x)=c(y) \\
		M	   & c(x)\neq c(y)
		\end{cases}
	$
for some positive constants $\delta$ and $M$. Then there exists a $\varphi_0$ for which eq.\ref{equ:qap_obj} is achieved, and $\varphi_0$ can be constructed from a solution to eq.\ref{equ:graph_part} denoted by $g_0$.
\end{theorem1}

This theorem assumes a strong condition that $(d_{ij})$ only has two distinct values, $\delta$ for locations within an interval and $M$ in between intervals. This is a simplification of the block structure and $(d_{ij})$, in reality, is usually more complex.

However, the concept of intervals can be generalised such that when $\delta$ and $M$ are non-constant, a function $c$ can still be defined on $V(H)$ such that the sum of distances within intervals is minimised. The motivation for such construct is that when a warehouse does not exhibit a clear-cut column structure, one can still think of an abstract ``column'' as a group of close locations. Note that this definition automatically specialises to the previous definition of $c$ when $\delta$ and $M$ are constant. This intuition can be formally expressed as 

\begin{align}
\text{Minimise} \quad & \sum_{i,j=1}^{n}\sum_{l=1}^{k}d_{ij}x_{il}x_{jl}  
\label{equ:graph_part_loc} \\
\text{Subject to} \quad & \sum_{l=1}^{k}x_{il} = 1 \quad i \in [1, n]
\label{equ:graph_part_loc_ct1} \\
\text{and} \quad & \sum_{i=1}^{n}x_{il} \leq s \quad l \in [1, k]
\label{equ:graph_part_loc_ct2} 
\end{align}

$(x_{ij})$ is an $n\times k$ decision matrix. $x_{ij}$ denotes whether item $i$ goes to location group $j$. Note that this graph partitioning problem can be thought of as the dual to eq.\ref{equ:graph_part}, with maximisation changed to minimisation. In effect, items that are frequently ordered together will be assigned to locations that are closer together.

Now both the set of locations and the set of items have been divided into subsets of equal size. Between subsets of locations, the distances are maximal, and between subsets of items, the interaction frequency is minimal. The final step is to produce a bijection from the set of subsets of items to the set of subsets of locations, and for each pair of subsets in the bijection $(items, locs)$, there is a sub-QAP of size $O(\sqrt{n})$, and therefore a QUBO of size $O(n)$, where $n$ is the total number of items.

Note that this is subject to $n$ being a perfect square; in practical situations where $n$ is not a perfect square, compromise has to be made in either finding the nearest smaller perfect square and do the optimisation on the smaller set of items and locations only, or use integer divisors of $n$ other than $\sqrt{n}$, and correspondingly deal with sub-QAPs of sizes other than $\sqrt{n}$. For example, if there are $n=3600$ locations, it is possible to divide it into 40 groups of 90, where each sub-QAP will have $n=90$.

The overall procedure runs the decomposition in Subsection \ref{sec:decomp} and then solves each individual sub-QAP using the exterior penalty method. Each sub-QAP utilises a conversion procedure to transform it into QUBO solvable by quantum (-inspired) annealer. Finally, the procedure forms global solutions by aggregating solutions to sub-problems. The details of the overall procedure, the exterior penalty method and the conversion procedure are available in \HT{Appendix F.6 and F.7}. 

\section{Experiments}
\label{sec:experiment}

The experiment has three objectives, namely 1) to evaluate the performance of the QAP decomposition heuristic in solving block-structural QAPs of various sizes, 2) to evaluate the effectiveness of the heuristic in minimising the warehouse picking distance, and 3) compare the performance of various computing hardware using different decomposition heuristics. In addition, we also included the heuristic algorithm ``Tabu search'' as a reference.

\subsection{Comparison on QAP}
\label{sec:warehouse_qap}

We compare the performance of decomposition on QAP. Block-structural QAP instances WH-8, WH-90, WH-180, WH-270, WH-3600, and WH-8100 are synthesised using randomly generated order sets. The number corresponds to the size of a problem. The results obtained for each of the datasets are average values over three runs. Please refer to \HT{Appendix F.3} for details of dataset and experimental settings.


\begin{table}[!t]
\centering
\caption{QAP time comparison}
\begin{tabular}{|l|c|c|c|c|c|c|}
\hline
Size              & 8     & 90     & 180    & 270    & 3600    & 8100     \\ \hline
$T_\text{QA}$     & 0.016 & -      & -      & -      & -       & -       \\ \hline
$T_\text{QBSolv}$ & 0.37  & 379.95 & 7454   & -      & -       & -       \\ \hline
$T_\text{DA}$     & 3.70  & 23     & -      & -      & -       & -       \\ \hline
$T_\text{decomp}$ & -     & -      & 60     & 68     & 983     & 2080    \\ \hline
$T_\text{Tabu}$   & 0.02  & 0.08   & 0.15   & 0.19   & 23.5    & 64.3    \\ \hline
$T_\text{SA}$     & 683   & 556    & 716    & 594    & 3447    & 6884    \\ \hline
\end{tabular}
\label{tab:warehouse_time}
\end{table}

QA can only handle QAP of size 8 due to the limitations in the D-Wave Chimera architecture. From Table \ref{tab:warehouse_time} we know that QA is the fastest among all others. Table \ref{tab:warehouse_energy} shows the energy of the solution found by QA is the highest. This matches with the results of experiments in Section \ref{sec:qap}.


DA can directly handle up to QAP of size 90. $T_\text{DA}$ 16 times faster than QBSolv on WH-90. In terms of the quality of the solutions, DA finds near-optimal solutions to WH-8. Although DA can handle a problem size of 90 directly without decomposition heuristics, the $E_\text{DA}$ is distinctly higher than that of Tabu and SA. 

For problems larger than 90, we use decomposition heuristics to tackle the problem. $T_\text{decomp}$ in Table \ref{tab:warehouse_time} is the time for DA solving all sub-problems. It scales linearly with the size of the problem. In terms of the quality of the solution, our decomposition heuristic can find a solution that is 14.3\% lower than that of the random solutions on WH-180. However, as the number of partitions increases, the optimality against the random solution vanishes. For example, on WH-8100, there is no difference in energy between random solution and the decomposition heuristic.

\begin{table}[!t]
\begin{center}
\caption{QAP energy comparison (lower is better)}
\begin{threeparttable}
\begin{tabular}{|l|c|c|c|c|c|c|}
\hline
Size              & 8    & 90      & 180     & 270     & 3600    & 8100    \\ \hline
$E_\text{QA}$     & 119  & -       & -       & -       & -       & -       \\ \hline
$E_\text{QBSolv}$ & 81   & $0.71m$ & NA      & -       & -       & -       \\ \hline
$E_\text{DA}$     & 81   & $0.67m$ & -       & -       & -       & -       \\ \hline
$E_\text{decomp}$ & -    & -       & $2.21m$ & $15.9m$ & $12.9b$ & $115b$  \\ \hline
$E_\text{Tabu}$   & 80   & $0.64m$ & $1.95m$ & $14.4m$ & $10.9b$ & $103b$  \\ \hline
$E_\text{SA}$     & 81   & $0.64m$ & $1.96m$ & $14.4m$ & $10.6b$ & $91.7b$ \\ \hline
$E_\text{Random}$ & -    & $0.75m$ & $2.58m$ & $17.5m$ & $13.1b$ & $115b$  \\ \hline
\end{tabular}
\begin{tablenotes}
\item `$m$' stands for `$\times 10^6$' while `$b$' means `$\times 10^9$'.
\end{tablenotes}
\end{threeparttable}
\label{tab:warehouse_energy}
\end{center}
\end{table}

Tabu is faster than all other methods on most datasets, except WH-8. 
In terms of the solution's quality, both Tabu and SA can find better solutions than other methods.
QBSolv did not find any feasible solution for WH-180.
 
\subsection{Comparison on Warehouse Management}
\label{sec:exp_warehouse}

We also generate a set of warehouse problems to test the effectiveness of the decomposition heuristic in reducing warehouse travel distance. The input data distribution is perturbed to provide modified versions of datasets in Section \ref{sec:warehouse_qap}. The name of the resulting problems has a tailing letter ``b''. This is to match the findings in \cite{tsige2013improving} that interaction-based methods such as QAP work well when 80\% of ordered items are concentrated in 20\% of the SKUs. In other words, there is a small set of commonly ordered products. The above assumptions are made about the order sets as well as the shape of the respective warehouses.

\begin{table}[htb]
\centering
\caption{Warehouse travel distance comparison}
\begin{tabular}{|c|c|c|c|c|c|}
\hline
Name    & ABC           & COI  & OOS           & Random        & decomp \\ \hline
WH-90b  & \textbf{733}  & 876  & 773           & 741           & 765    \\ \hline
WH-180b & \textbf{1418} & 1434 & 1425          & \textbf{1418} & 1422   \\ \hline
WH-270b & 3457          & 4524 & \textbf{3084} & 3619          & 3098   \\ \hline
\end{tabular}
\label{tab:solution_quality}
\end{table}

A simulation framework is built to calculate the distance travelled given order, and an assignment. We include three representative assignment heuristics, i.e., Cube-per-order Index (COI) \cite{malmborg1990revised}, class-based storage policy (ABC) \cite{petersen2004improving}, and Order-oriented-swapping (OOS) \cite{mantel2007order}, as references. A brief review of these heuristics are available in \HT{Appendix F.1}. Our decomposition heuristic is implemented based on DA, which solves a block-structured QAP generated from WH-270b. The numbers in Table \ref{tab:solution_quality} are the travelling distance of a picker, given different warehouse assignment policies. We run the experiment five times and get the averages to indicate the performance of these policies.

Table \ref{tab:solution_quality} shows that OOS is the best among the first three heuristics. The performance of our heuristics is very close to that of OOS. The objective function of QAP and the cost of warehouse assignment are not identical. Our decomposition heuristic does not perform as well as direct methods without decomposition from the point of view of QAP, but it performs well on warehouse assignment problems.

\section{Discussion}
\label{sec:discussion}

\subsection{Quantum(-inspired) Annealers: Pros, Cons and what's Missing}
\label{sec:pros_cons}

\textbf{Speed} The running time of D-Wave Quantum Annealer (QA) depends on the settings of its annealing process and is independent of the problem size. The time of Fujitsu Digital Annealer (DA) depends on the problem size. In our experiments, the problem-solving time of QA is generally fast on simple problem settings. DA has a good scalability in terms problem size.

To accelerate QA and DA from a problem perspective, we can reduce the size of the problem. 
For example, smart encoding technique \cite{tan2021qubit} uses fewer qubits to represent larger problems for gate-based models. The equivalent research efforts in the domain of annealing-based computers is missing.
On the other hand, problem simplification could also helps. For example, data scaling \cite{goh2021hybrid} smoothens the energy landscape of a permutation optimisation problem and helps DA find promising results with fewer iterations. A more general simplification method is missing to facilitate solver for other combinatorial optimisation problems.

To accelerate QA and DA from a method perspective, we can optimise the annealing process for a quantum annealer \cite{venturelli2019reverse} and a classical annealer \cite{isakov2015optimised}. This shortens the problem-solving time, as well as potentially improve of the optimality of the solutions. However, many works in this area are problem-specific and heavily parameterised. A general optimisation method is missing.

\textbf{Optimality} Overall, both QA and DA have a decrease in the quality of solutions as the problem size increases, although the reason behind are totally different. QA suffers from more errors in quantum mechanics \cite{pudenz2014error} and analog control \cite{pearson2019analog} as the problem size increases. DA has difficulty in exploring the solution space as it expends quickly along with the increase of problem size. 

QA outperforms DA in a simple problem setting, but loses its leading position when the problem setting gets complicated. Let $\mathbb{S}_p$ denotes the solution space to the original problem and $\mathbb{S}_v$ denotes the space expanded by the decision variables in the QUBO form. QA performs poorly when $\mathbb{S}_p$ is smaller than $\mathbb{S}_v$. In comparison, DA is less sensitive to this factor. For example, QA outperforms DA on max-cut in Section \ref{sec:maxcut_connect}, but underperforms DA on MVC in Section \ref{sec:mvc_connect}, despite the two experiments share the same dataset. We reach this conclusion by comparing QA and DA on different problems settings, which is missing in existing benchmark works.

There has been quite a lot of research in improving annealing-based computers. Techniques for improving speed from a problem perspective could also improve the optimality. \emph{Error Correction} \cite{pudenz2014error, vinci2016nested} is a popular research direction for QA, but not DA.
Furthermore, finetuning hyper-parameters such as the annealing process \cite{venturelli2019reverse}, or penalty weight \cite{huang2021qross} can also potentially improve its optimality.

\subsection{Decomposition on QAP}
\label{sec:disc_qap}

Decomposition improves the scalability of annealing-based computers. Experiments in Section \ref{sec:experiment} suggest our heuristic solves block-structural QAP with improved quality when the size is limited below 270. However, for larger problem sizes, our heuristic gradually loses its utility.

\emph{Cases where decomposition fail.} There could be a few reasons. First, solutions to sub-QAPs are not optimal. Second, the heuristic may be inaccurate as size increases. It is possible that when too many columns of 90 are stacked together, the effect of assigning item pairs in one column is superseded by the vast shape of the warehouse, such that even items from different columns, which have relatively low interaction frequency, contribute significantly to the overall traffic because the distance becomes large. Therefore, it is not advisable to split items and locations into too many subsets.

\emph{Matching the subsets.} In Section \ref{sec:decomp}, the heuristic partitions the items and locations according to an aggregate interaction-frequency measure and distance measure. This is a special case of the more generalised randomised decomposition technique in \cite{mihic2018randomized}, in which the items and locations are randomly taken to form sub-QAPs and lead to promising results. 
Therefore, one of the possibilities is to add a random layer on top of the current decomposition layer. 
In a similar spirit, the subsets are matched randomly in this project. Instead, there could be other organised ways as mentioned in Section \ref{sec:decomp}. When the number of bijections is small, the matching can be considered exhaustively to determine which one is the best. For example, for WH-180, there are only 2 item and location subsets. There are only 2 ways to match the subsets.  A more interesting but much more difficult question is how a matching would be a good seed.

The overall decomposition heuristic is only proven to sustain optimality for a very restricted simplification. It would be encouraging if further advancements can be made in this direction, expanding the solution quality guarantee to a larger class of block-structural QAPs. However, it is so far unclear how that could be done because that would involve analysing the distance matrix case-by-case.


\section{Conclusion}
\label{sec:conclusion}

Recent quantum(-inspired) annealers show promise in solving combinatorial optimisation problems. In this paper, we compared a true quantum annealer with a CMOS digital annealer on three problems ranging from simple to complex ones. We also compare them in the context of decomposition. 
Experiments suggest that the performance of quantum(-inspired) annealers is closely related to the problem settings. Decomposition techniques extend the scalability of quantum(-inspired) annealers. However, getting promising solutions is still challenging for computing devices with limited capability. Through experiments and analysis, we highlighted the research directions that can improve the utility and scalability of quantum(-inspired) annealers.

\section*{Acknowledgments}
We acknowledge the funding support from Agency for Science, Technology and Research (\#21709).

%

\bibliographystyle{unsrtnat}
\bibliography{reference}

\end{document}


\appendix
\appendixpage
\addappheadtotoc

\section{Datasets}

All datasets in this paper are available from \url{https://github.com/ianmalcolm/annealer-benchmark}. We are adding README.md, ground truth and python scripts to improve the usability of the repo.

\section{Backends}
\label{sec:appendix_backends}

A brief introduction of the backends are as follows.

D-Wave is a world-leading company that designs and builds quantum annealers. The Quantum Processing Unit (QPU) in the D-Wave's quantum annealing system is a lattice of interconnected qubits. Each qubit is made of a superconducting loop. D-Wave released the D-Wave 2000Q system in 2017, the QPU in which employs a Chimera graph architecture \cite{vert2019limitations}, equipped with 2048 qubits and 6016 couplers. Due to the variation in the manufacturing, the actual working resource in a quantum annealer varies.

\begin{figure}[htb]
\centering
\resizebox{0.66\columnwidth}{!}{%
\input{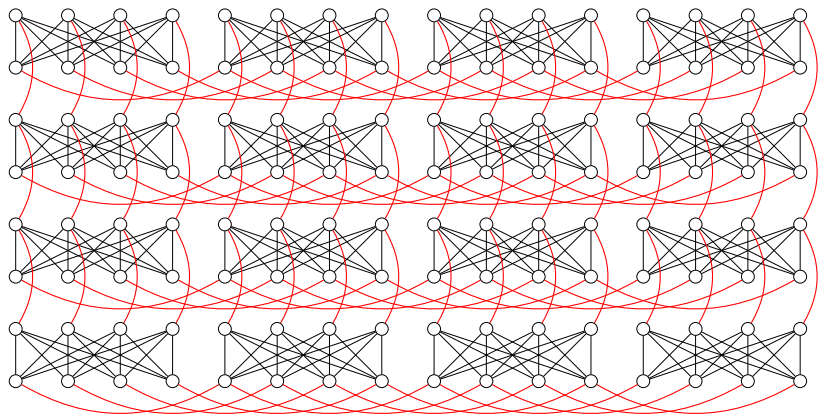}
}
\caption{Chimera graph architecture \cite{dwave_arch}. Circles are qubits. A black straight lines represents an internal coupler. The red curved lines are the external couplers.}
\label{fig:chimera_graph}
\end{figure}

Fig.\ref{fig:chimera_graph} shows a part of the \emph{Chimera} graph architecture. Each black circle represents a qubit. Every eight qubits form a complete bipartite graph. The connections within a bipartite graph are internal couplers. The connections across bipartite graphs are external couplers. A connection between two qubits corresponds to a quadratic term of two decision variables in a QUBO problem.

D-Wave released the Advantage system in 2019. One of the major differences between these two systems is the graph architecture of their QPUs. The Advantage system QPU employs a \emph{Pegasus} \cite{dwave_arch} graph architecture, which has 5640 qubits and 40,484 couplers. Generally speaking, we expect a quantum annealer to solve larger and denser problems if it has more qubits and couplers.

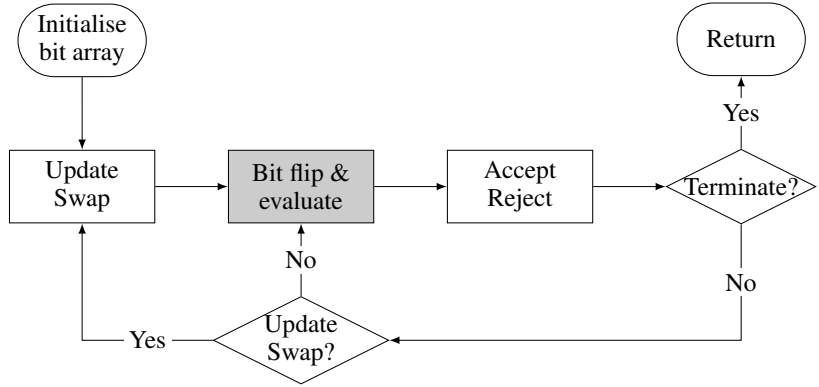
\begin{figure}[htb]
\centering
\resizebox{0.66\columnwidth}{!}{%
\begin{tikzpicture}

\node[draw,
    rounded rectangle,
    minimum width=2cm,
    minimum height=1cm,
    align=center,
] (bstart) {Initialise \\ bit array};
    
\node[draw,
    below=of bstart,
    minimum width=2cm,
    minimum height=1cm,
    align=center
] (butemp) {Update\\Swap};

\node[draw,
    right=of butemp,
    minimum width=2cm,
    minimum height=1cm,
    align=center,
    fill={rgb:black,1;white,4},
] (bpropose) {Bit flip \& \\ evaluate};

\node[draw,
    right=of bpropose,
    minimum width=2cm,
    minimum height=1cm,
    align=center
] (bselect) {Accept \\ Reject};

\node[draw,
    diamond,
    aspect=2,
    right=of bselect,
    minimum width=2cm,
    inner sep=0,
    align=center
] (bterminate) {Terminate?};

\node[draw,
    diamond,
    aspect=2,
    below=of bpropose,
    minimum width=2cm,
    inner sep=0,
    align=center
] (butempq) {Update\\Swap?};

\node[draw,
    rounded rectangle,
    above=of bterminate,
    minimum width=2cm,
    minimum height=1cm,
] (breturn) {Return};

\draw[-latex] (bstart)     -- (butemp);
\draw[-latex] (butemp)     -- (bpropose);
\draw[-latex] (bpropose)   -- (bselect);
\draw[-latex] (bselect)    -- (bterminate);
\draw[-latex] (bterminate) -- (breturn)  node[pos=0.5, fill=white,inner sep=2pt] {Yes};
\draw[-latex] (bterminate) |- (butempq)  node[pos=0.25,fill=white,inner sep=2pt] {No};
\draw[-latex] (butempq)    -- (bpropose) node[pos=0.5, fill=white,inner sep=2pt] {No};
\draw[-latex] (butempq)    -| (butemp)   node[pos=0.25,fill=white,inner sep=2pt] {Yes};

\end{tikzpicture}
}
\caption{Fujitsu Digital Annealer Workflow Diagram}
\label{fig:da}
\end{figure}

The Fujitsu \textbf{Digital Annealer} (DA) \cite{tsukamoto2017accelerator} is a hardware implementation of an enhanced variant of the Simulated Annealing (SA) algorithm. The enhancements include but are limited to parallel-trial, dynamic offset and Parallel Tempering (PT) with Isoenergetic Cluster Moves (ICM) \cite{zhu2020borealis}. The workflow of DA is shown in Fig.\ref{fig:da} \cite{aramon2019physics}.

DA employs Metropolis-Hastings update with Parallel Tempering. In each cycle, it adjusts the temperature, proposes bit flips, and accepts/rejects the proposals according to the temperature and periodically swaps configurations between systems. Apart from the sophisticated techniques in proposing and selecting updates, the competence of DA also comes from the efficient hardware implementation that exploits the parallelism in the algorithm. We include DA in the evaluation.

We include the D-Wave QPU of Chimera and Pegasus architecture, and Fujitsu Digital Annealer in the evaluation. We also include \textbf{Simulated Annealing} (SA) by D-Wave, which is an open-source CPU implementation based on the Metropolis-Hastings algorithm. We include QBSolv from D-Wave, which is a heuristic hybrid solver that incorporates classical computer and quantum annealer. We configure QBSolv to use Tabu search and use it as a software baseline in the experiments. We also include \textbf{Gurobi 9.1}, which represents the SOTA commercial optimisation solver. When we generate a synthetic dataset for the experiments, we use Gurobi to find the approximated optimal solutions for reference.

\begin{table*}[tb]
\centering
\caption{Solver settings}
\label{tab:solver_settings}
\begin{threeparttable}
\begin{tabular}{|p{0.15\linewidth}|p{0.2\linewidth}|p{0.1\linewidth}|p{0.20\linewidth}|p{0.1\linewidth}|p{0.1\linewidth}|}
\hline
\textbf{Solver}            & \textbf{Chimera} & \textbf{Pegasus} & \textbf{DA}  & \textbf{QBSolv} & \textbf{SA} \\ \hline
\textbf{Implementation} &
  D-Wave   DW\_2000Q\_6\tnote{1} &
  D-Wave   Advantage System 1.1\tnote{2} &
  Fujitsu   Quantum-inspired Computing Digital Annealer On-Premises Service &
  D-Wave Ocean Software\tnote{3} &
  D-Wave Ocean Software\tnote{4} \\ \hline
\textbf{Number of samples} & 100              & 100              & 128 replicas & 20              & 100         \\ \hline
\textbf{Annealing process} &
  Default annealing schedule\tnote{5} with annealing time ranging from 2 to 2000 $\mu s$, annealing offsets disabled, 1000us programming thermalisation and 0s readout thermalisation.\tnote{6} &
  Same as Chimera &
  Parallel tempering, with 128 replicas. The number of iterations ranges from $10^5$ to $10^9$. Offset increase rate is disabled \tnote{7} &
  Default solver Tabu search &
  Problem specific and geometry beta scheduling. Total number of sweeps ranges from $10^2$ to $10^4$ \tnote{8} \\ \hline
\textbf{Precision of QUBO coefficient} &
  Rescale fp32 and operates in anologue mode &
  Same as Chimera &
  Cast fp64 to 64/76-bit integer for quadratic/linear term for problem size below 4096 qubits, or to 16/76-bit integer for problem size over 4096 qubits &
  fp32 &
  fp32 \\ \hline
\textbf{Postprocess}       & Disabled         & Disabled         & Disabled     & N.A.            & N.A.        \\ \hline
\textbf{Random seed}       & N.A.             & N.A.             & N.A.         & 1234            & 1234        \\ \hline
\textbf{Initial state}     & None             & None             & None         & None            & None        \\ \hline
\end{tabular}
\begin{tablenotes}
\item[1] AWS Braket Device ARN: \url{arn:aws:braket:::device/qpu/d-wave/DW_2000Q_6}  \\
\item[2] AWS Braket Device ARN: \url{arn:aws:braket:::device/qpu/d-wave/Advantage_system1}  \\
\item[3] \url{https://github.com/dwavesystems/qbsolv}  \\
\item[4] \url{https://github.com/dwavesystems/dwave-neal}  \\
\item[5] QPU-specific anneal schedules documents: \url{https://docs.dwavesys.com/docs/latest/doc\_physical\_properties.html\#doc-qpu-characteristics}  \\
\item[6] Complete D-Wave QPU parameters: \url{https://docs.dwavesys.com/docs/latest/c_solver_parameters.html}  \\
\item[7] Complete DA annealing parameters: \url{https://portal.aispf.global.fujitsu.com/apidoc/da/jp/api-ref/da-qubo-en.html}  \\
\item[8] Temperature scheduling for SA \url{https://github.com/dwavesystems/dwave-neal/blob/a11e477c3a6b3585d75ee8b58be75a4127d0c17c/neal/sampler.py#L281}  \\
\end{tablenotes}
\end{threeparttable}
\end{table*}

The settings of the solvers are presented in Table \ref{tab:solver_settings}. We set the \emph{number of samples} (a.k.a number of shots, solutions, or readout) per task \#samples=100 for Chimera, Pegasus, SA. The annealing process is crucial to the quality of solutions. We are following the default annealing schedule provided by D-Wave, which empirically works on most of problems.  We carry out the experiments with different \emph{annealing time}, which determines the duration of annealing schedule, to understand if a problem is sensitive to it. For Chimera and Pegasus, we change the annealing time from 2 to 2000 $\mu s$. This range almost cover the full range allowed by the devices, which is 1 to 2000 $\mu s$.

In the implementation of SA, number of sweeps, or \#sweeps, is a hyper-parameter that is similar to the annealing time in D-Wave annealer. we change \#sweeps from $10^2$ to $10^4$. $10^3$ is the default setting, which lead to 1 sweep for each annealing temperature beta. \#sweeps higher than $10^3$ causes multiple sweeps per beta, lower causes skip of some beta.

For DA, we use Parallel Tempering mode with \#replicas=128. The \#replicas=128 means DA runs 128 copies of a problem instance in parallel, each of the instances is randomly initialized, at different temperatures. At the end of annealing, we get 128 samples from DA. This is different from the rest of the backends, where a problem instance is sampled repeatedly, in a sequential way. The \#replicas corresponds to the parallelism of DA. The upper limit of \#replicas is 128. We fix \#replicas=128 to make sure DA has its advantage in the comparison. The number of iterations, or \#iterations, is a hyper-parameter similar to the annealing time in D-Wave annealer. We are varying \#iterations from $10^5$ to $10^9$. $10^5$ is the lower limit of \#iterations. Experimental results suggest that \#iterations higher than $10^9$ does not make big difference in terms of the quality of solutions. Some problems are simple enough to be solved with fewer \#iterations. Having too large \#iterations puts DA in disadvantage in terms of speed.

We use randomly generated initial state for all backends. For SA and QBSolv, we use random seed=1234 to improve the reproducibility.

In the main text, we only include the results with the best hyper-parameters in the comparison. By saying best, we refer to the following rules: If a hyper-parameter outperforms others in terms of objective energy, it is the best. If all hyper-parameters are similar energy-wise, then the one with lowest time cost is the best. An optimal hyper-parameter on one problem setting is not necessarily optimal on other problems settings. 

\subsection{Gurobi}

The problem formulation for Gurobi goes like this. For max-cut problem, we formulate its QUBO form and solve it on Gurobi. For MVC and QAP, we formulate the objective as QUBO form and implement the problem constraints using Gurobi constraints. QUBO form is not in favour of Gurobi. There are other smart classical encoding methods for Gurobi. For QAP for example, instead of using one-hot encoding, one can use categorical variables to represent locations and factories and solve the problem more efficiently. But we choose QUBO form on Gubori because it provides a better comparability between annealing-based binary solvers and classical heuristic solvers.

The solving time of Gurobi presented in the comparison plot of the main text is the time for traversing the whole solution space, if not terminated by a pre-defined timeout threshold. It is possible that a global optima is discovered way before the solution space is traversed. Gurobi program maintains a ``best-so-far'' solution during the search. We plot time-to-solution for Gurobi, to understand how fast Gurobi approaches global optimal solution.

The Gurobi program is running on a server, which is equipped with an Intel Core i9-10900X CPU, 128GB DDR4 memory and 128GB HDD swap memory.   In our experimental settings, Gurobi always instantiate 20 threads and occupies almost all CPU time slices, DDR4 memory and swap memory, until it finishes the search. We pay efforts to minimise the interference from irrelevant processes on the same server, because some problem instances in this paper are really pushing the server toward limit and cause Gurobi to terminate with ``OUT\_OF\_MEMORY'' errors.

\clearpage

\section{max-cut}

\subsection{Pegasus-like max-cut problems}
\label{sec:appendix_max-cut_pegasus}

\textbf{Problem generation} We start the problem generation by first checking the working status of D-Wave QPU. The actual architecture of a QPU could be different from the description in the official documentation. This is because there is variation in manufacturing that puts some resources of the QPU in non-working conditions. In our case, we are accessing D-Wave from AWS Braket \cite{gonzalez2021cloud}. There are 5436 qubits and 37440 couplers available on D-Wave Advantage System 1.1. By accessing DWaveSampler.target\_structure, one can find the graph, $G_\textrm{pegasus} = (V_\textrm{pegasus}, E_\textrm{pegasus})$ representing the architecture of the target QPU. 

Based on $G_\textrm{pegasus}$, we generate random graphs with a specified number of nodes. Given $G_\textrm{pegasus}$ and a specific number of nodes $n$, we randomly choose $\left | V_\textrm{pegasus} \right |-n$ nodes from $G_\textrm{pegasus}$, remove these nodes and corresponding edges. The randomness in the graph generation is controlled by a random seed for good reproducibility. The resulting graph is a sub-graph of $G_\textrm{pegasus}$ with $n$ nodes. We can map this graph onto QPU without a resort to minor embedding effort.

In the pegasus-like max-cut experiment, we have ten random graphs of such. $\left | V \right |$ of these graphs ranges from 543 to 5430.

At the late stage of work, AWS Braket D-Wave Advantage System 1.1 is retired, and is replaced with 4.1 and 6.1. Although they follow exactly the same architecture, the generated problems based on 1.1 are not applicable to 4.1 and 6.1, because some qubits or couplers that were available on 1.1 are not necessarily in working condition on 4.1 or 6.1.

\begin{figure}[htb]
    \centering
    \begin{subfigure}[b]{0.45\linewidth}
        \centering
    	\includegraphics[width=\textwidth]{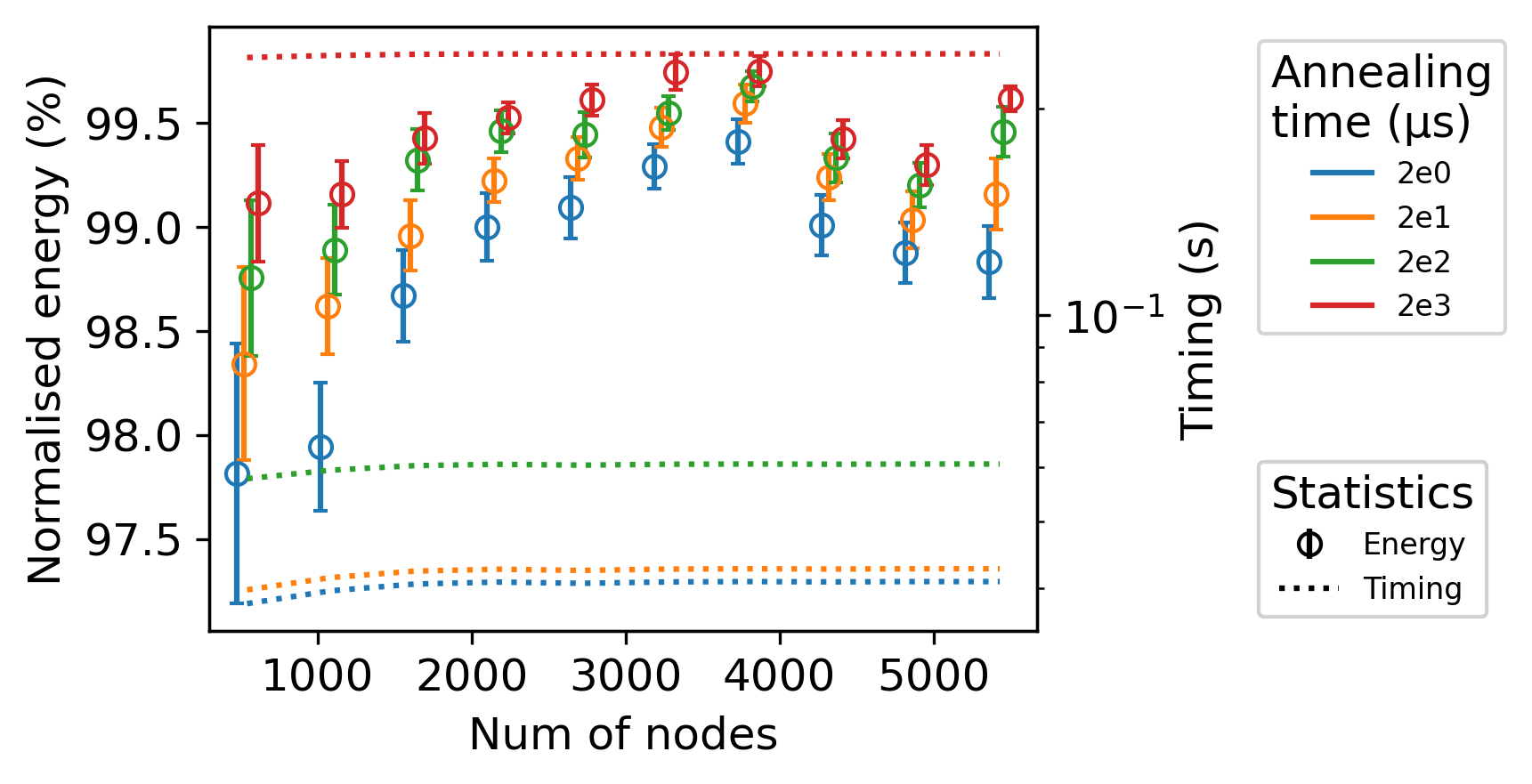}
    	\caption{Pegasus}
    	\label{fig:app_max-cut_pegasus_pegasus}
    \end{subfigure}
    \begin{subfigure}[b]{0.45\linewidth}
        \centering
    	\includegraphics[width=\textwidth]{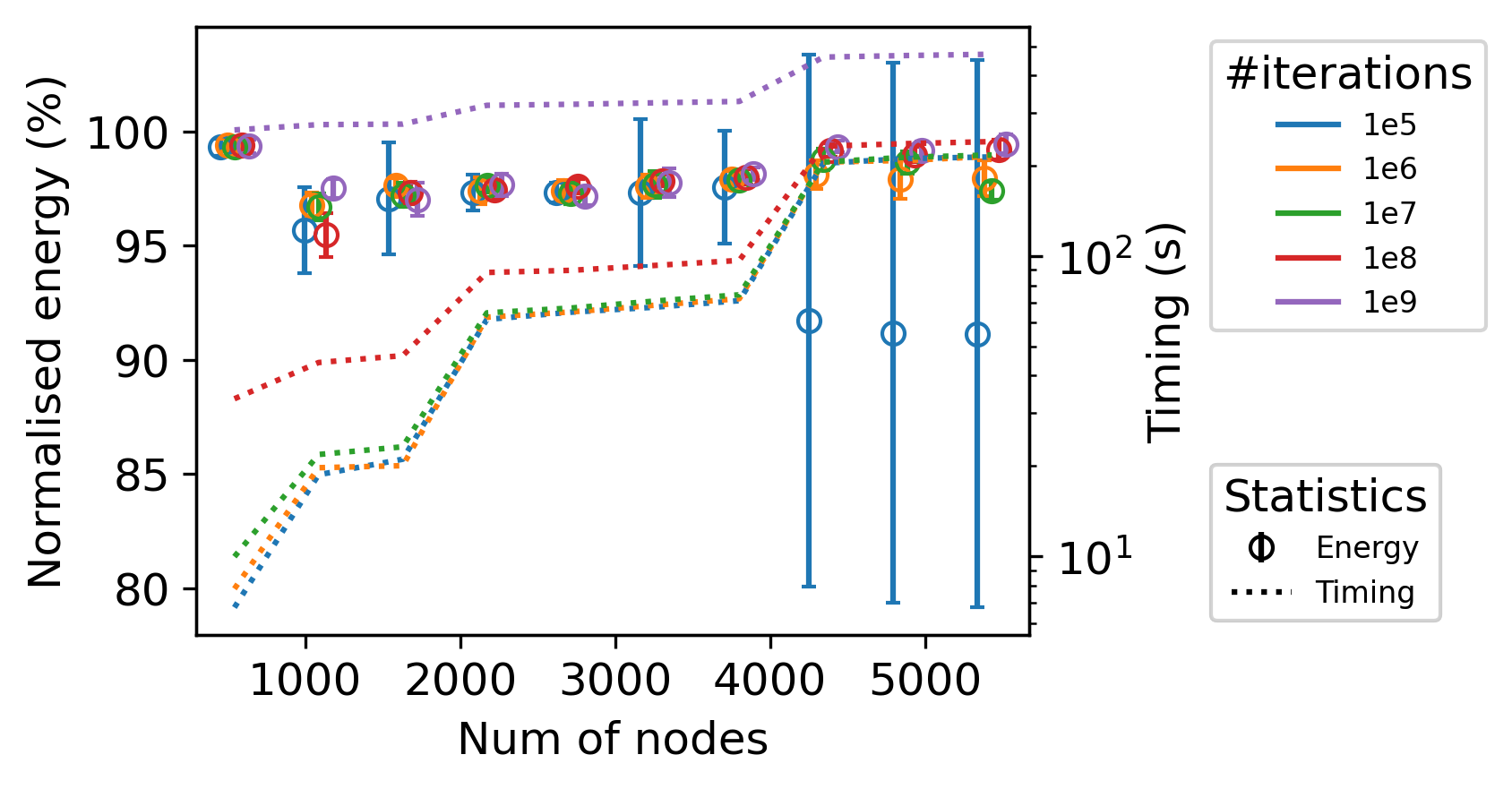}
    	\caption{DA}
    	\label{fig:app_max-cut_pegasus_da}
    \end{subfigure}
    \begin{subfigure}[b]{0.45\linewidth}
        \centering
    	\includegraphics[width=\textwidth]{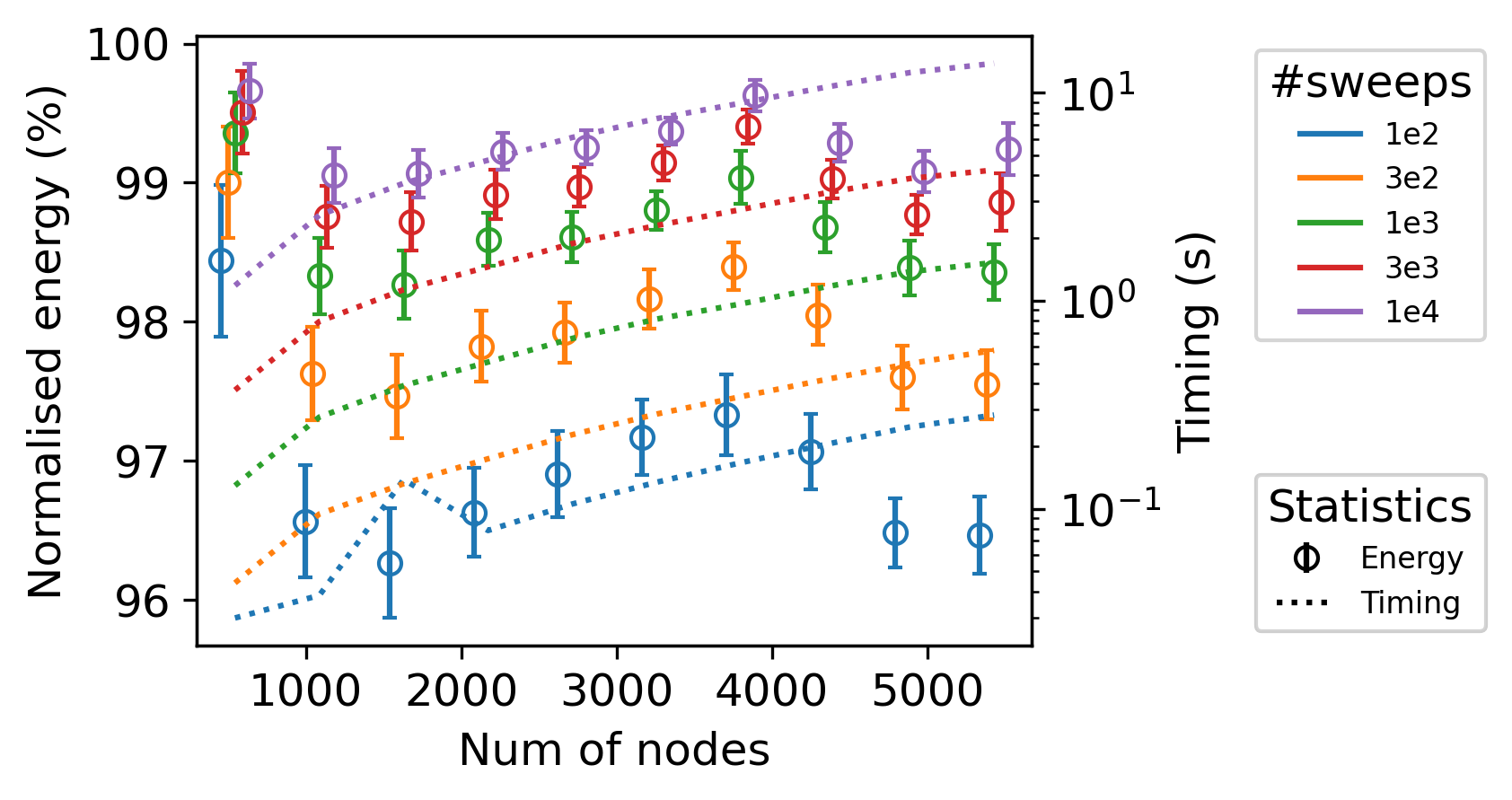}
    	\caption{SA}
    	\label{fig:app_max-cut_pegasus_sa}
    \end{subfigure}
    \caption{max-cut, Pegasus-like problems. The error bar of energy and timing. X axis is the number of nodes in a graph. The primary Y axis is energy normalised to that of Gurobi. The secondary Y axis corresponds to dotted curves and represents timing in seconds, in log scale. The circle of an error bar represents mean energy, while the top and bottom represent $\pm$ standard deviation. We shift the error bars a little bit to improve readability and avoid overlap. Since max-cut is a maximisation problem, higher energy is better.}
    \label{fig:app_max-cut_pegasus_hp}
\end{figure}

Figure \ref{fig:app_max-cut_pegasus_hp} shows the performance of Pegasus, DA and SA on max-cut Pegasus-like problems. For D-Wave Pegasus, figure \ref{fig:app_max-cut_pegasus_pegasus} suggest that annealing time of 2000 $\mu s$ achieves higher energy and outperforms shorter annealing time. Adiabatic theory \cite{born1928beweis,tong2005quantitative} suggests that longer annealing time will ensure a slower adiabatic process and a better result. In practice \cite{mcgeoch2014adiabatic,hauke2020perspectives}, D-Wave annealers operate in non-adiabatic mode. An open quantum system de-coheres and suffers from thermal noises, which sometimes could also improve the performance of a quantum system. In our case, we cannot distinguish if the improvement comes from the quantum mechanics or the thermal noises. We include annealing time of 2000 $\mu s$ for D-Wave Pegasus in the main text for comparison between solvers.

Figure \ref{fig:app_max-cut_pegasus_da} shows the performance of DA on the Pegasus-like max-cut problem. The runs with iterations larger than $10^6$ are relatively consistent in terms of the variance of energy. Further increase of the iterations does not make prominent difference. We present the results from \#iterations=$10^6$ in the main text and compare it with other solvers. DA with $10^5$ iteration on large problem size leads to $\pm 10\%$ variance in energy. We are not going investigate the reason, since in-depth analysis of a solver and its underlying algorithm is out of the scope of this work.

Figure \ref{fig:app_max-cut_pegasus_sa} shows the performance of SA on the Pegasus-like max-cut problem. The performance are relatively consistent on problems of different size. But higher number of sweeps provides significantly better results. We include the results of \#sweeps=$10^4$ in the main text for comparison.

\begin{figure}[htb]
    \centering
	\includegraphics[width=0.45\linewidth]{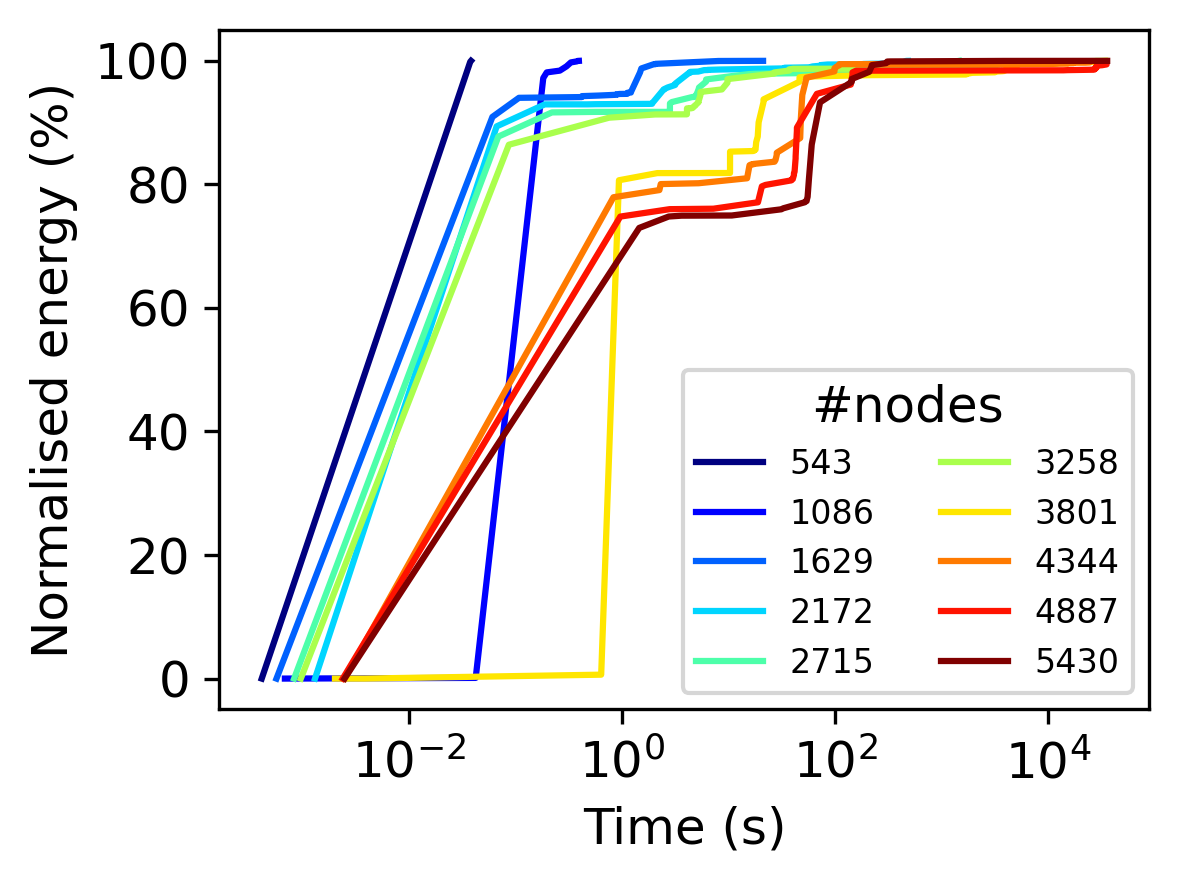}
	\caption{Gurobi on max-cut Pegasus-like problem. Time to solution. X axis is the time in seconds in log scale. Y axis is the energy normalised to optimal solution (found by Gurobi)}
	\label{fig:app_max-cut_pegasus_gurobi}
\end{figure}

Figure \ref{fig:app_max-cut_pegasus_gurobi} shows the time-to-solution of Gurobi on max-cut Pegasus-like problems. Although some large problem instances, e.g. the one with 5430 nodes, take quite a long time before the termination of the optimisation program (because of reaching pre-configured timeout), Gurobi can find very promising solutions within roughly 100 seconds.

Please note that a point in the figure marks the time Gurobi spends to discover the promising solution at that moment. Gurobi is not sure if the discovered the solution is globally optimal. It has to spend some more time to traverse the whole solution space with some smart pruning techniques to make sure it is not possible to find any better solutions. The time for traversing the solution space is reported in Table \ref{tab:max-cut_pegasus_gurobi_traverse}.

\begin{table}[htb]
\centering
\caption{Gurobi time-to-traverse on Pegasus-like max-cut problems}
\label{tab:max-cut_pegasus_gurobi_traverse}
\begin{tabular}{|l|l|l|l|}
\hline
\#nodes & Time (s) & \#nodes & Time (s) \\ \hline
543     & 0.041288 & 3258    & 36000.05 \\ \hline
1086    & 0.25132  & 3801    & 36000.03 \\ \hline
1629    & 14.66663 & 4344    & 36000.41 \\ \hline
2172    & 430.4116 & 4887    & 36000.05 \\ \hline
2715    & 1131.357 & 5430    & 36000.05 \\ \hline
\end{tabular}
\end{table}

To collect the intermediate results of Gurobi, we have to pass a callback to Gurobi program, which introduce extra time cost. For a problem instance that last for 36000 seconds, the time spent on the callback is over 1000 seconds.

\clearpage

\subsection{Chimera-like max-cut problems}
\label{sec:appendix_max-cut_chimera}

The chimera-like graphs are randomly generated, following the D-Wave Chimera architecture. The generation is similar to that of the Pegasus-like max-cut problem. Please refer to section \ref{sec:appendix_max-cut_pegasus} for the details of the problem generation.

\begin{figure}[htb]
    \centering
    \begin{subfigure}[b]{0.45\linewidth}
        \centering
        \includegraphics[width=\linewidth]{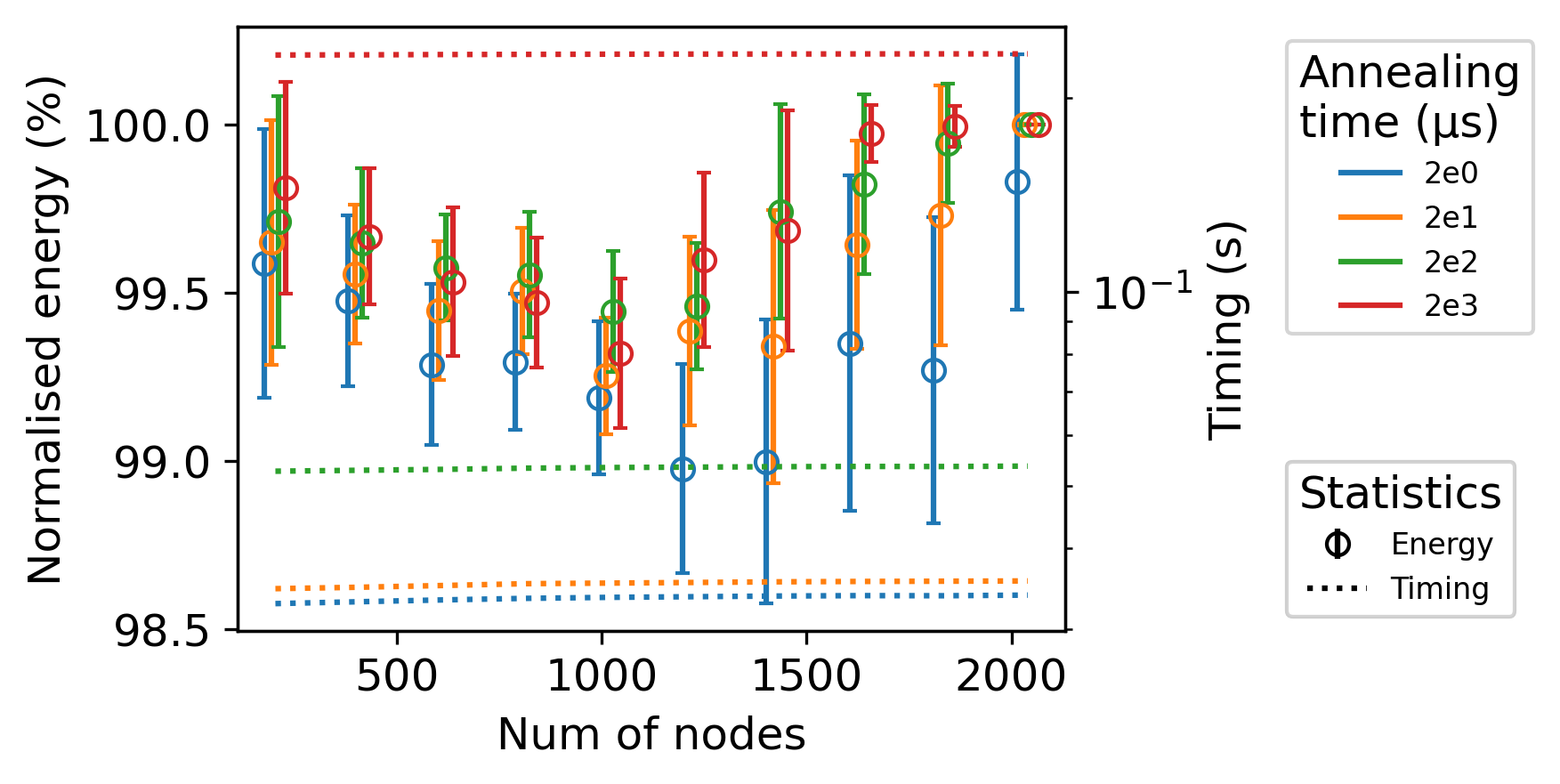}
        \caption{Chimera}
        \label{fig:app_max-cut_chimera_chimera}
    \end{subfigure}
    \begin{subfigure}[b]{0.45\linewidth}
        \centering
        \includegraphics[width=\linewidth]{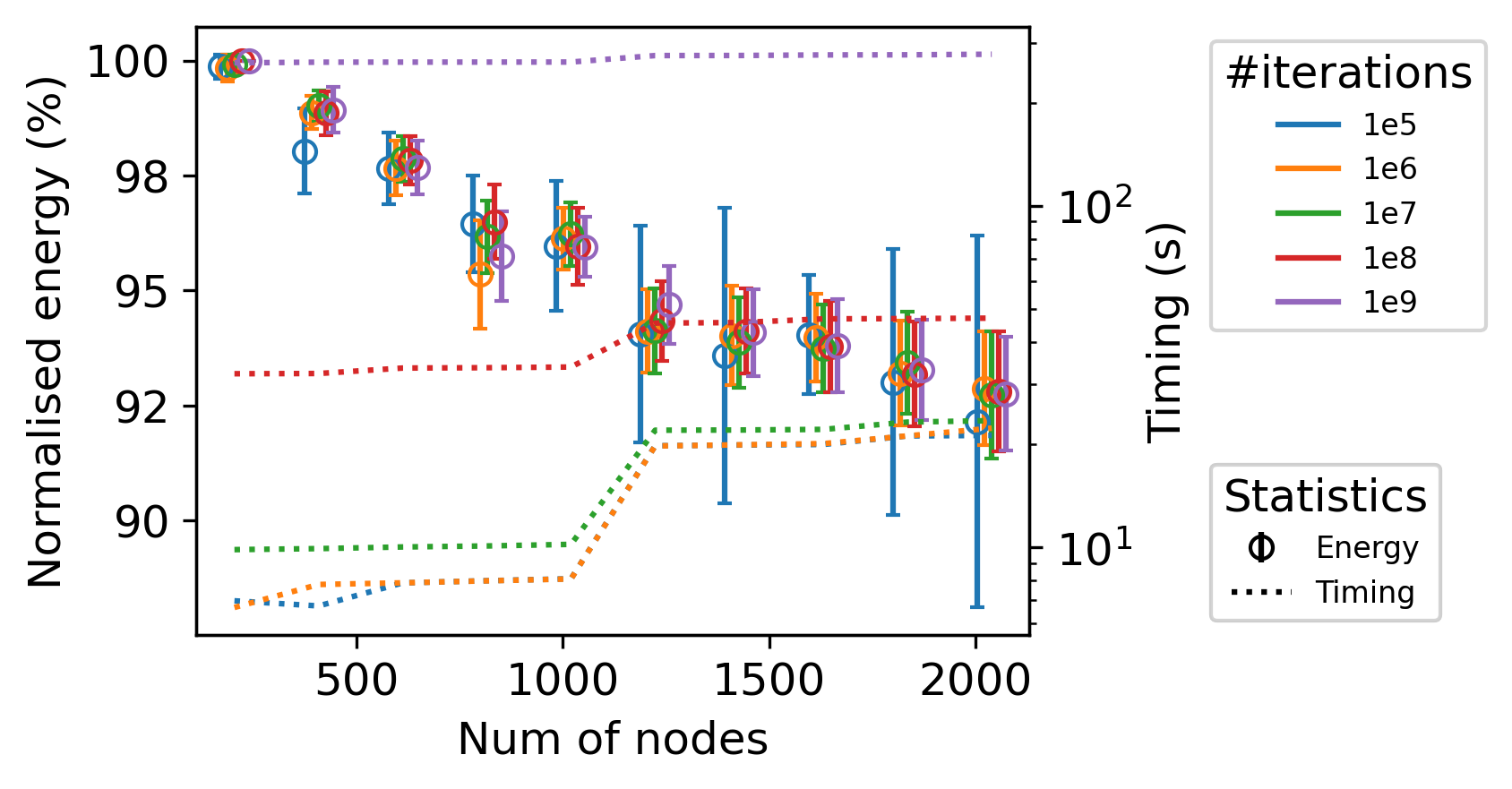}
        \caption{DA}
        \label{fig:app_max-cut_chimera_da}
    \end{subfigure}
    \begin{subfigure}[b]{0.45\linewidth}
        \centering
        \includegraphics[width=\linewidth]{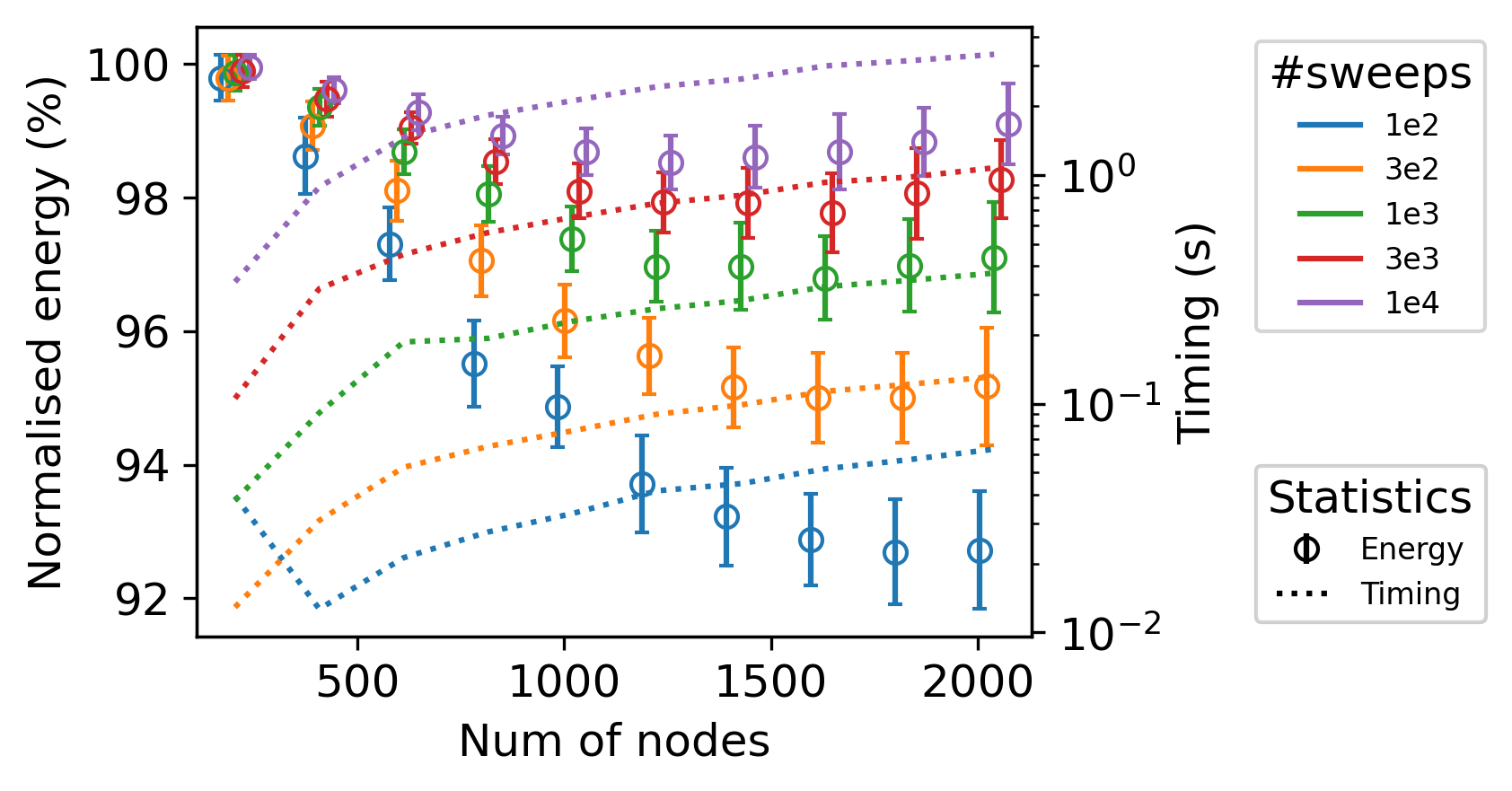}
        \caption{SA}
        \label{fig:app_max-cut_chimera_sa}
    \end{subfigure}
    \caption{max-cut, Chimera-like problems. The error bar of energy and timing. X axis is the number of nodes in a graph. The primary Y axis is energy normalised to that of Gurobi. The secondary Y axis corresponds to dotted curves and represents timing in seconds, in log scale. The circle of an error bar represents mean energy, while the top and bottom represent $\pm$ standard deviation. We shift the error bars a little bit to improve readability and avoid overlap. Since max-cut is a maximisation problem, higher energy is better.}
    \label{fig:app_max-cut_chimera_hp}
\end{figure}

Figure \ref{fig:app_max-cut_chimera_hp} shows the performance of D-Wave Chimera and the Digital Annealer on the max-cut Chimera-like problems. Through figure \ref{fig:app_max-cut_chimera_chimera} we understand that a longer annealing time, i.e. 2000 $\mu s$, leads to the best performance in energy. But 2000 $\mu s$ does not give Chimera prominent advantage over that of 200 $\mu s$. This result coincides with our observation with the Pegasus architecture. The performance of quantum and classical annealers in terms of energy on max-cut Chimera-like problems are generally improved given higher hyper-parameters. But for DA, the room of improvement is limited.

For Chimera, we only include the results of $200\mu s$ for comparison between solvers. For DA and SA We only include the results of \#iterations=$10^6$ and \#sweeps=$10^4$ for comparison between solvers. 
\begin{figure}[htb]
     \centering
     \begin{subfigure}[b]{0.45\linewidth}
        \centering
    	\includegraphics[width=\textwidth]{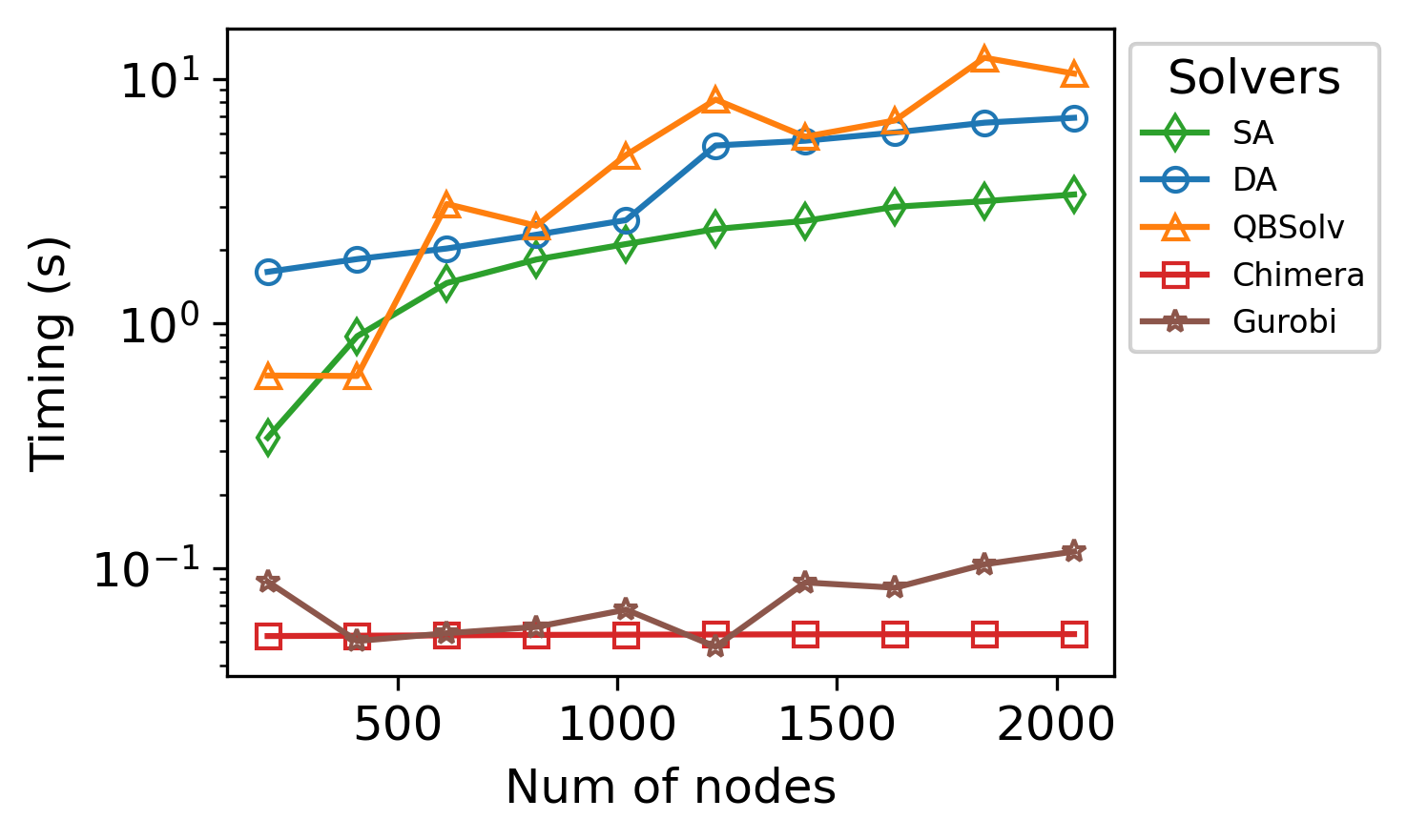}
    	\caption{Timing}
    	\label{fig:app_max-cut_chimera_timing}
     \end{subfigure}
     \begin{subfigure}[b]{0.45\linewidth}
        \centering
    	\includegraphics[width=\textwidth]{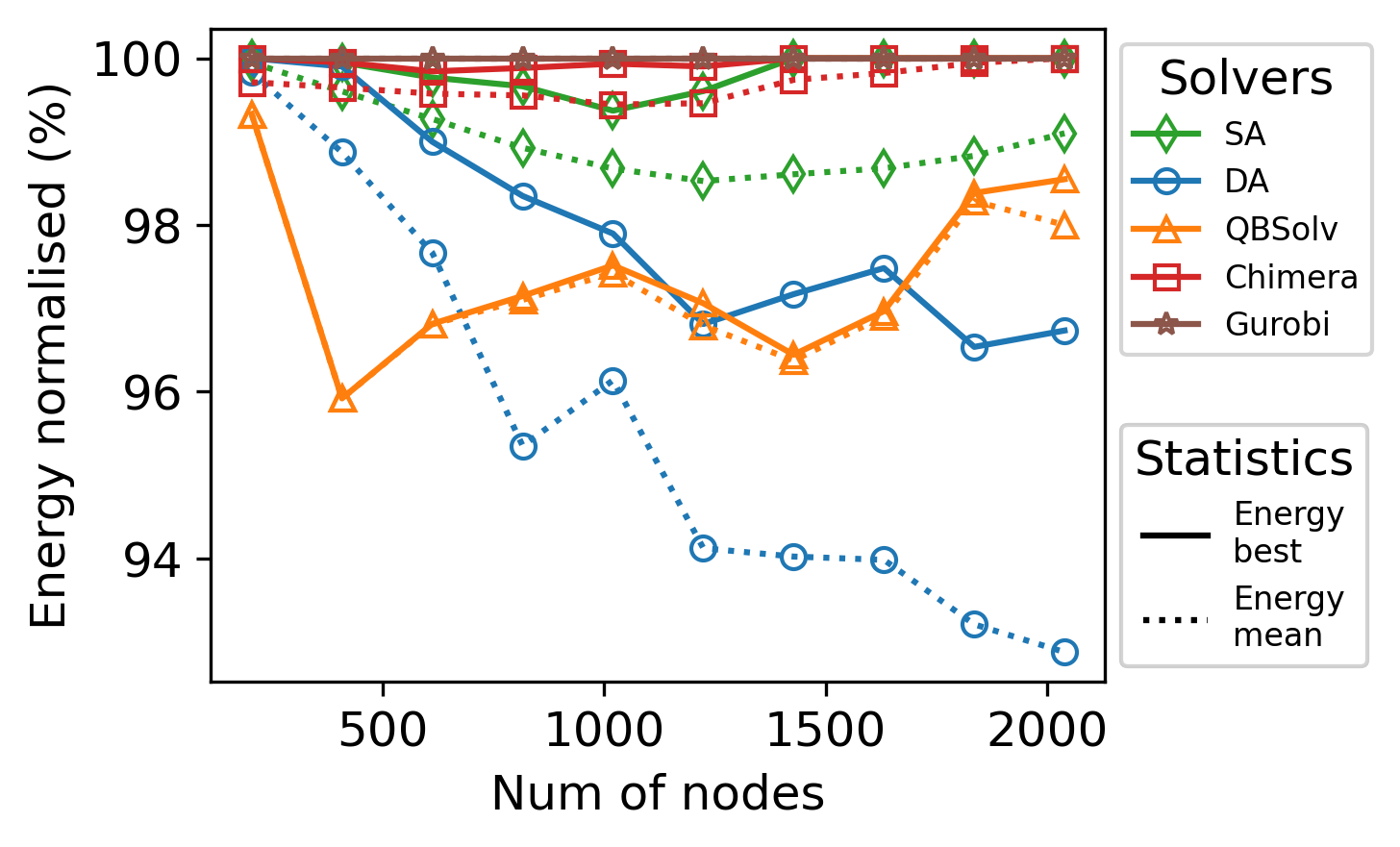}
    	\caption{Normalised energy}
    	\label{fig:app_max-cut_chimera_energy}
     \end{subfigure}
    \caption{Max-cut, Chimera-like problems. Comparison between solvers. a) X axis is the number of nodes. Y axis is the timing of the solvers, in log scale. b) X axis is the number nodes. Y axis is the energy normalised to that of Gurobi. We include mean energy (dotted lines) and best energy (solid lines) to represent the statistics of energy.}
    \label{fig:app_max-cut_chimera_performance}
\end{figure}

Figure \ref{fig:app_max-cut_chimera_performance} shows the performance of different solvers on max-cut Chimera-like problems. From figure \ref{fig:app_max-cut_chimera_timing} we understand that Chimera is the fastest solver for the most of the time. Gurobi is faster than classical annealing-based solvers and a little bit slower than Chimera. In terms of energy, according to figure \ref{fig:app_max-cut_chimera_energy} Chimera is the most promising solver among other annealing-based solvers. The gap between Chimera and DA is widening, as the problem size increases.

\begin{figure}[htb]
    \centering
	\includegraphics[width=0.45\linewidth]{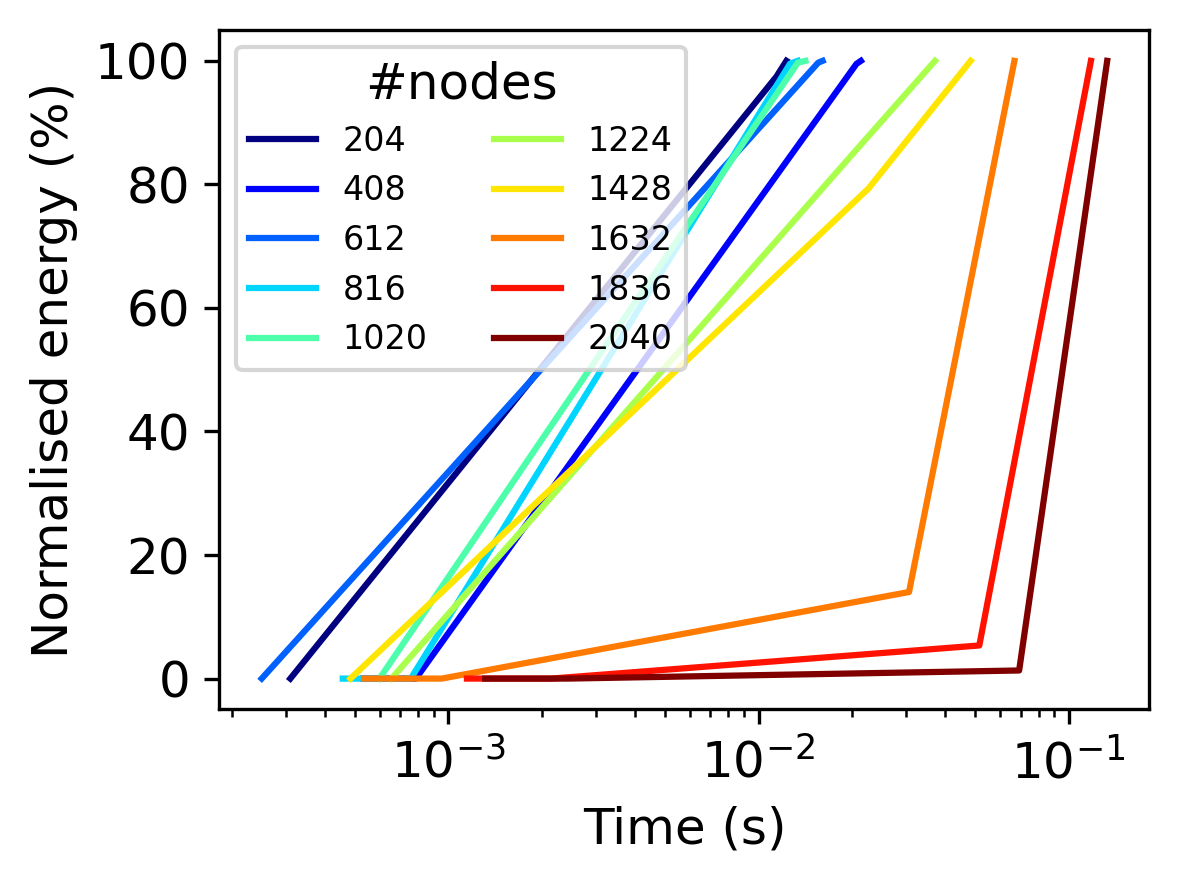}
	\caption{Gurobi on max-cut Chimera-like problems. Time to solution. X axis is the time in seconds in log scale. Y axis is the energy normalised to optimal solution (found by Gurobi)}
	\label{fig:app_max-cut_chimera_gurobi}
\end{figure}

Figure \ref{fig:app_max-cut_chimera_gurobi} shows the performance of Gurobi on max-cut Chimera-like problems. In this figure, most of the promising solutions can be found within 0.1 seconds. The time-to-traverse information is listed in Table \ref{tab:max-cut_chimera_gurobi_traverse}

\begin{table}[htb]
\centering
\caption{Gurobi time-to-traverse on Chimera-like max-cut problems}
\label{tab:max-cut_chimera_gurobi_traverse}
\begin{tabular}{|l|l|l|l|}
\hline
\#nodes & Time (s)    & \#nodes & Time (s)    \\ \hline
204     & 0.088003159 & 1224    & 0.047677994 \\ \hline
408     & 0.050300837 & 1428    & 0.087327957 \\ \hline
612     & 0.054126978 & 1632    & 0.083067894 \\ \hline
816     & 0.057450056 & 1836    & 0.103567839 \\ \hline
1020    & 0.067600012 & 2040    & 0.116798878 \\ \hline
\end{tabular}
\end{table}

\clearpage

\subsection{Connectivity-varied max-cut problems}
\label{sec:appendix_max-cut_connectivity}

\textbf{Problem generation} We use the ``dense\_gnm\_random\_graph'' method from networkx, i.e., the open-source python library, to generate random graphs. We set $n=145$, as this is the largest complete graph that can be mapped onto a Pegasus QPU. We sweep the averaged degree to create graphs with different connectivity. For example, to generate a graph with an averaged degree of 100, we set $m=\mathrm{floor}(145\times 100 \div 2)$, which represents the total number of edges in the graph.

\begin{figure}[htb]
    \centering
    \begin{subfigure}[b]{0.45\linewidth}
        \centering
    	\includegraphics[width=\textwidth]{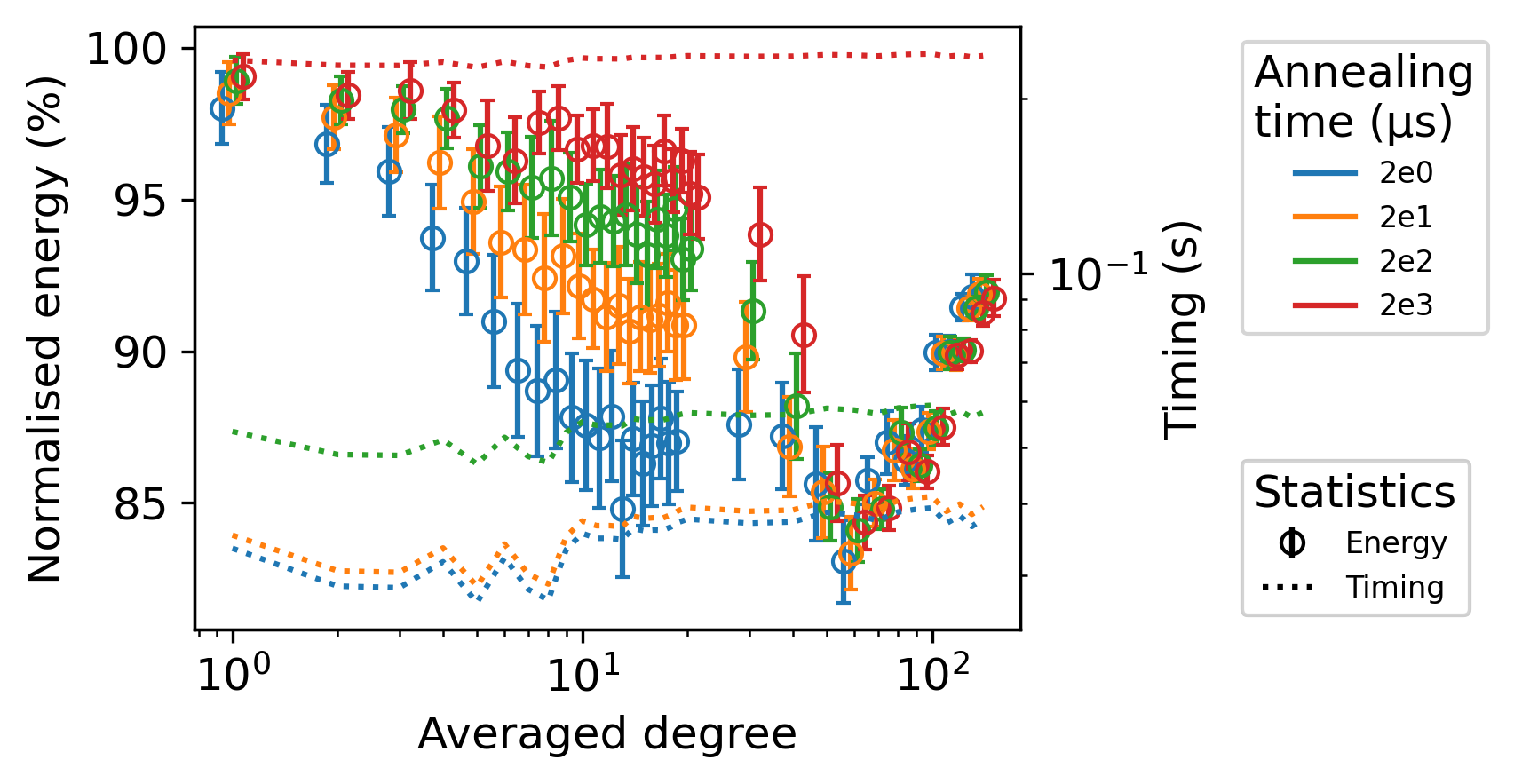}
    	\caption{Pegasus}
    	\label{fig:app_max-cut_connectivity_pegasus}
    \end{subfigure}
    \begin{subfigure}[b]{0.45\linewidth}
        \centering
        \includegraphics[width=\textwidth]{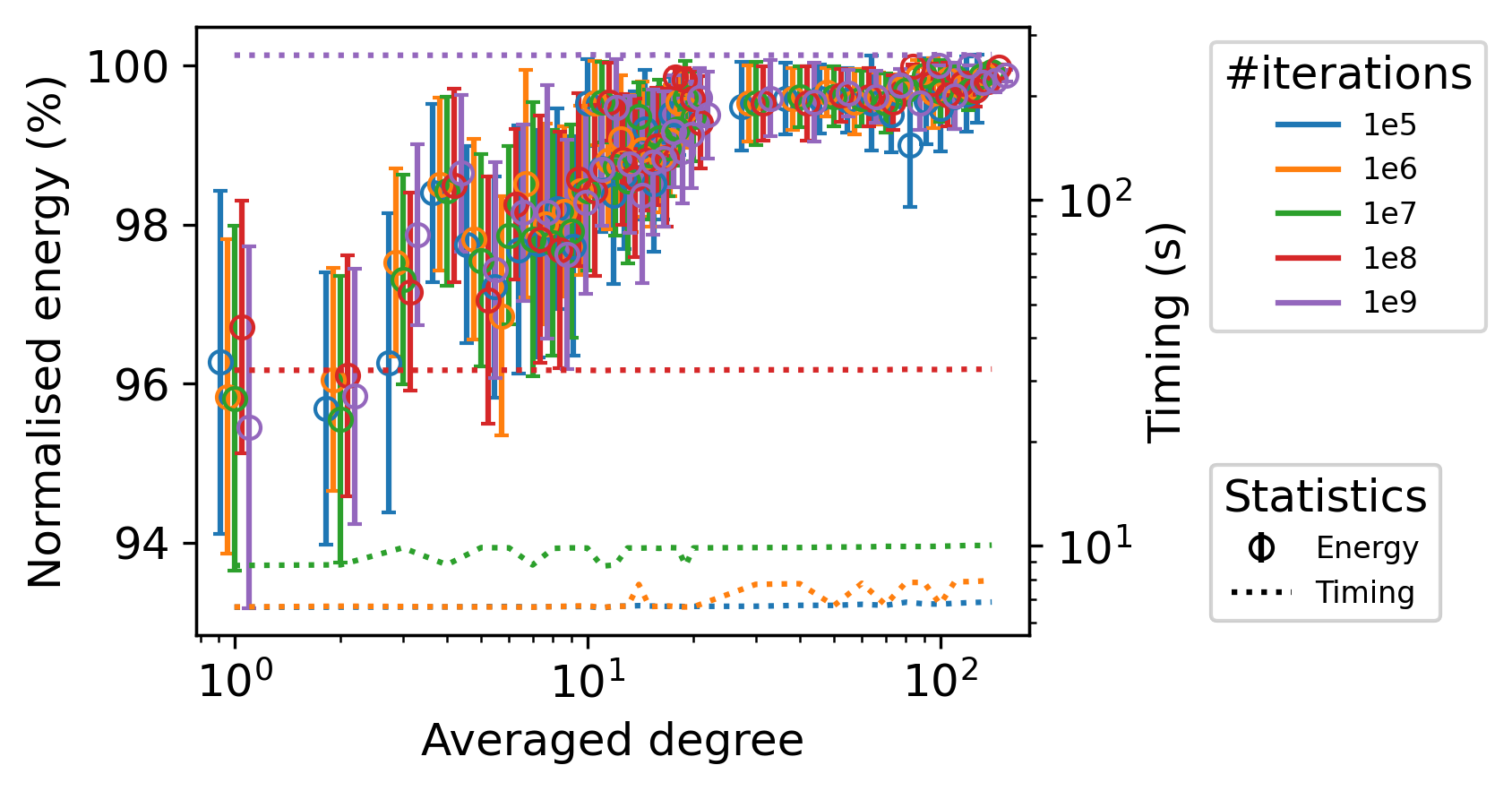}
        \caption{DA}
        \label{fig:app_max-cut_connectivity_da}
    \end{subfigure}
    \begin{subfigure}[b]{0.45\linewidth}
        \centering
        \includegraphics[width=\textwidth]{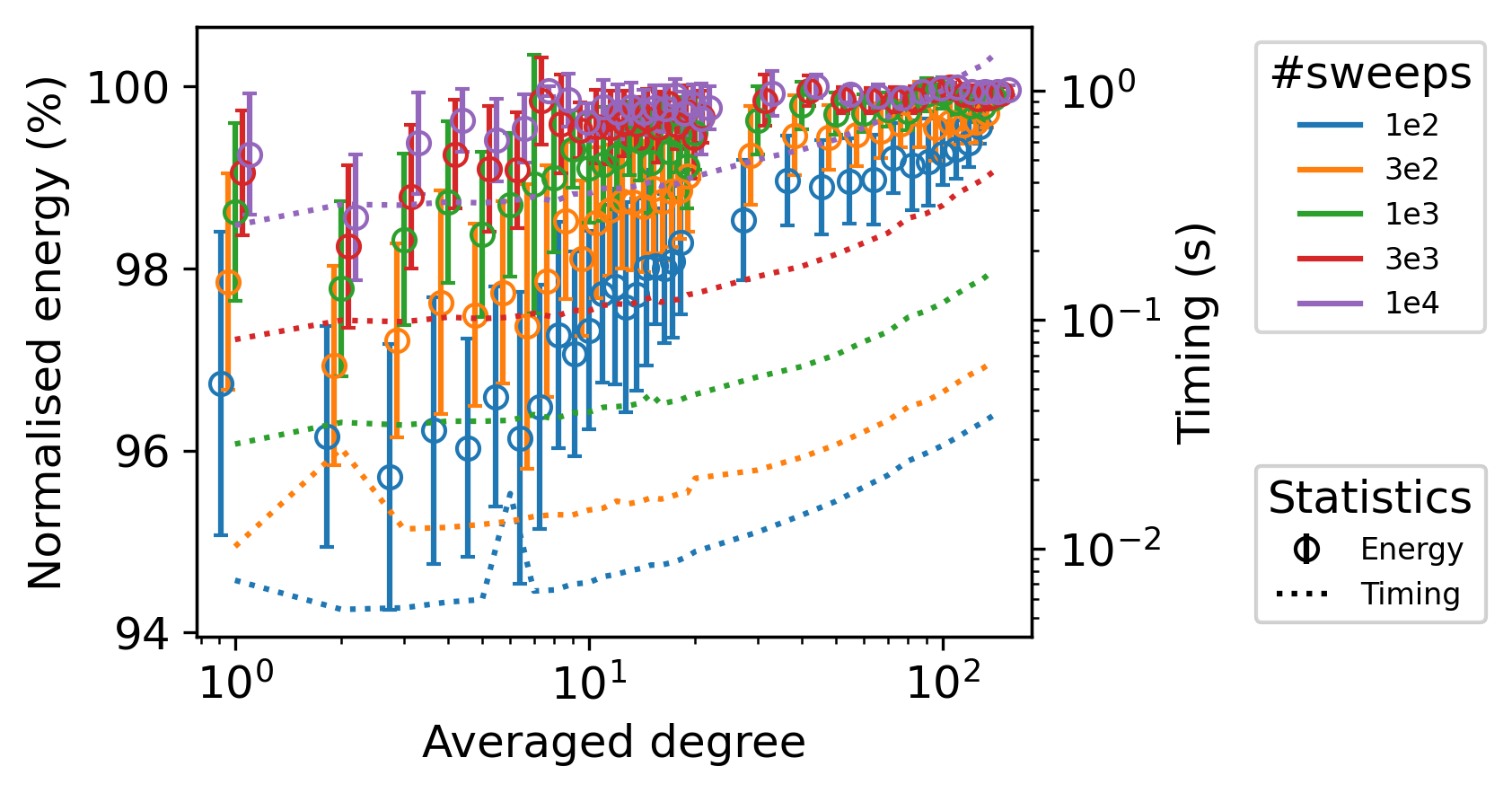}
        \caption{SA}
        \label{fig:app_max-cut_connectivity_sa}
    \end{subfigure}
    \caption{Max-cut, connectivity-varied problems. The error bar of energy and timing. X axis is the averaged degrees, in log scale. The primary Y axis is energy normalised to that of Gurobi. The secondary Y axis corresponds to dotted curves and represents timing in seconds, in log scale. The circle of an error bar represents mean energy, while the top and bottom represent $\pm$ standard deviation. We shift the error bars a little bit to improve readability and avoid overlap. Since max-cut is a maximisation problem, higher energy is better.}
    \label{fig:app_max-cut_connectivity_hp}
\end{figure}

Figure \ref{fig:app_max-cut_connectivity_hp} shows the performance of the solvers on the max-cut Chimera-like problems. Through figure \ref{fig:app_max-cut_connectivity_pegasus} we understand that a longer annealing time, i.e. 2000 $\mu s$, leads to the best performance in energy, when the averaged degree is below 12. But with higher averaged degrees, the advantage disappear. DA is not sensitive to its hyper-parameter. SA is more sensitive to its hyper-parameter, compared with DA. 

For Pegasus, we only include the results of $2000\mu s$ in the main text for comparison between solvers. For DA and SA We only include the results of \#iterations=$10^6$ and \#sweeps=$10^4$ in the main text for comparison between solvers. 

\begin{figure}[htb]
    \centering
	\includegraphics[width=0.45\linewidth]{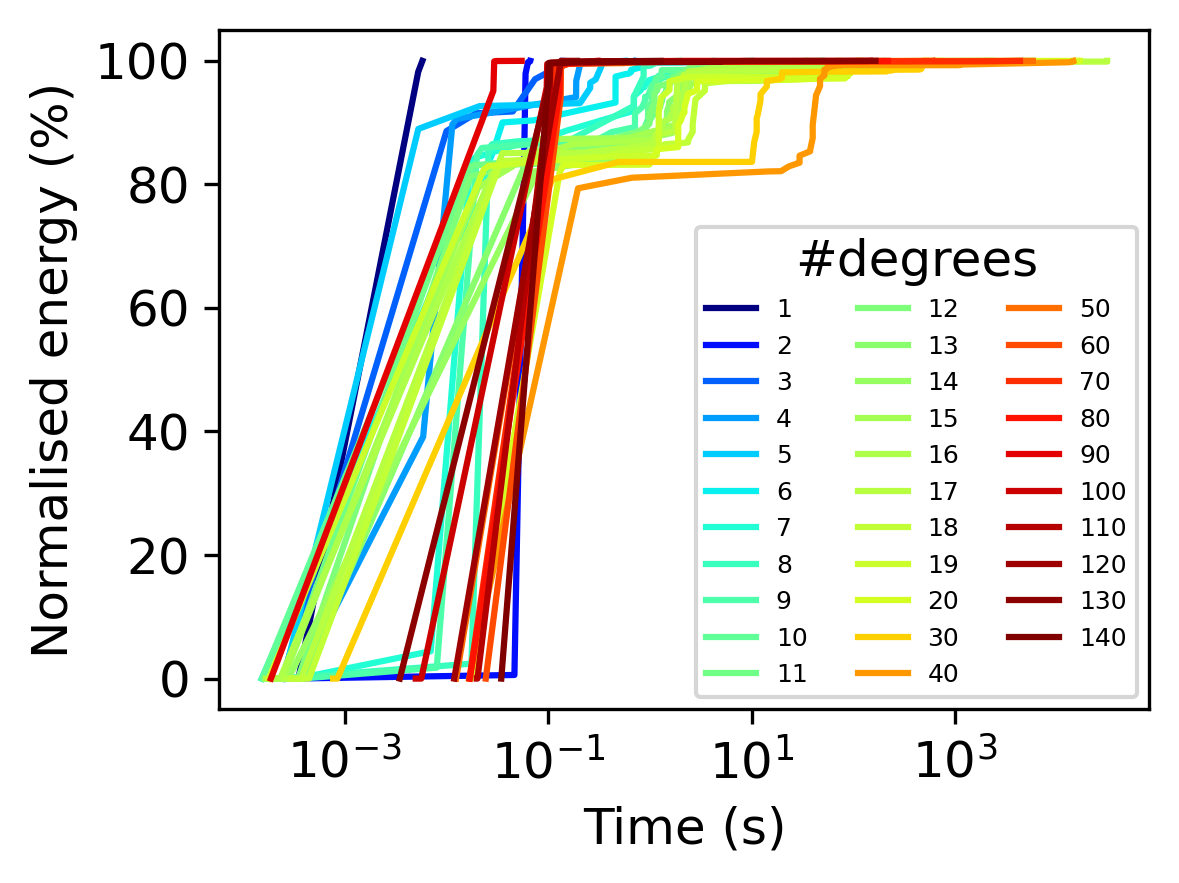}
	\caption{Gurobi on max-cut connectivity-varied problem. Time to solution. X axis is the time in seconds in log scale. Y axis is the energy normalised to optimal solution (found by Gurobi)}
	\label{fig:app_max-cut_connectivity_gurobi}
\end{figure}

Figure \ref{fig:app_max-cut_pegasus_gurobi} shows the time-to-solution of Gurobi on max-cut connectivity-varied problems. Gurobi can find very promising solutions within roughly 10 seconds. The time-to-traverse information is listed in Table \ref{tab:max-cut_connectivity_gurobi_traverse}.

\begin{table}[htb]
\centering
\caption{Gurobi time-to-traverse on connectivity-varied max-cut problems}
\label{tab:max-cut_connectivity_gurobi_traverse}
\begin{tabular}{|l|l|l|l|}
\hline
\#degrees & Time (s) & \#degrees & Time (s) \\ \hline
1         & 0.004468 & 17        & 19396.73 \\ \hline
2         & 0.034891 & 18        & 36000.02 \\ \hline
3         & 0.051368 & 19        & 36000.02 \\ \hline
4         & 0.252771 & 20        & 36000.01 \\ \hline
5         & 0.319412 & 30        & 36000.03 \\ \hline
6         & 3.782623 & 40        & 36000.01 \\ \hline
7         & 4.946026 & 50        & 36000.11 \\ \hline
8         & 7.411871 & 60        & 36000.13 \\ \hline
9         & 42.18519 & 70        & 36000.1  \\ \hline
10        & 224.94   & 80        & 36000.33 \\ \hline
11        & 348.8753 & 90        & 36000.26 \\ \hline
12        & 674.4734 & 100       & 36000.18 \\ \hline
13        & 1015.575 & 110       & 36000.79 \\ \hline
14        & 1766.733 & 120       & 36000.79 \\ \hline
15        & 3245.935 & 130       & 36000.14 \\ \hline
16        & 9300.061 & 140       & 36000.14 \\ \hline
\end{tabular}
\end{table}

\clearpage

\section{mvc}

\subsection{MVC Pegasus-like problems}
\label{sec:appendix_mvc_pegasus}

We reuse the graphs generated for Section \ref{sec:appendix_max-cut_pegasus} and formulate mvc problem based on these graphs.

\begin{figure}[htb]
     \centering
     \begin{subfigure}[b]{0.45\linewidth}
        \centering
    	\includegraphics[width=\textwidth]{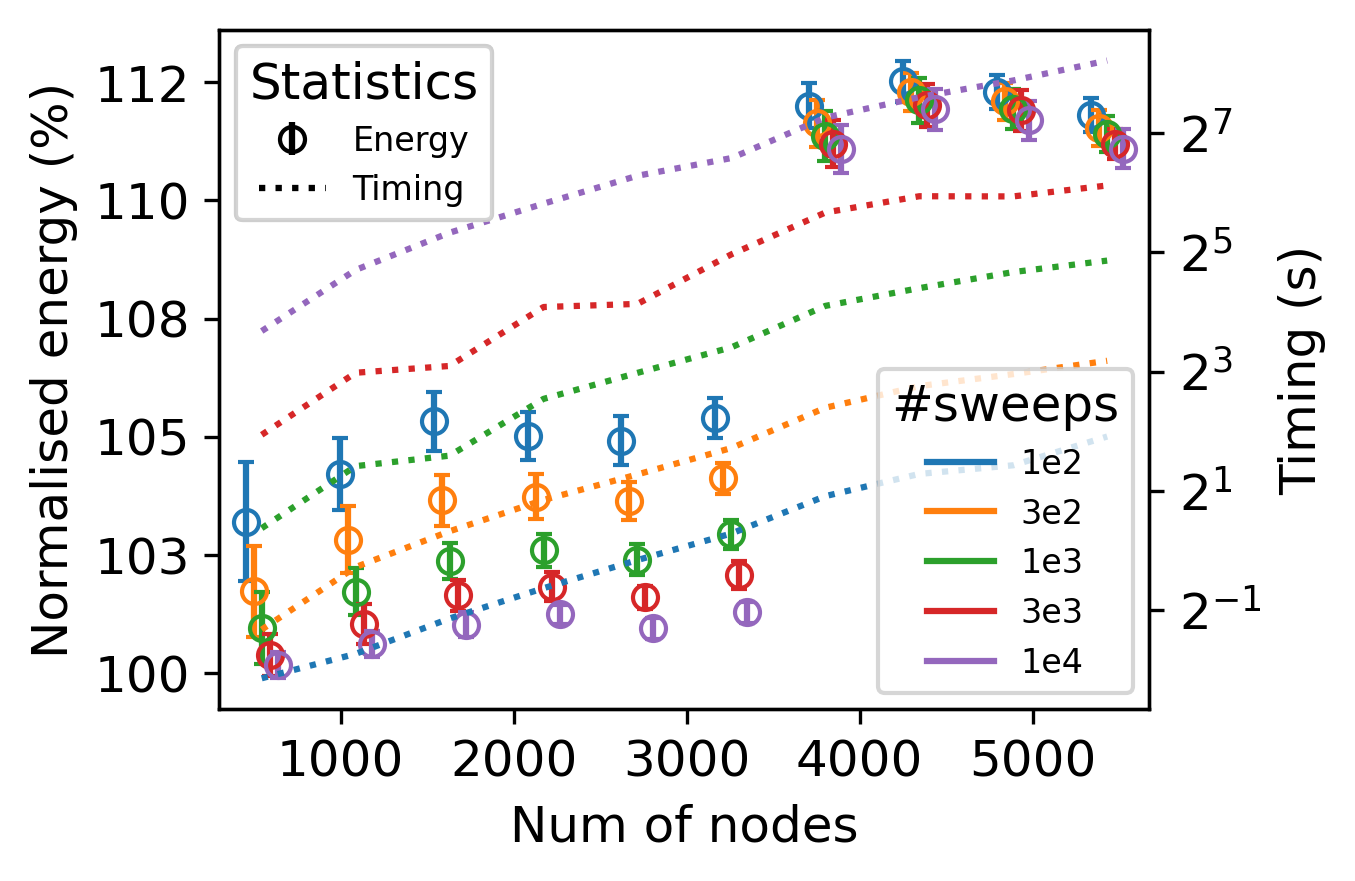}
    	\caption{Error bar of energy and timing}
    	\label{fig:app_mvc_pegasus_sa_energy}
     \end{subfigure}
     \begin{subfigure}[b]{0.45\linewidth}
        \centering
    	\includegraphics[width=\textwidth]{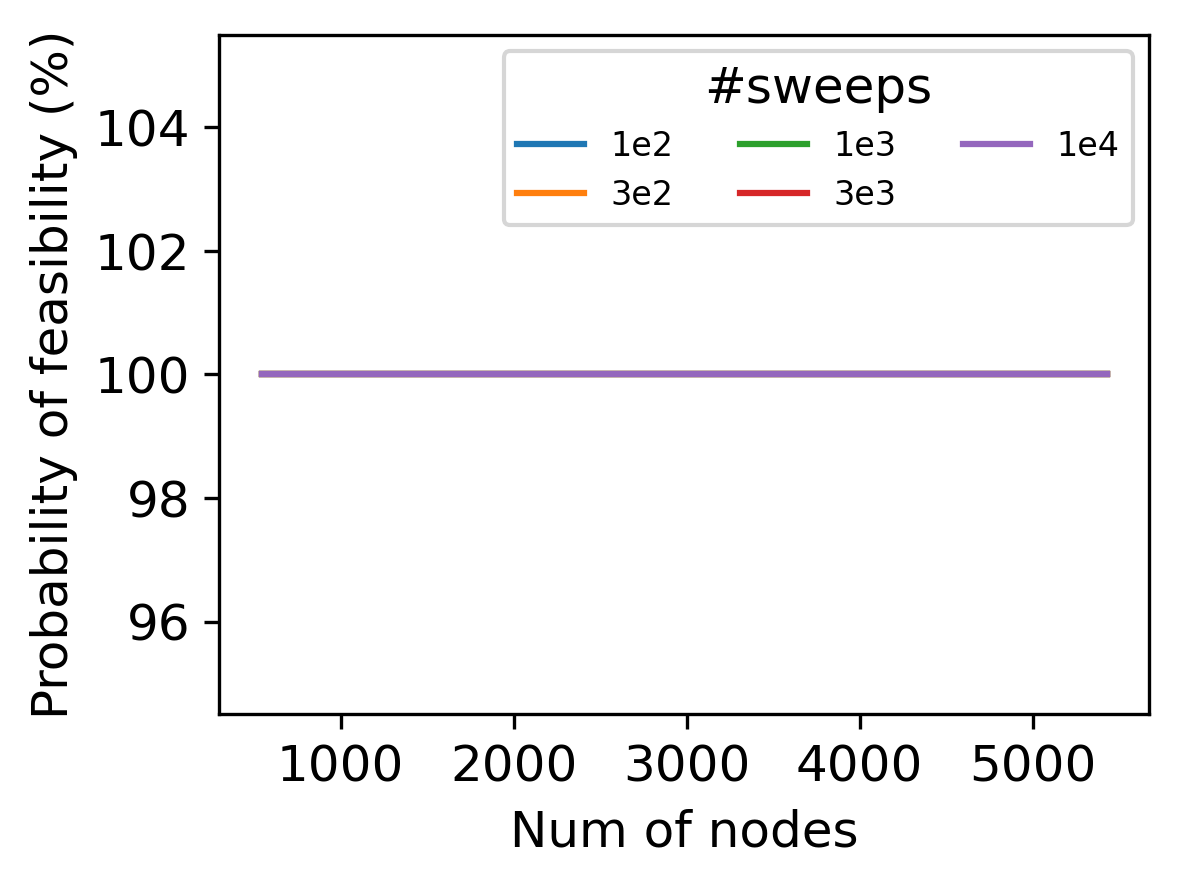}
    	\caption{Probability of feasibility}
    	\label{fig:app_mvc_pegasus_sa_pf}
     \end{subfigure}
    \caption{SA on Pegasus-like MVC problems. a) The error bar of energy and timing. X axis is the number of nodes in a graph. The primary Y axis is energy normalised to that of Gurobi. The secondary Y axis corresponds to dotted curves and represents timing in seconds, in log scale. The circle of an error bar represents mean energy, while the top and bottom represent $\pm$ standard deviation. We shift the error bars a little bit to improve readability and avoid overlap. Since MVC is a minimisation problem, lower energy is better. b) Probability of feasibility. Y axis represents the percentage of samples that meet the constraints of the problem.}
    \label{fig:app_mvc_pegasus_sa}
\end{figure}

Figure \ref{fig:app_mvc_pegasus_sa} shows the performance of SA on Pegasus-like MVC problems. For SA, large \#sweeps translates to better performance and longer time cost. This is consistent across a range of problem size and hyper-parameter settings.

\begin{figure*}[tb]
     \centering
     \begin{subfigure}[b]{0.38\textwidth}
        \centering
    	\includegraphics[width=\textwidth]{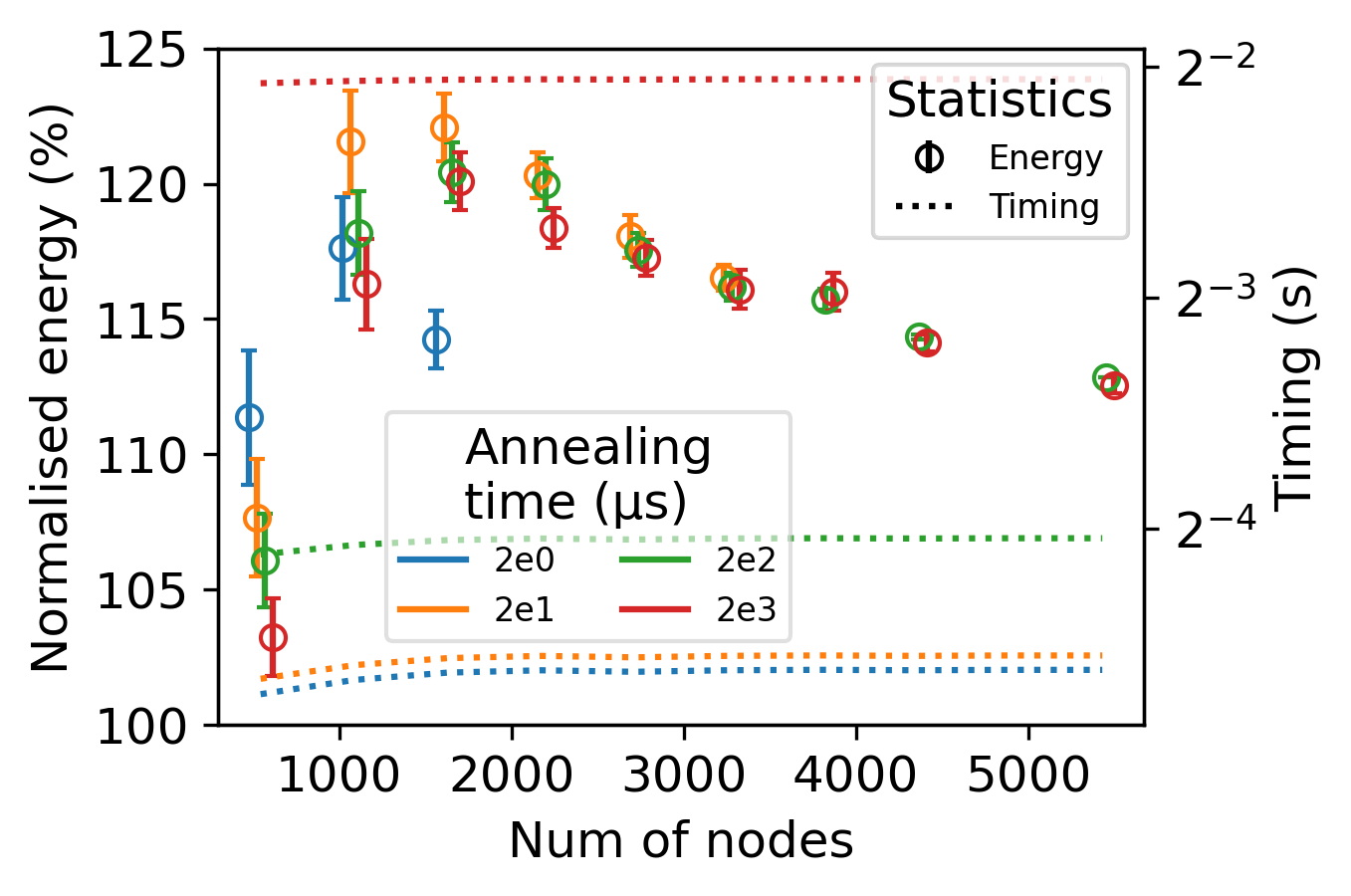}
    	\caption{Error bar of energy and timing}
    	\label{fig:app_mvc_pegasus_pegasus_energy}
     \end{subfigure}
     \begin{subfigure}[b]{0.33\textwidth}
        \centering
    	\includegraphics[width=\textwidth]{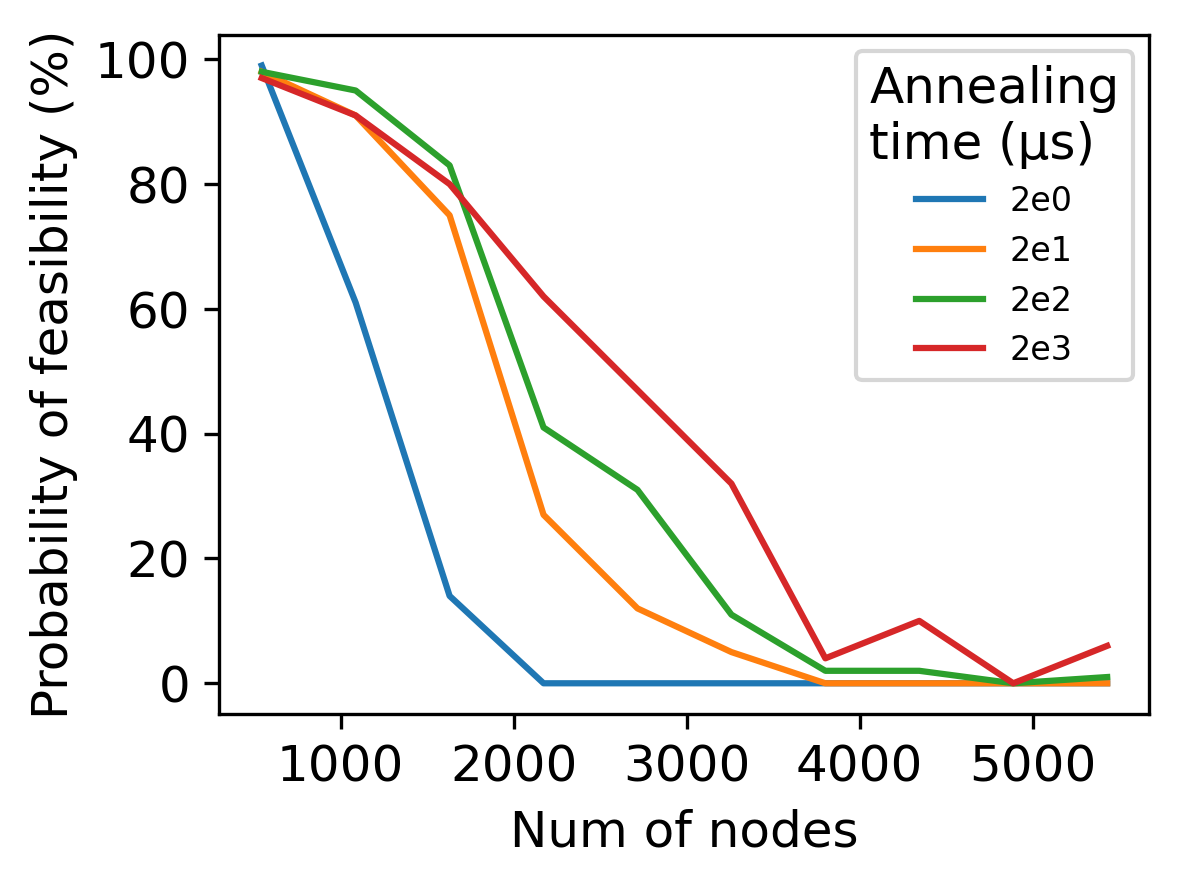}
    	\caption{Probability of feasibility}
    	\label{fig:app_mvc_pegasus_pegasus_pf}
     \end{subfigure}
     \begin{subfigure}[b]{0.95\textwidth}
        \centering
    	\includegraphics[width=\textwidth]{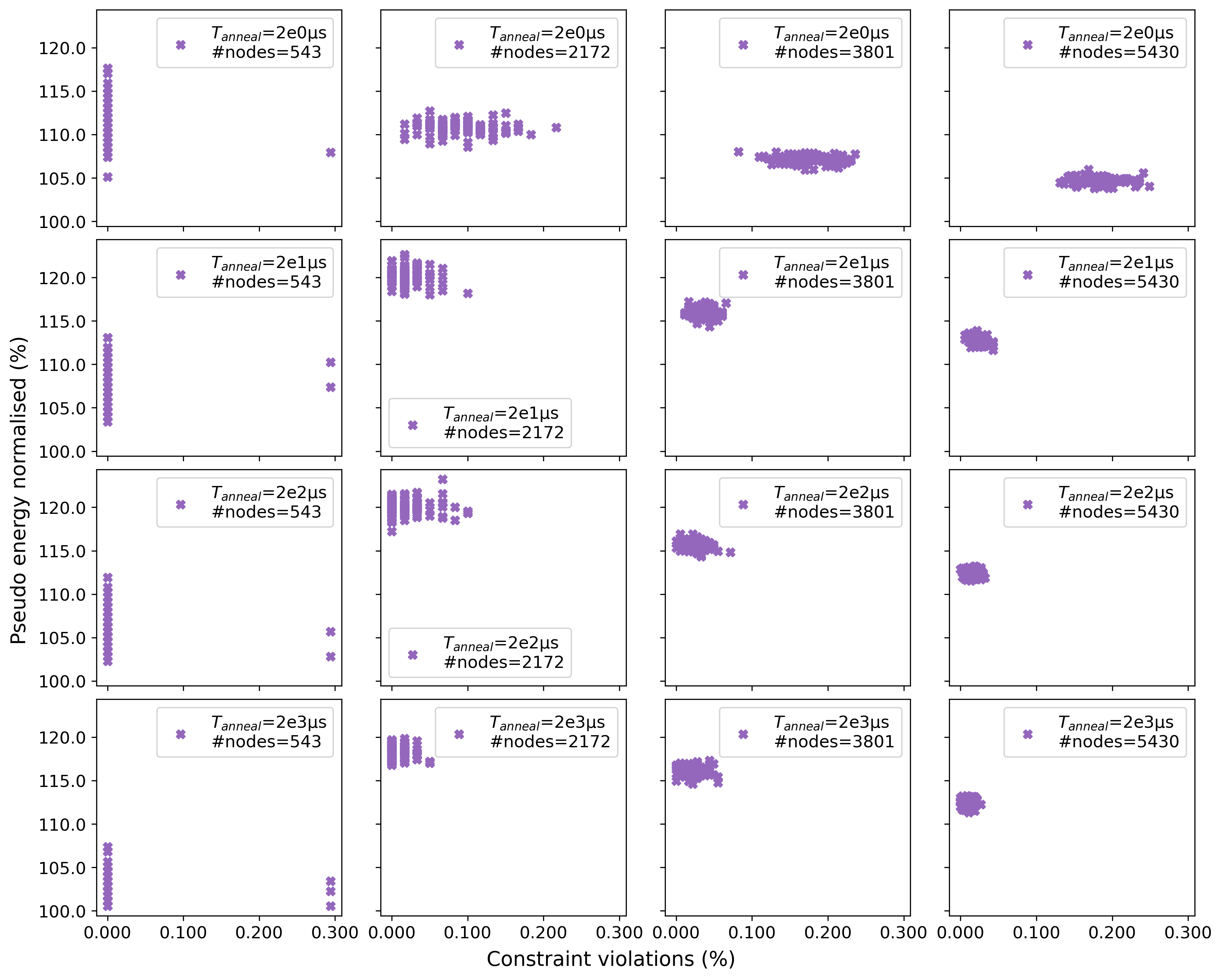}
    	\caption{Constraint violation}
    	\label{fig:app_mvc_pegasus_pegasus_violation}
     \end{subfigure}
    \caption{D-Wave Pegasus on Pegasus-like MVC problems. a) The error bar of energy and timing. X axis is the number of nodes in a graph. The primary Y axis is energy normalised to that of Gurobi. The secondary Y axis corresponds to dotted curves and represents timing in seconds, in log scale. The circle of an error bar represents mean energy, while the top and bottom represent $\pm$ standard deviation. We shift the error bars a little bit to improve readability and avoid overlap. Since MVC is a minimisation problem, lower energy is better. b) Probability of feasibility. Y axis represents the percentage of samples that meet the constraints of the problem. c) Constraint violations over a range of problem sizes and hyper-parameters. Each mark is a solution. X axis represents the percentage of violated constraints in a problem. Y axis represents pseudo energy, normalised to that of Gurobi.}
    \label{fig:app_mvc_pegasus_pegasus}
\end{figure*}

Figure \ref{fig:app_mvc_pegasus_pegasus} shows the performance of D-Wave Pegasus on Pegasus-like MVC problems. From figure \ref{fig:app_mvc_pegasus_pegasus_energy} we understand that longer annealing time usually produces the best solutions. From figure \ref{fig:app_mvc_pegasus_pegasus_pf} we can also see that longer annealing time produces the best probability of feasibility. Although Pegasus-like MVC problems have exactly the same graph topology as those of Pegasus-like max-cut problems, the optimal setting of the annealing time is on the different end of the allowed range. We reach to the conclusion that the optimal value of annealing time is closely related to the problem settings. However, hyper-parameter optimisation is not the focus of this paper. We only pick the hyper-parameters with the most promising results to be included for comparison between solvers.

Figure \ref{fig:app_mvc_pegasus_pegasus_violation} aims to compare between feasible and infeasible solutions, over a range of problem sizes and hyper-parameter settings. If you have a look at the first row, where the annealing time is 2 $\mu s$, as the problem size is getting larger, the cluster of solutions is moving away from 0\% violation. The pseudo energy of the solutions are lowering, meaning that the feasibility is traded for lower pseudo energy. If you have a look at the last row, where the annealing time is 2000 $\mu s$, the situation is the other way around. The pseudo energy is higher on larger problems, The lower pseudo energy is traded for higher feasibility. 

We know that quantum mechanics is not the only driving power for problem optimisation in a quantum annealer \cite{hauke2020perspectives}. Other things, like thermal noise, also play important roles in the optimisation process. It would be interesting if we can identify which part is dominating the performance in Pegasus-like max-cut and Pegasus-like MVC problems, and find a guideline on the proper setting of the annealing time. 

\begin{figure*}[tb]
     \centering
     \begin{subfigure}[b]{0.38\textwidth}
        \centering
    	\includegraphics[width=\textwidth]{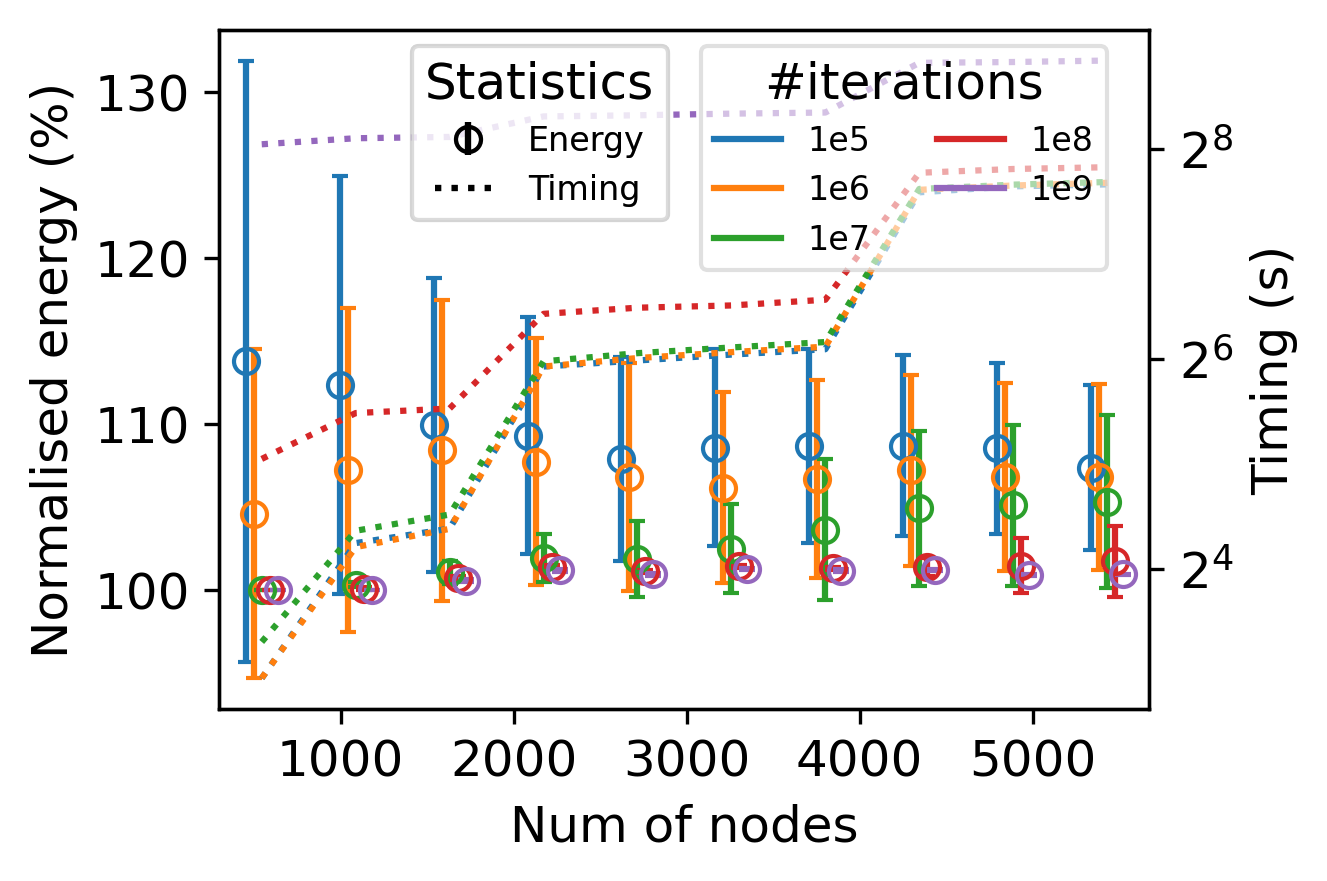}
    	\caption{Error bar of energy and timing}
    	\label{fig:app_mvc_pegasus_da_energy}
     \end{subfigure}
     \begin{subfigure}[b]{0.33\textwidth}
        \centering
    	\includegraphics[width=\textwidth]{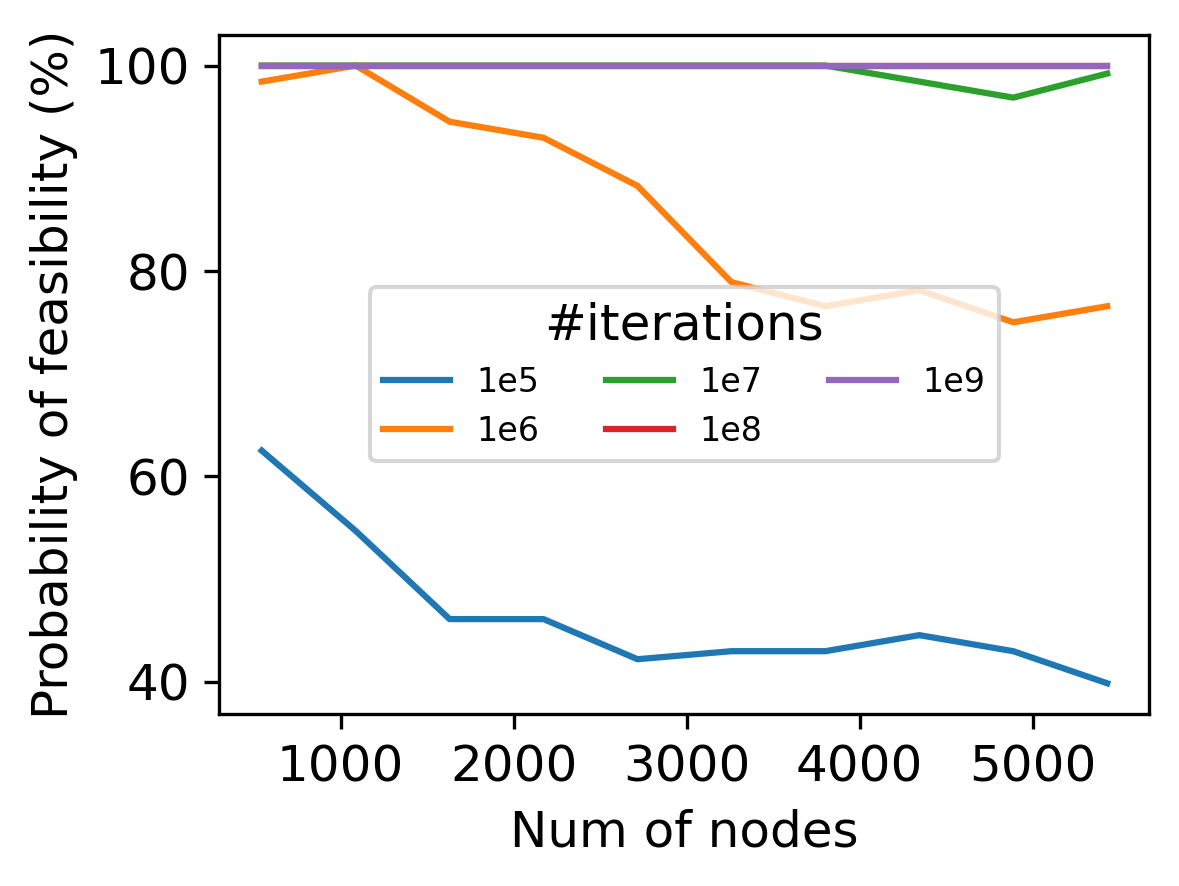}
    	\caption{Probability of feasibility}
    	\label{fig:app_mvc_pegasus_da_pf}
     \end{subfigure}
     \begin{subfigure}[b]{0.95\textwidth}
        \centering
    	\includegraphics[width=\textwidth]{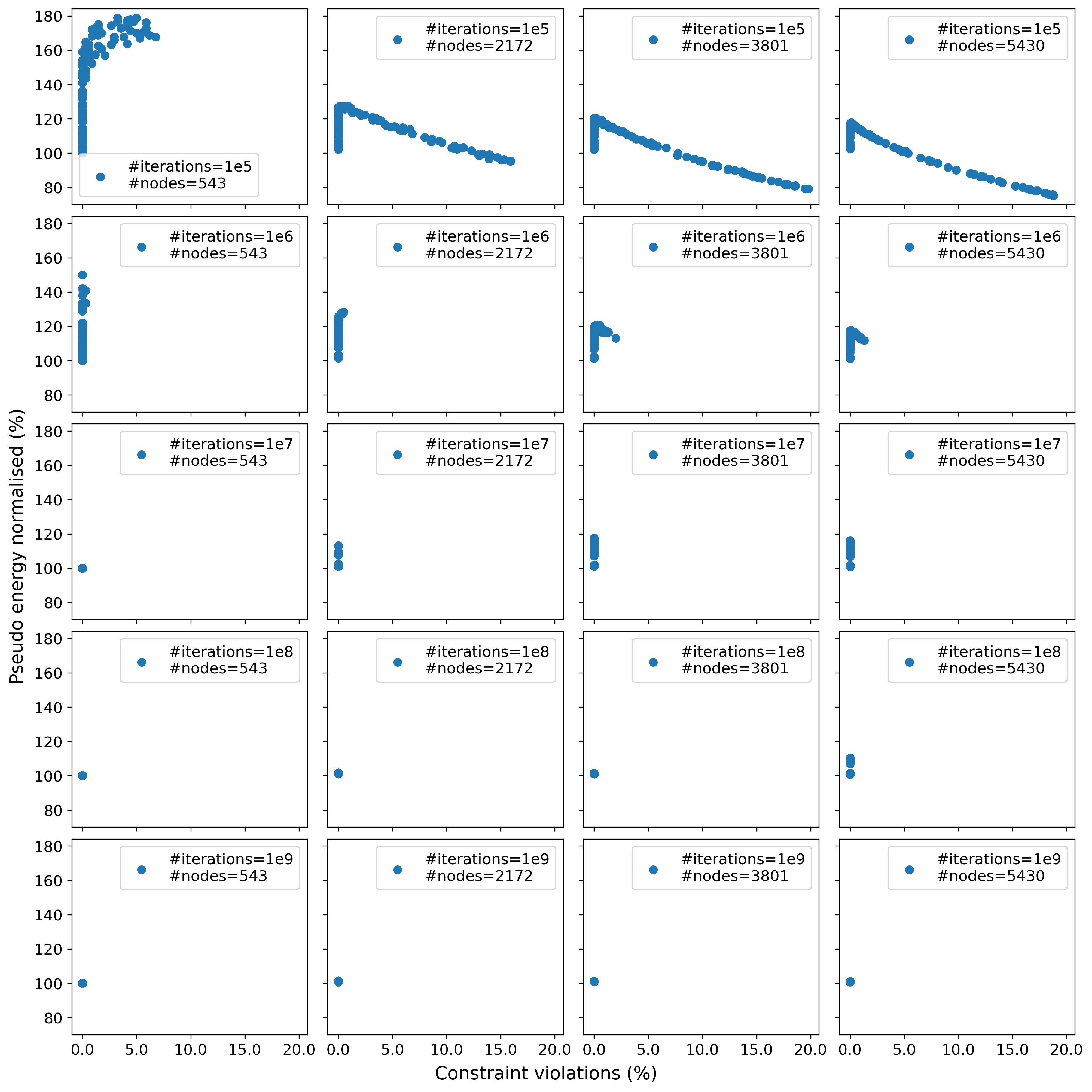}
    	\caption{Constraint violation}
    	\label{fig:app_mvc_pegasus_da_violation}
     \end{subfigure}
    \caption{DA on Pegasus-like MVC problems. The setting of the plots is the same as that of figure \ref{fig:app_mvc_pegasus_pegasus}}
    \label{fig:app_mvc_pegasus_da}
\end{figure*}

Figure \ref{fig:app_mvc_pegasus_da} shows the performance of DA on Pegasus-like MVC problems. Figure \ref{fig:app_mvc_pegasus_da_energy} and \ref{fig:app_mvc_pegasus_da_pf} suggest that when \#iterations is small, e.g. $10^5$, there is large variance in energy and low probability of feasibility. Figure \ref{fig:app_mvc_pegasus_da_violation} suggest that small \#iterations, like $10^5$, cannot effectively explore the solution space and leave some of the solution infeasible. the feasibility is traded for low pseudo energy.

For Pegasus, we only include the results of $2000\mu s$ for comparison between solvers. For DA and SA We only include the results of \#iterations=$10^7$ and \#sweeps=$10^4$ for comparison between solvers. 

\begin{figure}[htb]
    \centering
    \begin{subfigure}[b]{0.45\linewidth}
        \centering
    	\includegraphics[width=\textwidth]{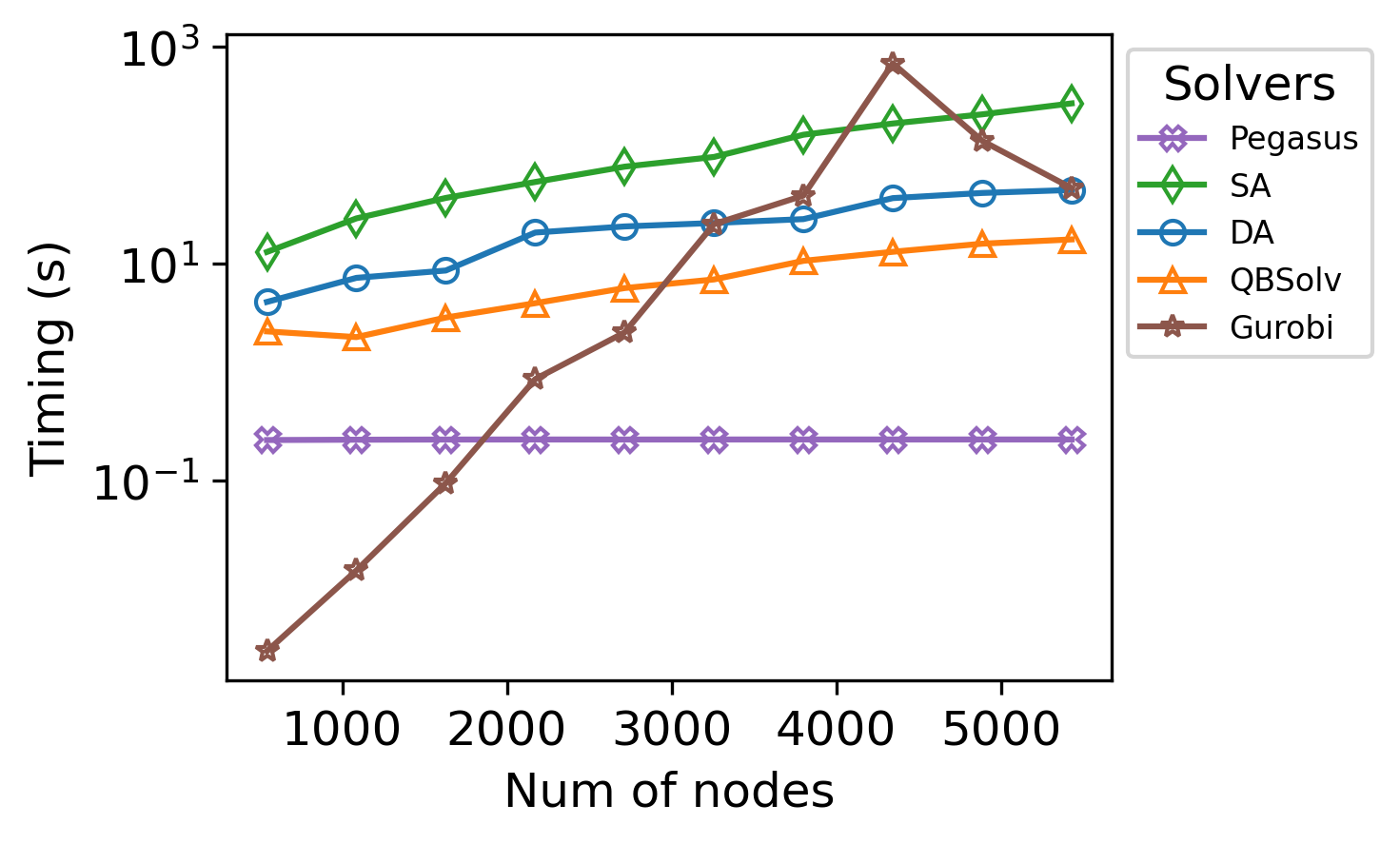}
    	\caption{Timing}
    	\label{fig:app_mvc_pegasus_timing}
    \end{subfigure}
    \begin{subfigure}[b]{0.45\linewidth}
        \centering
    	\includegraphics[width=\textwidth]{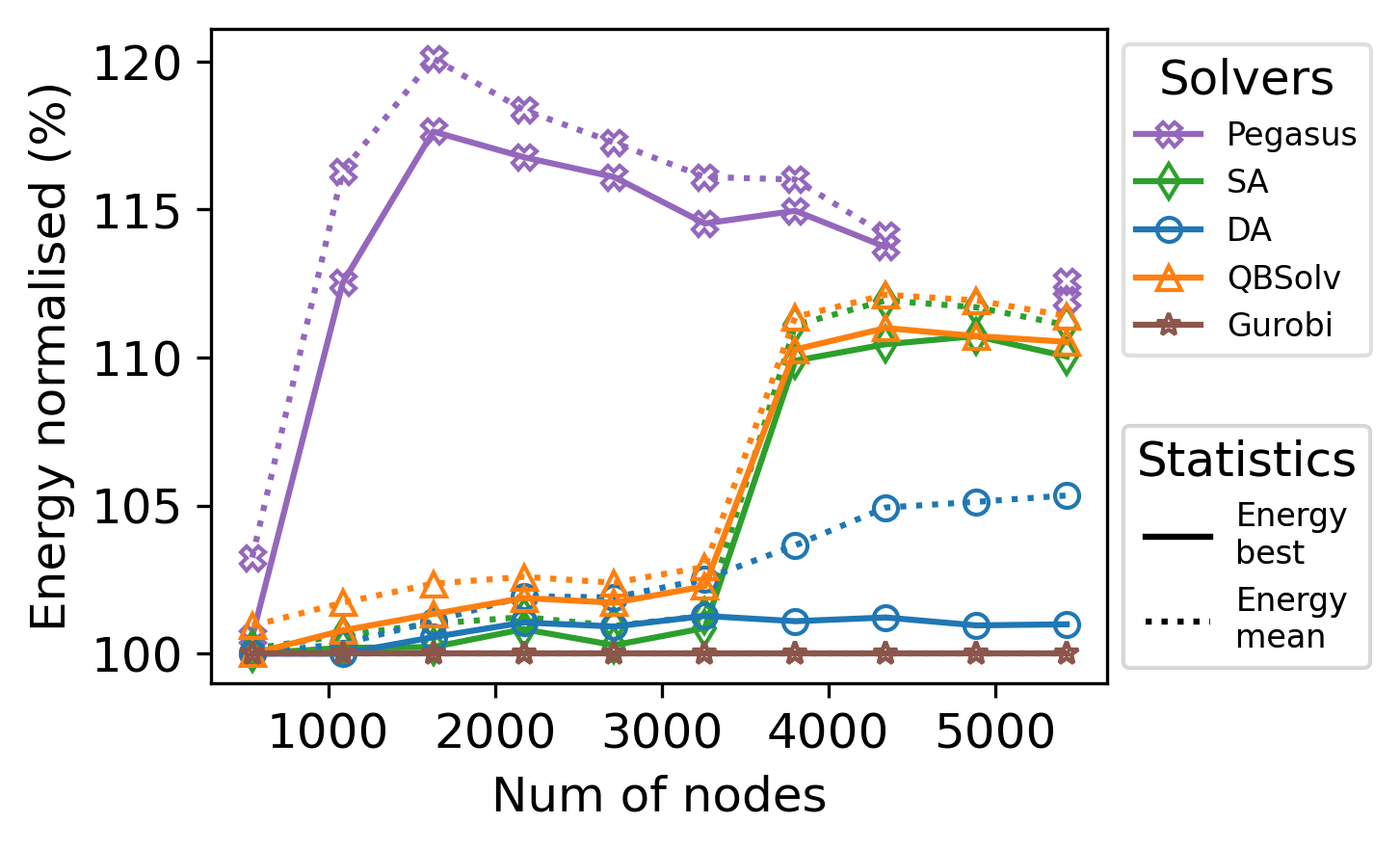}
    	\caption{Normalised energy}
    	\label{fig:app_mvc_pegasus_energy}
    \end{subfigure}
    \caption{Energy and timing of solvers on Pegasus-like MVC problems. We choose $T_{anneal} = 2000 \mu s$ for D-Wave Pegasus, \#iterations=$10^5$ for DA and \#sweeps=$10^4$ for SA. The setting of the figure is the same as that of figure \ref{fig:app_max-cut_chimera_performance}.}
    \label{fig:app_mvc_pegasus}
\end{figure}

Figure \ref{fig:app_mvc_pegasus} shows the energy and timing of the solvers on Pegasus-like MVC problems. D-Wave Pegasus is faster than the other annealing-based solvers, but is worst in terms of the energy of solutions. DA is faster than SA, and is the best among all annealing-based solvers in terms of energy.

\begin{figure}[htb]
    \centering
    \begin{subfigure}[b]{0.45\linewidth}
        \centering
    	\includegraphics[width=\textwidth]{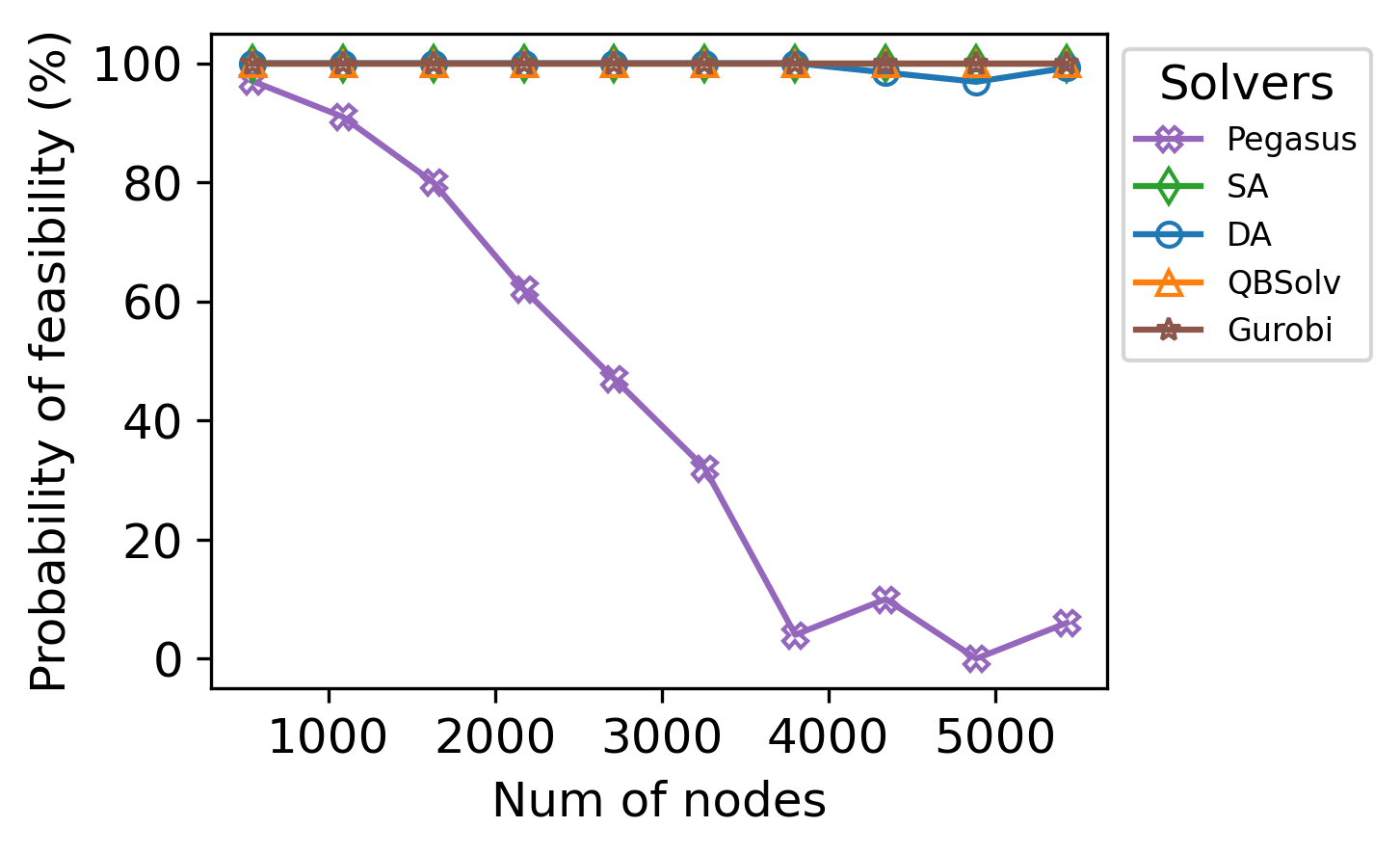}
    	\caption{Probability of feasibility}
    	\label{fig:mvc_pegasus_pf}
    \end{subfigure}
    \begin{subfigure}[b]{0.45\linewidth}
        \centering
    	\includegraphics[width=\textwidth]{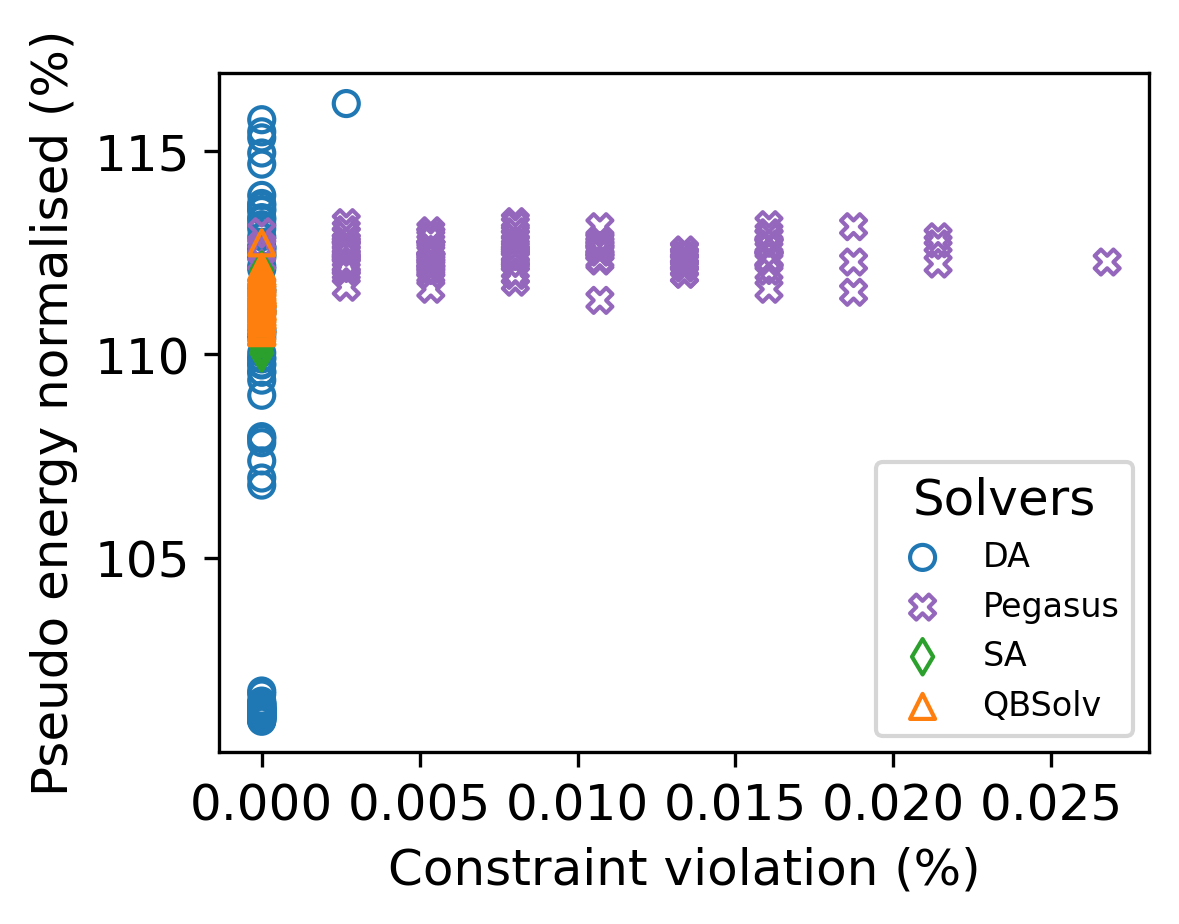}
    	\caption{Constraint Violation}
    	\label{fig:mvc_pegasus_violation}
     \end{subfigure}
    \caption{Feasibility/violation of solvers on Pegasus-like MVC problems. The hyper-parameter setting is the same as that in figure \ref{fig:app_mvc_pegasus}. a) X axis is the number of nodes. Y axis is percentage of samples that meet the constraints of the problem. b) constraint violation on pegasus\_node5430. X axis is the percentage of constraints being violated. Y axis is the pseudo energy, normalised to that of Gurobi.}
    \label{fig:mvc_pegasus_indepth}
\end{figure}

Figure \ref{fig:mvc_pegasus_indepth} shows the probability of feasibility, as well as the constraint violation, of solvers on Pegasus-like MVC problems. All solvers except Pegasus can find feasible solutions effectively on all problem instances. The curves of probability of feasibility are mostly overlapped with that of Gurobi in figure \ref{fig:mvc_pegasus_pf}. Figure \ref{fig:mvc_pegasus_violation} suggests that there is not much violation in the samples of Pegasus. One can fix the violations and get solutions of quality that is close to SA.

\begin{figure}[htb]
    \centering
	\includegraphics[width=0.45\linewidth]{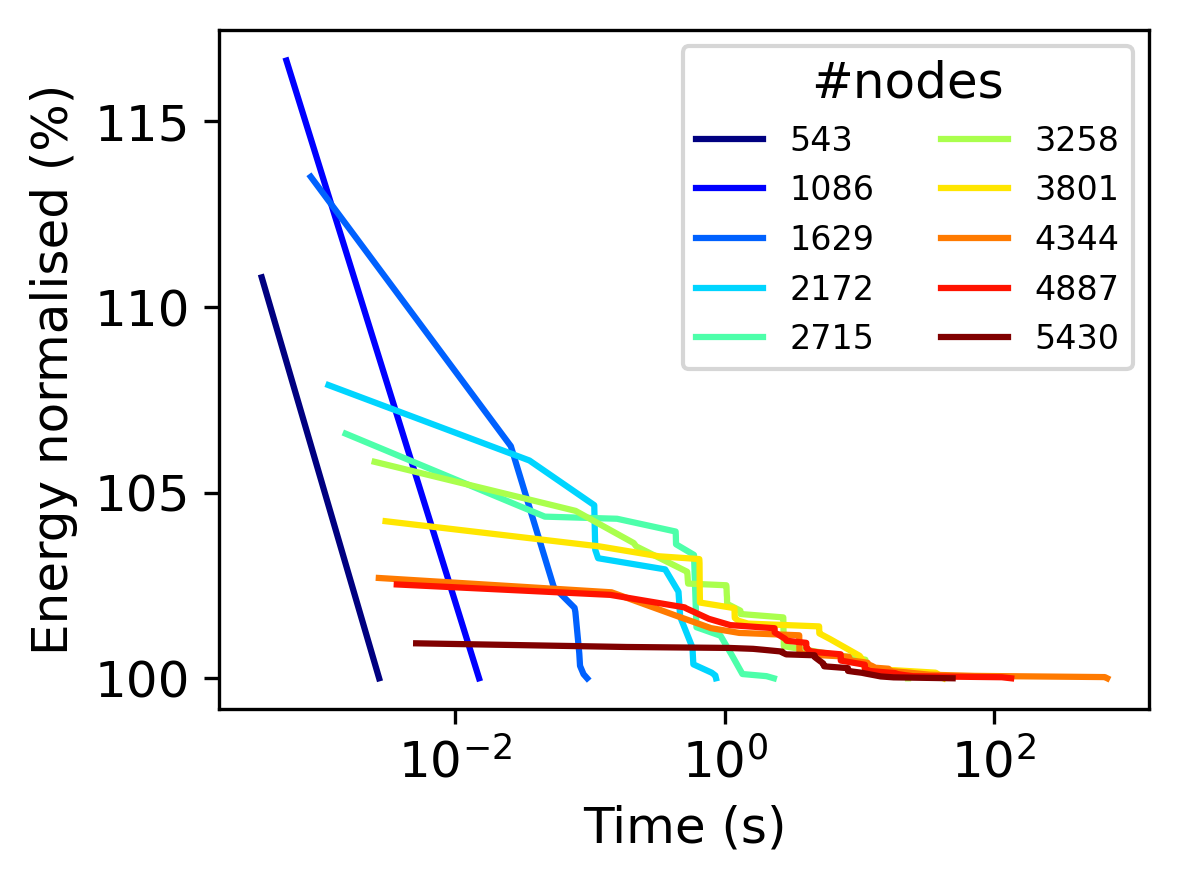}
	\caption{Gurobi time-to-solution on Pegasus-like MVC problems}
	\label{fig:app_mvc_pegasus_gurobi}
\end{figure}

Figure \ref{fig:app_mvc_pegasus_gurobi} shows the Gurobi time-to-solution plot. Gurobi can find promising solutions within a few seconds. The Gurobi time-to-traverse on Pegasus-like MVC problems is listed in Table \ref{tab:mvc_pegasus_gurobi_traverse}

\begin{table}[htb]
\centering
\caption{Gurobi time-to-traverse on Pegasus-like MVC problems}
\label{tab:mvc_pegasus_gurobi_traverse}
\begin{tabular}{|l|l|l|l|}
\hline

\#nodes & Time     & \#nodes & Time     \\ \hline
543     & 0.002709 & 3258    & 23.13427 \\ \hline
1086    & 0.014983 & 3801    & 42.36129 \\ \hline
1629    & 0.095287 & 4344    & 699.0575 \\ \hline
2172    & 0.864816 & 4887    & 135.1517 \\ \hline
2715    & 2.327012 & 5430    & 49.6165  \\ \hline
\end{tabular}
\end{table}

\clearpage

\subsection{MVC connectivity-varied problems}
\label{sec:appendix_mvc_connectivity}

We reuse the graphs generated for Section \ref{sec:appendix_max-cut_connectivity} and formulate mvc problem based on these graphs.

\begin{figure}[htb]
     \centering
     \begin{subfigure}[b]{0.45\linewidth}
        \centering
    	\includegraphics[width=\textwidth]{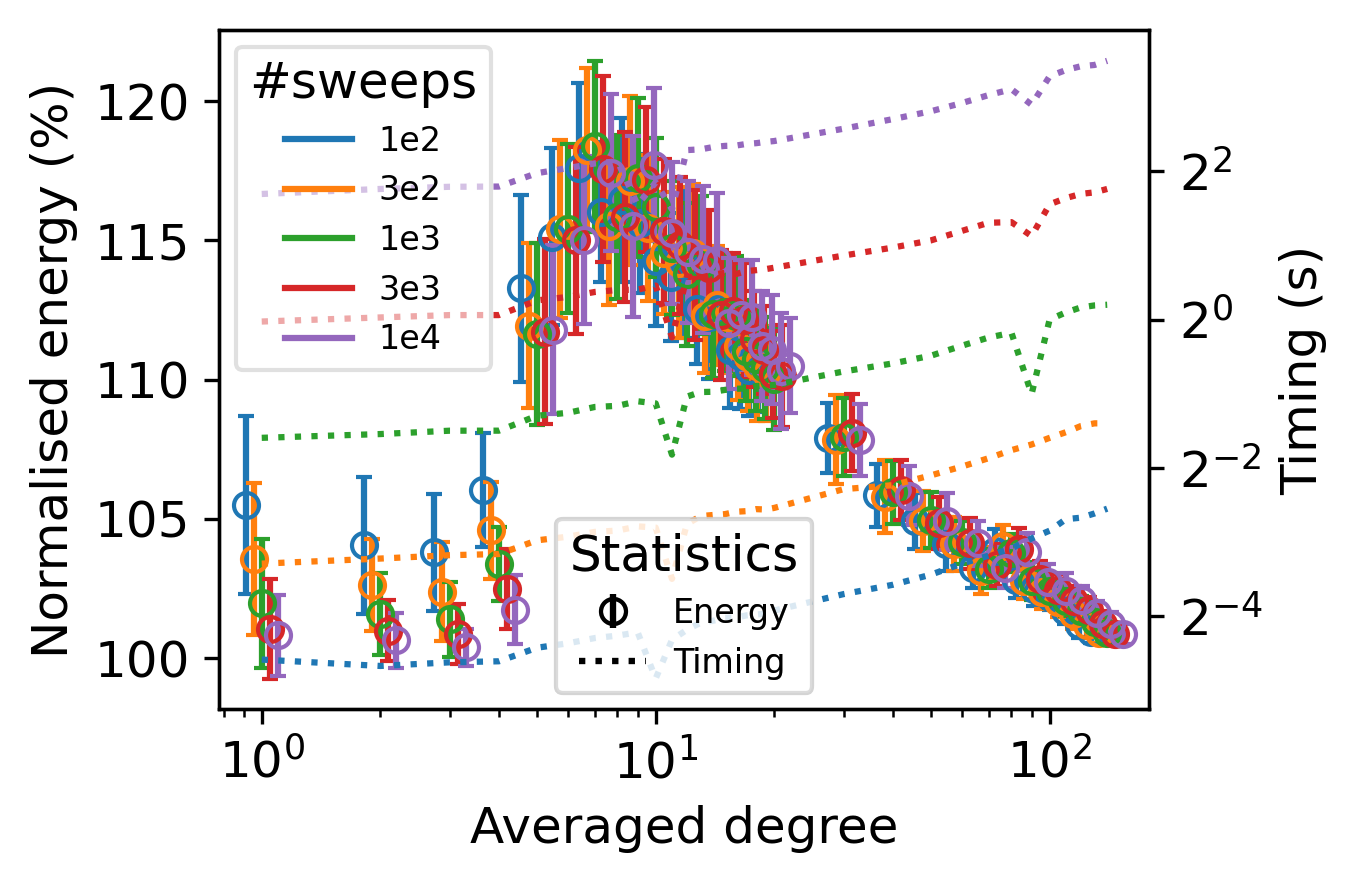}
    	\caption{Error bar of energy and timing}
    	\label{fig:app_mvc_connectivity_sa_energy}
     \end{subfigure}
     \begin{subfigure}[b]{0.45\linewidth}
        \centering
    	\includegraphics[width=\textwidth]{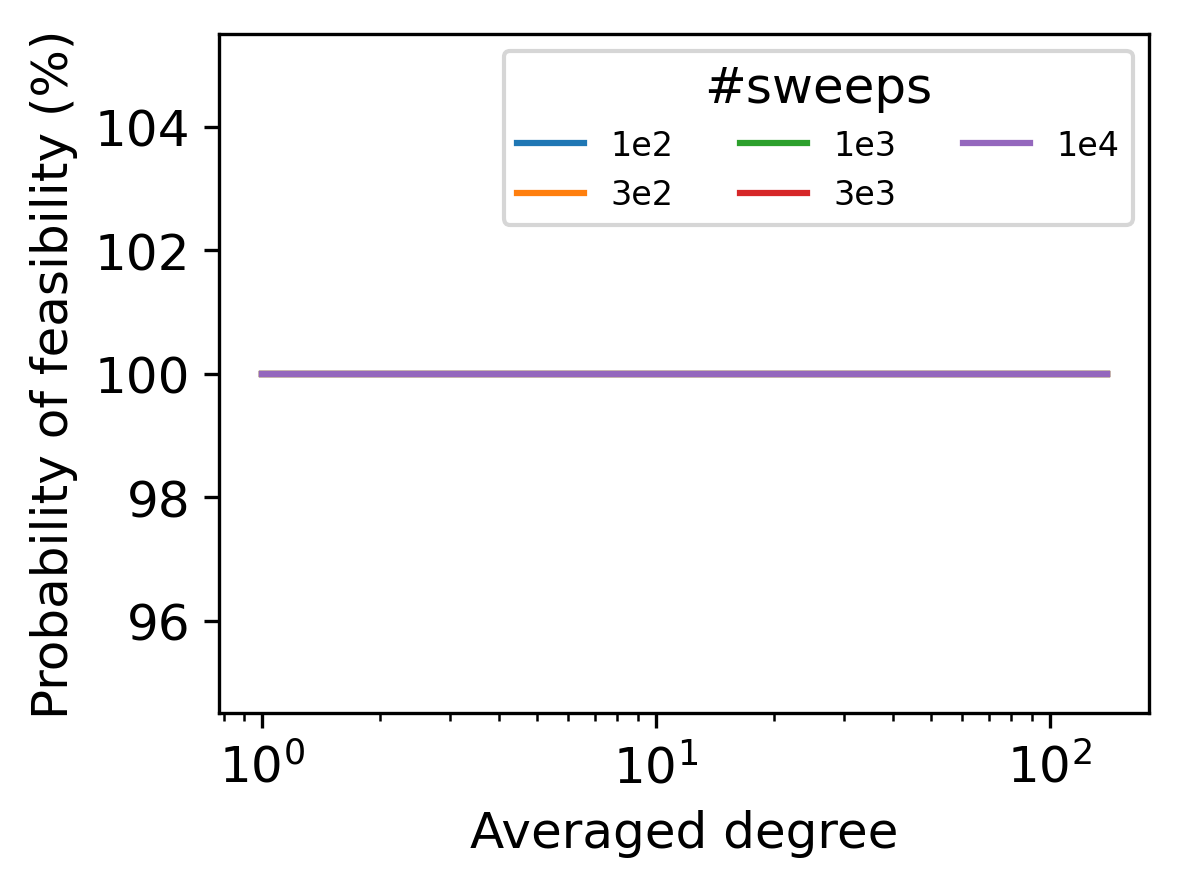}
    	\caption{Probability of feasibility}
    	\label{fig:app_mvc_connectivity_sa_pf}
     \end{subfigure}
    \caption{SA on Pegasus-like MVC problems. The setting of the plots is the same as that of figure \ref{fig:app_mvc_pegasus_pegasus}, except that we skip the plot of constraint violation, as the solutions returned by SA are all feasible.}
    \label{fig:app_mvc_connectivity_sa}
\end{figure}

Figure \ref{fig:app_mvc_connectivity_sa} shows the performance of SA on connectivity-varied MVC problems. For SA, large \#sweeps translates to better performance and longer time cost. This is consistent across a range of problem size and hyper-parameter settings.

\begin{figure*}[tb]
     \centering
     \begin{subfigure}[b]{0.38\textwidth}
        \centering
    	\includegraphics[width=\textwidth]{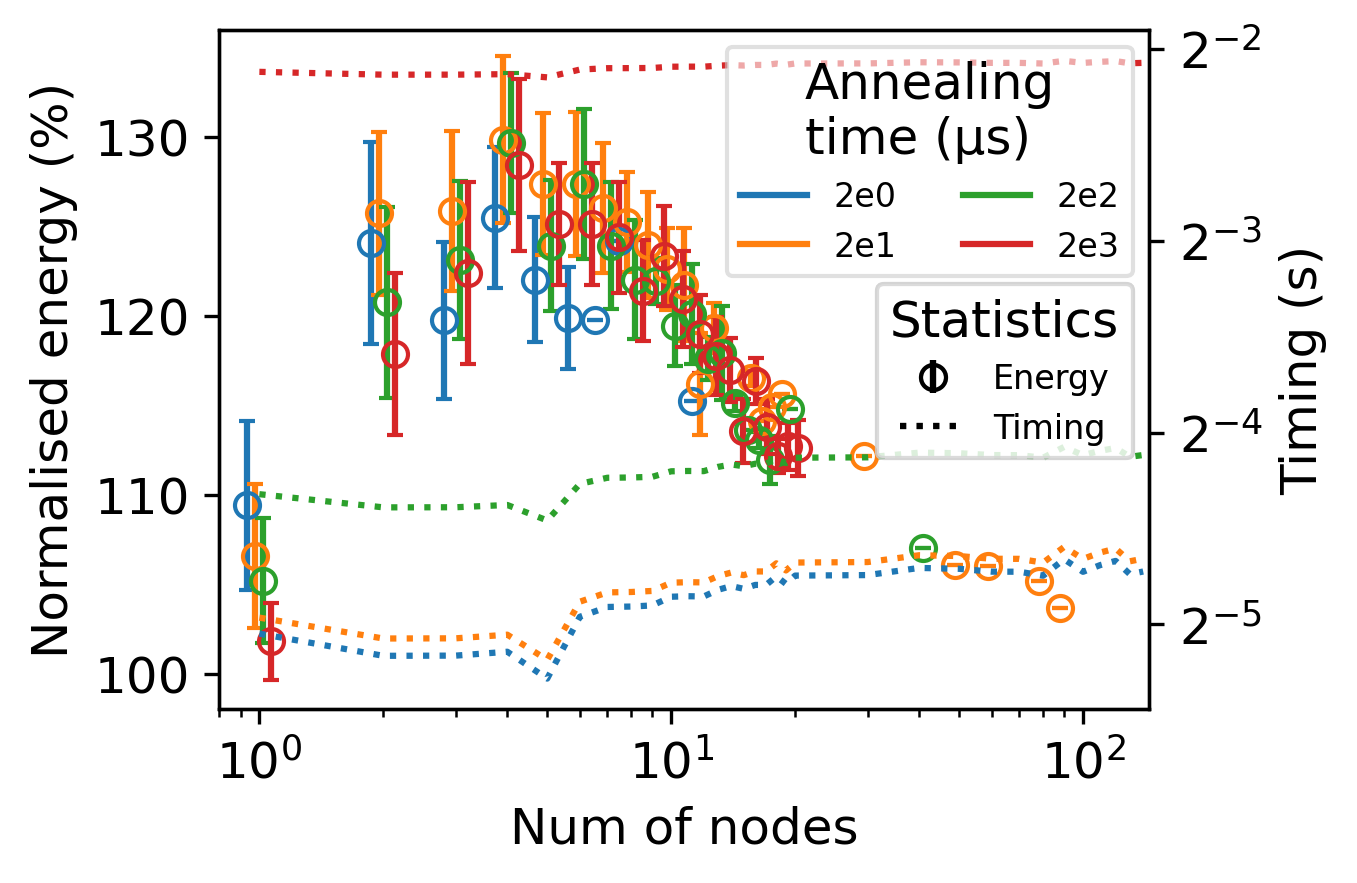}
    	\caption{Error bar of energy and timing}
    	\label{fig:app_mvc_connectivity_pegasus_energy}
     \end{subfigure}
     \begin{subfigure}[b]{0.33\textwidth}
        \centering
    	\includegraphics[width=\textwidth]{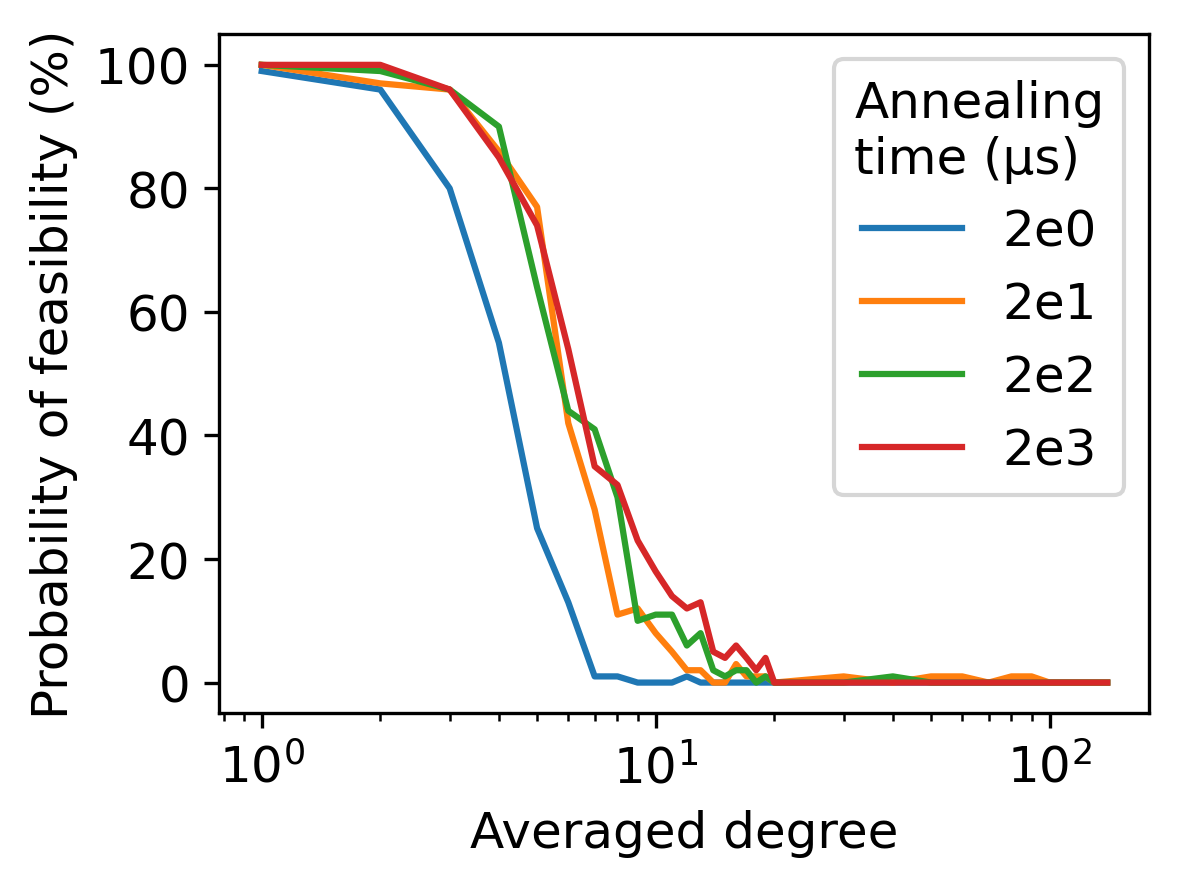}
    	\caption{Probability of feasibility}
    	\label{fig:app_mvc_connectivity_pegasus_pf}
     \end{subfigure}
     \begin{subfigure}[b]{0.95\textwidth}
        \centering
    	\includegraphics[width=\textwidth]{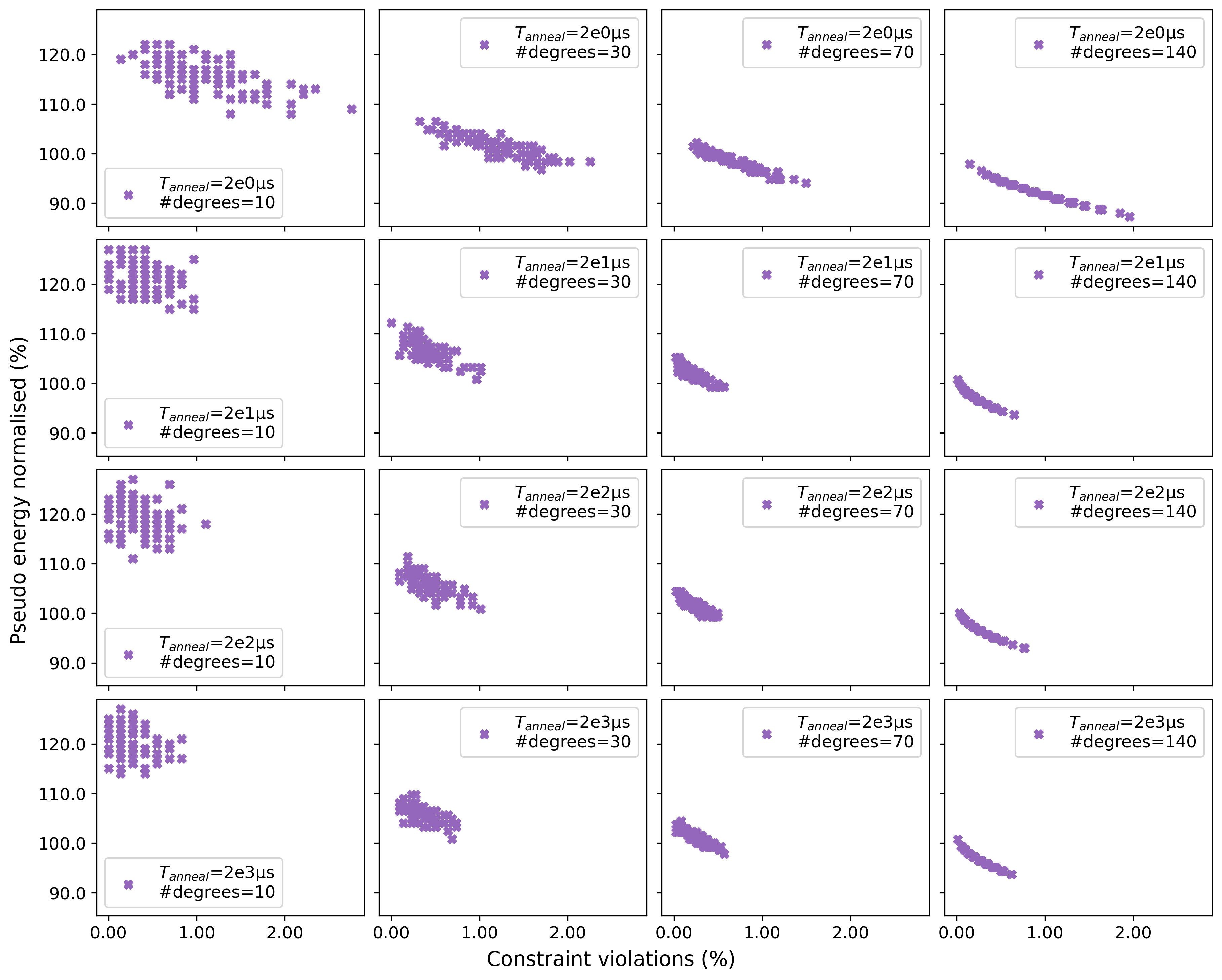}
    	\caption{Constraint violation}
    	\label{fig:app_mvc_connectivity_pegasus_violation}
     \end{subfigure}
    \caption{D-Wave Pegasus on connectivity-varied MVC problems. The plot setting is the same as that in figure \ref{fig:app_mvc_pegasus_pegasus}}
    \label{fig:app_mvc_connectivity_pegasus}
\end{figure*}

Figure \ref{fig:app_mvc_connectivity_pegasus} shows the performance of D-Wave Pegasus on connectivity-varied MVC problems. From figure \ref{fig:app_mvc_connectivity_pegasus_energy} we understand that longer annealing time usually produces the best solutions, but the annealing time of 20 $\mu s$ produces better probability of feasibility. From figure \ref{fig:app_mvc_pegasus_pegasus_pf} we known that the difference between 20$\mu s$ and 2000$\mu s$ is trivial. This observation suggest that optimality and feasibility can be detached. 

In figure \ref{fig:app_mvc_pegasus_pegasus_violation} first row last column sub-plot, the marker spreads widely and are mostly close to the bottom of the plot. This suggest that shorter annealing time ensure a lower (pseudo) energy, but is more likely to produce infeasible solutions. This is opposite to the sub-plot of last row first column, where markers has higher (pseudo) energy but closer to the 0\% violation.

\begin{figure*}[tb]
     \centering
     \begin{subfigure}[b]{0.38\textwidth}
        \centering
    	\includegraphics[width=\textwidth]{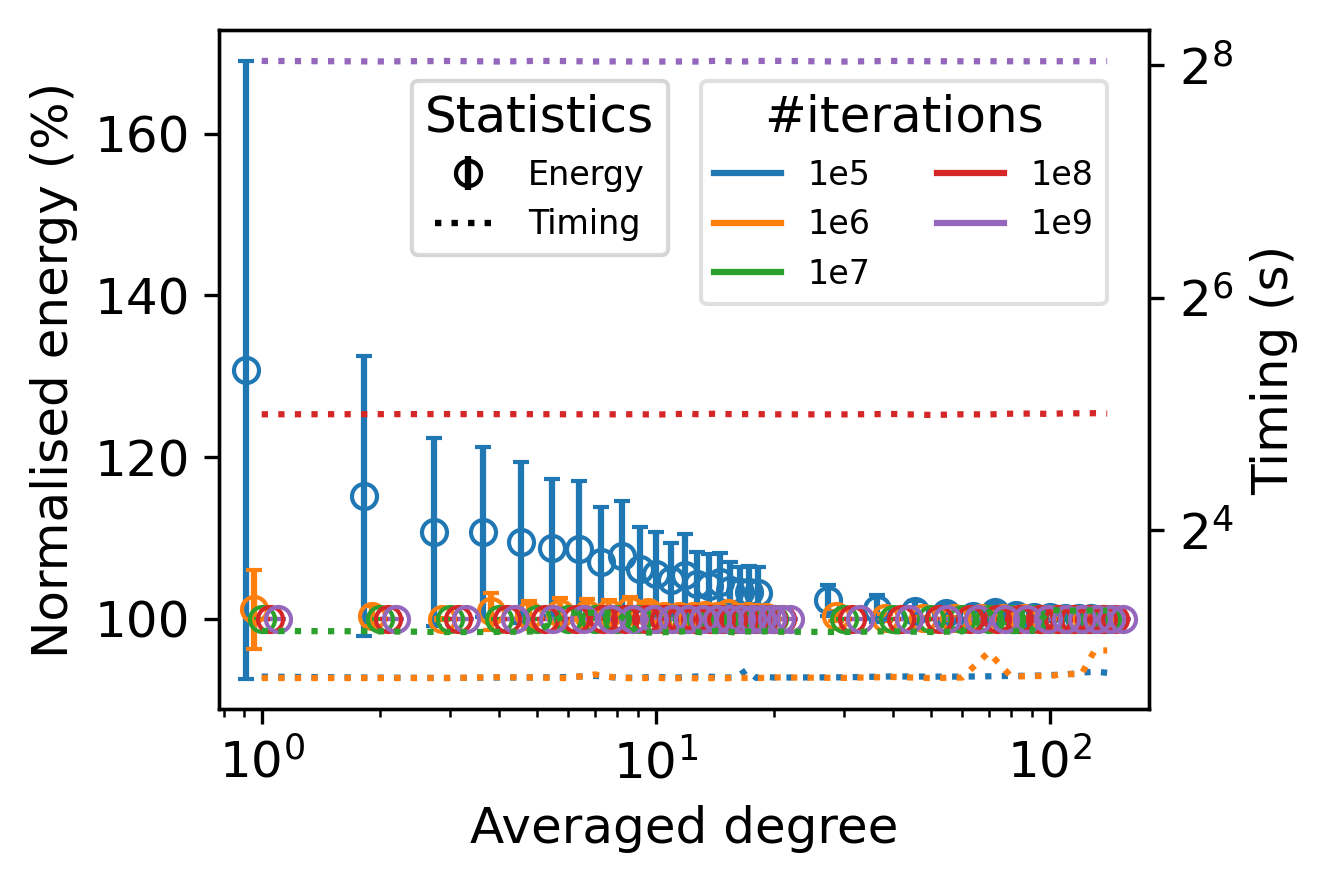}
    	\caption{Error bar of energy and timing}
    	\label{fig:app_mvc_connectivity_da_energy}
     \end{subfigure}
     \begin{subfigure}[b]{0.33\textwidth}
        \centering
    	\includegraphics[width=\textwidth]{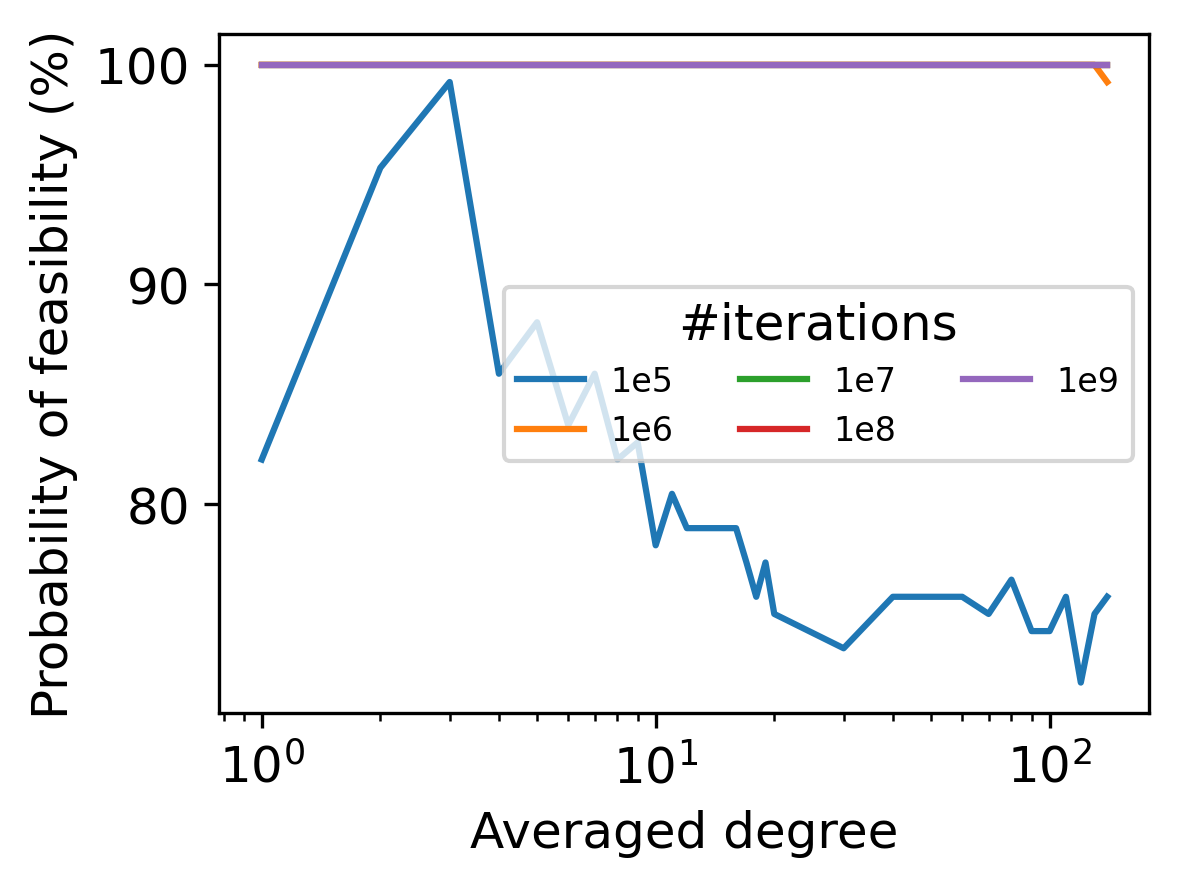}
    	\caption{Probability of feasibility}
    	\label{fig:app_mvc_connectivity_da_pf}
     \end{subfigure}
     \begin{subfigure}[b]{0.95\textwidth}
        \centering
    	\includegraphics[width=\textwidth]{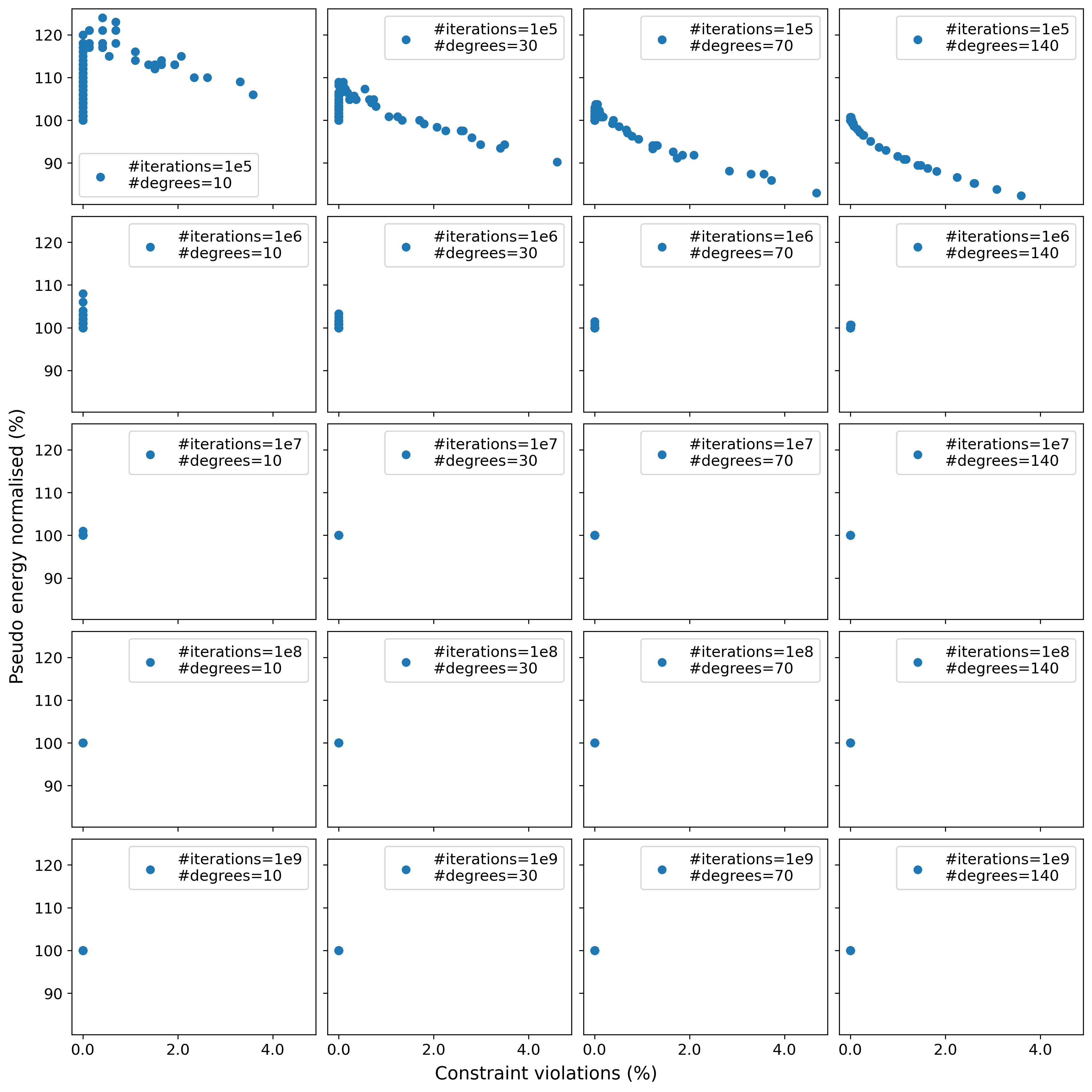}
    	\caption{Constraint violation}
    	\label{fig:app_mvc_connectivity_da_violation}
     \end{subfigure}
    \caption{DA on connectivity-varied MVC problems. The plot setting is the same as that in figure \ref{fig:app_mvc_pegasus_pegasus}}
    \label{fig:app_mvc_connectivity_da}
\end{figure*}

Figure \ref{fig:app_mvc_connectivity_da} shows the performance of DA on connectivity-varied MVC problems. This set of plots are similar to those in figure \ref{fig:app_mvc_pegasus_da}, except that the problem size in terms of number of nodes is much smaller and trivial to DA. DA can solve these problems more quickly.

For Pegasus, we only include the results of $2000\mu s$ in the main text for comparison between solvers. For DA and SA We only include the results of \#iterations=$10^5$ and \#sweeps=$10^2$ in the main text for comparison between solvers. 

\begin{figure}[htb]
    \centering
	\includegraphics[width=0.45\linewidth]{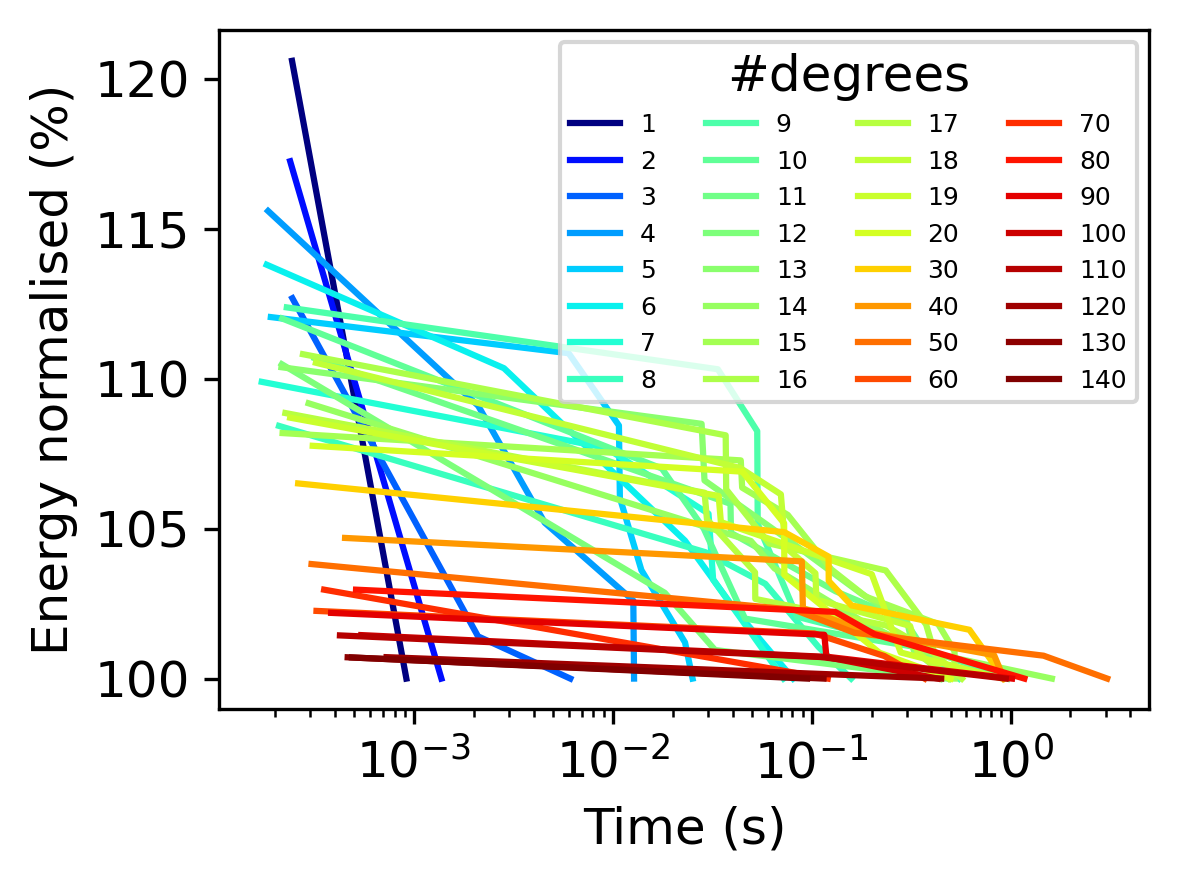}
	\caption{Gurobi time-to-solution on connectivity-varied MVC problems}
	\label{fig:app_mvc_connectivity_gurobi}
\end{figure}

Figure \ref{fig:app_mvc_connectivity_gurobi} shows the Gurobi time-to-solution plot. Gurobi can find promising solutions within a second. The Gurobi time-to-traverse on Pegasus-like MVC problems is listed in Table \ref{tab:mvc_connectivity_gurobi_traverse}

\begin{table}[htb]
\centering
\caption{Gurobi time-to-traverse on connectivity-varied MVC problems}
\label{tab:mvc_connectivity_gurobi_traverse}
\begin{tabular}{|l|l|l|l|}
\hline

\#degrees & Time   & \#degrees & Time   \\ \hline
1         & 0.0009 & 17        & 0.5694 \\ \hline
2         & 0.0014 & 18        & 0.4882 \\ \hline
3         & 0.0061 & 19        & 0.9114 \\ \hline
4         & 0.0128 & 20        & 0.4998 \\ \hline
5         & 0.0252 & 30        & 0.9279 \\ \hline
6         & 0.0806 & 40        & 0.9257 \\ \hline
7         & 0.0722 & 50        & 3.0588 \\ \hline
8         & 0.1579 & 60        & 0.3693 \\ \hline
9         & 0.5501 & 70        & 0.1204 \\ \hline
10        & 0.967  & 80        & 1.1727 \\ \hline
11        & 0.4539 & 90        & 1.0164 \\ \hline
12        & 0.4778 & 100       & 0.4307 \\ \hline
13        & 0.4948 & 110       & 0.9458 \\ \hline
14        & 1.6128 & 120       & 0.4478 \\ \hline
15        & 0.523  & 130       & 0.1147 \\ \hline
16        & 0.4513 & 140       & 0.0948 \\ \hline
\end{tabular}
\end{table}

\clearpage

\subsection{MVC Benchmark from DIMACS 10th Challenge}
\label{sec:appendix_mvc_dimacs10}

The graphs in the experiment are extracted from DIMACS 10th Challenge (\url{https://www.cc.gatech.edu/dimacs10/index.shtml}). Authors with interests can go to Github and find useful course works on the MVC benchmark in CSE-6410 from Georgia Tech. Students of this course have made a few comprehensive benchmarks of heuristic methods on these MVC problems. A few examples are listed below:

\begin{itemize}
    \item \url{https://github.com/Z-Jiang/CS-6140}
    \item \url{https://github.com/sangyh/minimum-vertex-cover}
    \item \url{https://github.com/ChujieChen/Minimum-Vertex-Cover}
    \item \url{https://github.com/xwave7/minimum-vertex-cover}
    \item \url{https://github.com/arjunchint/Minimum-Vertex-Cover}
\end{itemize}

\begin{figure}[htb]
     \centering
     \begin{subfigure}[b]{0.45\linewidth}
        \centering
    	\includegraphics[width=\textwidth]{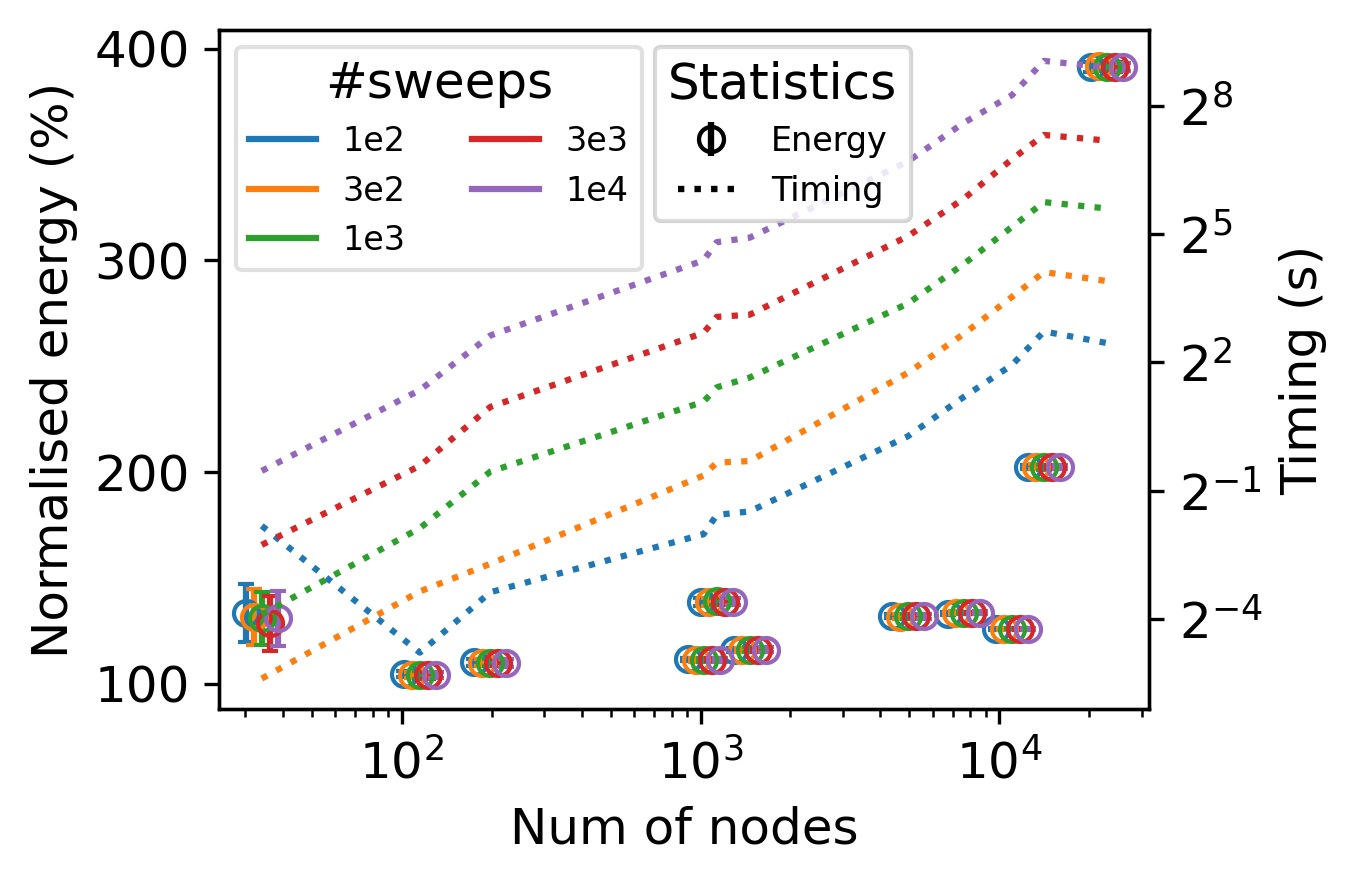}
    	\caption{Error bar of energy and timing}
    	\label{fig:app_mvc_dimacs10_sa_energy}
     \end{subfigure}
     \begin{subfigure}[b]{0.45\linewidth}
        \centering
    	\includegraphics[width=\textwidth]{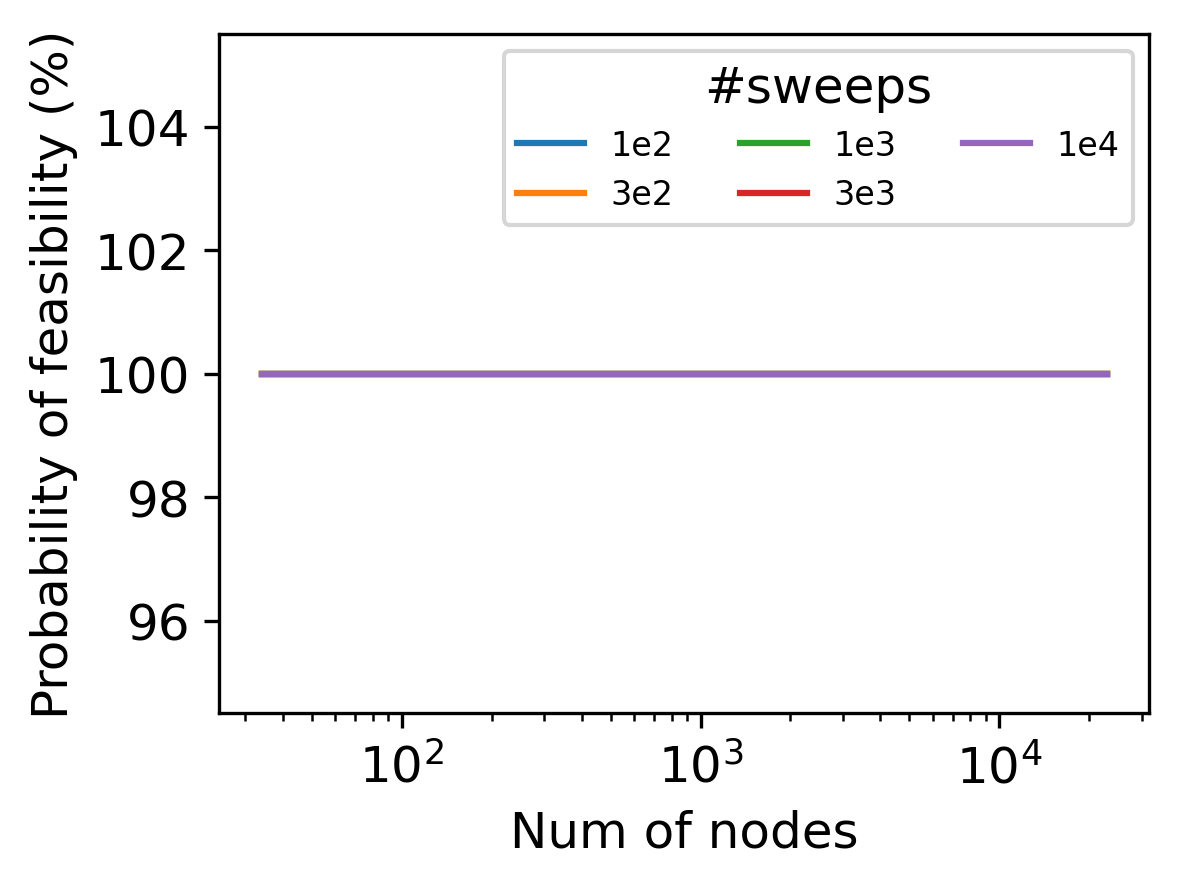}
    	\caption{Probability of feasibility}
    	\label{fig:app_mvc_dimacs10_sa_pf}
     \end{subfigure}
    \caption{SA on DIMACS10 MVC problems. The setting of the plots is the same as that of figure \ref{fig:app_mvc_pegasus_pegasus}, except that we skip the plot of constraint violation, as the solutions returned by SA are all feasible.}
    \label{fig:app_mvc_dimacs10_sa}
\end{figure}

Figure \ref{fig:app_mvc_dimacs10_sa} shows the performance of SA on DIMACS 10th Challenge MVC problems. For SA, large \#sweeps translates to longer time cost, but make no big difference in terms of energy.

\begin{figure*}[tb]
     \centering
     \begin{subfigure}[b]{0.38\textwidth}
        \centering
    	\includegraphics[width=\textwidth]{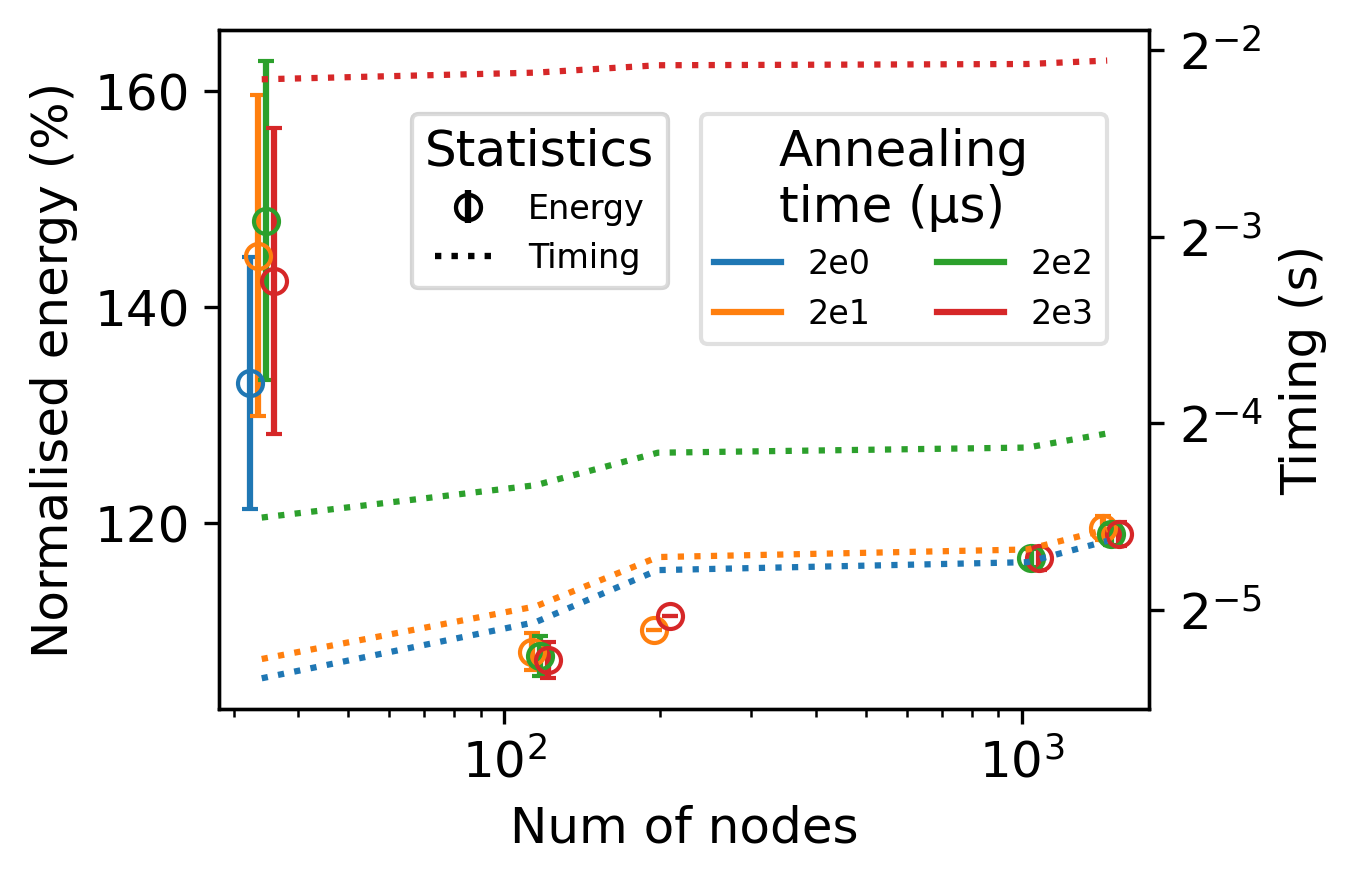}
    	\caption{Error bar of energy and timing}
    	\label{fig:app_mvc_dimacs10_pegasus_energy}
     \end{subfigure}
     \begin{subfigure}[b]{0.33\textwidth}
        \centering
    	\includegraphics[width=\textwidth]{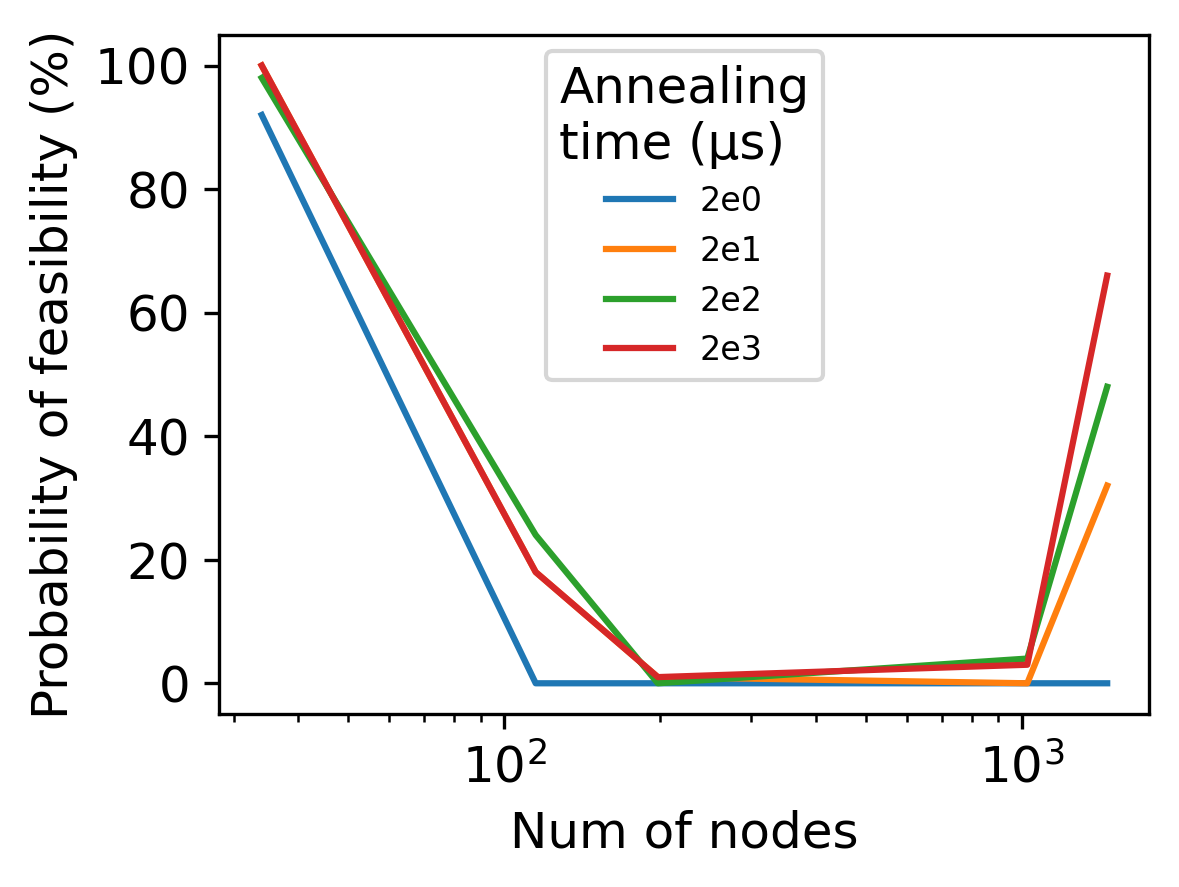}
    	\caption{Probability of feasibility}
    	\label{fig:app_mvc_dimacs10_pegasus_pf}
     \end{subfigure}
     \begin{subfigure}[b]{0.95\textwidth}
        \centering
    	\includegraphics[width=\textwidth]{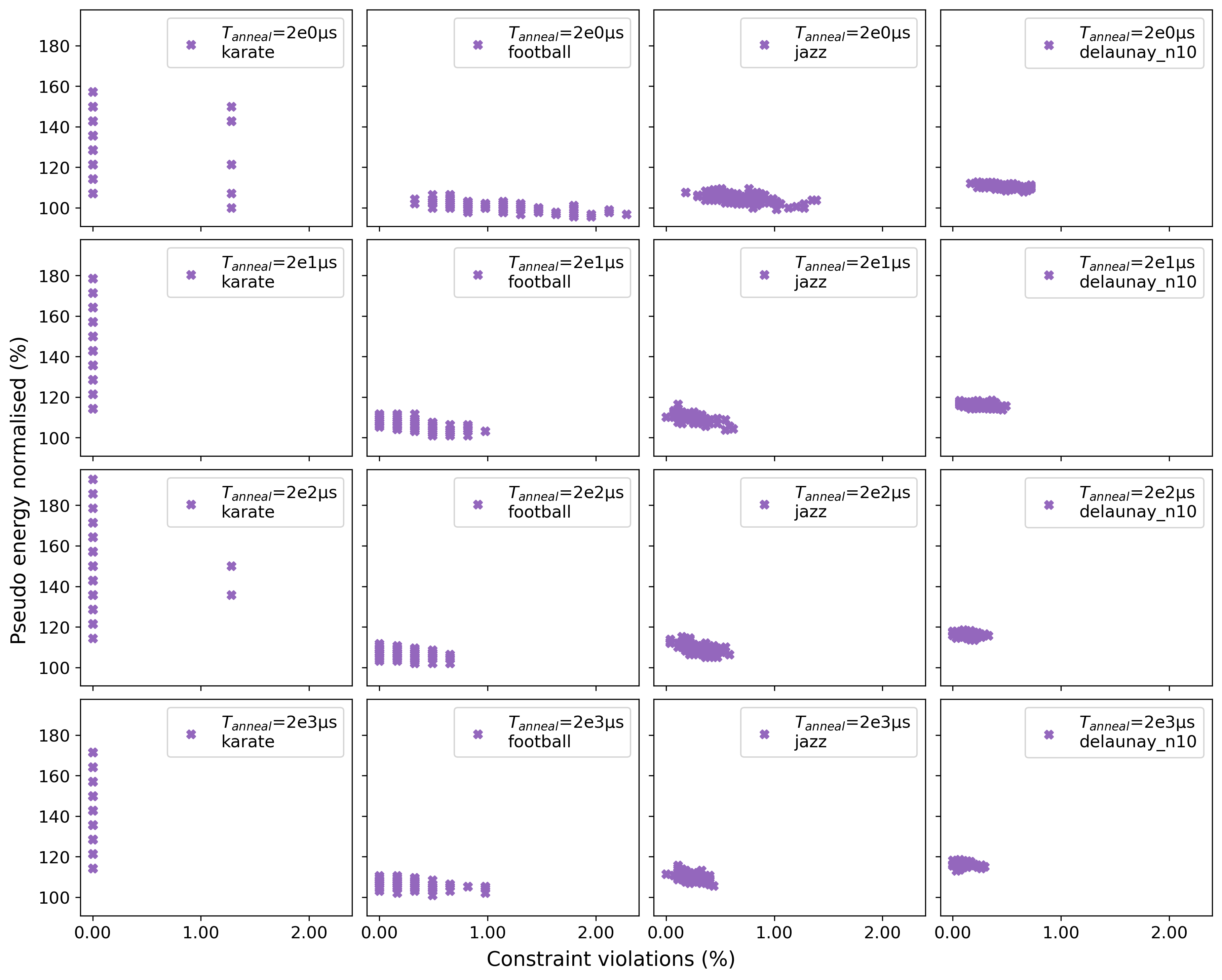}
    	\caption{Constraint violation}
    	\label{fig:app_mvc_dimacs10_pegasus_violation}
     \end{subfigure}
    \caption{D-Wave Pegasus on DIMACS10 MVC problems}
    \label{fig:app_mvc_dimacs10_pegasus}
\end{figure*}

Figure \ref{fig:app_mvc_dimacs10_pegasus} shows the performance of D-Wave Pegasus on DIMACS 10th Challenge MVC problems. As the topology of the graphs in the dataset varies largely, we cannot observe clear trend from figure \ref{fig:app_mvc_dimacs10_pegasus_energy} and \ref{fig:app_mvc_dimacs10_pegasus_pf}. But we can at least understand that Pegasus is able to find very limited number of feasible solutions, when the problem size is relatively. Annealing time as short as 2 $\mu s$ can hardly produce feasible solutions for graphs with over 100 nodes. This observation match with the previous synthetic datasets. 

From figure \ref{fig:app_mvc_pegasus_pegasus_violation} we understand there is limited amount of constraint violation in the solutions, as most of clusters are very close to the 0\% violation line. With a naive fixation step, one can save values from those broken solutions.

\begin{figure*}[tb]
     \centering
     \begin{subfigure}[b]{0.38\textwidth}
        \centering
    	\includegraphics[width=\textwidth]{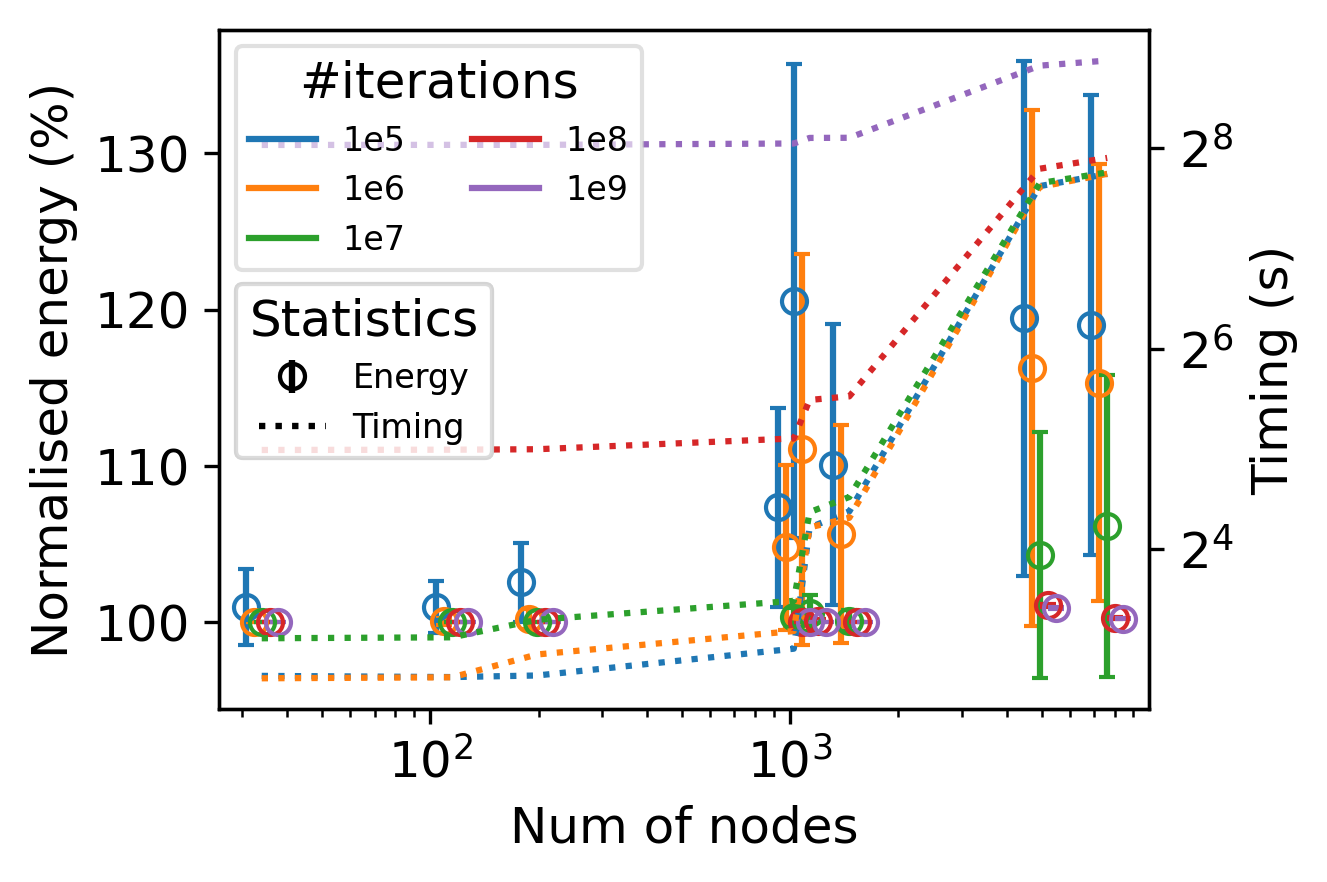}
    	\caption{Error bar of energy and timing}
    	\label{fig:app_mvc_dimacs10_da_energy}
     \end{subfigure}
     \begin{subfigure}[b]{0.33\textwidth}
        \centering
    	\includegraphics[width=\textwidth]{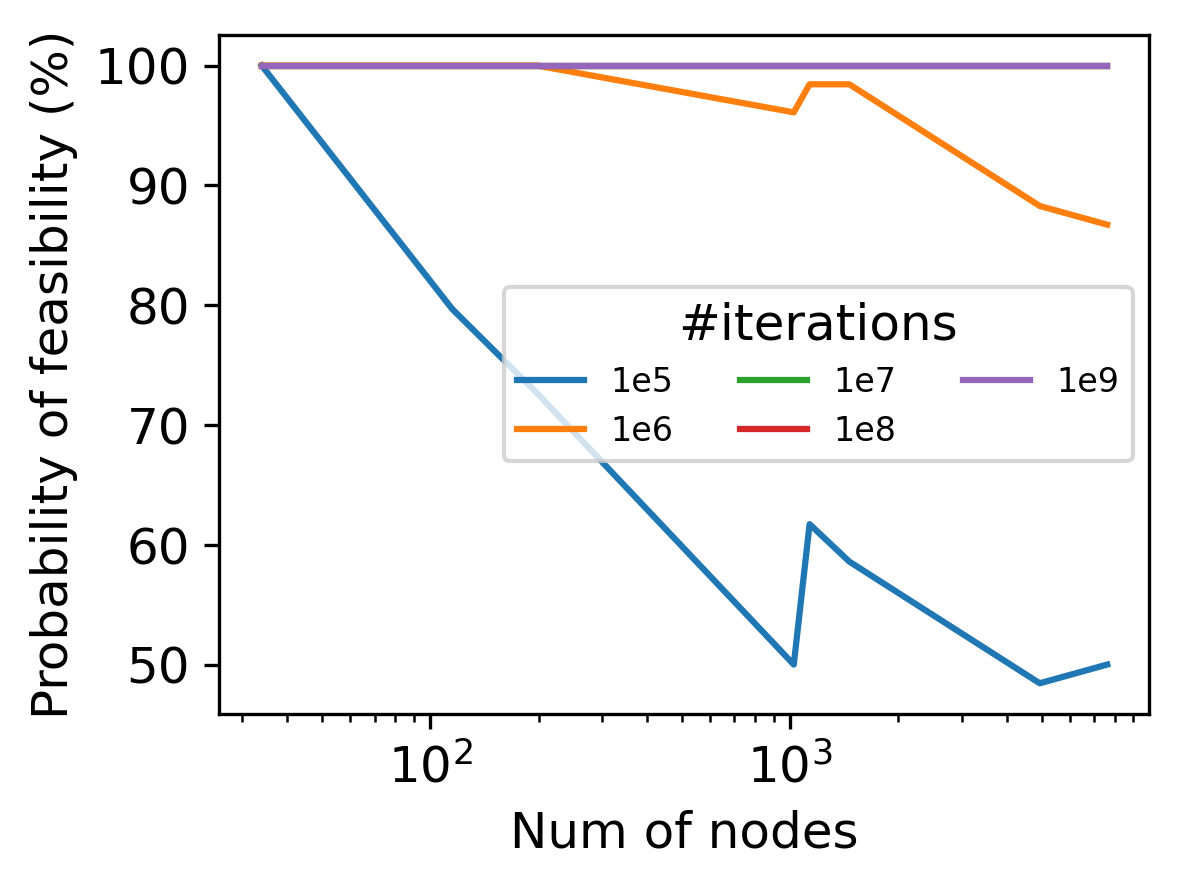}
    	\caption{Probability of feasibility}
    	\label{fig:app_mvc_dimacs10_da_pf}
     \end{subfigure}
     \begin{subfigure}[b]{0.95\textwidth}
        \centering
    	\includegraphics[width=\textwidth]{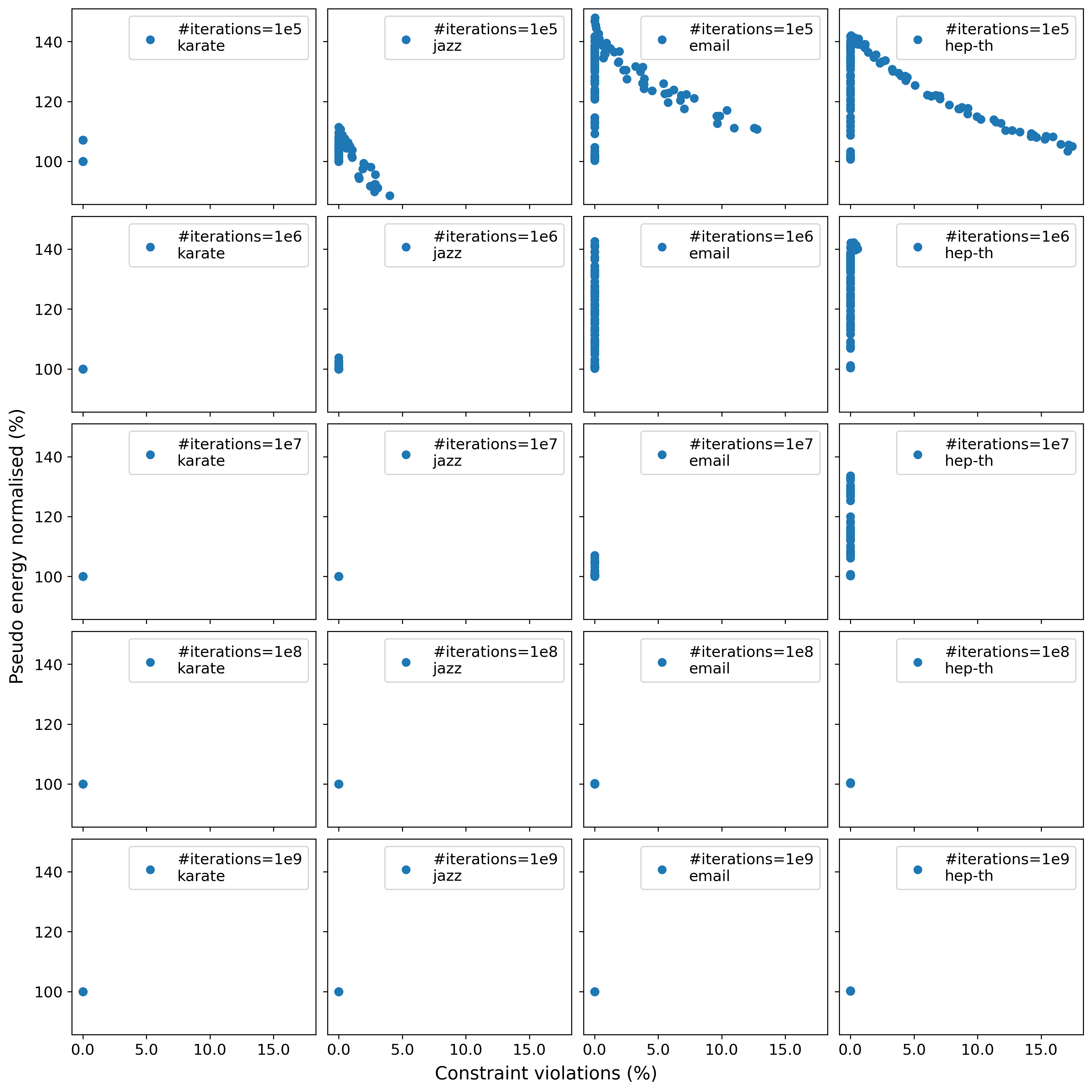}
    	\caption{Constraint violation}
    	\label{fig:app_mvc_dimacs10_da_violation}
     \end{subfigure}
    \caption{DA on DIMACS10 MVC problems}
    \label{fig:app_mvc_dimacs10_da}
\end{figure*}

Figure \ref{fig:app_mvc_dimacs10_da} shows the performance of DA on DIMACS 10th Challenge MVC problems. This set of plots are similar to those in figure \ref{fig:app_mvc_connectivity_da}, except that the problem size in terms of number of nodes is much smaller and trivial to DA. DA can solve these problems more quickly.

For Pegasus, we only include the results of $2000\mu s$ in the main text for comparison between solvers. For DA and SA We only include the results of \#iterations=$10^6$ and \#sweeps=$10^2$ in the main text for comparison between solvers.

\begin{figure}[htb]
    \centering
	\includegraphics[width=0.45\linewidth]{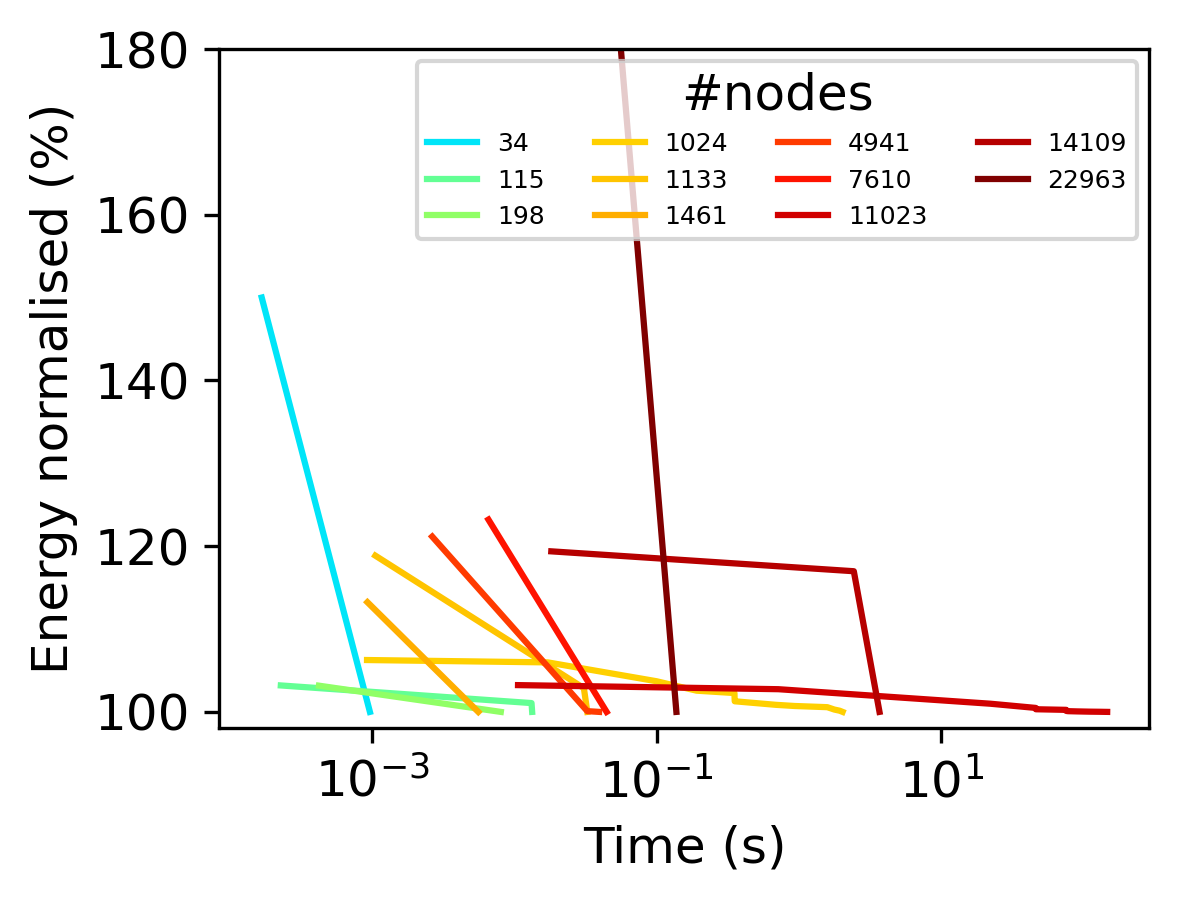}
	\caption{Gurobi time-to-solution}
	\label{fig:app_mvc_dimacs10_gurobi}
\end{figure}

Figure \ref{fig:app_mvc_dimacs10_gurobi} shows the Gurobi time-to-solution plot. Gurobi can find promising solutions within a second.

The ``dimacs10-netscience'' problem describes a network of co-authorships in the area of network science. It has 1461 nodes and 2742 edges. The more detailed description of this problem can be found via \url{http://konect.cc/networks/dimacs10-netscience/}.

%


We use the benchmark from DIMACS 10th Challenge in 2012, because we believe it closely follows the trend of requirements nowadays. However, most of the previous works on heuristic MVC methods are based on the benchmark from DIMACS 2nd Challenge in 1992. We evaluate this outdated benchmark as well.

\clearpage

\subsection{MVC Benchmark from DIMACS 2nd Challenge}
\label{sec:appendix_mvc_dimacs2}

The MVC dataset from DIMACS 2nd Challenge \url{http://archive.dimacs.rutgers.edu/pub/challenge/graph/benchmarks/volume/Clique/} is one of the most widely adopted \cite{wang2019exact, cai2013numvc, li2020numwvc} benchmarks for evaluating MVC related algorithms, despite of the fact that it is a 30-ish years old benchmark. We do not include quantum annealer in this experiment because most of the problem instances in the benchmark is too large or dense for quantum annealers. We set \#iterations=$10^8$ for DA and \#sweeps=$10^3$ for SA.

\begin{table*}[tb]
\centering
\caption{Performance on MVC benchmark from DIMACS 2nd Challenge. The performance is shown in Energy/Timing (s)/$P_f$ format.}
\label{tab:mvc_dimacs2}
\begin{tabular}{|l|c|c|c|c|c|}
\hline
Name             & \#nodes/\#edges  & Optimal & da            & qbsolv         & sa             \\ \hline
brock200\_2      & 200    / 9876    & 188     & 189/27.033/1  & 189/4.29/1     & 192/2.213/1    \\ \hline
brock200\_4      & 200    / 13089   & 183     & 192/27.025/1  & 192/1.384/1    & 193/2.398/1    \\ \hline
brock400\_2      & 400    / 59786   & 371     & 392/27.316/1  & 392/3.149/1    & 395/10.057/1   \\ \hline
brock400\_4      & 400    / 59765   & 367     & 393/27.317/1  & 393/2.544/1    & 395/10.093/1   \\ \hline
brock800\_2      & 800    / 208166  & 776     & 790/27.846/1  & 790/8.71/1     & 793/34.695/1   \\ \hline
brock800\_4      & 800    / 207643  & 774     & 790/27.834/1  & 790/9.48/1     & 792/35.435/1   \\ \hline
C125.9           & 125    / 6963    & 91      & 121/26.924/1  & 121/0.665/1    & 122/1.308/1    \\ \hline
C250.9           & 250    / 27984   & 206     & 245/27.107/1  & 245/1.659/1    & 245/4.678/1    \\ \hline
C500.9           & 500    / 112332  & 443     & 495/27.468/1  & 495/7.144/1    & 496/18.693/1   \\ \hline
C1000.9          & 1000   / 450079  & 932     & 994/28.183/1  & 995/14.26/1    & 996/74.469/1   \\ \hline
C2000.5          & 2000   / 999836  & 1984    & 1983/32.724/1 & 1985/42.001/1  & 1989/178.126/1 \\ \hline
C2000.9          & 2000   / 1799532 & 1920    & 1994/32.946/1 & 1995/55.965/1  & 1995/318.634/1 \\ \hline
C4000.5          & 4000   / 4000268 & 3982    & 3982/49.541/1 & 3985/147.304/1 & 3988/730.413/1 \\ \hline
DSJC500.5        & 500    / 62624   & 487     & 487/27.457/1  & 487/4.353/1    & 490/10.871/1   \\ \hline
DSJC1000.5       & 1000   / 249826  & 985     & 985/28.144/1  & 986/11.126/1   & 990/43.229/1   \\ \hline
gen200\_p0.9\_44 & 200    / 17910   & 156     & 195/27.028/1  & 195/1.01/1     & 196/3.105/1    \\ \hline
gen200\_p0.9\_55 & 200    / 17910   & 145     & 195/27.03/1   & 195/1.33/1     & 196/3.109/1    \\ \hline
gen400\_p0.9\_55 & 400    / 71820   & 345     & 392/27.314/1  & 392/2.408/1    & 392/11.949/1   \\ \hline
gen400\_p0.9\_65 & 400    / 71820   & 335     & 393/27.303/1  & 393/2.653/1    & 394/12.024/1   \\ \hline
gen400\_p0.9\_75 & 400    / 71820   & 325     & 394/27.315/1  & 394/2.599/1    & 395/11.878/1   \\ \hline
hamming8-4       & 256    / 20864   & 240     & 240/27.113/1  & 240/1.098/1    & 243/3.69/1     \\ \hline
hamming10-4      & 1024   / 434176  & 984     & 1004/28.229/1 & 1004/13.756/1  & 1008/71.61/1   \\ \hline
keller4          & 171    / 9435    & 160     & 156/26.995/1  & 156/0.808/1    & 159/1.797/1    \\ \hline
keller5          & 776    / 225990  & 749     & 745/27.805/1  & 745/8.611/1    & 754/38.329/1   \\ \hline
keller6          & 3361   / 4619898 & 3302    & 3298/47.664/1 & 3298/159.914/1 & 3318/835.81/1  \\ \hline
MANN\_a27        & 378    / 70551   & 252     & 375/27.277/1  & 375/2.402/1    & 375/11.581/1   \\ \hline
MANN\_a45        & 1035   / 533115  & 690     & 1032/30.363/1 & 1032/16.445/1  & 1032/88.17/1   \\ \hline
MANN\_a81        & 3321   / 5506380 & 2221    & 3318/47.809/1 & 3318/172.374/1 & 3318/971.542/1 \\ \hline
p\_hat300-1      & 300    / 10933   & 292     & 261/27.179/1  & 261/1.868/1    & 271/2.329/1    \\ \hline
p\_hat300-2      & 300    / 21928   & 275     & 273/27.181/1  & 273/1.816/1    & 281/4.106/1    \\ \hline
p\_hat300-3      & 300    / 33390   & 264     & 291/27.168/1  & 291/2.674/1    & 293/5.786/1    \\ \hline
p\_hat700-1      & 700    / 60999   & 689     & 635/27.798/1  & 637/4.187/1    & 660/11.511/1   \\ \hline
p\_hat700-2      & 700    / 121728  & 656     & 651/27.807/1  & 653/6.29/1     & 668/21.366/1   \\ \hline
p\_hat700-3      & 700    / 183010  & 638     & 690/27.708/1  & 690/8.831/1    & 692/30.694/1   \\ \hline
p\_hat1500-1     & 1500   / 284923  & 1488    & 1413/31.966/1 & 1417/13.831/1  & 1449/52.967/1  \\ \hline
p\_hat1500-2     & 1500   / 568960  & 1435    & 1438/31.598/1 & 1440/21.961/1  & 1462/100.844/1 \\ \hline
p\_hat1500-3     & 1500   / 847244  & 1406    & 1488/31.725/1 & 1489/29.186/1  & 1492/147.645/1 \\ \hline
\end{tabular}
\end{table*}

\clearpage

\section{qap}

\subsection{TinyQAP problems}
\label{sec:appendix_qap_tiny}

\textbf{Problem generation} We start the problem generation by randomly generating $n$ points in a two-dimensional space using ``np.random.rand'' from NumPy with a random seed of 1234. The $n$ points are saved for reserved usage. We then generate the distance adjacency matrix from the list of points. For the generation of the flow matrix, we generate $F_{n\times n}=(m+m.T) / 2$, where $m$ is an $n\times n$ matrix generated by the same random method. We set the diagonal of $F_{n\times n}$ to be zero to eliminate loops.


We evaluate Chimera, Pegasus, DA and SA on this dataset. Specifically, we restrict Chimera on problem of size below 8, since this is the largest QAP it can handle.

\begin{figure*}[tb]
     \centering
     \begin{subfigure}[b]{0.38\textwidth}
        \centering
    	\includegraphics[width=\textwidth]{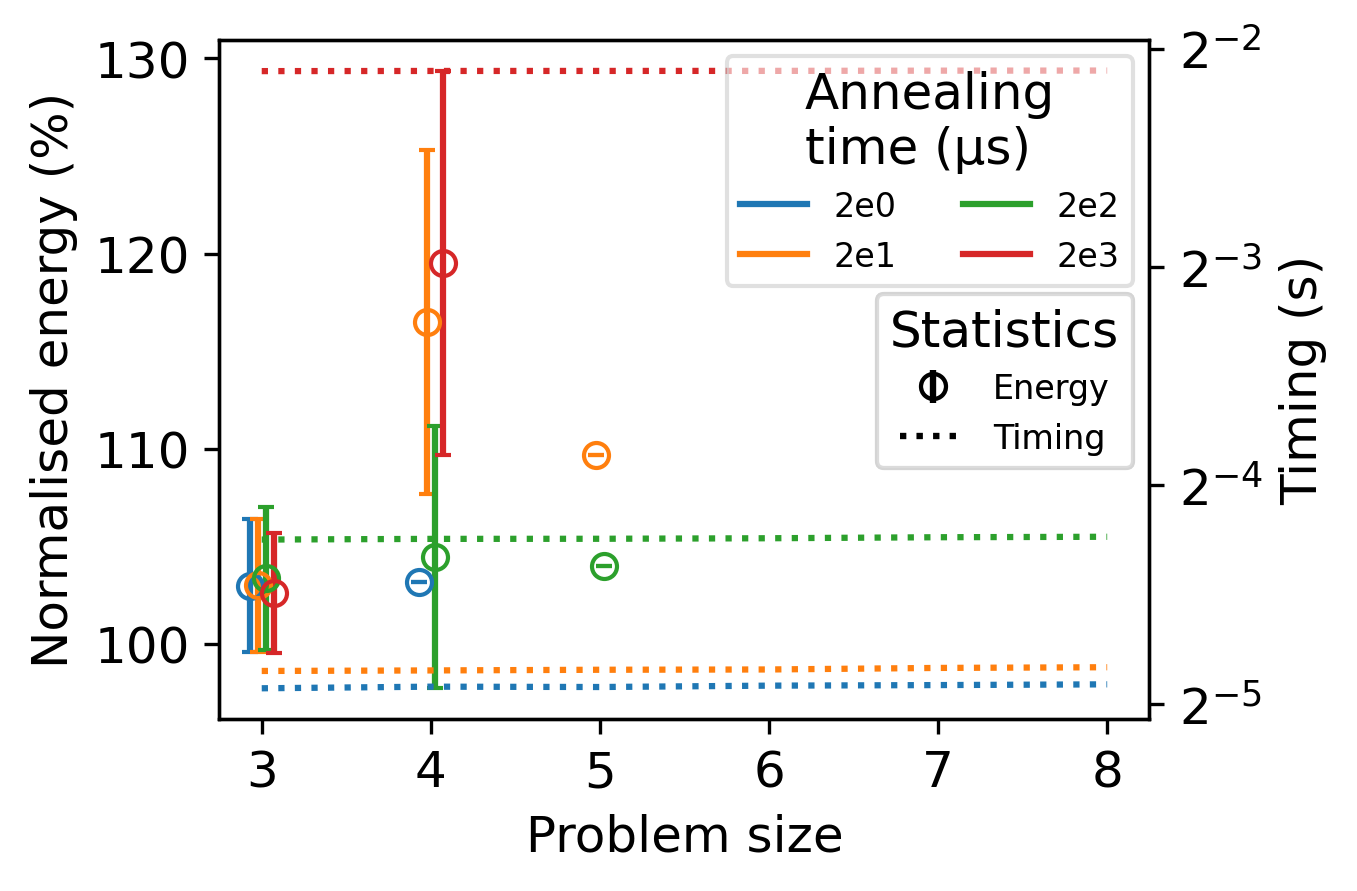}
    	\caption{Error bar of energy and timing}
    	\label{fig:app_qap_tinyqap_chimera_energy}
     \end{subfigure}
     \begin{subfigure}[b]{0.33\textwidth}
        \centering
    	\includegraphics[width=\textwidth]{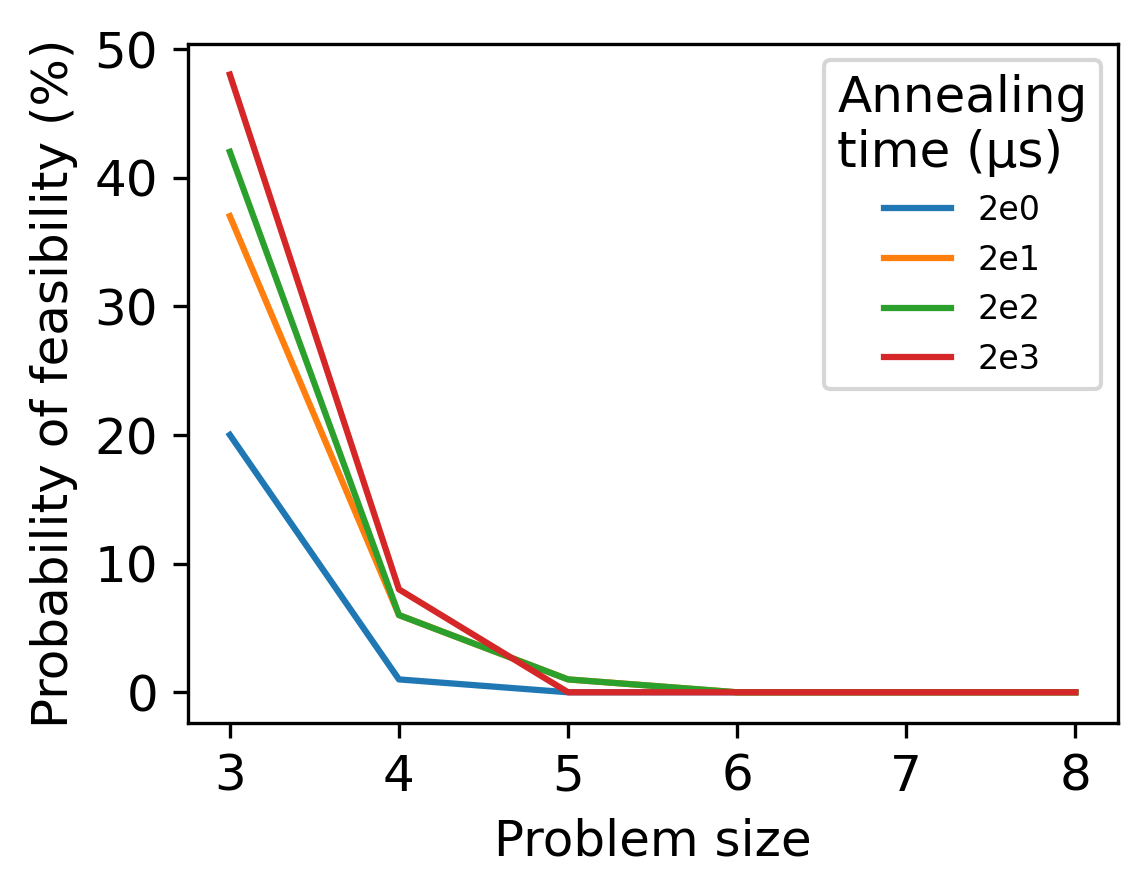}
    	\caption{Probability of feasibility}
    	\label{fig:app_qap_tinyqap_chimera_pf}
     \end{subfigure}
     \begin{subfigure}[b]{0.95\textwidth}
        \centering
    	\includegraphics[width=\textwidth]{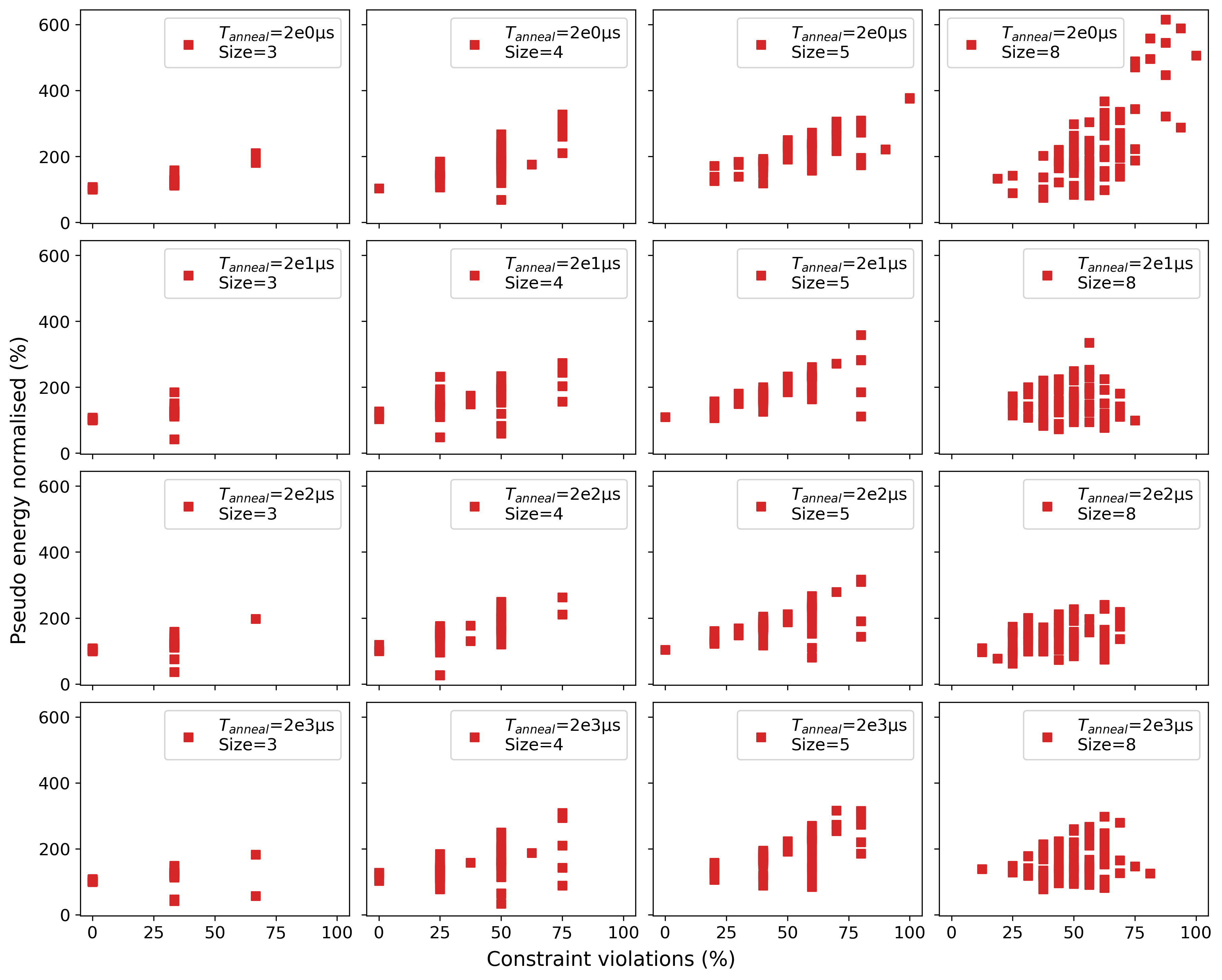}
    	\caption{Constraint violation}
    	\label{fig:app_qap_tinyqap_chimera_violation}
     \end{subfigure}
    \caption{D-Wave Chimera on Tinyqap. The plot setting is the same as that in figure \ref{fig:app_mvc_pegasus_pegasus}}
    \label{fig:app_qap_tinyqap_chimera}
\end{figure*}

Figure \ref{fig:app_qap_tinyqap_chimera} shows the performance of D-Wave Chimera on Tinyqap problems. Problem of size larger than 8 is not included because they are beyond the capacity of Chimera. From figure \ref{fig:app_qap_tinyqap_chimera_energy} we understand that longer annealing time does not have clear advantage over shorter ones in terms of energy as well as feasibility. This is also supported by figure \ref{fig:app_qap_tinyqap_chimera_pf}, which suggest that Chimera can hardly find any feasible solutions when the problem size is larger than 5. Figure \ref{fig:app_qap_tinyqap_chimera_violation}  shows that there is high level of constraints violation. The sub-plots in the upper right corner and lower left corner suggest there could be as much as 50\%-100\% constraints being violated in a solution. This translate to more efforts in fixing the broken constraints, compared with those solutions in MVC problems.

\begin{figure*}[tb]
     \centering
     \begin{subfigure}[b]{0.38\textwidth}
        \centering
    	\includegraphics[width=\textwidth]{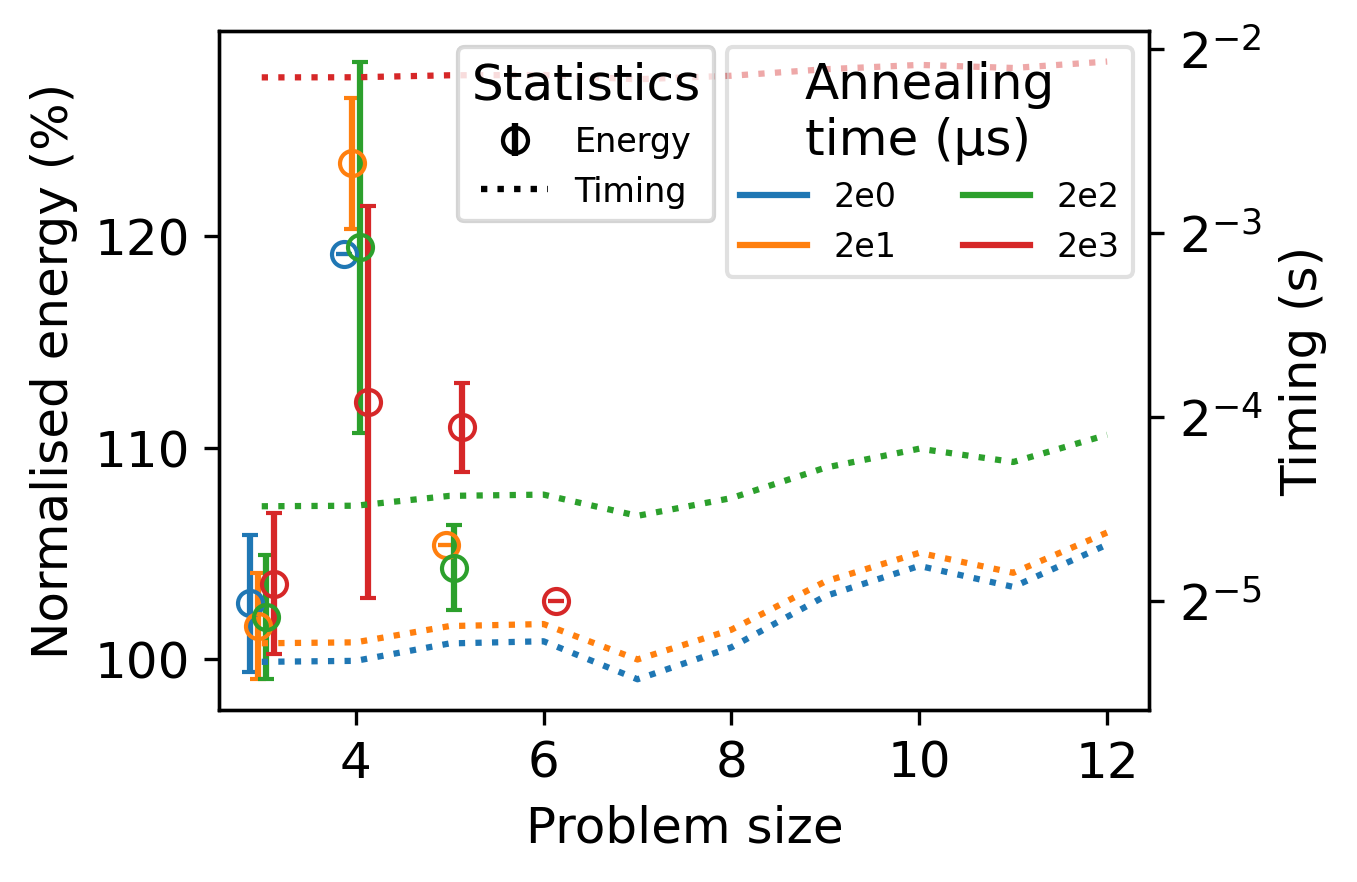}
    	\caption{Error bar of energy and timing}
    	\label{fig:app_qap_tinyqap_pegasus_energy}
     \end{subfigure}
     \begin{subfigure}[b]{0.33\textwidth}
        \centering
    	\includegraphics[width=\textwidth]{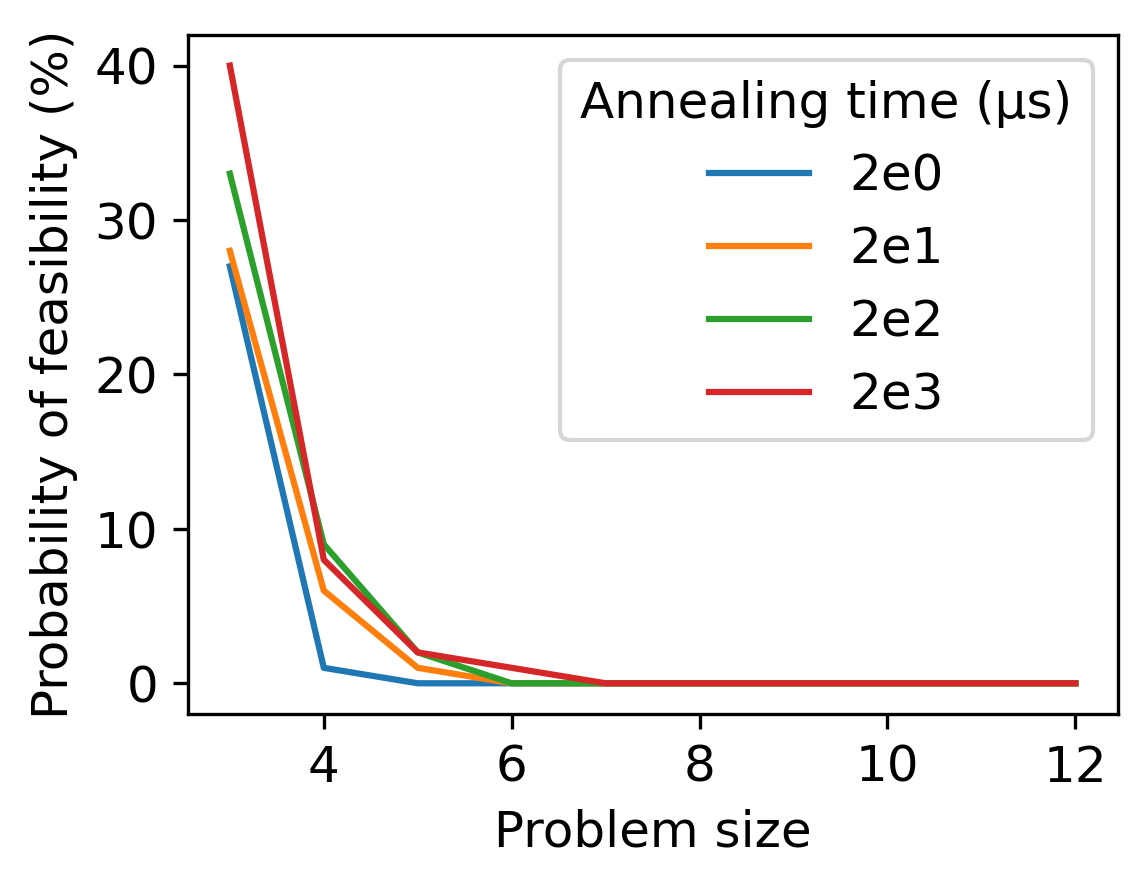}
    	\caption{Probability of feasibility}
    	\label{fig:app_qap_tinyqap_pegasus_pf}
     \end{subfigure}
     \begin{subfigure}[b]{0.95\textwidth}
        \centering
    	\includegraphics[width=\textwidth]{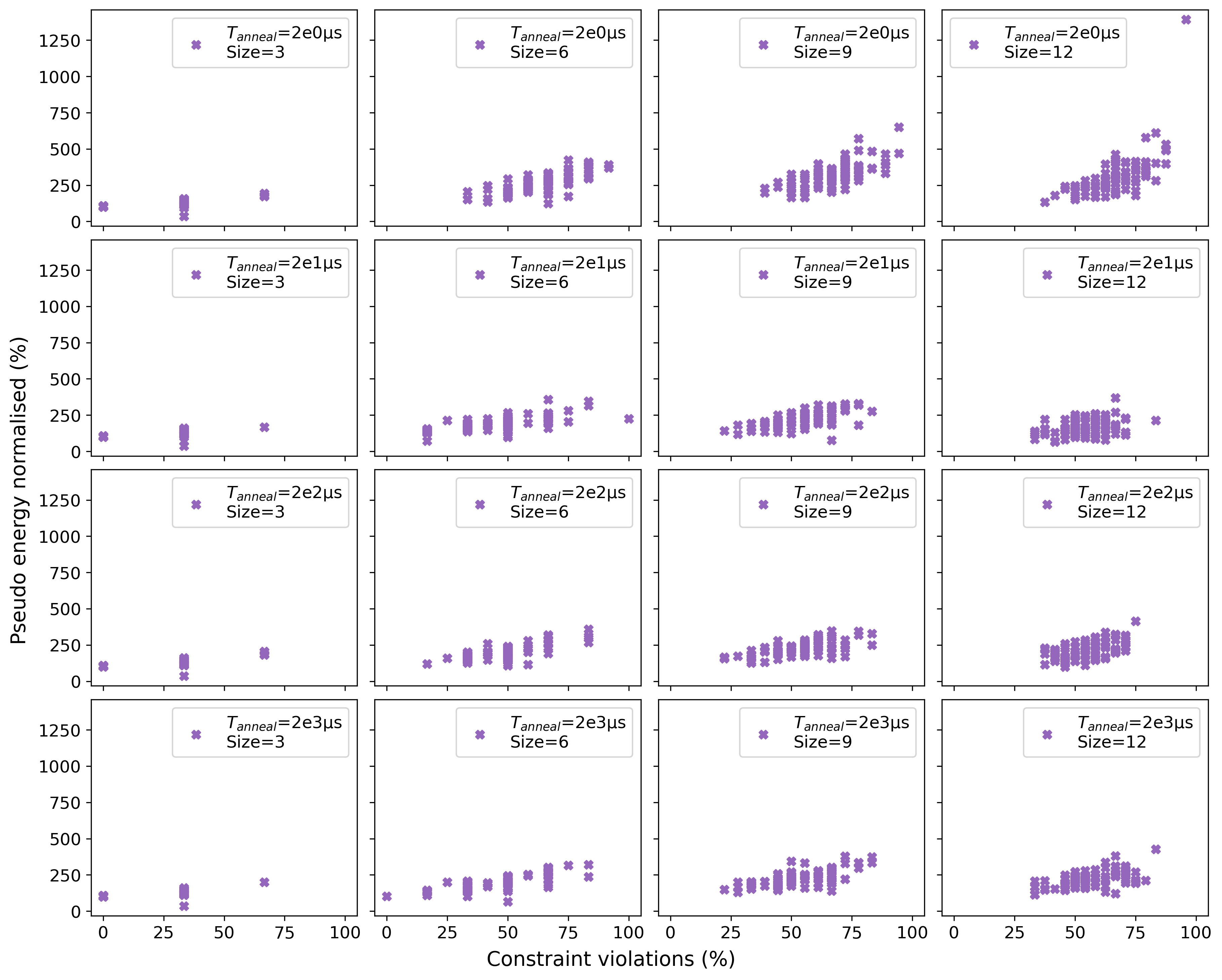}
    	\caption{Constraint violation}
    	\label{fig:app_qap_tinyqap_pegasus_violation}
     \end{subfigure}
    \caption{D-Wave Pegasus on Tinyqap. The plot setting is the same as that in figure \ref{fig:app_mvc_pegasus_pegasus}}
    \label{fig:app_qap_tinyqap_pegasus}
\end{figure*}

Figure \ref{fig:app_qap_tinyqap_pegasus} shows the performance of D-Wave Pegasus on Tinyqap problems. Comparing with figure \ref{fig:app_qap_tinyqap_chimera} we known that Pegasus has no clear advantage either in terms of energy of in feasibility. The solutions by Pegasus have high level of constraint violations, which also similar to that of Chimera.

\begin{figure*}[tb]
     \centering
     \begin{subfigure}[b]{0.38\textwidth}
        \centering
    	\includegraphics[width=\textwidth]{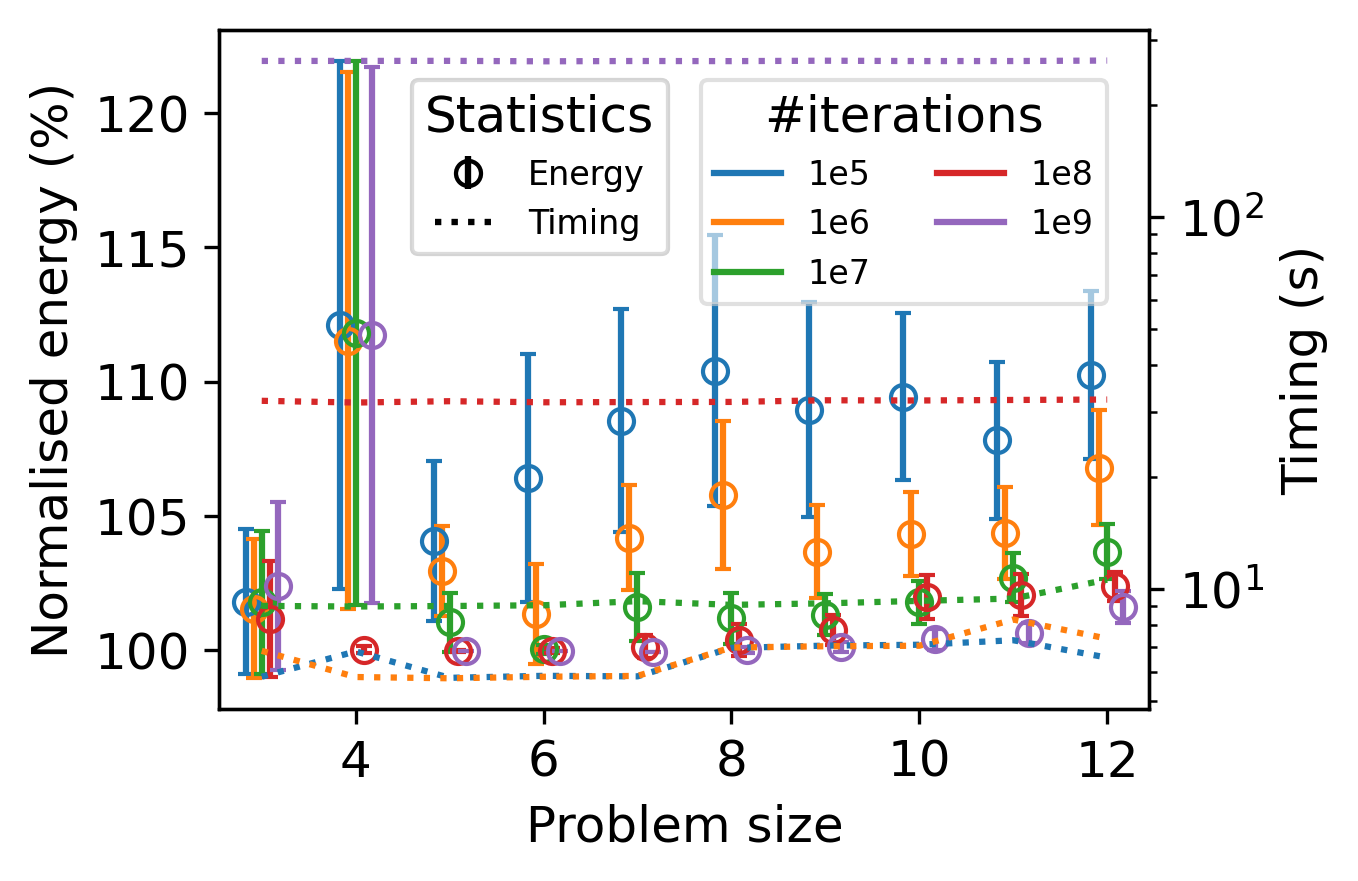}
    	\caption{Error bar of energy and timing}
    	\label{fig:app_qap_tinyqap_da_energy}
     \end{subfigure}
     \begin{subfigure}[b]{0.33\textwidth}
        \centering
    	\includegraphics[width=\textwidth]{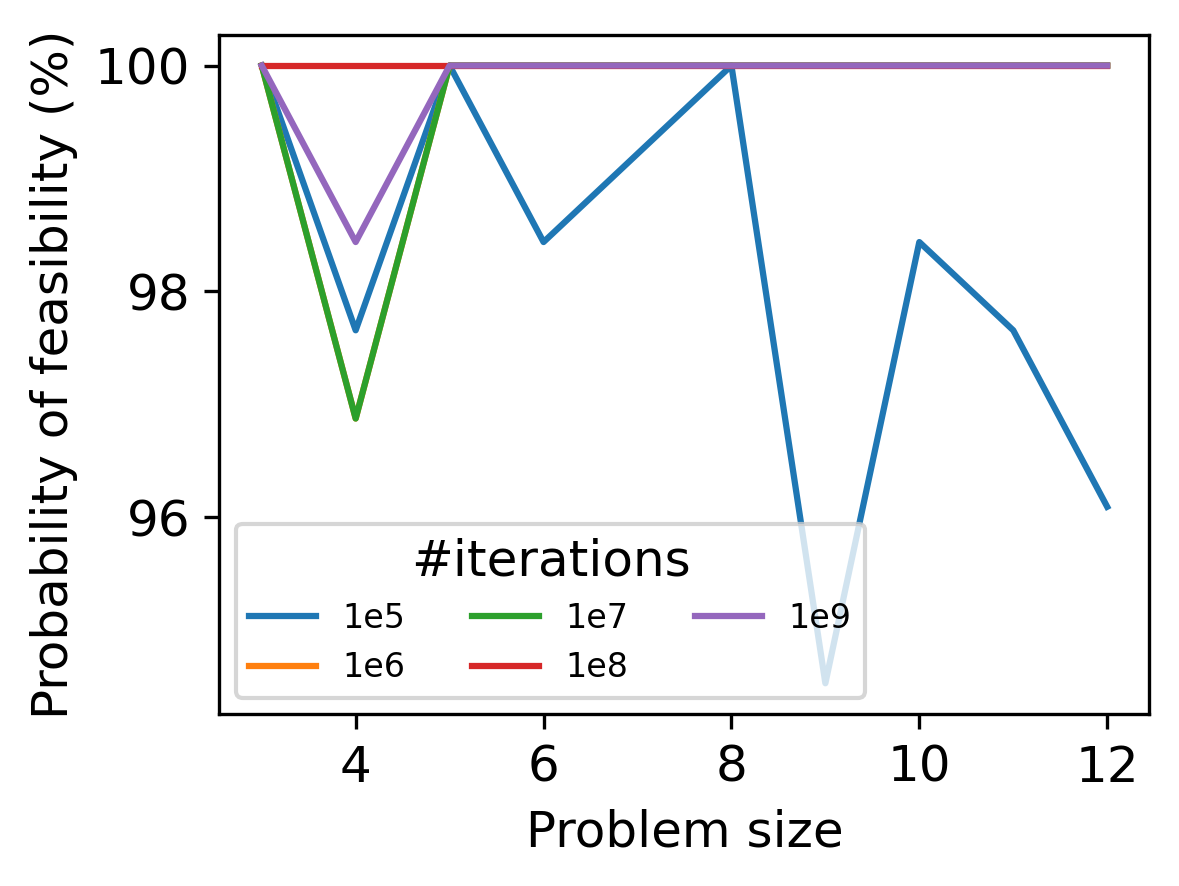}
    	\caption{Probability of feasibility}
    	\label{fig:app_qap_tinyqap_da_pf}
     \end{subfigure}
     \begin{subfigure}[b]{0.95\textwidth}
        \centering
    	\includegraphics[width=\textwidth]{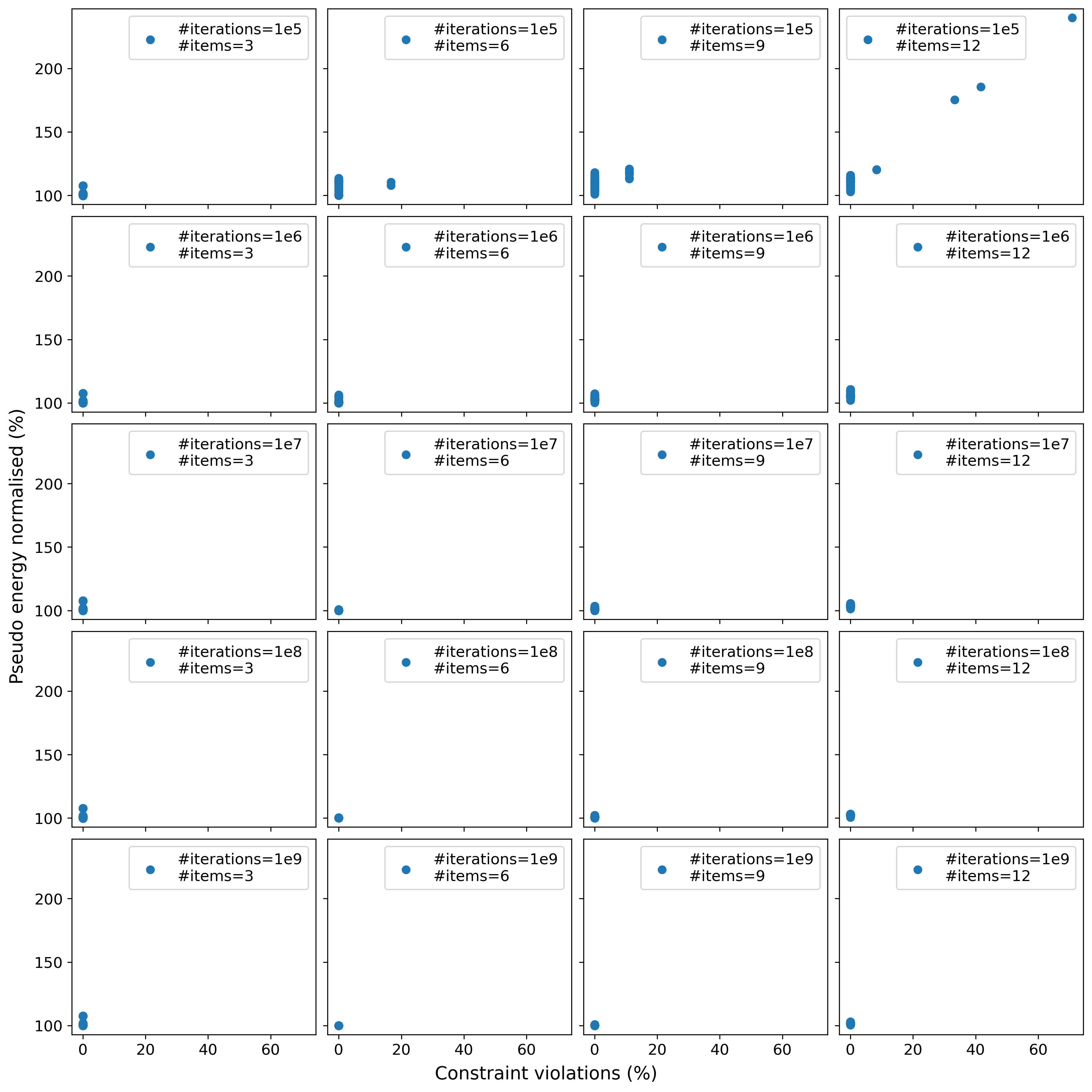}
    	\caption{Constraint violation}
    	\label{fig:app_qap_tinyqap_da_violation}
     \end{subfigure}
    \caption{DA on Tinyqap. The plot setting is the same as that in figure \ref{fig:app_mvc_pegasus_pegasus}}
    \label{fig:app_qap_tinyqap_da}
\end{figure*}

Figure \ref{fig:app_qap_tinyqap_pegasus} shows the performance of DA on Tinyqap problems. Figure \ref{fig:app_qap_tinyqap_da_energy} suggest that higher \#iterations can produce better results. This is consistent across all problem size. Figure \ref{fig:app_qap_tinyqap_da_pf} suggest that DA can find feasible solutions in most cases. \#iterations as low as $10^5$ only suffers a few percent loss in probability of feasibility. Figure \ref{fig:app_qap_tinyqap_da_violation} suggest that the infeasible cases has much higher level of constraint violations, compared with DA's performance on MVC problems.

\begin{figure*}[tb]
     \centering
     \begin{subfigure}[b]{0.38\textwidth}
        \centering
    	\includegraphics[width=\textwidth]{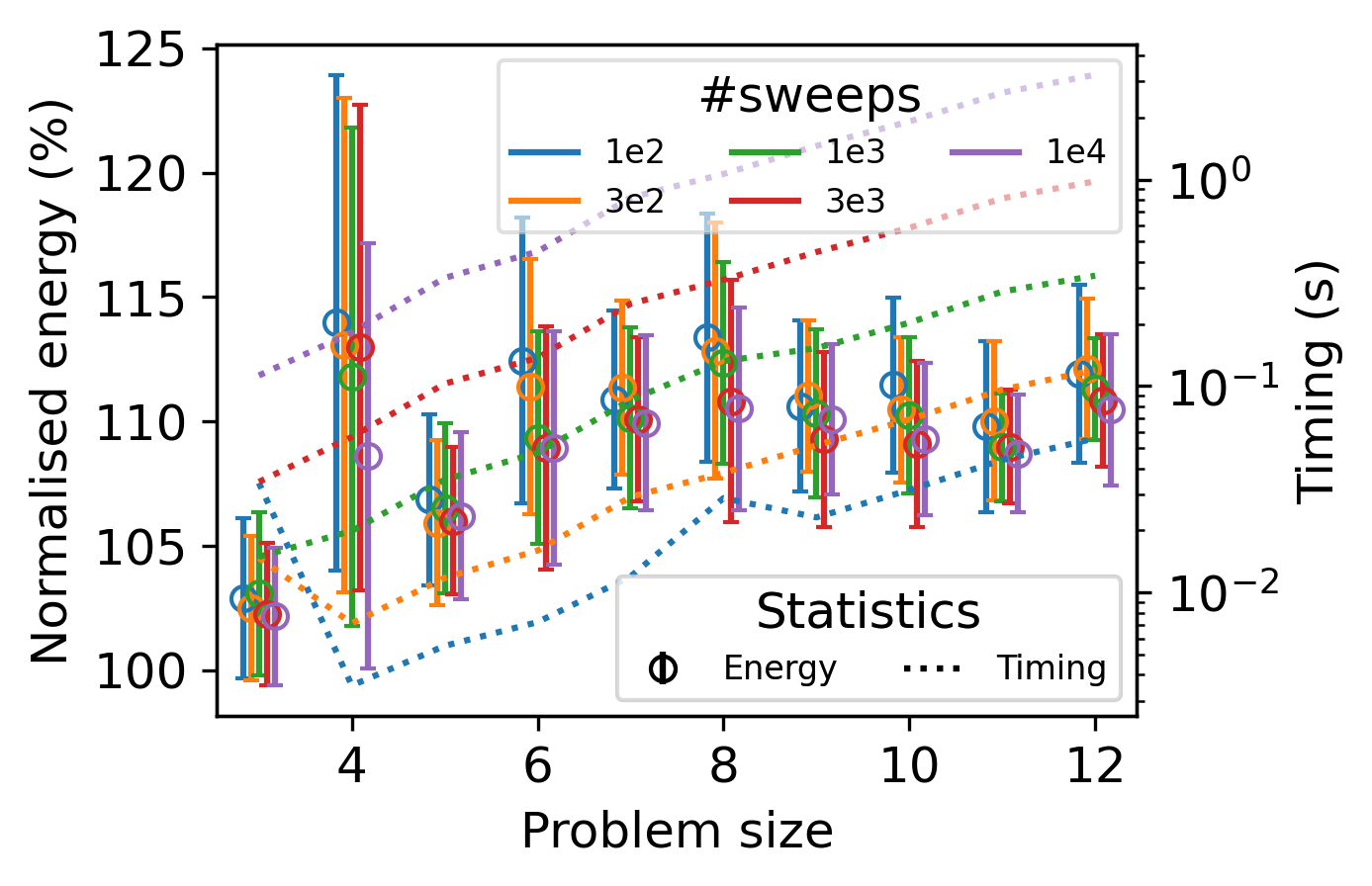}
    	\caption{Error bar of energy and timing}
    	\label{fig:app_qap_tinyqap_sa_energy}
     \end{subfigure}
     \begin{subfigure}[b]{0.33\textwidth}
        \centering
    	\includegraphics[width=\textwidth]{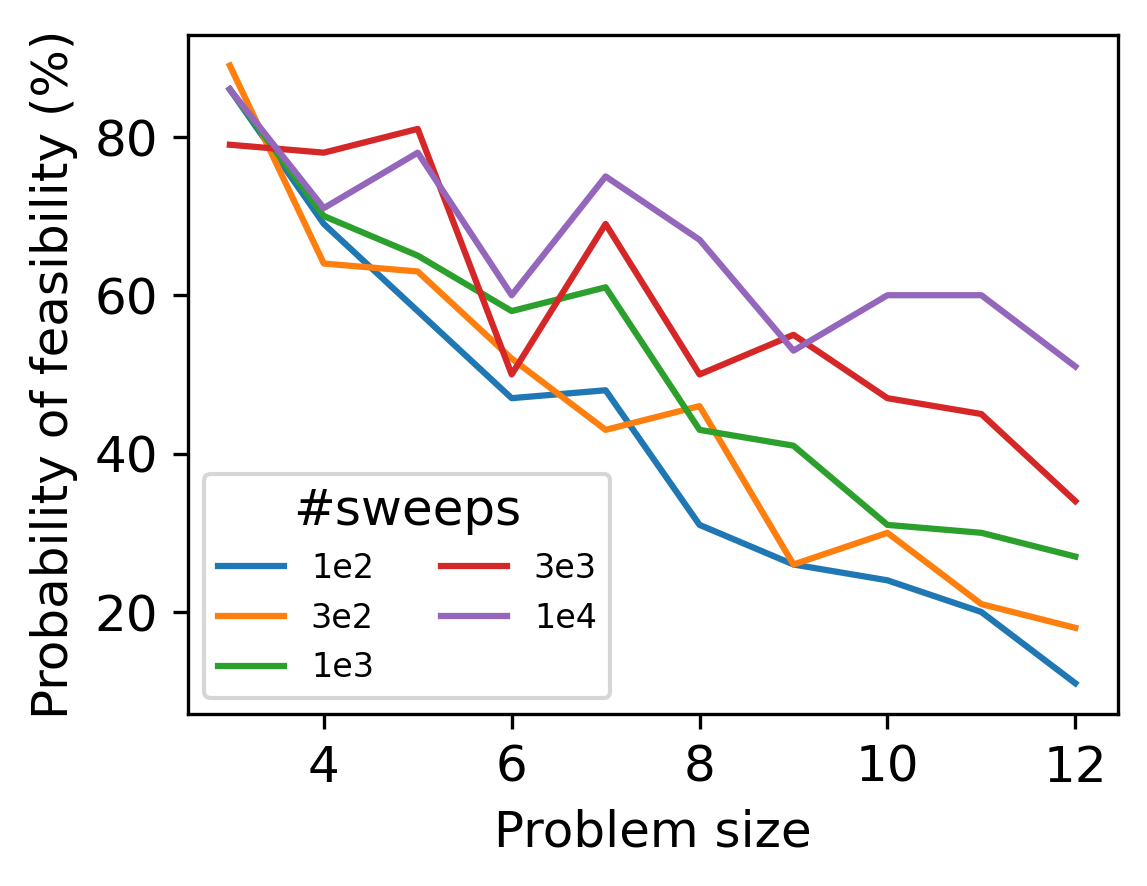}
    	\caption{Probability of feasibility}
    	\label{fig:app_qap_tinyqap_sa_pf}
     \end{subfigure}
     \begin{subfigure}[b]{0.95\textwidth}
        \centering
    	\includegraphics[width=\textwidth]{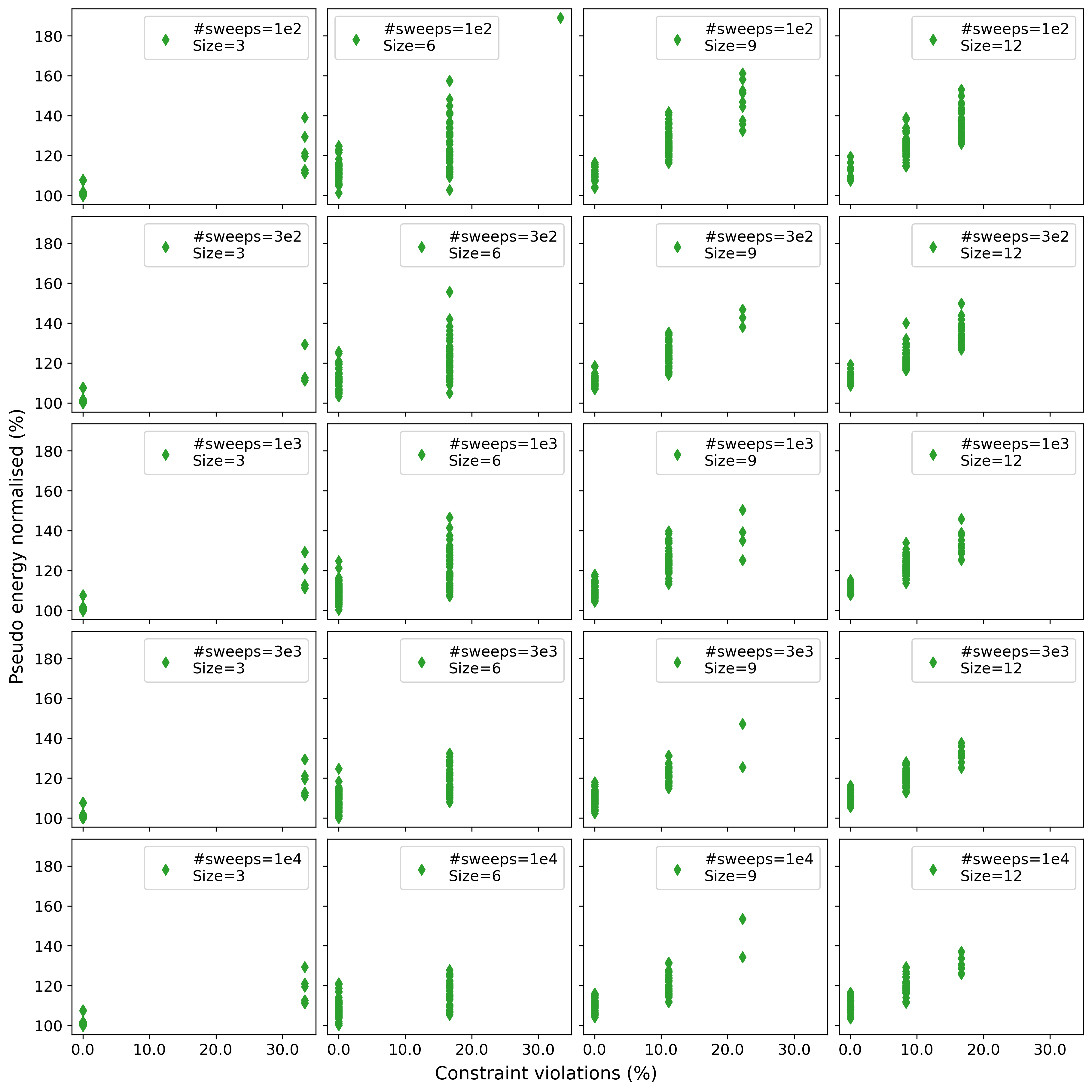}
    	\caption{Constraint violation}
    	\label{fig:app_qap_tinyqap_sa_violation}
     \end{subfigure}
    \caption{SA on Tinyqap. The plot setting is the same as that in figure \ref{fig:app_mvc_pegasus_pegasus}}
    \label{fig:app_qap_tinyqap_sa}
\end{figure*}

Figure \ref{fig:app_qap_tinyqap_sa} shows the performance of SA on Tinyqap problems. Figure \ref{fig:app_qap_tinyqap_sa_energy} suggest that higher \#sweeps can produce slightly better results. This is consistent across all problem size. Figure \ref{fig:app_qap_tinyqap_sa_pf} suggest that SA generally has difficulty in finding feasible solutions. Higher \#sweeps improves probability of feasibility. On all MVC problems SA can always find feasible solutions, but figure \ref{fig:app_qap_tinyqap_sa_violation} suggest that the infeasible cases have up to 30\% constraint violations.

For D-Wave annealer, we only include the results of $200\mu s$ for Chimera and $2000\mu s$ for Pegasus in the main text for comparison between solvers. For DA and SA We only include the results of \#iterations=$10^8$ and \#sweeps=$3\times 10^3$ in the main text for comparison between solvers. 

\begin{figure}[htb]
    \centering
	\includegraphics[width=0.45\linewidth]{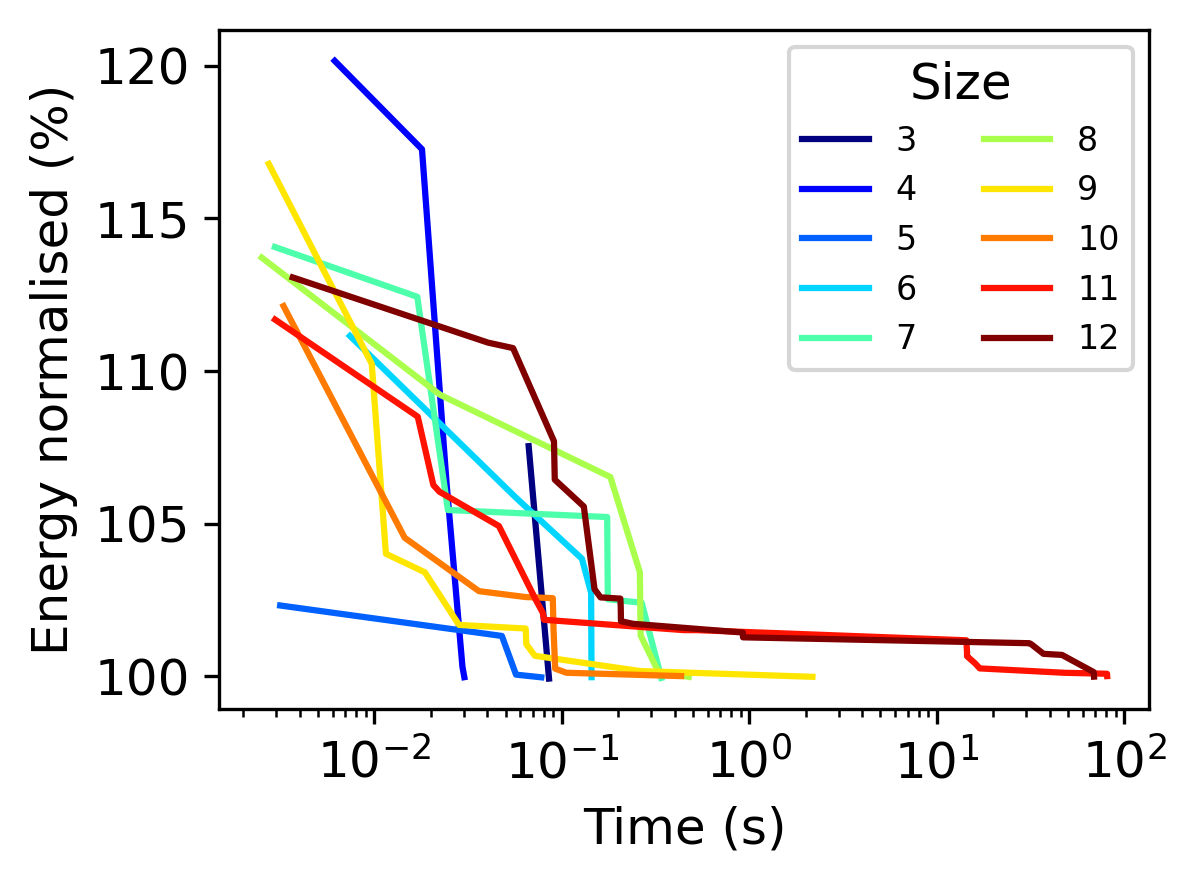}
	\caption{Gurobi time-to-solution on TinyQAP problems}
	\label{fig:app_qap_tiny_gurobi}
\end{figure}

Figure \ref{fig:app_qap_tiny_gurobi} shows the Gurobi time-to-solution plot. Gurobi can find promising solutions within a few seconds. The Gurobi time-to-traverse on TinyQAP is listed in Table \ref{tab:qap_tiny_gurobi_traverse}

\begin{table}[htb]
\centering
\caption{Gurobi time-to-traverse on TinyQAP problems}
\label{tab:qap_tiny_gurobi_traverse}
\begin{tabular}{|l|l|l|l|}
\hline
Size & Time     & Size & Time     \\ \hline
3    & 0.025959 & 8    & 1.3042   \\ \hline
4    & 0.037391 & 9    & 2.283769 \\ \hline
5    & 0.055867 & 10   & 30.85997 \\ \hline
6    & 0.095988 & 11   & 217.3864 \\ \hline
7    & 0.52411  & 12   & 1443.717 \\ \hline
\end{tabular}
\end{table}

\clearpage

\subsection{Tai Benchmark from QAPLIB}
\label{sec:appendix_qap_tai}

Tai Benchmark is available from \url{https://coral.ise.lehigh.edu/data-sets/qaplib/qaplib-problem-instances-and-solutions/#Ta}


\begin{figure*}[tb]
     \centering
     \begin{subfigure}[b]{0.38\textwidth}
        \centering
    	\includegraphics[width=\textwidth]{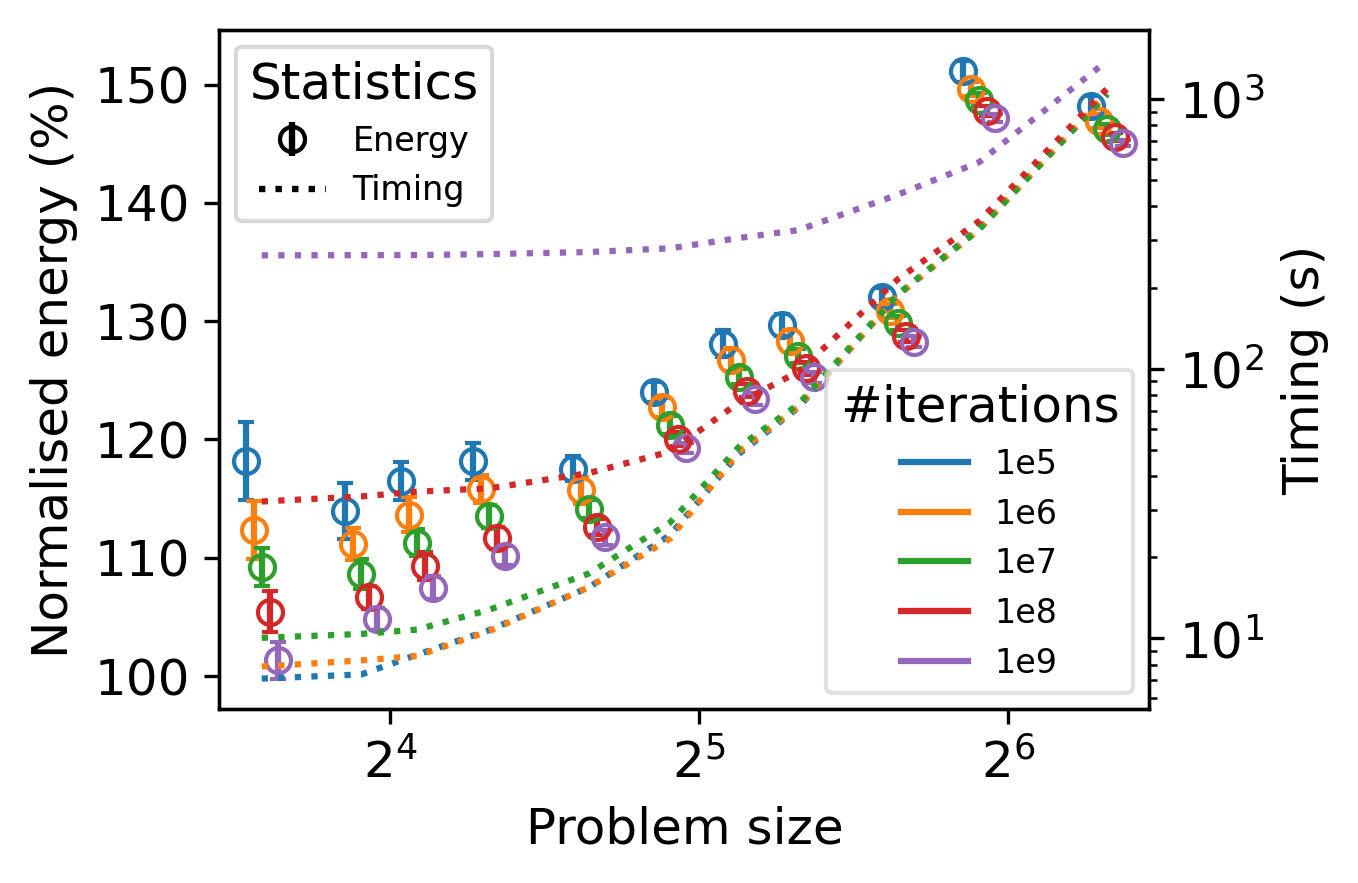}
    	\caption{Error bar of energy and timing}
    	\label{fig:app_qap_tai_da_energy}
     \end{subfigure}
     \begin{subfigure}[b]{0.33\textwidth}
        \centering
    	\includegraphics[width=\textwidth]{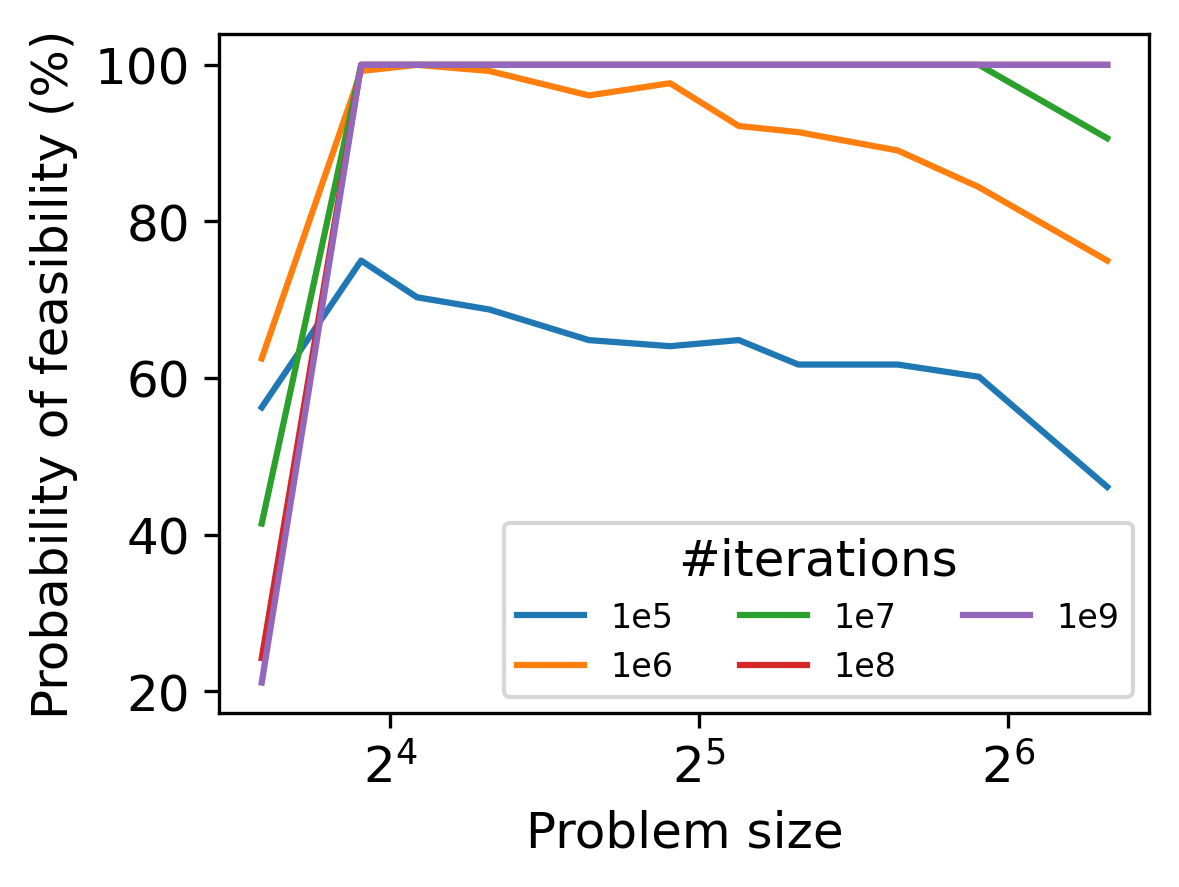}
    	\caption{Probability of feasibility}
    	\label{fig:app_qap_tai_da_pf}
     \end{subfigure}
     \begin{subfigure}[b]{0.95\textwidth}
        \centering
    	\includegraphics[width=\textwidth]{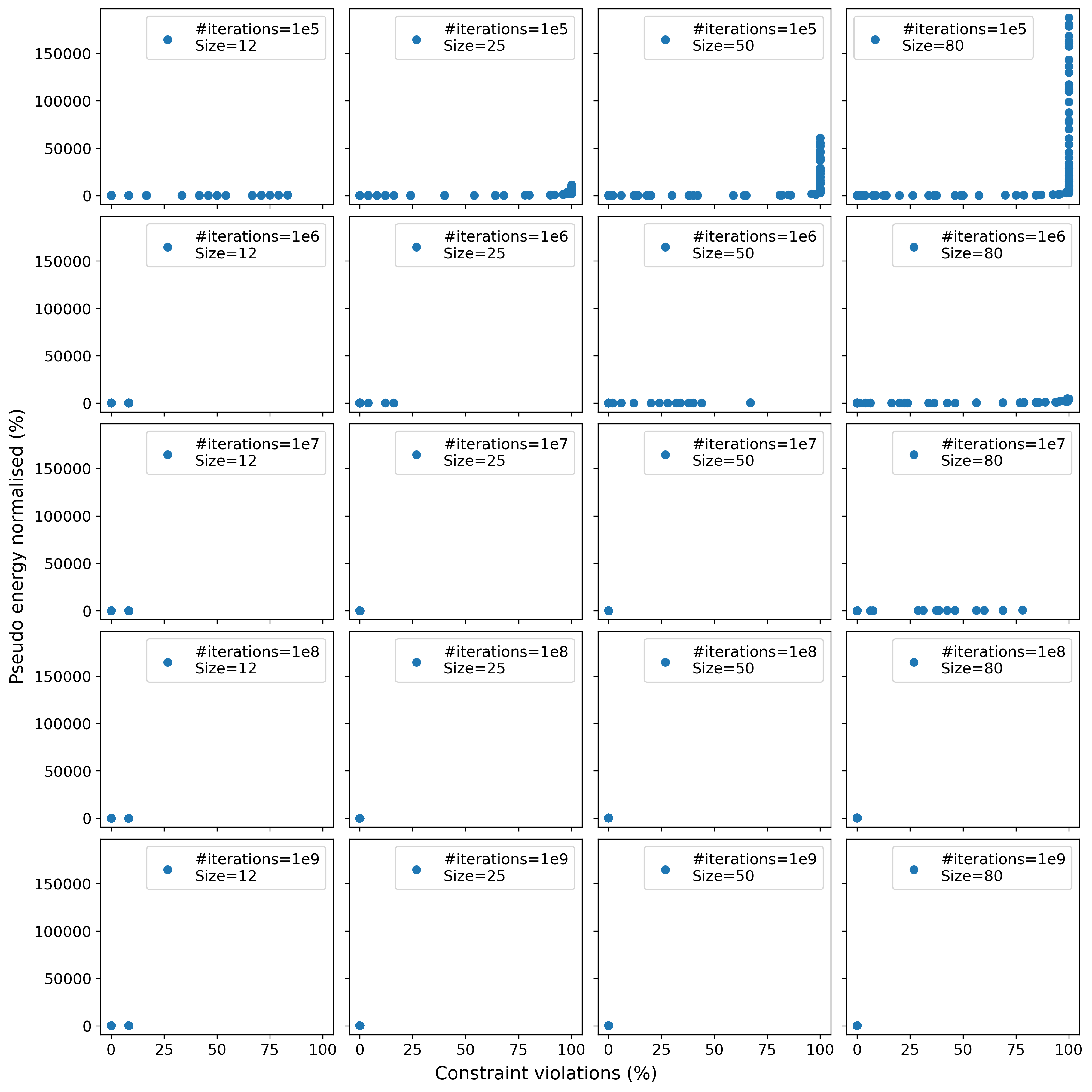}
    	\caption{Constraint violation}
    	\label{fig:app_qap_tai_da_violation}
     \end{subfigure}
    \caption{DA on Tai of QAPLIB. The plot setting is the same as that in figure \ref{fig:app_mvc_pegasus_pegasus}}
    \label{fig:app_qap_tai_da}
\end{figure*}

Figure \ref{fig:app_qap_tai_da} shows the performance of DA on Tai problems from QAPLIB. Problem of size larger than 80 is not included because they are beyond the capacity of DA. From figure \ref{fig:app_qap_tai_da_energy} we understand that higher \#iterations produces better solutions in terms of energy. This is also supported by figure \ref{fig:app_qap_tai_da_pf}, which suggest that lower \#iterations has difficulty in finding feasible solutions Figure \ref{fig:app_qap_tai_da_violation} also support our observation in figure \ref{fig:app_qap_tai_da_pf}.

\begin{figure*}[tb]
     \centering
     \begin{subfigure}[b]{0.38\textwidth}
        \centering
    	\includegraphics[width=\textwidth]{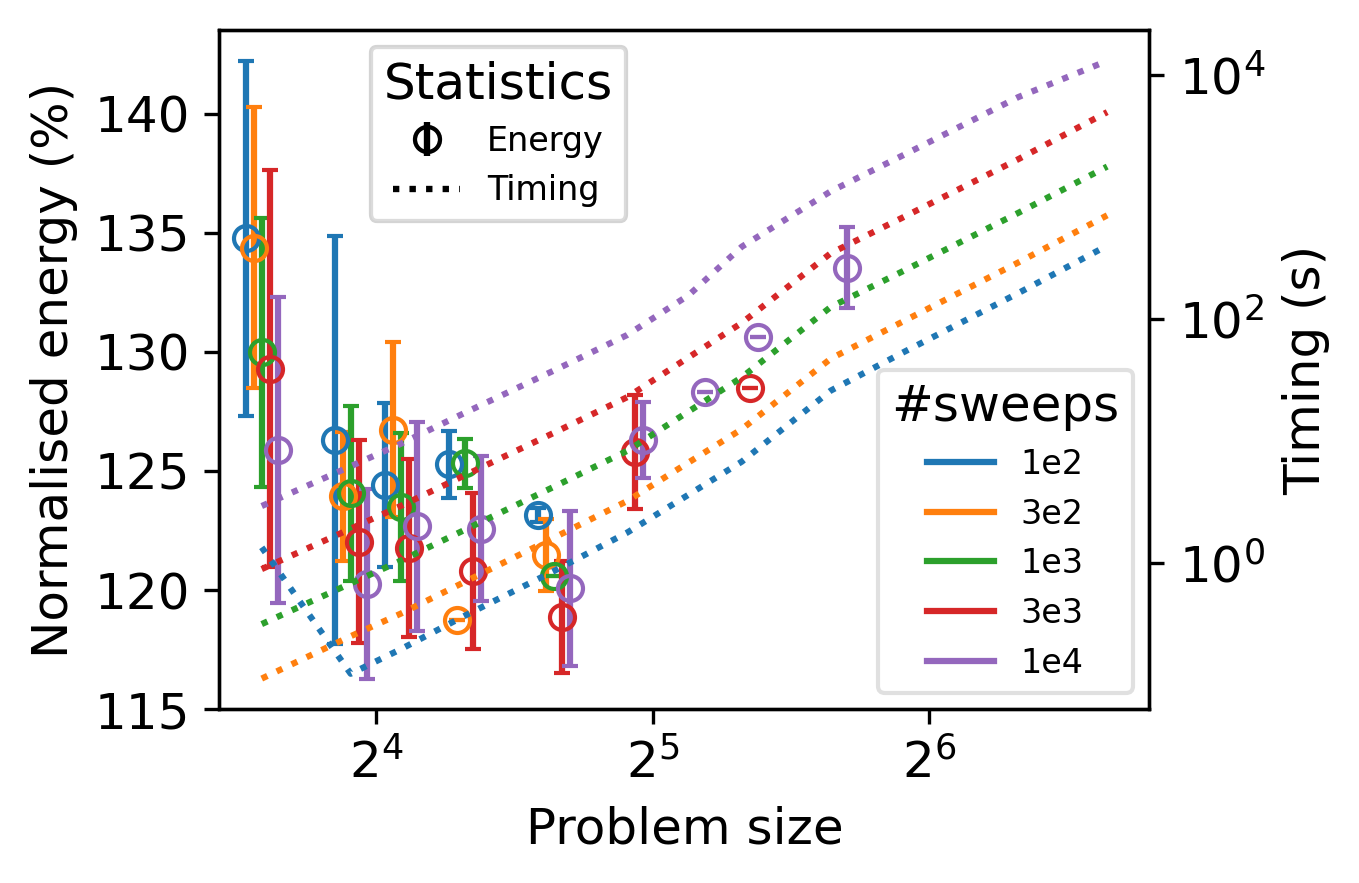}
    	\caption{Error bar of energy and timing}
    	\label{fig:app_qap_tai_sa_energy}
     \end{subfigure}
     \begin{subfigure}[b]{0.33\textwidth}
        \centering
    	\includegraphics[width=\textwidth]{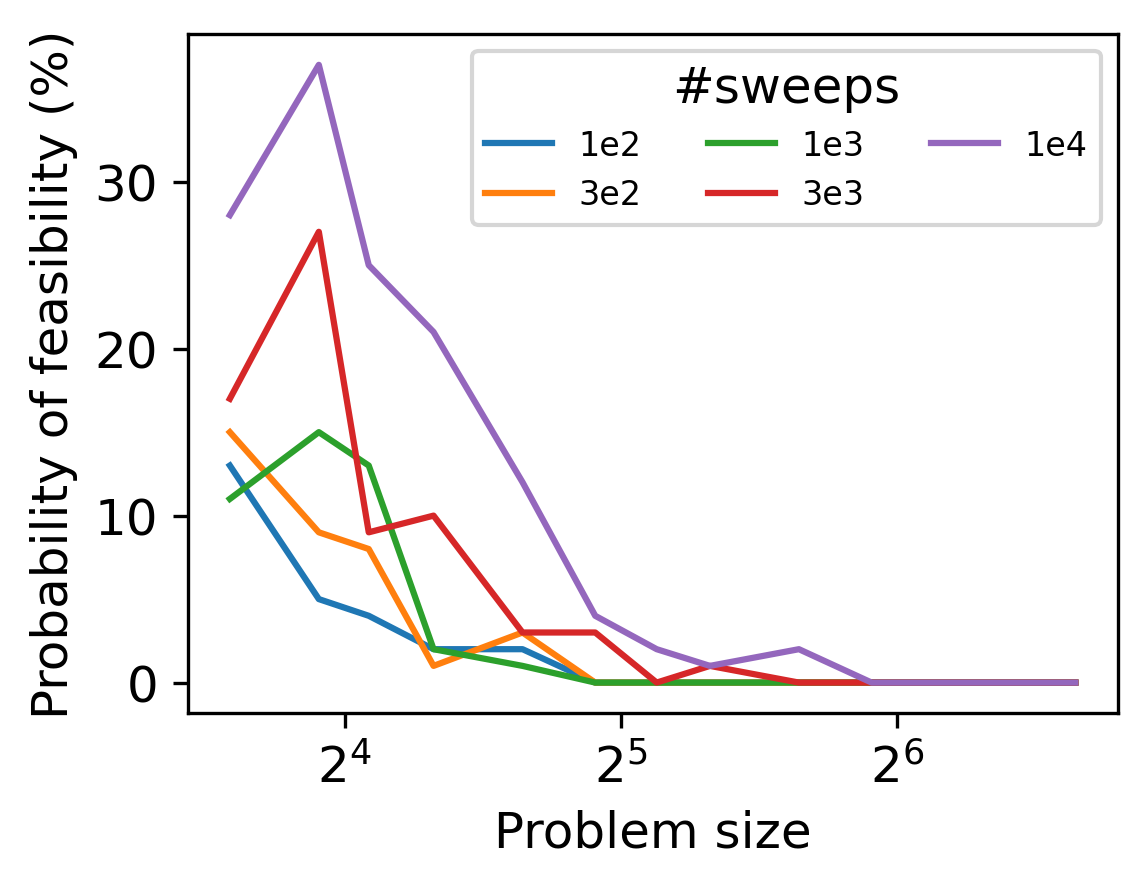}
    	\caption{Probability of feasibility}
    	\label{fig:app_qap_tai_sa_pf}
     \end{subfigure}
     \begin{subfigure}[b]{0.95\textwidth}
        \centering
    	\includegraphics[width=\textwidth]{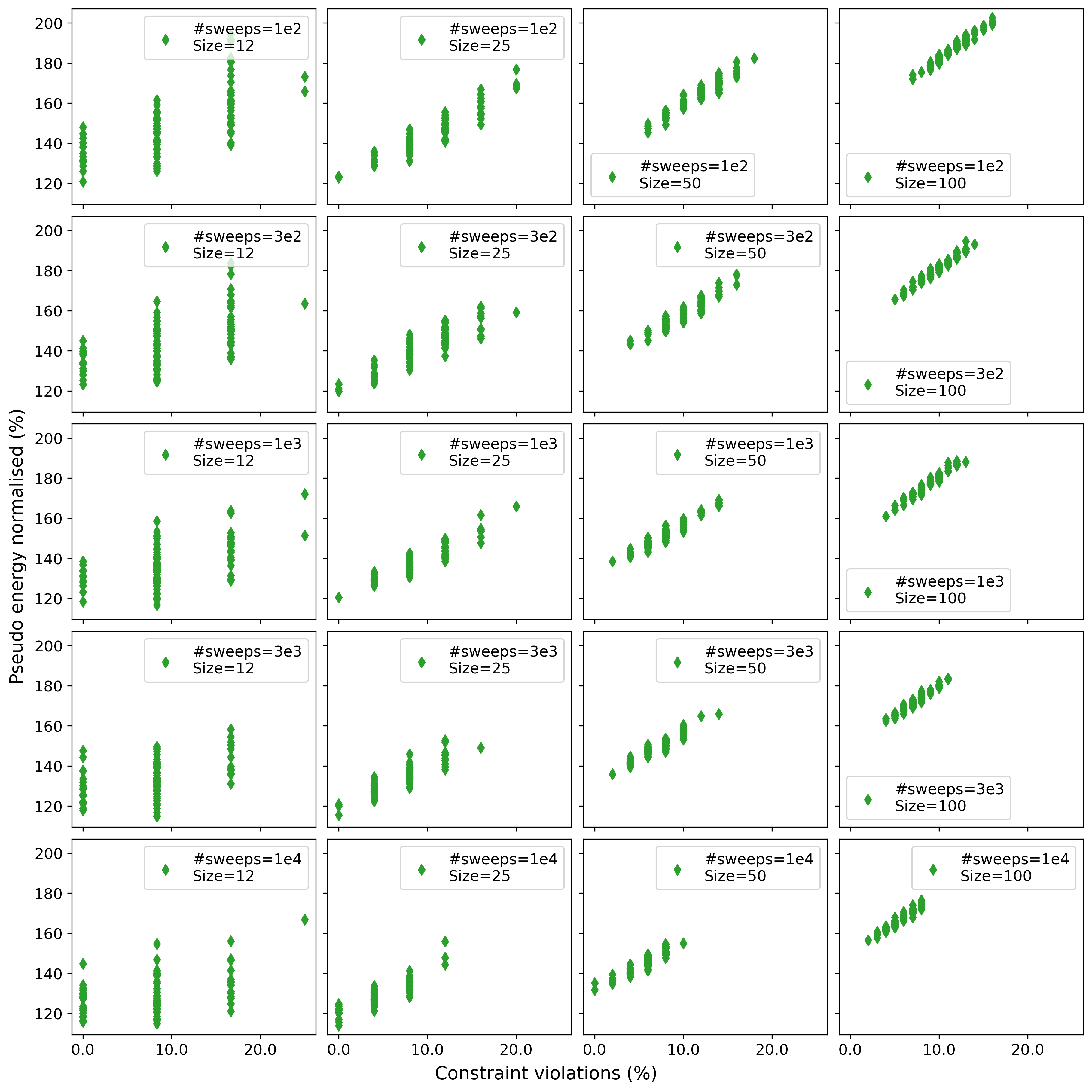}
    	\caption{Constraint violation}
    	\label{fig:app_qap_tai_sa_violation}
     \end{subfigure}
    \caption{SA on Tai of QAPLIB. The plot setting is the same as that in figure \ref{fig:app_mvc_pegasus_pegasus}}
    \label{fig:app_qap_tai_sa}
\end{figure*}

Figure \ref{fig:app_qap_tai_sa} shows the performance of SA on Tai problems from QAPLIB.  From figure \ref{fig:app_qap_tai_sa_energy} and \ref{fig:app_qap_tai_sa_pf} we understand that higher \#sweeps produces better solutions in terms of energy and feasibility. SA generally has difficulty in finding feasible solutions when problem size is over 60. Figure \ref{fig:app_qap_tai_sa_violation} also support our observation in figure \ref{fig:app_qap_tai_sa_pf}.

For Pegasus, we only include the results of $2000\mu s$ in the main text for comparison between solvers. For DA and SA We only include the results of \#iterations=$10^8$ and \#sweeps=$3\times 10^3$ in the main text for comparison between solvers. 

We do not run Gurobi over Tai dataset because we use ground truth from QAPLIB.

\clearpage
 
\section{Warehouse}

\subsection{Warehouse Assignment Heuristics}

Various storage assignment policies have been developed. 
For example, randomized storage policy (RSP) \cite{petersen1997evaluation}, assign items to storage randomly.
Dedicated storage policy (DSP) \cite{fumi2013minimizing} always assigns items to their own locations in the warehouse. This means every item has its dedicated location. This policy then relies on order pickers' memory power to minimise the travel distance.
Cube-per-order Index (COI) \cite{malmborg1990revised} assigns items according to their ratio between volume and popularity. For the interest of this project, we consider all items to have identical volume, so this reduces to arranging items according to popularity.  
Order-oriented-swapping (OOS) \cite{mantel2007order} It starts with a random assignment and runs a certain number of iterations. In each iteration, two items in the assignment are swapped. The distance is recalculated and the new assignment is accepted if it reduces the distance, and is accepted with some probability otherwise. Effectively, this is the technique of Simulated Annealing (SA).
class-based storage policy (ABC) \cite{petersen2004improving}, groups SKUs into classes according to popularity and allocates a dedicated area for each class. The number of popularity classes and the number of items in each class are arbitrary.
The QAP formulation bridges the gap between the warehouse assignment problem and various sophisticated solvers. Following this, there are successful attempts in implementing QAP \cite{ruijter2007improved} for warehouse assignment and favourable results are observed.

\subsection{Computing platforms}

\emph{Annealing-based computers} We include \emph{Chimera} Quantum Annealer (QA) and Fujitsu Digital Annealer (DA) in the experiment. The hyper-parameters are the same as that in Section \ref{sec:appendix_backends}, except we use $\#\text{shots}=10000$ for QA to increase the chance of finding feasible solutions. When problem size is above 90, we use our decomposition heuristic plus DA to solve the problem. Only annealer:solve() in Algorithm \ref{alg:overall} runs on the annealing hardware. The rest of the algorithm runs on the CPU of a server. The server has an Intel Xeon Silver 4116 at 2.1GHz with 128GB DDR4 memory and 960GB SSD storage.

\emph{Baselines} We include three classical methods based on CPU, i.e. \emph{QBSolv} from D-Wave, Simulated Annealing \cite{graidl2020pymhlib} (SA) and Tabu \cite{davidgasquez2015qap}, as baselines. SA (python) and Tabu (C++) are specialised for permutation optimization, meaning that the neighbouring function for QAP is implemented in programming language, such that constraints do not have to be incorporated in the objective function. We also include random solutions in the comparison, which serves as the bottom line of a QAP solver.

\subsection{Warehouse Dataset}

Block-structural QAP instances of sizes 8, 90, 180, 270, 3600, and 8100 are synthesised using randomly generated order sets. In addition, for objective 2), input data distribution is perturbed to provide modified versions of datasets of size 270. This is to match the findings in \cite{tsige2013improving} that interaction-based methods such as QAP work well when 80\% of ordered items are concentrated in 20\% of the SKUs. In other words, there is a small set of commonly ordered products.  The following assumptions are made about the order sets as well as the shape of the respective warehouses. WH-270b below is for objective 2), whereas others are for objective 1).

\begin{table}[htb]
\centering
\caption{Warehouse dataset}
\begin{tabular}{|c|c|c|c|c|}
\hline
Name & Problem Size & \#subsets &  \# rows &  \#columns \\ \hline
WH-8    & 8            & 1         & 4                 & 2                   \\ \hline
WH-90   & 90           & 1         & 45                & 2                   \\ \hline
WH-180  & 180          & 2         & 45                & 4                   \\ \hline
WH-270  & 270          & 3         & 45                & 6                   \\ \hline
WH-270b & 270          & 3         & 45                & 6                   \\ \hline
WH-3600 & 3600         & 40        & 45                & 80                  \\ \hline
WH-8100 & 8100         & 90        & 45                & 180                 \\ \hline
\end{tabular}
\label{tab:dataset}
\end{table}

In this dataset, the smallest size of the problem instance is 8, which is the maximum problem size directly handled by chimera. 90 is the biggest problem size supported by DA. Problem size bigger than 90 is divided into groups of 90 for decomposition. For example, the order set of size 3600 can be solved as 40 sub-QAPs, each of size 90.  All except size 8 warehouses are assumed to comprise columns of size 45, and there is an aisle running in between columns.

To ensure a fair comparison between the heuristic and the software library, critical parameters are set as identical. In particular, both are set to run for 10,000,000 iterations, which is the determining factor of the speed of any simulated annealing algorithm. Note that in the heuristic, our hardware has to perform 10,000,000 iterations for each of the sub-QAPs. Other annealing parameters such as maximum temperature and temperature interval are left to the default settings.

The results obtained for each of the datasets are average values over a certain number of runs.  The software library runs 3 times for each dataset. The number of runs for the decomposition heuristic varies across datasets due to overhead. For WH-8, WH-90, WH-180, and WH-270 the heuristic is run thrice, so the readings are averages of 3. For WH-3600 and WH-8100 the heuristic is run once.

\subsection{Formal description}

We can think of the the distance matrix and frequency matrix as complete graphs. The nodes of the graphs represent items and locations themselves. Formally, $$G=(V(G),E(G)) \; |V(G)|=n$$ is the graph of items. $$H=(V(H),E(H)) \; |V(H)|=n$$ is the graph of locations. Define the edge weights of $G$ to be interaction frequencies, and those of $H$ to be distances. Formally, $$f:E\left ( G \right )\rightarrow \mathbb{R} \qquad f(i,j)=f_{ij}$$ $$d:E\left ( H \right )\rightarrow \mathbb{R} \qquad d(i,j)=d_{ij}$$ An assignment is given by a bi-jection $\varphi: V(G) \rightarrow V(H)$. The goal is to solve the following minimisation:

\begin{equation}
    \min_{\varphi} \sum_{(i,j) \in E(G)}f(i,j)d(\varphi (i), \varphi(j))
    \label{equ:warehouse_obj}
\end{equation}

One can check that the objective function of QAP and \ref{equ:warehouse_obj} are equivalent.

\subsection{Decomposition}
\label{sec:appendix_decomp}

Next, we describe the decomposition formally and provide proof of the theoretical boundary. In order to minimise the travelling distance of an order picker, intuitively, we want to maximise the interaction frequencies of items within a block. This objective can be formally described as follows:

\begin{equation}
\text{Maximise} \quad \sum_{i,j=1}^{n}\sum_{l=1}^{k}f_{ij}x_{il}x_{jl}
\label{equ:graph_part}
\end{equation}

Equation \ref{equ:graph_part} denotes the sum of interaction frequencies among items within their respective subsets. $(x_{ij})$ is an $n \times k$ decision matrix. $x_{ij}$ denotes whether item $i$ goes to subset $j$. The objective comes with the following constraints:

\begin{align}
\sum_{l=1}^{k}x_{il} = 1 \quad \forall i, 1\leq i\leq n
\label{equ:graph_part_ct1}\\
\sum_{i=1}^{n}x_{il} \leq s \quad \forall l, 1\leq l \leq k
\label{equ:graph_part_ct2}
\end{align}

Equation \ref{equ:graph_part_ct1} means an item can belong to exactly one subset. Note that each subset can have at most $s=\frac{n}{k}$ items. Therefore, equation \ref{equ:graph_part_ct2} means each subset must not exceed its capacity $s$.

Note that equation \ref{equ:graph_part_ct1} is actually another QAP with a particular decision matrix (but with different constraints). Thus, it can be translated into a graph formulation. In the graph formulation, $(x_{ij})$ is equivalent to a function $g: V(G)\to K$, which maps an element of $V(G)$ into its subset, such that $g(i) = a \iff x_{ia} = 1$. This function is well-defined due to constraint equation \ref{equ:graph_part_ct1}. Equation \ref{equ:graph_part} can be converted to the equivalent graph formulation as follows:

\begin{align}
\sum_{i,j=1}^{n}\sum_{l=1}^{k}f_{ij}x_{il}x_{jl} 
&= \sum_{l=1}^{k}\sum_{\substack{i,j=1\\x_{il} = x_{jl} = 1}}^{n}f_{ij} \nonumber \\
&= \sum_{l=1}^{k}\sum_{\substack{i,j=1\\g(i) = g(j)=l}}^{n}f_{ij} \nonumber \\
&= \sum_{\substack{(i,j)\in E(G)\\g(i) = g(j)}}f(i,j) \label{equ:graph_part_g}
\end{align}

which, intuitively, is the sum of flows within subsets. This is the setup for the theorem below, which states that with a solution to partitioning problem, an optimal solution to the overall QAP objective can be constructed.

\newtheorem{theorem1}{Theorem}
\begin{theorem1}
Let $d(x,y) = \begin{cases}
		\delta & c(x)=c(y) \\
		M	   & c(x)\neq c(y)
		\end{cases}
	$
for some positive constants $\delta, M$. Then there exists a $\varphi_0$ for which the warehouse objective is achieved, and $\varphi_0$ can be constructed from a solution to partition objective denoted by $g_0$.
\end{theorem1}

\begin{proof}
Suppose $\varphi:V(G)\to V(H)$ is some assignment. Then the warehouse objective can be written as:

\begin{align*}
 & \qquad \sum_{(i,j)\in E(G)} f(i,j) d(\varphi(i),\varphi(j))\\
&= \sum_{\substack{(i,j)\in E(G) \\ c(\varphi(i))=c(\varphi(j))}} f(i,j) d(\varphi(i),\varphi(j)) \\
& \qquad + \sum_{\substack{(i,j)\in E(G) \\ c(\varphi(i))\ne c(\varphi(j))}} f(i,j) d(\varphi(i),\varphi(j))\\
&= \delta \left(\sum_{\substack{(i,j)\in E(G) \\ c(\varphi(i))=c(\varphi(j))}} f(i,j)\right) + M\left(\sum_{\substack{(i,j)\in E(G) \\ c(\varphi(i))\ne c(\varphi(j))}} f(i,j)\right) \\
&= \delta\mathcal{A} + M(\mathcal{F} - \mathcal{A})\\
&= M\mathcal{F} + (\delta - M)\mathcal{A}
\end{align*}

where $\mathcal{F} = \sum_{(i,j)\in E(G)} f(i,j)$, which is the sum of interaction frequencies between all items, and $\mathcal{A} = \sum_{\substack{(i,j)\in E(G) \\ c(\varphi(i))=c(\varphi(j))}} f(i,j)$ is the sum of interaction frequencies between items within partitions. Since $\delta -M <0$, the original objective is minimised if and only if $\mathcal{A}$ is maximised.

Note that $\varphi:V(G)\to V(H)$ and $c:V(H) \to K$, therefore we can view $c \circ \varphi: V(G) \to K$ and $\mathcal{A} = \sum_{\substack{(i,j)\in E(G) \\ c\circ \varphi(i))=c\circ \varphi(j))}} f(i,j)$. Then $g_0:V(G)\to K$ is by definition the solution to
$$\max_{g} \sum_{\substack{(i,j)\in E(G)\\g(i) = g(j)}}f(i,j)$$
which means
$$\sum_{\substack{(i,j)\in E(G)\\g_0(i) = g_0(j)}}f(i,j) \le \mathcal{A}, \quad \forall\varphi$$ 

Now we need to construct a $\varphi$ such that $c(\varphi(i)) = c(\varphi(j)) \iff g_0(i) = g_0(j)$. This is equivalent to saying that after splitting items into subsets according to $g_0$, assign the items such that items within the same subset are in the same interval. The relative position of items within an interval does not matter. Therefore, there are many possible answers.
\end{proof}
\bigskip

The above method assumes a strong condition that $(d_{ij})$ only has two distinct values, $\delta$ for locations within an interval and $M$ in between intervals. This is a simplification of the block structure and $(d_{ij})$, in reality, is usually more complex.

However, the concept of intervals can be generalised such that when $\delta$ and $M$ are non-constant, a function $c$ can still be defined on $V(H)$ such that the sum of distances within intervals is minimised. The motivation for such construct is that when a warehouse does not exhibit a clear-cut column structure, one can still think of an abstract ``column'' as a group of close locations. Note that this definition automatically specialises to the previous definition of $c$ when $\delta$ and $M$ are constant. This intuition can be formally expressed as 

\begin{equation}
\text{Minimise} \quad \sum_{i,j=1}^{n}\sum_{l=1}^{k}d_{ij}x_{il}x_{jl}
\label{equ:graph_part_loc}
\end{equation}

Subject to:

\begin{align}
\sum_{l=1}^{k}x_{il} = 1 \quad \forall i, 1\leq i\leq n
\label{equ:graph_part_loc_ct1}\\
\sum_{i=1}^{n}x_{il} \leq s \quad \forall l, 1\leq l \leq k
\label{equ:graph_part_loc_ct2}
\end{align}

$(x_{ij})$ is an $n\times k$ decision matrix. $x_{ij}$ denotes whether item $i$ goes to location group $j$. Note that this graph partitioning problem can be thought of as the dual to equation \ref{equ:graph_part}, with maximisation changed to minimisation. In effect, items that are frequently ordered together will be assigned to locations that are closer together.

Now both the set of locations and the set of items have been divided into subsets of equal size. Between subsets of locations, the distances are maximal, and between subsets of items, the interaction frequency is minimal. The final step is to produce a bijection from the set of subsets of items to the set of subsets of items, and for each pair of subsets in the bijection $(items, locs)$, there is a sub-QAP of size $O(\sqrt{n})$, and therefore a QUBO of size $O(n)$, where $n$ is the total number of items.

Note that this is subject to $n$ being a perfect square; in practical situations where $n$ is not a perfect square, compromise has to be made in either finding the nearest smaller perfect square and do the optimisation on the smaller set of items and locations only, or use integer divisors of $n$ other than $\sqrt{n}$, and correspondingly deal with sub-QAPs of sizes other than $\sqrt{n}$. For example, if there are $n=3600$ locations, it is possible to divide it into 40 groups of 90, where each sub-QAP will have $n=90$.

\subsection{QAP-to-QUBO conversion}
\label{sub:appendix_convert}

A QAP has to be converted to QUBO before it is solvable by a quantum annealer. This is also true of each of the sub-QAPs generated by the decomposition procedure . Here the procedures of QAP-to-QUBO conversion is described. Readers familiar with square penalty may safely skip this section.

For the linear constraint of the QAP objective, there is one constraint for each $i$. Thus, for each $i$ the quadratic penalty term $P(\sum_{k=1}^{n}x_{ik}-1)^2$ is added to the original objective function, where $P$ is some positive constant to be determined a-posteriori. Note that this penalty term, when expanded, is also a quadratic form, and thus can be represented as an addition to the QUBO matrix $Q$.

We can express this constraint in a more general form. For the decision vector $\bar{x}$, locate the position of $x_{ik}$ and instantiate a 0-1 bit-vector with length $n$, with the corresponding positions having 1 and the other positions having 0. For all $n$ constraints, collect the respective bit-vectors as rows of a $n^2\times n^2$ matrix $A$ (note that not all rows of A has to be used as a constraint, in which case the row is left as zero). All constraints are therefore in the form $A\bar{x}=\bar{b}$, where $\bar{b}$ is just a length $n^2$ vector of all 0's except $n+n$ 1's (corresponding to the $n+n$ constraints we have) in the corresponding rows of $A$. To convert the constraints into a sum of squares we take the inner product \[(Ax-b)^T(Ax-b)\] We can embed the penalty $P$ easily into $A$ and $b$ by scaling. Subsequently, the expanded QUBO coefficient matrix can be obtained by matrix manipulation. It could be shown that the final coefficient matrix to be augmented is $(A^TA-2D)$, where $D=diag(b^TA)$. The complete algorithm is described below. 

\begin{algorithm}
\caption{Conversion from QAP to QUBO}
\label{alg:qap2qubo}
\begin{algorithmic}[1]
\Statex {//Given: FindIndex$(i,j)$ that maps quadratic index of $x_{ij}$ to its corresponding position in the decision vector $\bar{x}$}
\Procedure{convert}{$F$,$D$,$P$}
	\State{$Q=\textrm{ZeroMatrix}(n^2\times n^2)$}
	\For {$i$ in $[1..n]$}
		\For {$j$ in $[1..n]$}
			\For {$k$ in $[1..n]$}
				\For {$l$ in $[1..n]$}
					\State {$x_{ik} = \textrm{FindIndex}(i,k)$}
					\State {$x_{jl} = \textrm{FindIndex}(j,l)$}
					\State {$Q[x_{ik}][x_{jl}] = F[i][j] \times D[k][l]$}
				\EndFor
			\EndFor
		\EndFor
	\EndFor
	\State {$Q'=$AtoQ(preparematrixA(), preparevectorB(), $P$)}
	\State {return $Q + Q'$}
\EndProcedure
\end{algorithmic}
\end{algorithm}

\begin{algorithm}
\caption{Prepare Maxtrix A}
\label{alg:preparematrixa}
\begin{algorithmic}[1]

\Statex
\Procedure{preparematrixA}{}
	\State {$A = \textrm{ZeroMatrix}(n^2\times n^2)$}
	\State {$i=1$}
	\For {$r$ in $[1..n]$}
		\For {$k$ in $1..n$}
			\State {$idx = \textrm{FindIndex}(i,k)$}
			\State {$A[r][idx] = 1$}
		\EndFor
		\State {$i=i+1$}
	\EndFor
	
	\State {$k=1$}
	\For {$r$ in $[(n+1)..2n]$}
		\For {$i$ in $1..n$}
			\State {$idx = \textrm{FindIndex}(i,k)$}
			\State {$A[r][idx] = 1$}
		\EndFor
		\State {$k=k+1$}
	\EndFor
	\State {return $A$}
\EndProcedure

\end{algorithmic}
\end{algorithm}

\begin{algorithm}
\caption{Prepare Vector B}
\label{alg:preparevectorb}
\begin{algorithmic}[1]

\Statex
\Procedure{preparevectorB}{}
	\State {$b=\textrm{ZeroVector}(n^2)$}
	\For {$r$ in $[1..2n]$}
		\State {$b[r] = 1$}
	\EndFor
	\State {return $b$}
\EndProcedure

\end{algorithmic}
\end{algorithm}

\begin{algorithm}
\caption{Convert Matrix A to QUBO}
\label{alg:atoq}
\begin{algorithmic}[1]

\Statex
\Procedure{AtoQ}{$A$,$b$,$P$}
	\State {Scale each row of $A$ by $\sqrt{P}$}
	\State {Scale each entry of $b$ by $\sqrt{P}$}
	\State {$D = b^TA$}
	\State {return $A^TA-2D$}
\EndProcedure

\end{algorithmic}
\end{algorithm}

In order to enable quantum annealing of a QAP, the problem must first be converted into a Quadratic Unconstrained Binary optimisation (QUBO) form, which effectively means constraints have to be subsumed as part of the objective function in some way. It is customary to do this by encoding the constraints as quadratic penalty terms which augment the objective function. It can be shown that for QUBOs, the optimal solution to the augmented objective function also minimises the original objective function.

\subsection{Exterior penalty method}
\label{sec:appendix_exterior}
We use exterior penalty method to iteratively solve a QUBO as a series of QUBO's with increasing penalty weights. Such is called the exterior penalty method, whose algorithm is described below. Readers familiar with this method in optimisation may safely skip this section.

The algorithm below starts with a random permutation matrix as the initial solution. Note that the procedure is equipped with a $isValid()$ function that tests if a solution satisfies constraints. 

\begin{algorithm}
\caption{Exterior Penalty Method}
\label{alg:exterior}
\begin{algorithmic}[1]
\Procedure{solve}{$F$,$D$,$\alpha_0$,$\beta$}
	\State {$\alpha = \alpha_0$}
	\State {solution $=$ randomPermutationMatrix$(n\times n)$}
	\Repeat
		\State {solution $=$ annealer.solve(convert($F,D,\alpha$), solution)}
		\State {$\alpha := \alpha \times \beta$}
	\Until {isValid(solution)}
\EndProcedure
\end{algorithmic}
\end{algorithm}

\subsection{Overall procedure}
\label{sec:appendix_procedure}

The overall procedure could be abstractly outlined in Algorithm \ref{alg:overall}.

\begin{algorithm}
\caption{Solving QAP with decomposition heuristics}
\label{alg:overall}
\begin{algorithmic}[1]
\Procedure{run}{$F$,$D$,$n$}
	\Statex \Comment{n-\#items/locations}
	\Statex \Comment{k-\#groups}
	\State{$s = \sqrt{n}$} \Comment{assume group size is $\sqrt{n}$}
	\State{$Fs,Ds,FtoD = \textrm{decompose}(F,D,s)$}
	\Comment{FtoD is the map between subsets and intervals}
	\State{$\alpha_0=10000$}
	\State{$\beta=1.5$}
	\State{$\textrm{subsolutions} = []$}
	\For {$i$ in $[1..k]$}
		\State{$\textrm{items}=F[i]$}
		\State{$\textrm{locations}=D[FtoD[i]]$}
		\State{$F'=\textrm{emptyMatrix}(s\times s)$}
		\State{$D'=\textrm{emptyMatrix}(s\times s)$}
		\For {$p,q$ in $items$}
			\State{$F'[\textrm{localindex}(p)][\textrm{localindex}(q)]=F[p][q]$}
		\EndFor
		\For {$k,l$ in $\textrm{locations}$}
			\State{$D'[\textrm{localindex}(k)][\textrm{localindex}(l)]=D[k][l]$}
		\EndFor
		\State{$\textrm{subsolutions.append}(\textrm{solve}(F',D',\alpha_0,\beta))$}
	\EndFor
	\State{combineSubsolutions(subsolutions)}
\EndProcedure
\end{algorithmic}
\end{algorithm}

Note that $\alpha_0$ and $\beta$ are hyper-parameters for penalty weight. Instead of statically determined, we iteratively adjust the penalty weight for the warehouse assignment problem to find the best solutions.

\subsection{Discussion on improving the solutions}

An important assumption in the comparison of the quality of solutions is that QAP is a good way of modelling warehouse assignments. 
With that assumption in mind, We observed that OOS gives a shorter picking distance than DA. This is to be expected since OOS is manually tuned and optimised for WH-270b, whereas QAP and its heuristic on DA are completely generic, modulo the initial estimation of penalty weights. A better solution to the QAP on DA will then yield a better solution for the warehouse assignment problem. 



\emph{Model for warehouse assignment} The current QAP model for warehouse assignment is the primitive one from \cite{mantel2007order}. There are many holes to plug in its assumptions.  In particular, the standard QAPs need to be solved with standard tools that give reliable solutions, which in turn can be simulated a-posteriori to ascertain their effectiveness in modelling the warehouse assignment problem. The problem of modelling warehouse assignment is also more subtle. There are other measures of interaction frequency other than the basic one, each of different numerical ranges and effectiveness, as reviewed in \cite{kofler2014optimising}. This is another aspect that carries the potential for improvement.

\bibliographystyle{unsrtnat}
\bibliography{reference}